\documentclass[12pt,oneside,english]{book}
\usepackage{mathptmx}

\usepackage[T1]{fontenc}
\usepackage[latin9]{inputenc}
\usepackage[a4paper]{geometry}
\usepackage{color}
\usepackage{array}
\usepackage{float}
\usepackage{textcomp}
\usepackage{amsmath}
\usepackage{amssymb}
\usepackage{stackrel}
\usepackage{graphicx}
\usepackage{setspace}
\makeatletter

\providecommand{\tabularnewline}{\\}

\newenvironment{lyxlist}[1]
	{\begin{list}{}
		{\settowidth{\labelwidth}{#1}
		 \setlength{\leftmargin}{\labelwidth}
		 \addtolength{\leftmargin}{\labelsep}
		 }}
	{\end{list}}

\usepackage{tocloft}

\usepackage{xpatch}
\xpatchcmd{\@chapter}{\addcontentsline{toc}{chapter}{\protect\numberline{\thechapter}#1}}{%
                      \addcontentsline{toc}{chapter}{\protect\numberline{}#1}}{\typeout{Success}}{\typeout{Failed!}}

\@ifundefined{definecolor}
 {\usepackage{color}}{}
\usepackage{array}\usepackage{amsthm}\usepackage{float} \makeatletter

\newcommand{\lyxmathsym}[1]{\ifmmode\begingroup\def\b@ld{bold}
  \text{\ifx\math@version\b@ld\bfseries\fi#1}\endgroup\else#1\fi}
\providecommand{\tabularnewline}{\\}

 \@ifundefined{definecolor}{

}{\makeatletter} \AtBeginDocument{
  
}

\providecommand{\tabularnewline}{\\}

\usepackage{fancyhdr}\usepackage{titlesec}\titleformat{\chapter}[display]
  {\normalfont\huge\bfseries\centering}{\MakeUppercase\chaptertitlename\ \thechapter}{20pt}{\LARGE}

\usepackage[square, numbers, comma, sort&compress]{natbib}
\newcommand{\ket}[1]{\vert{#1}\rangle}
\newcommand{\bra}[1]{\langle{#1}\vert}

\newcommand{\inpr}[2]{\langle{#1}\vert{#2}\rangle}

\newcommand{\abs}[1]{\vert{#1}\vert}

\fancypagestyle{plain}{
  \fancyhead{}
  \renewcommand{\headrulewidth}{0pt}
}

\usepackage[english]{babel}

\addto\captionsenglish{
  \renewcommand{\contentsname}%
    {\Large{TABLE OF CONTENTS}}%
}

\usepackage[labelfont={bf}, tableposition=top]{caption}
\addto\captionsenglish{%
   }
\addto\captionsenglish{%
   }

\makeatother

\usepackage{babel}
\begin{document}
\pagestyle{fancy} \cfoot{\thepage}\rhead{}

\begin{center}
\textbf{\Large{}DESIGN AND ANALYSIS OF}{\Large\par}
\par\end{center}

\begin{center}
\textbf{\Large{}COMMUNICATION PROTOCOLS}{\Large\par}
\par\end{center}

\begin{center}
\textbf{\Large{}USING QUANTUM RESOURCES}{\Large\par}
\par\end{center}

~

\begin{center}
\emph{\large{}Thesis submitted in fulfillment of the requirements
for the Degree of}{\large\par}
\par\end{center}

~

\begin{center}
\textbf{\large{}DOCTOR OF PHILOSOPHY}\\
\par\end{center}

\begin{center}
By\\
\par\end{center}

\begin{center}
\textbf{MITALI SISODIA}\\
\textbf{ENROLLMENT NO.}\\
\textbf{15410006}
\par\end{center}
\begin{quotation}
\begin{figure}[H]
\centering{}\includegraphics[scale=1.2]{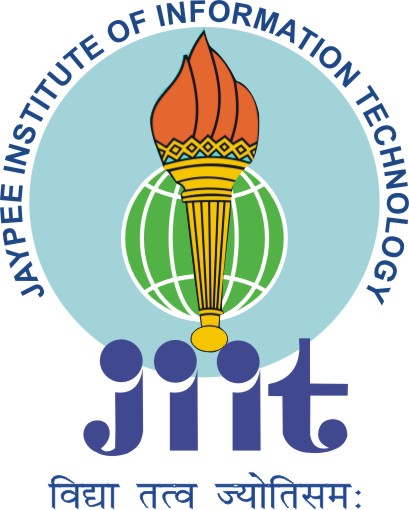}
\end{figure}
\end{quotation}
\begin{center}
Department of Physics and Materials Science and Engineering
\par\end{center}

\begin{center}
JAYPEE INSTITUTE OF INFORMATION TECHNOLOGY\\
(Declared Deemed to be University U/S 3 of UGC Act)\\
A-10, SECTOR-62, NOIDA, INDIA\\
\par\end{center}

\begin{center}
July, 2019
\par\end{center}

\thispagestyle{empty}

\renewcommand{\headrulewidth}{0pt}

\pagebreak{} {\large{}\pagenumbering{roman}\addcontentsline{toc}{chapter}{INNER
FIRST PAGE}} ~

 ~\

~\\

~

~

~~

~

~

~~

~

~

~\\
\\

~~

~~

~

~

~

~~

~~~

~~

\begin{center}
{\small{}@ Copyright JAYPEE INSTITUTE OF INFORMATION TECHNOLOGY}\\
{\small{}(Declared Deemed to be University U/S 3 of UGC Act)}\\
{\small{}NOIDA, INDIA.}\\
{\small{}July, 2019}\\
{\small{}ALL RIGHTS RESERVED}\\
\par\end{center}

\begin{center}
{\large{}$\pagestyle{plain}$}{\small{}\newpage}{\small\par}
\par\end{center}

\textbf{\emph{\large{}This thesis is dedicated to my parents and my
husband for their constant support and unconditional love. I love
you all dearly.}}{\large\par}

{\large{}$\thispagestyle{empty}$}{\large\par}

\textbf{\emph{\large{}\newpage}}{\large{}\lhead{}\addcontentsline{toc}{chapter}{TABLE
OF CONTENTS}}\tableofcontents{}\newpage{}
\begin{center}
{\large{}\addcontentsline{toc}{chapter}{DECLARATION BY THE SCHOLAR}}{\large\par}
\par\end{center}

\begin{center}
\textbf{\Large{}DECLARATION BY THE SCHOLAR}{\LARGE{} {}}{\LARGE\par}
\par\end{center}

~

~\\
~

~

\begin{doublespace}
I hereby declare that the work reported in the Ph.D. thesis entitled{\LARGE{}
}\textbf{\large{}\textquotedblleft Design and Analysis of Communication
Protocols Using Quantum Resources\textquotedblright{}}\textbf{ }submitted
at \textbf{\large{}Jaypee Institute of Information Technology, Noida,
India,} is an authentic record of my work carried out under the supervision
of \textbf{\large{}Prof. Anirban Pathak}. I have not submitted this
work elsewhere for any other degree or diploma. I am fully responsible
for the contents of my Ph.D. Thesis.
\end{doublespace}

~

~

~

~~\\

{\large{}(Signature of the Scholar)}{\large\par}

~

{\large{}(Mitali Sisodia)}{\large\par}

~

{\large{}Department of Physics and Materials Science and Engineering}{\large\par}

~

{\large{}Jaypee Institute of Information Technology, Noida, India}{\large\par}

~

{\large{}Date:}{\large\par}
\begin{flushleft}
{\large{}{}}{\large\par}
\par\end{flushleft}

{\large{}$\thispagestyle{empty}$}{\large\par}

\textbf{\emph{\large{}\newpage}}{\large{} $\thispagestyle{empty}$$\mbox{}$}\textbf{\emph{\large{}\newpage}}{\large\par}
\begin{flushleft}
{\large{}
\addcontentsline{toc}{chapter}{SUPERVISOR'S CERTIFICATE}}{\large\par}
\par\end{flushleft}

\begin{center}
\textbf{\Large{}SUPERVISOR'S CERTIFICATE }{\LARGE{}{}}{\LARGE\par}
\par\end{center}

~

~~

~\\

\begin{doublespace}
This is to certify that the work reported in the Ph.D. thesis entitled{\LARGE{}
}\textbf{\large{}\textquotedblleft Design and Analysis of Communication
Protocols Using Quantum Resources\textquotedblright{}} submitted by
\textbf{\large{}Mitali Sisodia} at \textbf{\large{}Jaypee Institute
of Information Technology, Noida, India,}{\large{} }is a bonafide
record of her original work carried out under my supervision. This
work has not been submitted elsewhere for any other degree or diploma.
\end{doublespace}

~

~

~

~

~

{\large{}(Signature of Supervisor)}{\large\par}

~

{\large{}(Prof. Anirban Pathak)}{\large\par}

~

{\large{}Department of Physics and Materials Science and Engineering}{\large\par}

~

{\large{}Jaypee Institute of Information Technology, Noida, India}{\large\par}

~

{\large{}Date:}{\large\par}
\begin{flushleft}
{\large{}{}}{\large\par}
\par\end{flushleft}

{\large{}$\thispagestyle{empty}$}{\large\par}

\textbf{\emph{\large{}\newpage}}{\large{} $\thispagestyle{empty}$$\mbox{}$}\textbf{\emph{\large{}\newpage}}{\large\par}

{\large{}$\thispagestyle{plain}$\addcontentsline{toc}{chapter}{ACKNOWLEDGEMENT}}{\large\par}
\begin{center}
\textbf{\Large{}ACKNOWLEDGEMENT }{\LARGE{}{}}{\LARGE\par}
\par\end{center}

~

Completion of this doctoral dissertation has been a truly life-changing
experience for me and it would not have been possible to do without
the support and guidance that\textcolor{black}{{} I received from many
people. }Now, I got a chance to express thanks from my heart through
this acknowledgment.

First and foremost, \textcolor{black}{I would like to start with the
person wh}o made the biggest difference in my life, my mentor, my
guide, my\textcolor{black}{{} supervisor Prof. A}nirban Pathak. I would
like to extend my sincere gratitude to him for the continuous support
of my doctoral research and introducing me to this exciting field
of science and for his dedicated help, advice, inspiration, encouragement
and continuous support, throughout my Ph.D. Without his guidance and
constant feedback this Ph.D. would not have been achievable. I have
learnt extensively from him, including how to raise new possibilities,
how to regard an old question from a new perspective, how to approach
a problem by systematic thinking, data-driven decision making and
exploiting serendipity.\emph{ }I would like to say\emph{- As you walk
with the GURU, you walk in the LIGHT of existence, away from the darkness
of IGNORANCE, you leave behind all the problems of your life, and
move towards the peak experiences of life.}

\textcolor{black}{I wish to re}spectfully express my gratitude to
Dr. Papia Chowdhury for all her kind support and a\textcolor{black}{ffections}.
Prof. \textcolor{black}{D. K. Rai} for their timely support provided
in successful completion of the thesis work. My DPMC members Prof.
S.P. Purohit and Prof. B. P. Chamola for their fruitful suggestions
during end semester\textcolor{black}{{} Ph.D. presen}tations and all\textcolor{black}{{}
faculty members} of the Department of Physics and Materials Science
and Engineering for their valuable suggestions during my studies.

Further, my heartfelt thanks to my super seniors Dr. Amit Verma, Dr.
Anindita Banerjee for their kind \textcolor{black}{supports. }My heartfelt
and special thanks to my senior who is my inspiration Dr. Chitra\textcolor{red}{{}
}Shukla for all her kind and continuous support, encouragement, discussions
over the phone and chat regarding research work. I always feel free
to discuss the work and share many other things any time with her.
Actually, I feel very happy and friendly with her and I love talking
to her very much. During my end semester presentations, I always called
her and asked many queries, after my p\textcolor{black}{resentations,
sh}e used to call me back and ask- hey Mitali, how was your presentation?
thank you Chitra Mam for everything. \textcolor{black}{Further, a
sp}ecial thanks to my senior Dr. Ki\textcolor{black}{shore Thapliyal,}
who has played a very important role in my \textcolor{black}{Ph.D.
duration}\textcolor{blue}{.}\textcolor{black}{{} I really thank to }him
for his great help, encouragement, continuous support, guidance, cooperation\textcolor{black}{{}
and motivation which have a}lways kept me going ahead. I owe a lot
of gratitude to\textcolor{black}{{} him for} always being there for
me and I feel privileged to be associated with a person like \textcolor{black}{him
during my research life, thanks for everything Kishore Sir, professionally,
you are \textquotedblleft perfection personified\textquotedblright .
I am also thankful to Dr. Abhishek Shukla for helping me doing work
on the IBM quantum computer}, Dr. Nasir Alam for helping me to clear
the basic concepts of physics and Dr. Meenakshi Rana for helping me
each and every stage of \textcolor{black}{Ph.D. and my of my personal
life,} too.

I would also like to thank to my seniors, juniors, friends Dr. Vikram
Verma, Dr. Rishi Dutt Sharma, Dr. Animesh, Ashwin Saxena, Priya Malpani
(near and dear friend), Kathakali Mandal, Vikas Deep, Ankit Pandey,
Vinne Malik for listening, advising, and supporting me through this
entire process, other research scholars and all the technical staff
Mr. \textcolor{black}{T. K. }Mishra, Mr. Kailash Chandra and Mr. Munish
Verma of Materials Science Laboratory, JIIT Noida, for their timely
support, kind help and cooperation during the smooth conduction of
B. Tech. Labs. I am indebted to the Jaypee Institute of Information
technology, Noida for providing me teaching assistantship during Ph.D.
work.

I cannot forget to mention the name of my \textcolor{black}{Bhaiya
(Dr. Manoj Kumar Chauhan), bhabi (Mrs. Rama Chauhan) and their kids
(Nonu and Pihu) for their support and lots of love. }

\textcolor{black}{It is my fortune to gratefully acknowledge the support
of m}y family members, respected parents who are my God, Mummy (Mrs.
Saroj Sisodia), Papa (Mr. Ashok Kumar Sisodia) for showing faith in
me and giving me liberty to choose what I desired, Sister (Neha Sisodia),
Brother (Vipul Kumar Sisodia) for their selfless love. Words are short
to express my deep sense of gratitude towards my in-laws, Mother-in-law
(Mrs. Renu Rajput), Father-in-law (Mr. Akhilesh Rajput) and Brother-in-law
(Priyanshu Rajput) who gave me the love, strength, blessings and patience
to work through all these years so that today I can stand proud with
my head held high.

\textcolor{black}{Finally, my special acknowledgment goes to a special
person who mean world to me, my better half my husband (Mr. Sudhanshu
Rajput) for his unconditional love, eternal and moral support and
understanding of my goals and aspirations. He was always around at
times, I thought that it is impossible to continue, he helped me to
keep things in perspective. These past several years have not been
an easy ride, both academically and personally. I truly thank Sudhanshu
for sticking by my side, even when I was irritable and depressed.
Thank you for all the little things you have done for me. Thanks for
listening to me always at late nights and really sorry for the sleepless
nights. I consider myself the luckiest in the world to have such a
supportive husband, standing behind me with his love and support.}

~
\begin{flushright}
(Mitali Sisodia)\pagebreak{}
\par\end{flushright}

\begin{flushleft}
{\large{}$\thispagestyle{plain}$\addcontentsline{toc}{chapter}{ABSTRACT}}{\large\par}
\par\end{flushleft}

\begin{center}
\textbf{\Large{}ABSTRACT}{\Large\par}
\par\end{center}

~

\textcolor{black}{This thesis is focused on the design and analysis
of quantum communication protocols. }Several schemes for quantum communication
have been introduced in the recent past. For example, quantum teleportation,
dense coding, quantum key distribution, quantum secure direct communication,
etc., have been rigorously studied in the last 2-3 decades. \textcolor{black}{Specifically,
a specific attention of the present thesis is to study the quantum
teleportation schemes with entangled orthogonal and nonorthogonal
states and their experimental realization, but not limited to it.
We have also studied some aspects of quantum cryptography. The present
thesis contains 7 chapters. In Chapter \ref{cha:Introduction1}, we
have introduced about the basic concepts related to quantum communication
schemes with a specific attention on quantum teleportation and quantum
cryptography and some specific examples of the quantum communication
schemes, which are rigorously studied in the next chapter of this
thesis. Chapters \ref{cha:resourceopt}-\ref{cha:nonorthogonal} are
dedicated to the quantum teleportation schemes using different type
of quantum resources (entangled orthogonal state and entangled nonorthogonal
state) and their experimental realization using superconductivity-based
IBM quantum computer. In these chapters, we have shown a perfect teleportation
of multi-qubit quantum states can be done using an optimal amount
of quantum resources and also shown a proof-of-principle experimental
realization of our optimal quantum teleportation scheme using IBMQX2
processor of five-qubit IBM quantum computer. We have also proved
that which quasi-Bell state (Bell-type entangled nonorthogonal states)
as a quantum channel is perfect for the teleportation scheme in the
absence and presence of noise. In Chapter \ref{cha:BellIBM}, we have
reported an experimental realization of a scheme for nondestructive
Bell state discrimination using the newest popular experimental platform,
i.e., superconductivity-based five-qubit IBM quantum computer, which
is recently placed in the cloud in 2016 and easily available free
of cost through the internet. In Chapter \ref{chap:OPTICAL-DESIGNS-FOR},
optical circuits for a set of quantum cryptographic schemes have been
designed using available optical elements, like a laser, beam splitter,
polarizing beam splitter, half wave plate. Finally, the thesis work
is concluded in Chapter \ref{cha:Conclusions-and-Scope} with a brief
discussion on the limitations of the present work and the scope for
future work.}

~

{\large{}$\thispagestyle{plain}$}\\

\textbf{\emph{\large{}\newpage}}{\large{} $\thispagestyle{empty}$$\mbox{}$}\textbf{\emph{\large{}\newpage}}{\large\par}
\begin{flushleft}
{\large{}$\thispagestyle{plain}$\addcontentsline{toc}{chapter}{LIST
OF ACRONYMS AND ABBREVIATIONS}}{\large\par}
\par\end{flushleft}

\begin{center}
\textbf{\Large{}LIST OF ACRONYMS \& ABBREVIATIONS}{\Large\par}
\par\end{center}

~
\begin{center}
~%
\begin{tabular}{ll}
AD & Amplitude Damping\tabularnewline
BB84 & Bennett and Brassard 1984\tabularnewline
BST & Bidirectional State Teleportation\tabularnewline
BS & Beam Splitter\tabularnewline
CDSQC & Controlled Deterministic Secure Quantum Communication\tabularnewline
CNOT & Controlled-NOT\tabularnewline
CQD & Controlled Quantum Dialogue\tabularnewline
DSQC & Deterministic Secure Quantum Communication\tabularnewline
EPR & Einstein-Podolsky-Rosen\tabularnewline
GHZ & Greenberger-Horne-Zeilinger\tabularnewline
HWP & Half Wave Plate\tabularnewline
IBM & International Business Machine\tabularnewline
MFI & Minimum Fidelity\tabularnewline
MASFI & Minimum Assured Fidelity\tabularnewline
MAVFI & Minimum Average Fidelity\tabularnewline
MDI & Measurement-Device-Independent\tabularnewline
NMR & Nuclear Magnetic Resonance\tabularnewline
OD & Optical Delay\tabularnewline
OS & Optical Switch\tabularnewline
PBS & Polarizing Beam Splitter\tabularnewline
PC & Polarization Controller\tabularnewline
PD & Phase Damping\tabularnewline
PM & Polarizing Modulator\tabularnewline
QD & Quantum Dialogue\tabularnewline
QT & Quantum Teleportation\tabularnewline
QIS & Quantum Information Splitting\tabularnewline
QKD & Quantum Key Distribution\tabularnewline
QSDC & Quantum Secure Direct Communication\tabularnewline
QST & Quantum State Tomography\tabularnewline
SQUID & Superconducting Quantum Interference Device\tabularnewline
SPDC & Spontaneous Parametric Down Conversion\tabularnewline
\end{tabular}
\par\end{center}

\begin{center}
{\large{}$\thispagestyle{plain}$}\newpage{}
\par\end{center}

{\large{}\addcontentsline{toc}{chapter}{LIST OF FIGURES}}\listoffigures
\textbf{\emph{\large{}\newpage}}{\large{} $\thispagestyle{empty}$$\mbox{}$}{\large\par}

{\large{}\addcontentsline{toc}{chapter}{LIST OF TABLES}}{\large\par}

\listoftables

\begin{center}
\textbf{\Large{}\newpage}{\Large\par}
\par\end{center}

\addtocontents{toc}{
\setlength{\cftchapindent}{-18pt}
}

\chapter[CHAPTER \thechapter \protect\newline INTRODUCTION]{INTRODUCTION \label{cha:Introduction1}}

{\large{}\pagenumbering{arabic}\lhead{}}{\large\par}

\section{What is quantum communication? \label{sec:Basic-ideas-of-QC}}

\textcolor{black}{Communication is the act of conveying an intended
message to another party through the use of mutually understood signs
and rules. Effective communication has played a pivotal role in the
development of civilization, and often the ability to communicate
effectively distinguishes the human being from the other living species.
With time, we have learned many techniques of communication, and our
dependence on the communication schemes have increased. In fact, in
a modern society communication plays a crucial role, and our dependence
on the communication technologies is increasing continuously with
the rapid development and enhanced uses of e-banking, mobile phones,
internet, IoT, etc. Motivated by this fact, the present thesis is
focused on a set of modern techniques of communication.}

\textcolor{black}{On the basis of the nature of the physical resources
used, this important aspect of modern life (communication) can be
broadly categorized into two classes: (1) classical communication
and (2) quantum communication. Classical communication includes, all
the traditional modes of communication, like internet, mobile, post,
where classical resources are used to perform the communication tasks.
In contrast, when nonclassical features of quantum mechanics (e.g.,
nonlocality or noncommutativity) and/or nonclassical states like entangled
states or squeezed states are used to perform a communication task,
it is referred to as quantum communication. This thesis is focused
on some aspects of quantum communication schemes which are not possible
in the classical domain. Some of the communication schemes studied
here require security, whereas others don't. Specifically, in what
follows, we study schemes for quantum teleportation and its variants
where security is not required and schemes related quantum cryptography
where security of the scheme is the primary concern.}

\textcolor{black}{In classical communication, the security of the
transferred information is not unconditional. As the security of the
communicated information in every public key cryptography system is
ensured via the computational complexity of the task used for creating
a key which is used for encryption of the message. In contrast, in
quantum communication, it is possible to attain the unconditional
security as the security of the schemes for quantum cryptography are
independent of computational complexity of a task and is obtained
using the laws of nature. In addition, there exist a few quantum communication
tasks such as teleportation and dense coding, which can be realized
only in the quantum domain. These communication tasks (teleportation,
dense coding and most of their variants) do not require security,
but they may be used as primitives for secure quantum communication.}

\textcolor{black}{In this thesis, we have worked on both types of
quantum communication schemes (i.e., quantum teleportation and quantum
cryptography), but with a greater stress on quantum teleportation.
From last 3-4 decades, several cryptographic and non-cryptographic
quantum communication tasks have been studied rigorously. Historically,
the journey of quantum communication formally began with the publication
of the pioneering work of S. Wiesner \cite{wiesner1983conjugate}
in 1983}\footnote{\textcolor{black}{A version of this paper was prepared and communicated
in 1970, but it was not accepted for publication at that time (for
detail see \cite{pathak2013elements}).}}\textcolor{black}{. In the next year (i.e., in 1984), Bennett and
Brassard \cite{bennett1984quantum} proposed the first ever protocol
of quantum key distribution (QKD). This pioneering work which is now
known as BB84 protocol, changed the entire notion of cryptographic
security. The security of this protocol was based on our inability
to perform simultaneous measurements in two nonorthogonal bases (non-commutativity),
and the protocol in its original form was described using the polarization
states of single photons. Later in 1991, Ekert introduced another
interesting protocol of QKD, which is now known as E91 protocol \cite{ekert1991quantum}.
In contrast to single particle states based BB84 protocol, E91 protocol
used the properties of entangled states. This was probably the first
occasion when entanglement was used for quantum communication. Soon,
it was realized that entanglement is one of the most important resources
for quantum communication, as in 1992 and 1993, the concept of dense
coding \cite{bennett1992communication}, and teleportation \cite{bennett1993teleporting}
were introduced, respectively, and entanglement was found to be essential
(may not be sufficient) for both of these communication schemes having
no classical analogue. Later, a stronger version of entanglement (Bell
nonlocality) has been found to be essential for device independent
quantum cryptography \cite{acin2007device}. Among these nonclassical
schemes of communication, teleportation can be viewed as one of the
most important schemes of quantum communication. It deserves special
attention, as a large number of other quantum communication schemes
can be viewed as variants of teleportation. For example, quantum information
splitting (QIS) \cite{zheng2006splitting,pathak2011efficient}, hierarchical
QIS \cite{wang2010hierarchical,shukla2013hierarchical}, quantum secret
sharing (QSS) \cite{hillery1999quantum}, quantum cryptography based
on entanglement swapping \cite{shukla2014orthogonal}, remote state
preparation \cite{pati2000minimum,sharma2015controlled} may be viewed
as variants of teleportation.}

\textcolor{black}{The first quantum teleportation scheme was introduced
by Bennett et al., in 1993 \cite{bennett1993teleporting}. }This scheme
was designed for the transmission of an unknown quantum state (a qubit)
from Alice (sender) to Bob (receiver) using two bits of classical
communication and a pre-shared maximally entangled state (see Section
\ref{subsec:Quantum-Teleportation-(QT)} for details).\textcolor{black}{{}
Dense coding (or super dense coding) is a closely related scheme for
quantum communication, where two classical bits of information is
transferred by using a single-qubit and prior shared entanglement.
Bennett et al., introduced this scheme in 1992 \cite{bennett1992communication}.
After these pioneering works, several quantum communication schemes
have been proposed \cite{cao2005teleportation,cao2013deterministic,chen2006general,chen2008quantum,choudhury2016teleportation,choudhury2017asymmetric,choudhury2017teleportation,da2007teleportation,dai2004probabilistic,deng2003two,deng2005comment,dong2008controlled,dong2011controlled,duan2014bidirectional,fang2000experimental,fang2003probabilistic,fu2014general,furusawa1998unconditional,gorbachev2000quantum,guan2014joint,hao2001controlled,hassanpour2015efficient,hassanpour2016bidirectional,henderson2000two,hillery1999quantum,hong2001probabilistic,huang2017performance,huelga2001quantum,jain2009secure,joo2003quantum,joy2017efficient,karlsson1998quantum,li2002quantum,li2007states,li2013bidirectional,li2016asymmetric,li2016quantum,li2016quantum1,li2017quantum,liu2002general,lo1999unconditional,lutkenhaus1999bell,ma2012quantum,majumder2017experimental,man2007genuine,mattle1996dense,mozes2005deterministic,muralidharan2008perfect,muralidharan2008quantum,nandi2014quantum,nguyen2004quantum,nie2009non,nie2011quantum,nielsen1998complete,pathak2011efficient,pati2005probabilistic,pati2007probabilistic,prakash2007improving,prakash2008effect,prakash2011increasing,prakash2012minimum,riebe2004deterministic,rigolin2005quantum,sang2016bidirectional,sharma2015controlled,sharma2016comparative,sharma2016verification,shi2000probabilistic,shukla2013bidirectional,shukla2013hierarchical,shukla2013group,shukla2013improved,shukla2014orthogonal,sisodia2016teleportation,sisodia2017design,sisodia2018comment,situ2010simultaneous,situ2014controlled,situ2015secure,song2008controlled,srinatha2014quantum,sun2016quantum,tan2016deterministic,thapliyal2015applications,thapliyal2015general,thapliyal2017quantum,ting2005controlled,tsai2010teleportation,tsai2011dense,van2001entangled,van2001quantum,wang2010hierarchical,wang2015bidirectional,weedbrook2004quantum,wei2016comment,xi2007controlled,xia2006quantum,xia2008generalized,xia2010teleportation,xiao2007controlled,xin2010bidirectional,xiu2005probabilistic,ya2008faithful,yadav2014two,yan2004scheme,yan2013bidirectional,yang2000multiparticle,yang2004efficient,yang2009quantum,yeo2003quantum,yeo2006teleportation,yin2012quantum,yong2001probabilistic,yu2013teleportation,zha2013bidirectional,zhang2006experimental,zhang2007perfect,zhang2009schemes,zhang2013quantum,zhang2016efficient,zhao2004experimental,zhao2017efficient,zhao2018quantum,zheng2006splitting,zheng2016complete,zhou2000controlled,zhou2007multiparty,shukla2017hierarchical}
which can be classified into following two classes.}
\begin{description}
\item [{\textcolor{black}{Class~1:}}] \textcolor{black}{Quantum communication
protocols without security: where security is not relevant (e.g.,
dense coding, teleportation and its variants).}
\item [{\textcolor{black}{Class~2:}}] \textcolor{black}{Quantum communication
protocols with security: where security is relevant. All schemes of
secure quantum communication, including the protocols for QKD and
secure direct quantum communication belong to this class.}
\end{description}
\textcolor{black}{In the following section, we have briefly discussed
the history of protocols of quantum communication belonging to the
above classes with a focus on the schemes which are relevant for this
thesis.}

\section{\textcolor{black}{A chronological history of protocols of quantum
communication\label{sec:A-chronological-history}}}

\textcolor{black}{In this section, we aim to discuss the historical
development of quantum communication schemes belonging to both the
classes mentioned above}\footnote{\textcolor{black}{As the existing literature is huge, some important
works on quantum communication may be excluded by us. Any such omission
is unintentional.}}\textcolor{black}{. To begin with, we discuss the schemes belonging
to Class 1. However, we have given more stress to the schemes that
can be viewed as variants of teleportation in comparison to the schemes
of dense coding as this thesis is focused on teleportation.}

\subsection{\textcolor{black}{A chronological history of quantum communication
protocols of Class 1\label{subsec:A-history-of}}}
\begin{description}
\item [{\textcolor{black}{1992:}}] \textcolor{black}{In 1992, the concept
of dense coding was proposed by Charles H Bennett and Stephen J Wiesner
\cite{bennett1992communication}. This scheme allowed Alice to send
2 bits of classical information to Bob by sending only one qubit provided
they already share an entangled state. This was exciting as the communication
capacity of this scheme was higher than the maximum possible classical
value (as maximum information that can be communicated classically
by sending a particle in one bit).}
\item [{\textcolor{black}{1993:}}] \textcolor{black}{In 1993, Charles H
Bennett et al., proposed the first quantum teleportation scheme \cite{bennett1993teleporting},
which does not have any classical analogue. This was an exciting development
as in this scheme (to be elaborated in Section \ref{subsec:Quantum-Teleportation-(QT)})
the quantum state to be teleported does not travel through the channel,
and at the end of the scheme all information about the state is lost
at the sender's end and the quantum state is obtained at the receiver's
end.}
\item [{\textcolor{black}{1997:}}] \textcolor{black}{Dik Bouwmeester et
al., presented the first experimental demonstration of the quantum
teleportation scheme \cite{bouwmeester1997experimental} using entangled
photons generated by type II spontaneous parametric down-conversion
(SPDC) process.}
\item [{\textcolor{black}{1998:}}] \textcolor{black}{Anders Karlsson and
Mohamed Bourennane proposed the first scheme for QIS or controlled
quantum teleportation  scheme \cite{karlsson1998quantum}, where a
sender teleports an unknown quantum state to two receivers. Due to
no-cloning theorem only one of them can reproduce the state with the
help of other receiver who may be treated as a controller.}
\item [{\textcolor{black}{2000:}}] \textcolor{black}{Arun K Pati proposed
the first scheme of remote state preparation, which can be viewed
as a scheme for QT of a known quantum state \cite{pati2000minimum}.
After that, several experimental realizations of remote state preparation
have been reported.}
\item [{\textcolor{black}{2001:}}] \textcolor{black}{Susana F Huelga et
al., \cite{huelga2001quantum} proposed a first scheme for bidirectional
(sender $\Longleftrightarrow$receiver) quantum teleportation or bidirectional
state teleportation (BST).}
\item [{\textcolor{black}{2004:}}] \textcolor{black}{Various facets of
teleportation have been experimentally demonstrated. For example,
a proof of principle experimental realizations of a scheme for quantum
information splitting was reported by Zhi Zhao et al., by preparing
a five-qubit entangled state of photon \cite{zhao2004experimental},
MD Barretti et al., demonstrated teleportation of massive particle
(atomic) qubits using ($^{9}$Be$^{+}$) ions confined in an ion trap
\cite{barrett2004deterministic} and Yun-Feng Huang et al., performed
experimental teleportation of a CNOT gate \cite{huang2004experimental}. }
\item [{\textcolor{black}{2005:}}] \textcolor{black}{Gustavo Rigolin proposed
a scheme \cite{rigolin2005quantum} for QT of an arbitrary two-qubit
state and shown that all the multi-qubit states can be teleported
using Bell states.}
\item [{\textcolor{black}{2006:}}] \textcolor{black}{Qiang Zhang et al.,
\cite{zhang2006experimental} experimentally realized the quantum
teleportation of a two-qubit composite system.}
\item [{\textcolor{black}{2010:}}] \textcolor{black}{The 16 kilometer free
space QT was achieved by Xian-Min Jin et al., in China \cite{jin2010experimental}.}
\item [{\textcolor{black}{2011:}}] \textcolor{black}{In 2011, Anirban Pathak
and Anindita Banerjee proposed an efficient and economical (as far
as the amount of quantum resource requirement is concerned) scheme
for the perfect quantum teleportation and controlled quantum teleportation
\cite{pathak2011efficient}.}
\item [{\textcolor{black}{2012:}}] \textcolor{black}{Xiao-Song Ma et al.,
reported QT over 143 kilometers between the Canary Islands of La Palma
and Tenerife \cite{ma2012quantum} using optical fiber, and Juan Yin
et al., also reported the free space QT and entanglement distribution
over 100 kilometers in China \cite{yin2012quantum}.}
\item [{\textcolor{black}{2012:}}] \textcolor{black}{Above mentioned QT
schemes were based on orthogonal-state-based entangled channels. To
the best of our knowledge the first nonorthogonal-state-based QT scheme
was proposed by Satyabrata Adhikari et al. \cite{adhikari2012quantum}.}
\item [{\textcolor{black}{2005-16:}}] \textcolor{black}{Various interesting
schemes of QT (two party and three party scheme) have been reported.
Here, we may mention a few of them, such as protocols designed by
Zhuo-Liang Cao and Wei Song in 2005 \cite{cao2005teleportation},
Li Da-Chuang and Cao Zhuo-Liang in 2007 \cite{da2007teleportation},
Li Song-Song et al., in 2008 \cite{song2008controlled}, Chia-Wei
Tsai and Tzonelih Hwang in 2010 \cite{tsai2010teleportation}, Yuan-hua
Li et al., in 2016 \cite{li2016quantum}. Authors of all these works
used costly (in terms of preparation and maintenance) quantum resources
(multi-qubit state) to teleport unknown quantum state.}
\item [{\textcolor{black}{2017:}}] \textcolor{black}{In 2017, we proposed
a general scheme of QT with a reduced amount of quantum resources
\cite{sisodia2017design} and their experimental realization. We have
also proposed a nonorthogonal-state-based QT scheme \cite{sisodia2016teleportation}.}
\item [{\textcolor{black}{2018:}}] \textcolor{black}{Many people are found
to propose QT schemes with higher amount of quantum resource even
after the introduction of our optimal scheme in 2017. So, in 2018,
we have written a specific comment \cite{sisodia2018comment} to specifically
show how our optimal scheme can be realized in a particular case.}
\end{description}

\subsection{\textcolor{black}{A chronological history of quantum communication
protocols of Class 2 \label{subsec:A-chronological-historyclass2}}}
\begin{description}
\item [{\textcolor{black}{1983:}}] \textcolor{black}{Stephen Wiesner \cite{wiesner1983conjugate}
published a paper entitled, ``Conjugate coding'' which inherently
contained many concepts of secure quantum communication. The formal
journey of quantum cryptography began with this paper.}
\item [{\textcolor{black}{1984:}}] \textcolor{black}{First protocol for
QKD was proposed by Charles H Bennett and G Brassard \cite{bennett1984quantum}.
The protocol is now known as BB84 protocol. This protocol requires
four states.}
\item [{\textcolor{black}{1991:}}] \textcolor{black}{Artur K Ekert proposed
a QKD protocol, which was based on Bell's theorem \cite{ekert1991quantum}.
Six states were used in this protocol.}
\item [{\textcolor{black}{1992:}}] \textcolor{black}{Charles H Bennett
proposed a new protocol to establish that two states selected from
two nonorthogonal bases would be sufficient for QKD \cite{bennett1992quantum}.
The protocol is now known as B92 protocol.}
\item [{\textcolor{black}{2002:}}] \textcolor{black}{The popular Ping-pong
(PP) protocol for quantum secure direct communication (QSDC) introduced
by Kim Bostr�m and Timo Felbinger \cite{bostrom2002deterministic}.}
\item [{\textcolor{black}{2004:}}] \textcolor{black}{The first scheme for
bidirectional secure quantum communication, i.e., a scheme for quantum
dialogue (QD) introduced by Ba An Nguyen \cite{nguyen2004quantum}
in 2004.}
\item [{\textcolor{black}{2005:}}] \textcolor{black}{Marco Lucamarini and
Stefano Mancini presented a protocol, which is known as LM05 scheme
\cite{lucamarini2005secure}. This is a PP- type protocol which can
be realized using single photon states, i.e., without using entanglement.}
\item [{\textcolor{black}{2009:}}] \textcolor{black}{Tae-Gon Noh proposed
a counterfactual protocol for QKD \cite{noh2009counterfactual}. The
protocol is now known as N09 protocol. This protocol is recently tested
experimentally \cite{brida2012experimental,liu2012experimental,ren2011experimental},
it's also interesting because it motivated researchers to introduce
many counterfactual schemes of secure quantum communication.}
\item [{\textcolor{black}{2006-2018:}}] \textcolor{black}{Several quantum
cryptographic protocols have been reported and a few of them have
been realized experimentally \cite{zhou2016measurement,zhou2018measurement,zhang2017quantum,khan2018satellite,liao2018satellite,zhu2017experimental,yang2009quantum}.
During this period, many other facets of QKD have also been explored.
For example, the notion of semi-quantum key distribution \cite{krawec2015security,yu2014authenticated},
counterfactual key distribution \cite{sun2010counterfactual,salih2014tripartite},
device independent QKD \cite{gisin2010proposal,lim2013device,pironio2009device},
semi-device independent QKD and measurement device independent QKD
have also been evolved and gradually matured (for a review see \cite{shenoy2017quantum}).
Here, it would be interesting to note that in a semi-quantum scheme,
the end user can be classical, and in a device independent scheme
a completely secure key distribution is possible even when the devices
used are faulty. However, to achieve complete device independence,
we would generally require 100\% efficient photon-detector, which
is not achievable at the moment. In contrast, measurement device independent
schemes allow one to distribute secure key when only the measurement
devices are faulty. The realization of this state of art scheme of
QKD requires Bell nonlocal state, but it's realized that one-way device
independence (measurement device independence can be realized using
steering). Further, schemes of secure direct communication have been
realized in the laboratories \cite{hu2016experimental,zhang2017quantum}.
In this decade, exciting works have been done in many facets of cryptography
beyond QKD and secure direct quantum communication. In particular,
many protocols of quantum voting \cite{hillery2006towards,vaccaro2007quantum,wen2011secure,jiang2012quantum,bao2017quantum,xue2017simple,thapliyal2017protocols,thapliyal2016analysis},
quantum auction \cite{hogg2007quantum,piotrowski2008quantum,qin2009cryptanalysis,zhao2010secure,sharma2017quantum},
quantum e-commerce, etc. have been studied in detail. Finally, various
commercial products for performing QKD have been launched and marketed
\cite{IDQ,mgQ}, some of them have also been successfully used in
providing information security to mega events like 2010 Soccer World
Cup. Any discussion on this topic would remain incomplete without
a mention of the recent successful satellite-based experimental quantum
communication \cite{khan2018satellite} between China and Austria
\cite{liao2018satellite}.}
\end{description}

\section{\textcolor{black}{Qubit and measurement basis \label{sec:Qubit-and-measurement}}}

\textcolor{black}{Before introducing quantum communication schemes,
we would like to introduce here the basic building block of quantum
information processing which is known as a ``qubit''. A qubit can
be viewed as a quantum analogue of a bit. As we know, classical information
is measured in bit, which is either in the state $0$ or in $1,$
similarly quantum information is represented by qubit, which is allowed
to exist simultaneously in the states $0$ and $1$. These states
or vectors are denoted as $|0\rangle=\left(\begin{array}{c}
1\\
0
\end{array}\right)$ and $|1\rangle=\left(\begin{array}{c}
0\\
1
\end{array}\right),$ which are written in the Dirac's notation or bra-ket notation. In
this notation, $|0\rangle$ and $|1\rangle$ are pronounced as $ket0$
and $ket1$, respectively. Transpose conjugate of a state vector described
as $|\psi\rangle$ is described as $\langle\psi|$ and is pronounced
as bra $\psi.$ Thus, the state vectors are usually described as a
column matrix representing $|\psi\rangle,$whereas $\langle\psi|$
is described by a row matrix. Now, a qubit is represented by a state
vector $|\psi\rangle=\alpha|0\rangle+\beta|1\rangle,$ where $\alpha$
and $\beta$ are the probability amplitudes and complex numbers, which
satisfies the condition $|\alpha|^{2}+|\beta|^{2}=1.$ $|\alpha|^{2}$
and $|\beta|^{2}$ are the probabilities of obtaining the qubit in
the state $|0\rangle$ and $|1\rangle$, respectively on performing
a measurement in the computational basis \cite{gruska1999quantum,pathak2013elements}.
The last statement and the fact that a quantum state collapses to
a basis state on performing a measurement in a particular basis will
be further clarified in the following paragraphs, but before we do
so, we need to elaborate on the basis sets used in this work.}

\textcolor{black}{A set of vectors is $\left\{ |v_{1}\rangle,|v_{2}\rangle,|v_{3}\rangle,.......|v_{n}\rangle\right\} $.
If the elements of this set are linearly independent (condition of
linear independence of a set of vectors is $\stackrel[i=1]{n}{\sum}a_{i}|v_{i}\rangle=0$
iff all $a_{i}=0$) to each other, satisfy the condition of orthogonality
$\langle v_{i}|v_{j}\rangle=\delta_{ij}$ and show a completeness
relation given by $\stackrel[i=1]{n}{\sum}|v_{i}\rangle\langle v_{i}|=1,$
then the set is known as a basis set. In this thesis, we have mostly
used three basis sets- computational basis, diagonal basis and Bell
basis. These basis sets can be defined as follows}

\textcolor{black}{
\[
\boldsymbol{\text{Computational basis: }}\left\{ |0\rangle,|1\rangle\right\} .
\]
In the context of quantum communication the basis elements (states
to which an arbitrary quantum state is projected on measurement) are
usually viewed as horizontal ($|0\rangle$) and vertical ($|1\rangle$)
polarization states of photon. In two-qubit scenario computational
basis refers to $\left\{ |00\rangle,|01\rangle,|10\rangle,|11\rangle\right\} .$
Similarly, we can easily extend it to multi-qubit scenario as one
can write a multi-qubit state as tensor product of states of individual
qubits. In general, in this basis set, basis elements are not expressed
in a superposition
\[
\begin{array}{c}
.\\
\boldsymbol{\text{Diagonal basis: }}\left\{ |+\rangle=\frac{|0\rangle+|1\rangle}{\sqrt{2}},|-\rangle=\frac{|0\rangle-|1\rangle}{\sqrt{2}}\right\} .
\end{array}
\]
In QKD, we usually view $|+\rangle$ as a photon polarized at $45^{{\rm o}}$
w.r.t. horizontal direction and $|-\rangle$ as a photon polarized
at $135^{{\rm o}}$  w.r.t. horizontal direction.
\[
\begin{array}{c}
.\\
\boldsymbol{\text{Bell basis: }}\left\{ |\psi^{+}\rangle=\frac{|00\rangle+|11\rangle}{\sqrt{2}},|\psi^{-}\rangle=\frac{|00\rangle-|11\rangle}{\sqrt{2}},|\phi^{+}\rangle=\frac{|01\rangle+|10\rangle}{\sqrt{2}},|\phi^{-}\rangle=\frac{|01\rangle-|10\rangle}{\sqrt{2}}\right\} .
\end{array}
\]
Here, the basis elements are maximally entangled and these are known
as Bell states.}

\section{\textcolor{black}{Various facets of quantum communication schemes
of Class 1 \label{sec:Various-facets-of-insecure}}}

\textcolor{black}{The pioneering work of Bennett et al., on QT drew
considerable attention of the quantum communication community since
its introduction in 1993 \cite{bennett1993teleporting}. As a consequence,
a large number of modified QT schemes have been proposed.  To elucidate
on this point, we may mention a few schemes of quantum communication
tasks, which can be viewed as modified schemes for QT. This set of
quantum communication tasks includes\textendash remote state preparation
\cite{pati2000minimum}, controlled quantum teleportation \cite{karlsson1998quantum,pathak2011efficient},
or equivalently quantum information splitting, \cite{hillery1999quantum,nie2011quantum}
bidirectional quantum teleportation \cite{huelga2001quantum}, bidirectional
controlled quantum teleportation \cite{shukla2013bidirectional,thapliyal2015general,thapliyal2015applications},
quantum secret sharing \cite{hillery1999quantum}, hierarchical quantum
secret sharing \cite{shukla2013hierarchical,mishra2015integrated},
and many more. The interconnection among these variants of QT can
be understood easily if we note that QT of a known state (i.e., when
the probability amplitudes (coefficients) of the state to be teleported
are known to the sender (usually referred to as Alice) but not to
the receiver (usually referred to as Bob)) is called remote state
preparation, whereas simultaneous QT of a quantum state each by Alice
and Bob is known as bidirectional state teleportation. Similarly,
controlled teleportation and bidirectional controlled state teleportation
schemes are controlled variants of QT and bidirectional state teleportation
schemes, where a third party (Charlie) supervises the whole proceeding
by preparing a quantum channel to be used for the task and withholding
part of the useful information. In other words, in controlled teleportation
(bidirectional controlled state teleportation), Alice and Bob can
execute a scheme for QT (bidirectional state teleportation) if Charlie
allows them to do so.}

\textcolor{black}{Many of the above-mentioned schemes have very interesting
applications (for details see Ref. \cite{pathak2013elements}). For
example, we may mention that every scheme of BST can be used to design
quantum remote control \cite{huelga2001quantum}. It is also worth
noting here that although teleportation (and most of its variants)
in its original form is not a secure communication scheme, it can
be used as a primitive for secure quantum communication. As a major
part of this thesis is focused around quantum teleportation and its
modified version, in the following section, we will briefly describe
the concept of QT.}

\subsection{\textcolor{black}{Quantum teleportation \label{subsec:Quantum-Teleportation-(QT)}}}

\textcolor{black}{Teleportation is a quantum task and it can be achieved
only by using a shared entangled state as a quantum channel and some
classical communication. The standard quantum teleportation scheme
uses a Bell state as quantum channel \cite{bennett1993teleporting}.
The basic idea of the original quantum teleportation scheme was to
teleport an unknown single-qubit quantum state $|\psi\rangle=\alpha|0\rangle+\beta|1\rangle$
from Alice (sender) to spatially separated Bob (receiver) by using
2 bits of classical communication using a shared entangled state (Bell
state). After the pioneering work of Bennett et al., a large number
of quantum teleportation schemes and their modifications have been
reported by using Bell state, GHZ state, and other multi-partite entangled
states as a quantum channel \cite{li2016quantum,hassanpour2016bidirectional,da2007teleportation,li2016asymmetric,song2008controlled,cao2005teleportation,muralidharan2008quantum,tsai2010teleportation,nie2011quantum,tan2016deterministic,wei2016comment,li2016quantum1,yu2013teleportation,nandi2014quantum}.
Several experimental realizations of quantum teleportation schemes
have also been reported (\cite{bouwmeester1997experimental,barrett2004deterministic,huang2004experimental,zhao2004experimental,zhang2006experimental,jin2010experimental,ma2012quantum,yin2012quantum}
and references therein).}

\begin{figure}
\begin{centering}
\textcolor{black}{\includegraphics[scale=0.6]{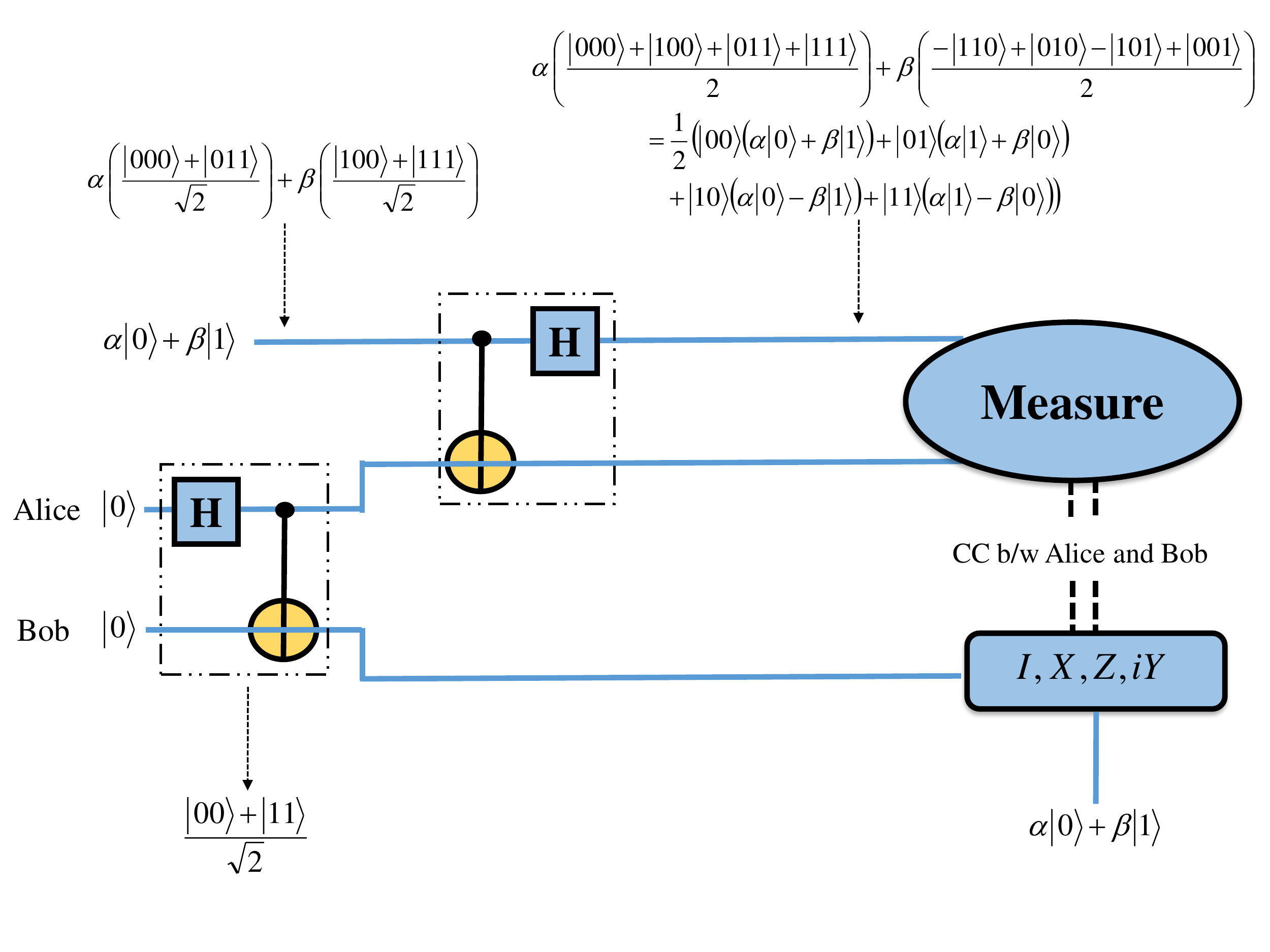}}
\par\end{centering}
\textcolor{black}{\caption{\label{fig:Standard-quantum-teleportation}Standard quantum teleportation
circuit\textcolor{black}{.}\textcolor{red}{{} }\textcolor{black}{CC
stands for classical communication. The quantum gates shown here are
described in Section \ref{sec:Brief-discussion-of}.}}
}
\end{figure}

\textcolor{black}{All these schemes and their modified version may
be classified into two classes- (1) Perfect teleportation- where an
unknown quantum state is transmitted with unit fidelity. A perfect
teleportation scheme is also called deterministic if the success rate
of the teleportation scheme is found to be unity, such a deterministic
perfect teleportation scheme requires a maximally entangled state
as a quantum channel. (2)}\textbf{\textcolor{black}{{} }}\textcolor{black}{Probabilistic
teleportation- After the Bennett et al.\textquoteright s scheme, it
was shown that the teleportation is possible with unit fidelity even
when a non-maximally entangled state is used as a quantum channel.
In that case, the success rate of teleportation will not be unity
such a scheme of teleportation is referred to as probabilistic teleportation
scheme. In what follows, we briefly describe the original scheme of
Bennett et al., as an example of deterministic perfect teleportation
scheme.}
\begin{description}
\item [{\textcolor{black}{Deterministic~perfect~teleportation:}}] \textcolor{black}{A
standard quantum circuit for deterministic perfect teleportation is
shown in Figure \ref{fig:Standard-quantum-teleportation}. According
to this figure, initially Alice and Bob share a maximally entangled
Bell state $|\psi^{+}\rangle=\frac{1}{\sqrt{2}}\left(|00\rangle{\rm {\rm +}}|11\rangle\right)$
of which the first qubit is with Alice and the second qubit is with
Bob. In the left most block of the circuit shown in Figure \ref{fig:Standard-quantum-teleportation},
this Bell state is created from the separable states ($|0\rangle\otimes|0\rangle$)
by applying a Hadamard gate }\footnote{\textcolor{black}{All quantum gates are explained in detail in Section\ref{sec:Brief-discussion-of}
.}}\textcolor{black}{{} on the first qubit followed by a CNOT gate (EPR
circuit) which uses the first qubit as the control qubit and the second
qubit as the target qubit. In the second box from left, a reverse
EPR circuit is shown. It helps Alice to entangled the state to be
teleported using the shared entangled state as follows. First Alice
combines her unknown quantum state $|\psi\rangle=\alpha|0\rangle+\beta|1\rangle$
to be teleported with her shared entangled qubit. The combined state
is}
\end{description}
\textcolor{black}{
\[
\begin{array}{llc}
|\psi\rangle_{combined} & = & |\psi\rangle\otimes|\psi^{+}\rangle\\
 & = & \left(\alpha|0\rangle+\beta|1\rangle\right)_{A}\otimes\frac{1}{\sqrt{2}}\left(|00\rangle+|11\rangle\right)_{AB}\\
 & = & \left(\alpha\frac{\left(|000\rangle+|011\rangle\right)}{\sqrt{2}}+\beta\frac{\left(|100\rangle+|111\rangle\right)}{\sqrt{2}}\right)_{AAB}
\end{array}.
\]
Then she applies a CNOT gate on the first two qubits (indexed as $AA$)
available with her. Where the first $A$ corresponds to a control
qubit and the second $A$ corresponds to a target qubit. This operation
transforms the combined state $|\psi\rangle_{combined}$ to $|\psi\rangle_{1}$}

\textcolor{black}{
\[
\begin{array}{llc}
|\psi\rangle_{1} & = & \left(\alpha\frac{\left(|000\rangle+|011\rangle\right)}{\sqrt{2}}+\beta\frac{\left(|110\rangle+|101\rangle\right)}{\sqrt{2}}\right)_{AAB}.\end{array}
\]
}

\textcolor{black}{After that Alice applies Hadamard operation on the
first qubit to yield}

\textcolor{black}{
\[
\begin{array}{lcl}
|\psi\rangle_{2} & = & \left(\alpha\frac{\left(|0\rangle+|1\rangle\right)}{\sqrt{2}}\frac{\left(|00\rangle+|11\rangle\right)}{\sqrt{2}}+\beta\frac{\left(|0\rangle-|1\rangle\right)}{\sqrt{2}}\frac{\left(|10\rangle+|01\rangle\right)}{\sqrt{2}}\right)_{AAB}\\
 & = & \frac{1}{2}\left(|00\rangle\left(\alpha|0\rangle+\beta|1\rangle\right)+|01\rangle\left(\alpha|1\rangle+\beta|0\rangle\right)\right.\\
 & + & \left.|10\rangle\left(\alpha|0\rangle-\beta|1\rangle\right)+|11\rangle\left(\alpha|1\rangle-\beta|0\rangle\right)\right)_{AAB}
\end{array}.
\]
}

\textcolor{black}{Subsequently, Alice measures her both qubits $AA$
in computational basis \{$|00\rangle,\,|01\rangle,$ $|10\rangle,|11\rangle$\}.
Subsequently, she sends her measurement result to Bob by classical
communication. At the end of Bob, he applies unitary operation on
his qubit to reconstruct the unknown quantum state. Pauli operation
applied by Bob depends on the classical communication of Alice as
shown in Table \ref{tab:Table-of-Alice's}. In all cases Bob reconstructs
$\alpha|0\rangle+\beta|1\rangle.$ Thus, the perfect teleportation
task is accomplished with unit fidelity. It may be noted that the
teleportation scheme described above is a scheme for teleportation
of an unknown single-qubit quantum state using a Bell state. Extending
this idea, in Chapters \ref{cha:resourceopt} and \ref{cha:comment},
we will present a scheme for the teleportation of the multi-qubit
unknown quantum states using minimal number of Bell states.}

\textcolor{black}{}

\subsubsection{\textcolor{black}{Entangled orthogonal and nonorthogonal state based
quantum teleportation \label{subsec:Orthogonal-and-nonorthogonal}}}

\textcolor{black}{Usually, standard entangled states, which are inseparable
states of orthogonal states, are used to implement teleportation based
schemes. However, entangled nonorthogonal states do exist, and they
may be used to implement some of these teleportation-based protocols
\cite{adhikari2012quantum}. Specifically, entangled coherent states
\cite{sanders1992entangled,mann1995bell,van2001entangled,peres1978unperformed,mann1992unique,wang2001bipartite,prakash2007improving,mishra2010teleportation,prakash2011increasing}
and Schr{\"o}dinger cat states prepared using $SU(2)$ coherent states
\cite{wang2000entangled} are the typical examples of entangled nonorthogonal
states. Such a state was first introduced by Sanders in 1992 \cite{sanders1992entangled}.}

\textcolor{black}{}
\begin{table}
\begin{centering}
\textcolor{black}{\caption{\label{tab:Table-of-Alice's}Table of Alice's measurement results
and Bob's appropriate choices of Pauli gates to reconstruct the unknown
state $\left(\alpha|0\rangle+\beta|1\rangle\right).$\textbf{ }}
}
\par\end{centering}
~

~
\begin{centering}
\textcolor{black}{}%
\begin{tabular}{|>{\centering}p{2cm}|>{\centering}p{3cm}|>{\centering}p{2cm}|>{\centering}p{3cm}|}
\hline 
\textcolor{black}{Alice measures}

~ & \textcolor{black}{State of Bob's qubit} & \textcolor{black}{Bob's operation} & \textcolor{black}{Bob's state after operation}\tabularnewline
\hline 
\textcolor{black}{$00$} & $\alpha|0\rangle+\beta|1\rangle$ & \textcolor{black}{$I$} & $\alpha|0\rangle+\beta|1\rangle$\tabularnewline
\hline 
\textcolor{black}{$01$} & $\alpha|0\rangle-\beta|1\rangle$ & \textcolor{black}{$Z$} & $\alpha|0\rangle+\beta|1\rangle$\tabularnewline
\hline 
\textcolor{black}{$10$} & $\alpha|1\rangle+\beta|0\rangle$ & \textcolor{black}{$X$} & $\alpha|0\rangle+\beta|1\rangle$\tabularnewline
\hline 
\textcolor{black}{$11$} & $\alpha|1\rangle-\beta|0\rangle$ & \textcolor{black}{$iY$} & $\alpha|0\rangle+\beta|1\rangle$\tabularnewline
\hline 
\end{tabular}
\par\end{centering}
\end{table}
\textcolor{black}{Since then several investigations have been performed
on the properties and applications of the entangled nonorthogonal
states. The investigations have yielded a handful of interesting results.
To be precise, in Ref. \cite{prakash2009swapping}, Prakash et al.,
have provided an interesting scheme for entanglement swapping for
a pair of entangled coherent states; subsequently, they investigated
the effect of noise on the teleportation fidelity obtained in their
scheme \cite{kumar2013noise}, and Dong et al., \cite{dong2014continuous}
showed that this type of entanglement swapping schemes can be used
to construct a scheme for continuous variable quantum key distribution
(QKD); the work of Hirota et al., \cite{hirota2001entangled} has
established that entangled coherent states, which constitute one of
the most popular examples of the entangled nonorthogonal states, are
more robust against decoherence due to photon absorption in comparison
to the conventional bi-photon Bell states; another variant of entangled
nonorthogonal states known as squeezed quasi-Bell state has recently
been found to be very useful for quantum phase estimation \cite{de2016quantum};
in \cite{adhikari2012quantum}, Adhikari et al., have investigated
the merits and demerits of an entangled nonorthogonal state-based
teleportation scheme analogous to the standard teleportation scheme.
In brief, these interesting works have established that most of the
quantum computing and communication tasks that can be performed using
usual entangled states of orthogonal states can also be performed
using entangled nonorthogonal states. Interestingly, nonorthogonal-based
QT scheme will be discuss in  Chapter \ref{cha:nonorthogonal}.}

\section{\textcolor{black}{Various facets of quantum communication schemes
of Class 2 \label{sec:Various-facts-of}}}

\textcolor{black}{It is well known that all the classical cryptographic
protocols are secure under some assumptions related to the complexity
of a computational task, and consequently security of classical schemes
are conditional. In contrast, secure communication is possible in
the quantum domain without any such assumption. So most of the quantum
cryptographic protocols or secure quantum communication protocols
are unconditionally secure, which is never achievable in the classical
regime. We have already mentioned that in 1984, Bennett and Brassard
\cite{bennett1984quantum} introduced the first QKD protocol, which
is now known as BB84 protocol. This unconditionally secure quantum key can be used for encryption in one-time pad \cite{vernam}. This pioneering work was followed by
several new }different protocols for QKD and other tasks related to
secure quantum communication\textcolor{black}{{} (see \cite{ekert1991quantum,bennett1992quantum,goldenberg1995quantum,shenoy2017quantum}
and references therein). Most of these protocols have not yet been
realized experimentally. The protocols which have not yet been reported
experimentally, require some modifications for the realizations using
the existing technology. Such a possibility is discussed in the present
thesis. This section aims to introduce a few protocols of secure quantum
communication that are relevant to the present thesis. Such protocols
are briefly discussed below.}

\subsection{\textcolor{black}{BB84 and other protocols of QKD \label{subsec:BB84-Protocol-of}}}

\textcolor{black}{Detailed description of BB84 and other protocols
of QKD and their inter-relations can be found in Chapter 8 of Ref.
\cite{pathak2013elements} and in the other text books, so we are
not going to provide detailed description of any scheme here. We just
wish to note that in contrast to classical schemes of secure communication
(where the security arises from the complexity of a computational
task), in QKD the security arises from the fundamental laws of physics.
In BB84 protocol, initially, Alice prepares a random sequence of quantum
states $\left\{ |0\rangle,|1\rangle,|+\rangle,|-\rangle\right\} $
or equivalently a random sequence of single photon states prepared
in horizontal, vertical, $45^{0}$, and $135^{0}$ polarized states.
After receiving the sequence, Bob measures each qubit randomly using
$\{|0\rangle,|1\rangle\}$ basis or $\{|+\rangle,|-\rangle\}$ basis
and subsequently announces which qubit is measured by him in which
basis. Alice checks in which cases the basis used for preparing the
state coincides with the basis used by Bob to measure the states.
She informs Bobs to keep those cases and discard the rest. In all
these cases, in absence of Eve the Bob's measurement should reveal
the same state as was prepared by Alice. To check that they compare
half of these cases, to check the presence of Eve (an authorized party
who wishes obtain the key). In the absence of Eve they use rest of
the bits as key by considering $|0\rangle$ and $|+\rangle$ as 0
and $|1\rangle$ and $|-\rangle$ as 1.}

\textcolor{black}{Here, it's important to note that as Eve does not
know which qubit is prepared in which basis, she cannot perform a
measurement without disturbing the state of the qubit. So her efforts
to obtain information by performing any type of measurement leaves
detectable trace which is revealed in the comparison step performed
by Alice and Bob. This primarily happens because of Eve's inability
to perform simultaneous measurement in two nonorthogonal bases (i.e.,
in $\{|0\rangle,|1\rangle\}$ basis or $\{|+\rangle,|-\rangle\}$
basis). Further, nocloning theorem states that an unknown quantum
state cannot be copied perfectly. As a consequence, Eve cannot keep
a copy of the transmitted qubit and perform measurement using appropriate
basis at a later time when Bob discloses the basis used by him to
measure a particular qubit. Thus, we can clearly see that the security
arises from noncommutativity (our inability to perform simultaneous
measurement in two or more nonorthogonal bases) and noncloning principle,
and it does not depend on the complexity of a computational task.
This is why schemes of quantum cryptography in general and schemes
of QKD in particular are considered as unconditionally secure.}

\textcolor{black}{In the above, we have elaborated what kind of laws
of physics lead to the security of QKD in view of BB84 protocol. After
this pioneering work many other schemes of QKD have been proposed.
Some of them (e.g., \cite{bennett1992quantum}) use nonorthogonal
states and thus the same principle as described above.}

~\\
~\\
~

\textcolor{black}{However, there exist various orthogonal states based
schemes, too (e.g., \cite{goldenberg1995quantum}). Here, we would
not elaborate on the working of these schemes as they are not used
in this work.}

\subsection{\textcolor{black}{Secure direct quantum communication \label{subsec:Secure-direct-quantum}}}

\textcolor{black}{Most of the initial proposals on quantum cryptography
were limited to QKD \cite{bennett1984quantum,ekert1991quantum,bennett1992quantum,goldenberg1995quantum}.
Later on, it was observed that secure quantum communication is possible
without prior generation of key. In other words, it is possible to
design schemes of direct secure quantum communication circumventing
prior distributed key. In last two decades, many schemes for secure
direct quantum communication have been proposed. All the schemes of
secure direct quantum communication can be classified into two categories
\cite{long2007quantum}: (1) deterministic secure quantum communication
scheme (DSQC) \nocite{bostrom2002deterministic,long2007quantum,yuan2011high,hai2006quantum,zhu2006secure,hwang2011quantum,zhong2005deterministic,yan2004scheme},
(2) quantum secure direct communication scheme (QSDC) \cite{long2002theoretically,bostrom2002deterministic,lucamarini2005secure}.
In DSQC protocol, Alice needs to send some classical information (at
least 1 bit for each qubit) to Bob, who can't decode the information
encoded by Alice in the absence of this classical information. In
QSDC, no such classical information is required for decoding.}

\subsection{\textcolor{black}{Quantum dialogue (QD) \label{subsec:Quantum-dialogue-(QD):}}}

\textcolor{black}{In all protocols of QSDC and DSQC (\cite{bostrom2002deterministic,long2007quantum,hai2006quantum,zhu2006secure,hwang2011quantum,zhong2005deterministic,yan2004scheme,braunstein2000dense,lucamarini2005secure}
and references therein), a secret message is transmitted from Alice
(sender) to Bob (receiver). Thus, these schemes are one way (unidirectional)
schemes for communication. In other words, Alice and Bob cannot exchange
their messages to each other at the same time. Of course they can
use two independent QSDC or DSQC schemes to do that. However, that
would not be considered as simultaneous communication. Keeping this
in mind, a notion of two-way communication using Bell states was proposed
by Ba An \cite{nguyen2004quantum} in 2004. The two-way communication
scheme which allows both Alice and Bob to simultaneously send message
to each other in a secure manner is referred to as quantum dialogue
(QD). A very important feature of this scheme is that in this scheme
information encoded by Alice and Bob simultaneously exists in the
channel. Almost immediately after Ba An's work, in 2005, Man et al.,
\cite{man2005quantum} found that Ba An's original protocol for QD
is not secure against intercept-resend attack and he proposed a modified
scheme. However, eventually, it was found that there were some issues
with Man et al.'s scheme, too. Finally, in 2005, Ba An proposed an
improved version of his original QD protocol and addressed all the
issues raised until then \cite{an2005secure}. After this work of
Ba An, several protocols of QD have been reported \cite{tan2008classical,gao2010two,xia2006quantum,li2009quantum,shukla2013group}.
The reason behind the interest on QD is obvious, as ability to perform
QD ensure the ability to perform QSDC/DSQC and QKD. This point will
be further elaborated in the next section.}

\subsection{\textcolor{black}{Controlled quantum dialogue (CQD) \label{subsec:Controlled-quantum-dialogue}}}

\textcolor{black}{Controlled QD corresponds to a novel scheme for
bidirectional secure quantum communication which is essentially a
scheme for QD controlled by a controller or a supervisor. It is a
three party scheme involving sender (Alice), receiver (Bob) and supervisor
(Charlie). In this two way communication scheme, Alice and Bob can
execute a scheme for QD provided the controller Charlie allows them
to do so. This scheme is interesting for various reasons. Firstly,
ability to perform CQD ensures the ability to perform QD. In term
if one can perform two way direct communication he can also perform
one way direct communication (we may assume that Alice encodes a message,
but Bob always encodes zero). Now ability to communicate a message
in a secure manner implies the ability to distribute a key in a secure
manner. Thus, if we can do CQD, we can do QD, DSQC, QKD, CDSQC, etc.}

\section{\textcolor{black}{Effect of noise on the protocols of quantum communication
\label{sec:Effect-of-noise-1}}}

\textcolor{black}{We have already introduced various quantum communication
schemes. Now in this section, we will discuss the effect of noise
on these quantum communication schemes. Generally, dynamics of a system
can be defined as unitary evolution neglecting the effect of surrounding
on it, which is also called closed system description. In contrast,
considering the effect due to ambient environment, the system dynamics
cannot always be described by a unitary evolution studied as open
quantum system. In open quantum systems, a pure state can evolve into
a mixed state. The effect of environment on the system can be studied
by describing unitary dynamics of the composite system-environment
state. Let's consider initial state of the system and environment
as $\rho_{s}$ and $\rho_{R}$ and consider evolution of the state
of the composite system, i.e., $\rho=\rho_{s}\otimes\rho_{R}$, under
unitary $U$ as}

\textcolor{black}{
\begin{equation}
\begin{array}{lcl}
\rho(t) & = & U(\rho_{s}\otimes\rho_{R})U^{\dagger}.\end{array}\label{eq:evolve}
\end{equation}
}

\textcolor{black}{The system evolution can be obtained by tracing
over the reservoir (environment) state as $\rho_{s}(t)=Tr_{R}\left\{ \rho(t)\right\} $.
For the sake of simplicity, we can assume environment to be in the
pure state, say $|e_{i}\rangle$. Then Eq. (\ref{eq:evolve}) can
be rewritten as 
\begin{equation}
\epsilon(\rho_{s})=Tr_{R}\left\{ U(\rho_{s}\otimes|e_{i}\rangle\langle e_{i}|)U^{\dagger}\right\} .\label{eq:2evolve}
\end{equation}
Here, $\epsilon$ is a completely positive and trace preserving dynamical
map, completely positive and trace preserving conditions are imposed
by the density matrix definition employing positivity and unit trace
condition. Here, we will briefly discuss a representation in which
the effect of environment on the system can be encapsulated nicely
as completely positive and trace preserving dynamical map. The representation
is known as operator sum representation or Kraus operator representation.
The system state from the above equation can be derived by tracing
out the environment in basis $|e_{k}\rangle$ thus Eq. (\ref{eq:2evolve})
becomes}

\textcolor{black}{
\begin{equation}
\epsilon(\rho_{s})=Tr_{R}\left\{ U(\rho_{s}\otimes|e_{i}\rangle\langle e_{i}|)U^{\dagger}\right\} =\sum_{k}\langle e_{k}|U(\rho_{s}\otimes|e_{i}\rangle\langle e_{i}|)U^{\dagger}|e_{k}\rangle=\sum_{k}E_{k}\rho_{S}E_{k}^{\dagger}.\label{eq:3-2}
\end{equation}
}

\textcolor{black}{The equation in the last row elucidate mathematical
statement of operator sum representation and ${E_{k}}$ are known
as Kraus operator which satisfy a completeness relation $\underset{i}{\sum}E_{k}E_{k}^{\dagger}=I.$}

\textcolor{black}{There are various types of noise which have been
well established in the past \cite{banerjeeopen,mcmahon2007quantum,nielsen2000quantum}.}

\textcolor{black}{In this thesis, we have mainly studied two types
of Markovian noise, amplitude damping (AD) noise and phase damping
(PD) noise. These two types of noise are briefly described below.}

\subsection{\textcolor{black}{Amplitude damping noise \label{subsec:Amplitude-damping-noise}}}

A large number of investigations considering different noise models
have been performed recently. The amplitude damping (AD) noise model
is one the most important noise models that have been studied recently.
This noise model has been rigorously studied  in the recent past (\cite{sharma2016verification,huang2007necessary,thapliyal2015quasiprobability,kim2012protecting,turchette2000decoherence,myatt2000decoherence,marques2015experimental,banerjeeopen,chuang1998bulk}
and references therein) because of the fact that it can mimic (simulate)
the dissipative interaction between a quantum system and a vacuum
bath. 

\textcolor{black}{The Kraus operators for AD noise are \cite{nielsen2000quantum}}

\textcolor{black}{
\begin{equation}
E_{0}=|0\rangle\langle0|+\sqrt{1-\eta}|1\rangle\langle1|,\,\,\,\,\,\,\,\,\,\,\,E_{1}=\sqrt{\eta}|0\rangle\langle1|.\label{eq:Kraus-damping-1}
\end{equation}
}

\textcolor{black}{where $\eta\,\left(0\leq\eta\leq1\right)$ is the
decoherence rate for the AD channel, which determines the effect of
the noisy channel on the quantum system.}

\subsection{\textcolor{black}{Phase damping noise \label{subsec:Phase-damping-noise}}}

\textcolor{black}{The phase damping (PD) noise involves information
loss about relative phases in a quantum state. The PD noise model
is characterized by the following Kraus operators.
\begin{equation}
\begin{array}{c}
E_{0}=|0\rangle\langle0|+\sqrt{1-\eta}|1\rangle\langle1|,\,\,\,\,\,\,\,\,\,\,\,E_{1}=\sqrt{\eta}|1\rangle\langle1|.\end{array}\label{eq:Kraus-dephasing-1}
\end{equation}
For PD noise, $\eta\,\left(0\leq\eta\leq1\right)$ is the decoherence
rate, which describes the probability of error due to PD channel.}

\subsection{\textcolor{black}{Fidelity as a quantitative measure of the effect
of noise \label{subsec:Fidelity-as-a}}}

\textcolor{black}{The concept of fidelity is a basic ingredient in
quantum communication scheme. Let us consider $\rho^{1}=|\psi_{1}\rangle\langle\psi_{1}|$
and $\rho^{2}=|\psi_{2}\rangle\langle\psi_{2}|$ are the two quantum
states on a finite dimensional Hilbert space, and we want to know
the closeness of these quantum states. So, one measure of distinguishability
or closeness between two quantum states is known as the fidelity.
In other words, fidelity is used to describe the closeness between
two quantum states ($\rho^{1}$ and $\rho^{2}$).}

\textcolor{black}{Given two states $\rho^{1}$ and $\rho^{2}$, generally
the fidelity is defined as the quantity}\footnote{\textcolor{black}{In some literature for example Ref \cite{jozsa1994fidelity},
the formula of fidelity $\left(Tr[\sqrt{\sqrt{\rho^{1}}\rho^{2}\sqrt{\rho^{1}}}]\right)^{2}$
has been used. Both definitions yield the same physical meaning.}}

\textcolor{black}{
\begin{equation}
F\left(\rho^{1},\rho^{2}\right)=Tr[\sqrt{\sqrt{\rho^{1}}\rho^{2}\sqrt{\rho^{1}}}].\label{eq:fidelity-1}
\end{equation}
}

\textcolor{black}{There are some special cases where we can acquire
more useful and easily understood expressions of fidelity. For simplicity,
here we consider two pure states. In this particular case, we may
consider $\rho^{1}=|\psi_{1}\rangle\langle\psi_{1}|$ and $\rho^{2}=|\psi_{2}\rangle\langle\psi_{2}|$
are both pure states for which $\left(\rho^{i}\right)^{2}=\rho^{i}.$
Consequently, $F\left(|\psi_{1}\rangle,\rho^{2}\right)=Tr[\sqrt{\sqrt{\rho^{1}}\rho^{2}\sqrt{\rho^{1}}}]=Tr[\sqrt{|\psi_{1}\rangle\langle\psi_{1}|\rho^{2}|\psi_{1}\rangle\langle\psi_{1}|}]=\sqrt{\langle\psi_{1}|\rho^{2}|\psi_{1}\rangle}.$
In another case,}

\textcolor{black}{$F\left(|\psi_{1}\rangle,|\psi_{2}\rangle\right)=Tr[\sqrt{\sqrt{\rho^{1}}\rho^{2}\sqrt{\rho^{1}}}]=Tr[\sqrt{|\psi_{1}\rangle\langle\psi_{1}|\psi_{2}\rangle\langle\psi_{2}|\psi_{1}\rangle\langle\psi_{1}|}]=|\langle\psi_{2}|\psi_{1}\rangle|.$}

\textcolor{black}{There are a few general properties of fidelity.}

\textbf{\textcolor{black}{(i) }}\textcolor{black}{The first is that
fidelity ranges between $0\leq F\left(\rho^{1},\rho^{2}\right)\leq1,$
with unity iff $\rho^{1}=\rho^{2}$ and $0$ if there is no overlap
whatsoever.}

\textbf{\textcolor{black}{(ii)}}\textcolor{black}{{} $F\left(\rho^{1},\rho^{2}\right)=F\left(\rho^{2},\rho^{1}\right)$.}

\textbf{\textcolor{black}{(iii)}}\textcolor{black}{{} Fidelity is invariant
under unitary transformations $F\left(U\rho^{1}U^{\dagger},U\rho^{2}U^{\dagger}\right)=F\left(\rho^{2},\rho^{1}\right)$.}

\textcolor{black}{We can also see fidelity as a quantitative measure
of the effect of noise. For example, a pure quantum state is $|\psi\rangle$
(initial state) so the density matrix of this initial state is $\rho=|\psi\rangle\langle\psi|.$
Now, if the pure state interacts with the environment which converts
it into mixed state $\rho'=\sum p_{i}|\psi_{i}\rangle\langle\psi_{i}|$
($p_{i}$ is the probability with $\sum p_{i}=1$). To quantify how
much the state is affected due to noise, we calculate fidelity between
the initial state and noise affected state by using above formula
\ref{eq:fidelity-1}.}

\section{\textcolor{black}{Quantum gates and quantum circuits \label{sec:Brief-discussion-of}}}

\textcolor{black}{Except NOT and Identity gates, all the other conventional
classical gates, such as AND, OR, NOR, NAND are irreversible $\left(\rightarrow\right)$,
in the sense that we cannot uniquely obtain the input state from the
output state. During operation of each of the above mentioned irreversible
gates, one bit of information is erased as these gates always map
a 2-bit input state into a 1-bit output state. It was Landauer's pioneering
work \cite{landauer1961irreversibility}, which led to the observation
that erasure of a bit involves a loss of energy amounting at least
$kT\log2$, and such an energy loss due to logical operation can be
circumvented by using reversible logic gates. It's technically possible
to make classical reversible gates. However, they are not of much
interest as they cannot be used to perform any task that cannot be
performed by the irreversible circuits. In contrast, all the quantum
gates are reversible $\left(\leftrightarrow\right)$ in nature, implying
that we can uniquely reconstruct the input states from the output
states (bijective mapping). Here, it may be noted that all the quantum
gates are reversible, but all the reversible gates are not referred
to as the quantum gates. Usually, quantum gates are called just quantum
gates and classical reversible gates are referred to as reversible
gates. In what follows, we will also follow the same convention. Basically,
a quantum gate or a quantum logic gate corresponds to a unitary operation
that maps a quantum state into another quantum state in a unique manner.
The map is always bijective as expected. In matrix notation, an $N$
qubit quantum gate is represented by $2^{N}\times2^{N}$ matrix. In
what follows, we will discuss important single-qubit and two-qubit
gates and their combinations, which lead to quantum circuit. Some
quantum gates, which are not used in this thesis will also be summarized.
To begin with, in the next section, we will introduce some single-qubit
gates.}

\subsection{\textcolor{black}{Single-qubit quantum gates \label{subsec:Single-qubit}}}

\begin{figure}
\centering{}\textcolor{black}{\includegraphics[scale=1.5]{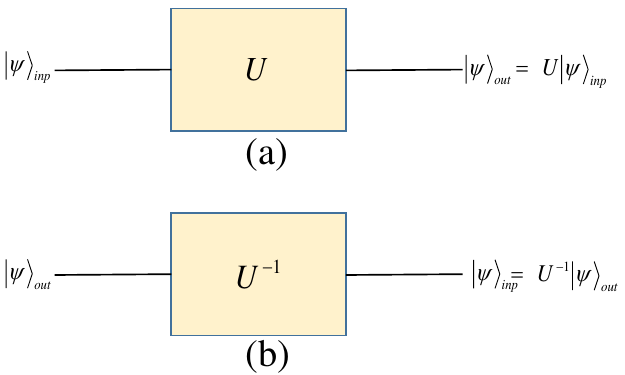}\caption{\label{fig:General-circuit-representation}General circuit representation
of an arbitrary single-qubit gate $U$ which is essentially reversible
(i.e., existence of $U$ implies the existence of $U^{-1}$). }
}
\end{figure}

\textcolor{black}{The vector representation of a single-qubit is $\left(\begin{array}{c}
\alpha\\
\beta
\end{array}\right)$ and a single-qubit quantum gate is represented by a $2\times2$ matrix.
Single-qubit gate is a unitary operator $U$, which transforms a single-qubit
state $|\psi\rangle_{inp}$ to another single-qubit state $|\psi\rangle_{out}=U|\psi\rangle_{inp}.$
Further, the existence of $U$ ensures the existence of $U^{-1}.$
Thus the feasibility of the inverse operation $U^{-1}|\psi\rangle_{out}=|\psi\rangle_{inp}$(see
Figure \ref{fig:General-circuit-representation}). In this section,
single-qubit gates are introduced.}

\textcolor{black}{}
\begin{description}
\item [{\textcolor{black}{1.}}] \textbf{\textcolor{black}{Pauli-X gate
or NOT gate:}}\textcolor{black}{{} The matrix representation of unitary
operator $X$ gate is}
\end{description}
\textcolor{black}{
\[
\begin{array}{ccc}
X & = & \left(\begin{array}{cc}
0 & 1\\
1 & 0
\end{array}\right).\end{array}
\]
}

\textcolor{black}{Operation of $X$ gate can be described by the following
input-output maps.}

\textcolor{black}{
\[
\begin{array}{cccccc}
X|0\rangle & = & |1\rangle,\, & X|1\rangle & = & |0\rangle,\end{array}
\]
}

\textcolor{black}{
\[
X(\alpha|0\rangle+\beta|1\rangle)=(\alpha|1\rangle+\beta|0\rangle).
\]
}
\begin{description}
\item [{\textcolor{black}{2.}}] \textbf{\textcolor{black}{Pauli-Z gate:
}}\textcolor{black}{$Z$ gate is represented in matrix form as follows}
\end{description}
\textcolor{black}{
\[
\begin{array}{ccc}
Z & = & \left(\begin{array}{cc}
0 & 1\\
-1 & 0
\end{array}\right).\end{array}
\]
}

\textcolor{black}{It is easy to see that $Z$ gate transforms the
single-qubit state as}

\textcolor{black}{
\[
\begin{array}{cccccc}
Z|0\rangle & = & |0\rangle,\, & Z|1\rangle & = & -|1\rangle,\end{array}
\]
}

\textcolor{black}{
\[
Z(\alpha|0\rangle+\beta|1\rangle)=(\alpha|0\rangle-\beta|1\rangle).
\]
}

\textcolor{black}{Thus, $Z$ gate just flips the phase. So, it is
also known as phase flip operator as well as $\sigma_{z}$.}
\begin{description}
\item [{\textcolor{black}{3.}}] \textbf{\textcolor{black}{Pauli-Y gate:
}}\textcolor{black}{Similarly, the matrix representation of $Y$ gate
or $\sigma_{y}$ is}
\end{description}
\textcolor{black}{
\[
\begin{array}{ccc}
Y & = & \left(\begin{array}{cc}
0 & -i\\
i & 0
\end{array}\right).\end{array}
\]
}

\textcolor{black}{It will transform the single-qubit states as}

\textcolor{black}{
\[
\begin{array}{cccccc}
Y|0\rangle & = & |1\rangle,\, & Y|1\rangle & = & |0\rangle,\end{array}
\]
}

\textcolor{black}{
\[
Y(\alpha|0\rangle+\beta|1\rangle)=(\alpha|1\rangle-\beta|0\rangle).
\]
}

\textcolor{black}{$Y$ gate is the combination of $X$ gate and $Z$
gate. It is a bit flip and phase flip operator.}
\begin{description}
\item [{\textcolor{black}{4.}}] \textbf{\textcolor{black}{Hadamard gate:}}\textcolor{black}{{}
The other single-qubit gate is Hadamard gate $\left(H\right)$, which
is represented by a matrix}
\end{description}
\textcolor{black}{
\[
\begin{array}{ccc}
H & = & \frac{1}{\sqrt{2}}\left(\begin{array}{cc}
1 & 1\\
1 & -1
\end{array}\right).\end{array}
\]
}

\textcolor{black}{The Hadamard transformations is defined by}

\textcolor{black}{
\[
\begin{array}{ccc}
H|0\rangle & = & \frac{1}{\sqrt{2}}\left(|0\rangle+|1\rangle\right),\end{array}
\]
}

\textcolor{black}{
\[
H|1\rangle=\frac{1}{\sqrt{2}}\left(|0\rangle-|1\rangle\right).
\]
}
\begin{description}
\item [{\textcolor{black}{5.}}] \textbf{\textcolor{black}{Phase gate: }}\textcolor{black}{We
can represent the phase gate by a matrix form is}
\end{description}
\textcolor{black}{
\[
\begin{array}{ccc}
P & = & \left(\begin{array}{cc}
1 & 0\\
0 & e^{i\theta}
\end{array}\right).\end{array}
\]
Since $\theta$ can have infinitely many values, but here we are taking
two values of $\theta.$ For $\theta=\frac{\pi}{2},$ we obtain $S$
gate described as}

\textcolor{black}{
\[
\begin{array}{ccc}
S=P\left(\frac{\pi}{2}\right) & = & \left(\begin{array}{cc}
1 & 0\\
0 & i
\end{array}\right).\end{array}
\]
}

\textcolor{black}{It is usually referred to as $S$ gate, and it maps}

\textcolor{black}{
\[
S|0\rangle=|0\rangle,\,S|1\rangle=i|1\rangle.
\]
}

\textcolor{black}{If $\theta=\frac{\pi}{4},$ then we obtain a $T$
gate which is described in matrix form as}

\textcolor{black}{
\[
\begin{array}{ccc}
T=P\left(\frac{\pi}{4}\right) & = & \left(\begin{array}{cc}
1 & 0\\
0 & \frac{1}{\sqrt{2}}\left(1+i\right)
\end{array}\right).\end{array}
\]
}

\textcolor{black}{It may be noted that $P$ gate is not self-inverse
in general.}

\subsection{\textcolor{black}{Two-qubit quantum gates \label{subsec:Two-qubit}}}

\textcolor{black}{A two-qubit state is represented by the column matrix
$\left(\begin{array}{c}
\alpha\\
\beta\\
\gamma\\
\delta
\end{array}\right)$ and a two-qubit quantum gate is represented by a $2^{2}\times2^{2}=4\times4$
matrix. Now, when a two-qubit quantum gate acts on a two-qubit state
it maps the input state into another two-qubit output state in a well-defined
manner which characterizes that particular gate.}
\begin{description}
\item [{\textcolor{black}{1}}] \textbf{\textcolor{black}{Controlled-NOT
gate-}}\textcolor{black}{{} Controlled NOT gate, i.e., CNOT gate, which
is represented by in a matrix form as,}
\end{description}
\textcolor{black}{
\[
\begin{array}{ccc}
{\rm CNOT} & = & \left(\begin{array}{cccc}
1 & 0 & 0 & 0\\
0 & 1 & 0 & 0\\
0 & 0 & 0 & 1\\
0 & 0 & 1 & 0
\end{array}\right),\end{array}
\]
 and the bra-ket notation of CNOT gate is}

\begin{figure}
\begin{centering}
\textcolor{black}{\includegraphics[scale=0.5]{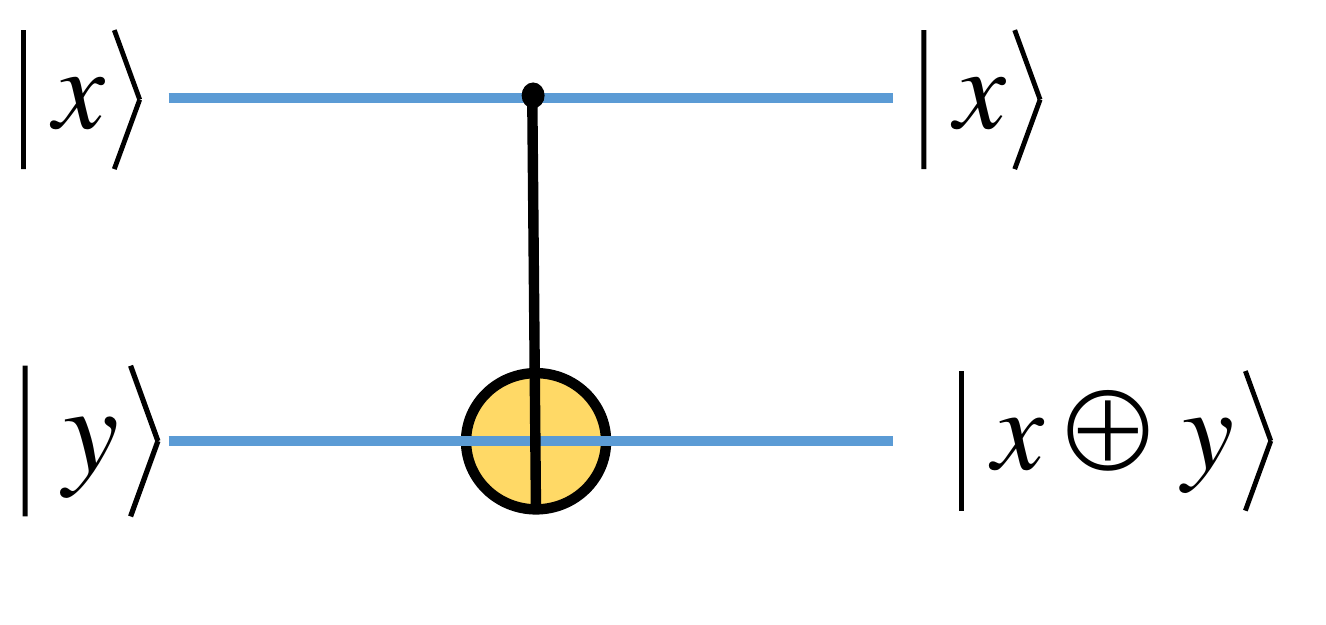}}
\par\end{centering}
\textcolor{black}{\caption{\label{fig:CNOT-gate-circuit}Circuit representation of a CNOT gate.}
}
\end{figure}

\textcolor{black}{
\[
\begin{array}{ccc}
{\rm CNOT} & = & |00\rangle\langle00|+|01\rangle\langle01|+|11\rangle\langle10|+|10\rangle\langle11|.\end{array}
\]
This gate is a two-qubit gate, in which first qubit is control and
second qubit is target as shown in Figure \ref{fig:CNOT-gate-circuit}.
According to the figure, if the first qubit as control qubit is $|0\rangle$
then target will not be flip if control qubit is $|1\rangle$ then
target will be flip. So, the CNOT mapping is}

\textcolor{black}{
\[
\begin{array}{ccc}
|00\rangle & \rightarrow & |00\rangle\\
|01\rangle & \rightarrow & |01\rangle\\
|10\rangle & \rightarrow & |11\rangle\\
|11\rangle & \rightarrow & |10\rangle
\end{array}.
\]
}

\textcolor{black}{}
\begin{description}
\item [{\textcolor{black}{2}}] \textbf{\textcolor{black}{SWAP gate:}}\textcolor{black}{{}
Another two-qubit gate is SWAP gate. The matrix of SWAP gate is}
\end{description}
\textcolor{black}{
\[
\begin{array}{ccc}
{\rm SWAP} & = & \left(\begin{array}{cccc}
1 & 0 & 0 & 0\\
0 & 0 & 1 & 0\\
0 & 1 & 0 & 0\\
0 & 0 & 0 & 1
\end{array}\right),\end{array}
\]
}

\textcolor{black}{and the same can be expressed in bra-ket notation
as}

\textcolor{black}{
\[
\begin{array}{ccc}
{\rm SWAP} & = & |00\rangle\langle00|+|10\rangle\langle01|+|01\rangle\langle10|+|11\rangle\langle11|.\end{array}
\]
}

\textcolor{black}{A SWAP gate maps a two-qubit state $|ab\rangle$
to $|ba\rangle.$ Specifically, it maps}

\textcolor{black}{
\[
\begin{array}{ccc}
|00\rangle & \rightarrow & |00\rangle\\
|01\rangle & \rightarrow & |10\rangle\\
|10\rangle & \rightarrow & |01\rangle\\
|11\rangle & \rightarrow & |11\rangle
\end{array}.
\]
}

\begin{figure}
\centering{}\textcolor{black}{\includegraphics[scale=0.52]{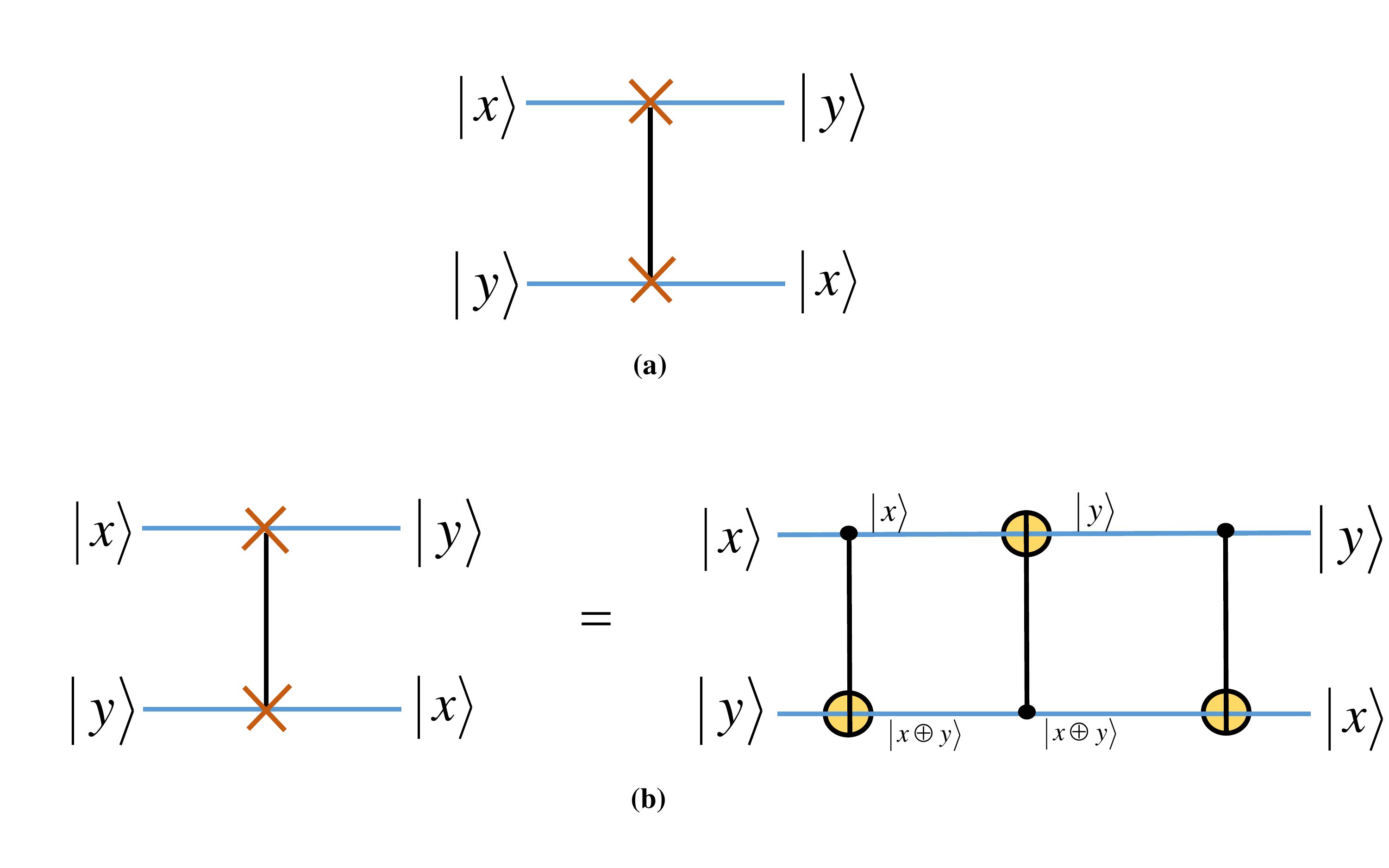}\caption{\textcolor{black}{\label{fig:Swap-gate}(a) Symbolic representation
of a SWAP gate (b) how to realize a SWAP gate using three CNOT gates.}}
}
\end{figure}

\textcolor{black}{A SWAP gate can be constructed by using three CNOT
gates as shown in Figure \ref{fig:Swap-gate}.}

\textcolor{black}{}

\subsection{\textcolor{black}{Other quantum gates\label{subsec:Other-quantum-gates:}}}

\textcolor{black}{Here, we would like to briefly mention about a few
other quantum gates which are often used. One such single qubit gate
is a square-root of NOT gate $\left(\sqrt{\text{NOT}}\right)$ which
can be represented by the following matrix $\frac{1}{2}\left(\begin{array}{cc}
1+i & 1-i\\
1-i & 1+i
\end{array}\right).$ It's action and the origin of the name can be visualized easily if
we note that $\sqrt{\text{NOT}}\sqrt{\text{NOT}}|0\rangle=|1\rangle$
and $\sqrt{\text{NOT}}\sqrt{\text{NOT}}|1\rangle=|0\rangle$. A controlled
version of this gate is also often used as a two-qubit gate. In fact,
in the domain of classical reversible circuits, this gate plays a
crucial role. Finally, there are 2 three-qubit gates which (or their
decomposition into single-qubit and two-qubit gates) are often used.
These gates are referred to as Toffoli and Fredkin gates and they
may be viewed as CCNOT and CSWAP gates, respectively. In the Toffoli
gate (first two qubits are control and the third qubit is target),
if first two qubits are found to be in state $|1\rangle$ then only
the target (third qubit) flips, nothing happens otherwise. Similarly,
in a Fredkin gate, if first qubit is found to be at state $|1\rangle$
then a SWAP operation is performed between second and third qubits,
and nothing happens otherwise. }

\subsection{\textcolor{black}{EPR circuit \label{subsec:EPR-circuit:}}}

\begin{figure}
\begin{centering}
\textcolor{black}{\includegraphics[scale=0.4]{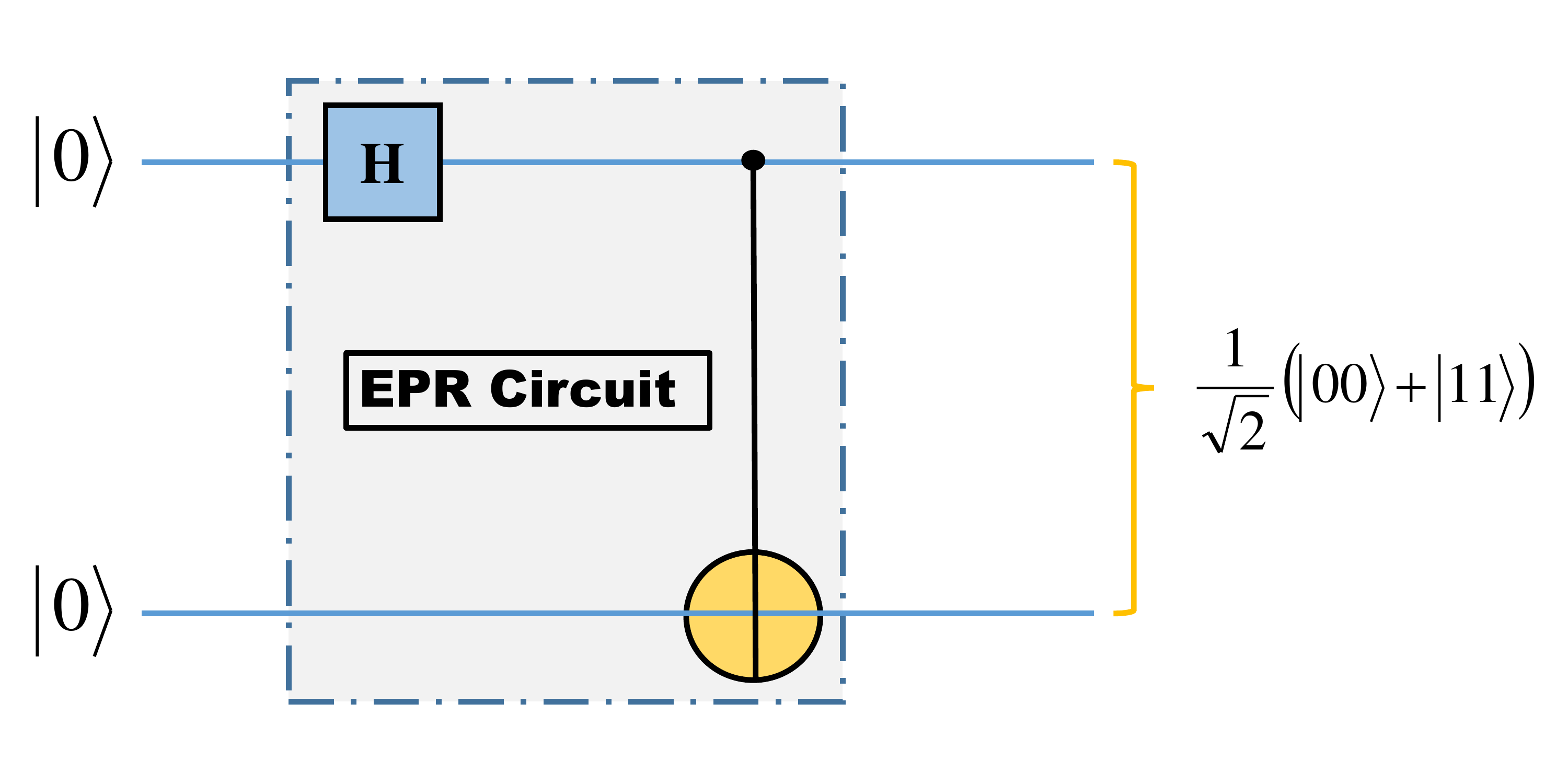}}
\par\end{centering}
\textcolor{black}{\caption{\label{fig:EPR circuit}Diagram for a quantum circuit. This particular
circuit is known as EPR circuit. It transforms a separable input states
to a Bell state as shown. Different inputs (e.g., $|01\rangle,|10\rangle,|11\rangle$)
will generate different Bell states.}
}
\end{figure}

\textcolor{black}{EPR circuit is the combination of single-qubit and
two-qubit quantum gate, i.e., Hadamard gate followed by CNOT gate,
which generate the maximally entangled state (see the circuit shown
in Figure \ref{fig:EPR circuit}). The mathematical operation of this
circuit is ${\rm CNOT\left(H\otimes I\right)}|00\rangle.$We start
with two separable states $|00\rangle.$ Now, first of all ${\rm H}$
applied on the first qubit and $I$ on second qubit then the state
maps into $\begin{array}{c}
\frac{1}{\sqrt{2}}\left(|00\rangle+|10\rangle\right).\end{array}$Subsequently, two-qubit gate CNOT operates on it, where the first
qubit works as control qubit and the second qubit works as target
qubit. According to the CNOT gate, second qubit will be flip when
first qubit is $|1\rangle.$ Now, the state is, ${\rm CNOT}\left(\frac{1}{\sqrt{2}}\left(|00\rangle+|10\rangle\right)\right)=\frac{1}{\sqrt{2}}\left(|00\rangle+|11\rangle\right),$which
is a maximally entangled state.}

\textcolor{black}{}

\subsection{\textcolor{black}{Useful optical components and how to realize quantum
gates optically? \label{subsec:Useful-optical-components}}}

\begin{figure}
\centering{}\textcolor{black}{\includegraphics[scale=0.55]{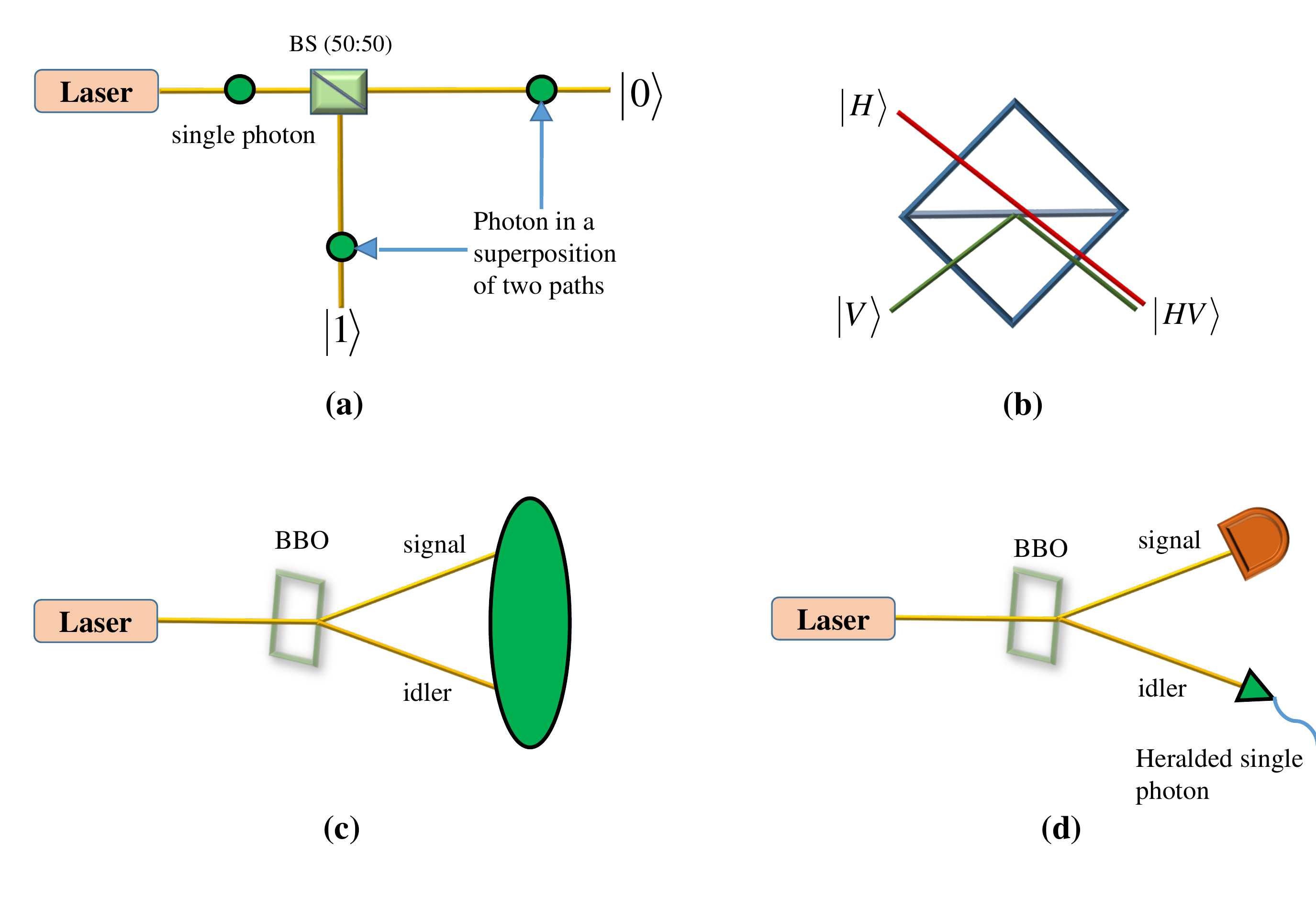}\caption{\label{fig:optical-elements} Schematic diagrams of optical elements.
(a) optically realization of qubit. (b) PBS, which transmits horizontally
polarized photon and reflects vertically polarized photon. (c) entanglement
generation through the SPDC process (d) A heralded single photon from
the SPDC process.}
}
\end{figure}

\textcolor{black}{In Chapter \ref{chap:OPTICAL-DESIGNS-FOR}, we wish
to present some optical setups, which can be used to realize various
schemes for secure quantum communication. As a background to that
here we plan to discuss how to realize a qubit optically and how to
use basic optical elements as quantum gates. In Figure \ref{fig:optical-elements}
(a), we use an attenuated laser (approximate single photon source)
to generated the single photons which incident on the symmetric beam
splitter. As we know the basic concept of symmetric (50:50) BS that
transmits half of the incident light and reflects rest half. Thus,
when a single photon is incident on a (50:50) BS then with half probability
the photon would be found on the transmitted path say $|0\rangle$
and with half probability it would be found on the reflected path
say $|1\rangle.$ This happens after measurement. Prior to that the
photon simultaneously exists in both paths as shown in Figure \ref{fig:optical-elements}
(a). Then a single photon emerges in a superposition state of the
photon $\alpha|0\rangle+\beta|1\rangle,$ for a (50:50) BS we obtain
$|\alpha|^{2}=|\beta|^{2}=\frac{1}{2}$, where $|\alpha|^{2}$ and
$|\beta|^{2}$ are the probabilities of getting the photon on transmitted
and reflected path. }

\textcolor{black}{In the above, a photonic qubit is introduced using
their position (path). A photonic qubit can also be defined using
other degrees of freedom of the photon (say orbital angular momentum
or polarization state). In this thesis, we will primarily deal with
the polarization based photonic qubits, where a single photon having
arbitrary linear polarization can be viewed to be in the state $|\psi\rangle=\alpha|H\rangle+\beta|V\rangle$.
If this photon is made to incident on a PBS (see Figure \ref{fig:optical-elements}
(b)), it will reflect with the probability $|\beta|^{2}$, and transmit
with the probability $|\alpha|^{2}.$ Thus, in other words, if this
qubit is measured in $\{|H\rangle,|V\rangle\}$ by applying two single
photon detectors at the two output ports of the PBS, then the qubit
will be found in the state $|H\rangle$ with probability $|\alpha|^{2}$
and in the state $|V\rangle$ with probability $|\beta|^{2}$. }

\textcolor{black}{Similarly, quantum gates can be realized by using
optical elements. There are two types of optical elements linear and
non-linear. Beam splitter (BS), half wave plate (HWP), etc., are linear
optical elements and down converters are non-linear optical elements.
Each optical element is represented by a square matrix, i.e., Jones
matrix. The Jones matrix of HWP with fast axis at an angle $\theta$
with respect to horizontal is $\left(\begin{array}{cc}
{\rm cos}2\theta & {\rm sin}2\theta\\
{\rm sin}2\theta & -{\rm cos}2\theta
\end{array}\right)$, if $2\theta=90^{0}$ then the matrix of HWP will be $\left(\begin{array}{cc}
0 & 1\\
1 & 0
\end{array}\right)$ and it would convert the horizontally polarized photon $\left(|H\rangle\right)$
into vertically polarized photon $\left(|V\rangle\right)$ and vice
versa. Consequently, a ${\rm HWP}_{\left(2\theta=90^{0}\right)}$
would work as a quantum NOT gate for the qubits based on polarization
states of a single photon. If $2\theta=45{}^{0}$, then the matrix
of HWP would become $\frac{1}{\sqrt{2}}\left(\begin{array}{cc}
1 & 1\\
1 & -1
\end{array}\right)$ and it would transforms the basis states as ${\rm HWP}_{\left(2\theta=45^{0}\right)}|H\rangle=\frac{1}{\sqrt{2}}\left(|H\rangle+|V\rangle\right)$
and ${\rm HWP_{\left(2\theta=45^{0}\right)}}|V\rangle=\frac{1}{\sqrt{2}}\left(|H\rangle-|V\rangle\right).$
Thus, in this case HWP would work as a Hadamard gate. Another optical
element of particular use is polarizing BS (PBS) which always transmit
the horizontally polarized light and reflect the vertically polarized
light as shown in Figure \ref{fig:optical-elements} (b). In Chapter
\ref{chap:OPTICAL-DESIGNS-FOR}, we will show that these optical components
play a crucial role in the experimental realization of the schemes
of secure quantum communication.}

\textcolor{black}{Further, we know that entangled states play a very
important role in quantum communication schemes. To optically generate
entangled states we need an interaction between photons which is possible
only by using nonlinear optical components. In this thesis, we have
proposed to use a nonlinear optical process called spontaneous parametric
down conversion (SPDC) to generate the entangled photons. SPDC process
is a second order nonlinear optical process in which a high frequency
photon (pump) gets converted into two correlated photons (signal and
idler)}\footnote{\textcolor{black}{The photons are correlated in momentum, frequency
and polarization.}}\textcolor{black}{{} of lower frequencies simultaneously in accordance
with the laws of conservation of momentum and energy}\footnote{\textcolor{black}{The combined energy and momentum of the generated
photons is equal to the energy and momentum of the incident photon.}}\textcolor{black}{{} after passing through the nonlinear crystal, like
barium borate  as shown in Figure \ref{fig:optical-elements} (c).
Now, if we put a detector on the signal photon's path and got it to
click, that means photon will exist on the idler's path. Thus, the
detection of one photon heralds the production of the other. This
is how through the SPDC process and heralding a single photon source
can be created for the use in designing experimentally feasible schemes
of quantum cryptography. Such schemes will be discussed in Chapter
\ref{chap:OPTICAL-DESIGNS-FOR}.}

\textcolor{black}{}

\section{\textcolor{black}{SQUID-based quantum computer \label{sec:SQUID-based}}}

\textcolor{black}{Quantum information processing can be done using
various experimental platforms, such as experimental architecture
based on NMR, ion-trap, silicon, Nitrogen-vacancy center. These experimental
facilities are not easily available to all the researchers. Only a
few researchers had adequate access to such quantum computing facilities.
Surprisingly, in 2016, the scenario has been changed considerably
after the introduction of a set of SQUID-based quantum computers by
IBM as these computers were placed on cloud and free access to these
computers were given to every researcher and students. This experimental
platform has attracted the attention of the entire quantum information
community because everyone can access it freely and easily through
IBM Quantum Experience. More specifically, IBM has designed several,
five-qubit, sixteen-qubit and twenty-qubit quantum computers. Interestingly,
several quantum communication tasks \cite{alsina2016experimental,fedortchenko2016quantum,devitt2016performing,behera2017experimental,hebenstreit2017compressed,linke2017experimental,majumder2017experimental,wootton2017demonstrating,ghosh2018automated,singh2018demonstration,hegade2017experimental,behera2018designing,dash2018exact,behera2018simulational,kapil2018quantum,malik2019first,pathak2018experimental}
have already been verified and tested by using these SQUID-based quantum
computers}\footnote{\textcolor{black}{Interested readers may find a detailed user guide
on how to use this superconducting-based quantum computer at \cite{IBMQE},
and a lucid explanation of the working principle of a IBM quantum
computer in Ref. \cite{steffen2011quantum}}}\textcolor{black}{. The first IBM quantum computer was made-up of
five superconducting transmon qubits. There are two processors of
five-qubit quantum computer (i.e., IBM QX2 see Figure \ref{fig:IBMQX2}
and IBMQX4 see Figure \ref{fig:IBM QX4}), one quantum computer (IBM
QX5) of sixteen-qubit see Figure \ref{fig:16-qubit} and one processor
(QS1\_1) of twenty-qubit IBM computer. In all the IBM processors,
gates from the Clifford+T gate library can be implemented directly.
Single-qubit gates can be applied anywhere on the qubit lines. However,
application of two-qubit gate (CNOT) is given by the architecture
as shown in Figures \ref{fig:IBMQX2}, \ref{fig:IBM QX4} and \ref{fig:16-qubit}.
In this thesis work, we have extensively used IBM QX2 processor (old
version) of five-qubit SQUID-based quantum computer to realize quantum
communication schemes experimentally.}

\textcolor{black}{Keeping the above in mind in this section, we will
discuss the technical aspects of the old version of the IBM QX2 processor
of five-qubit IBM quantum computer with focus on the architecture
of the computer, nature of the qubits realized in it and their manipulation,
and readout. In the subsections, we will briefly discuss IBM QX4 and
IBM QX5, too.}

\subsection{\textcolor{black}{IBM QX2 \label{subsec:IBM-QX2}}}

\begin{figure}
\begin{centering}
\textcolor{black}{\includegraphics{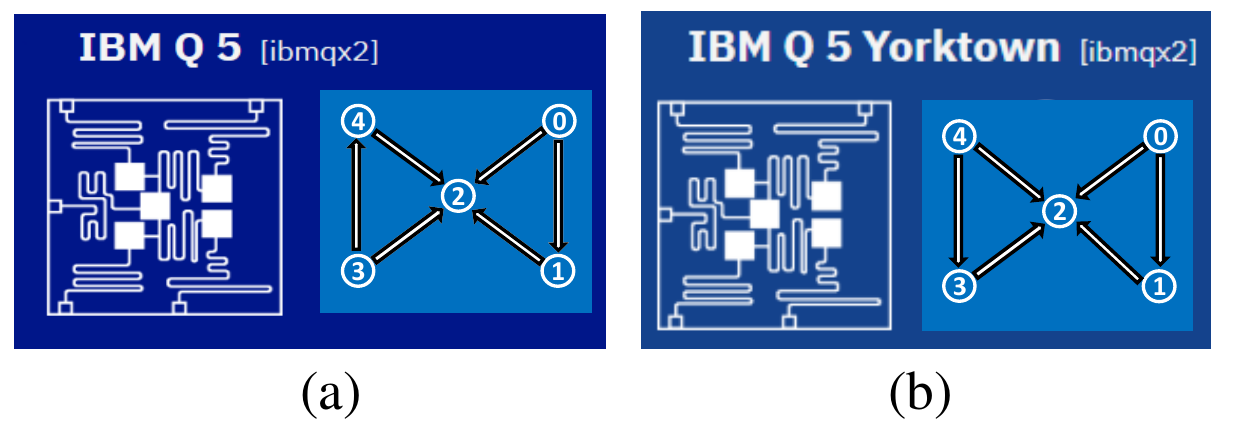}}
\par\end{centering}
\textcolor{black}{\caption{\label{fig:IBMQX2}Figures (a) and (b) are the old and new versions
of IBM QX2 processor.}
}
\end{figure}

\textcolor{black}{IBM QX2 is one of the processors of five-qubit quantum
computer placed in cloud by IBM. This processor has two versions.
In Figures \ref{fig:IBMQX2} (a) and (b) the topology and schematic
diagram of the processor is shown for the old version and new version
of IBM QX2, respectively. These two versions are different in the
sense that their topology is different which implies that the CNOT
gates that can be implemented directly in a particular version is
different from the other version.}

\textcolor{black}{Specifically, the allowed CNOT operations for the
old version are q{[}0{]} $\rightarrow$ {} q{[}1{]}, q{[}0{]} $\rightarrow$  q{[}2{]},
q{[}4{]}$\rightarrow$ {} q{[}2{]}, q{[}3{]} $\rightarrow$  q{[}2{]},
q{[}3{]} $\rightarrow$  q{[}4{]}, q{[}1{]} $\rightarrow$ {} q{[}2{]}
and for new version are q{[}0{]} $\rightarrow$ {} q{[}1{]}, q{[}0{]}
$\rightarrow$  q{[}2{]}, q{[}4{]}$\rightarrow$ {} q{[}2{]}, q{[}4{]}$\rightarrow$  q{[}3{]},
q{[}3{]} $\rightarrow$  q{[}2{]}, q{[}1{]} $\rightarrow$ {} q{[}2{]}.
Here q{[}i{]} $\rightarrow$ {} q{[}j{]} means q{[}i{]} is the control
bit and q{[}j{]} is the target bit.}

\begin{figure}
\begin{centering}
\textcolor{black}{\includegraphics[scale=0.52]{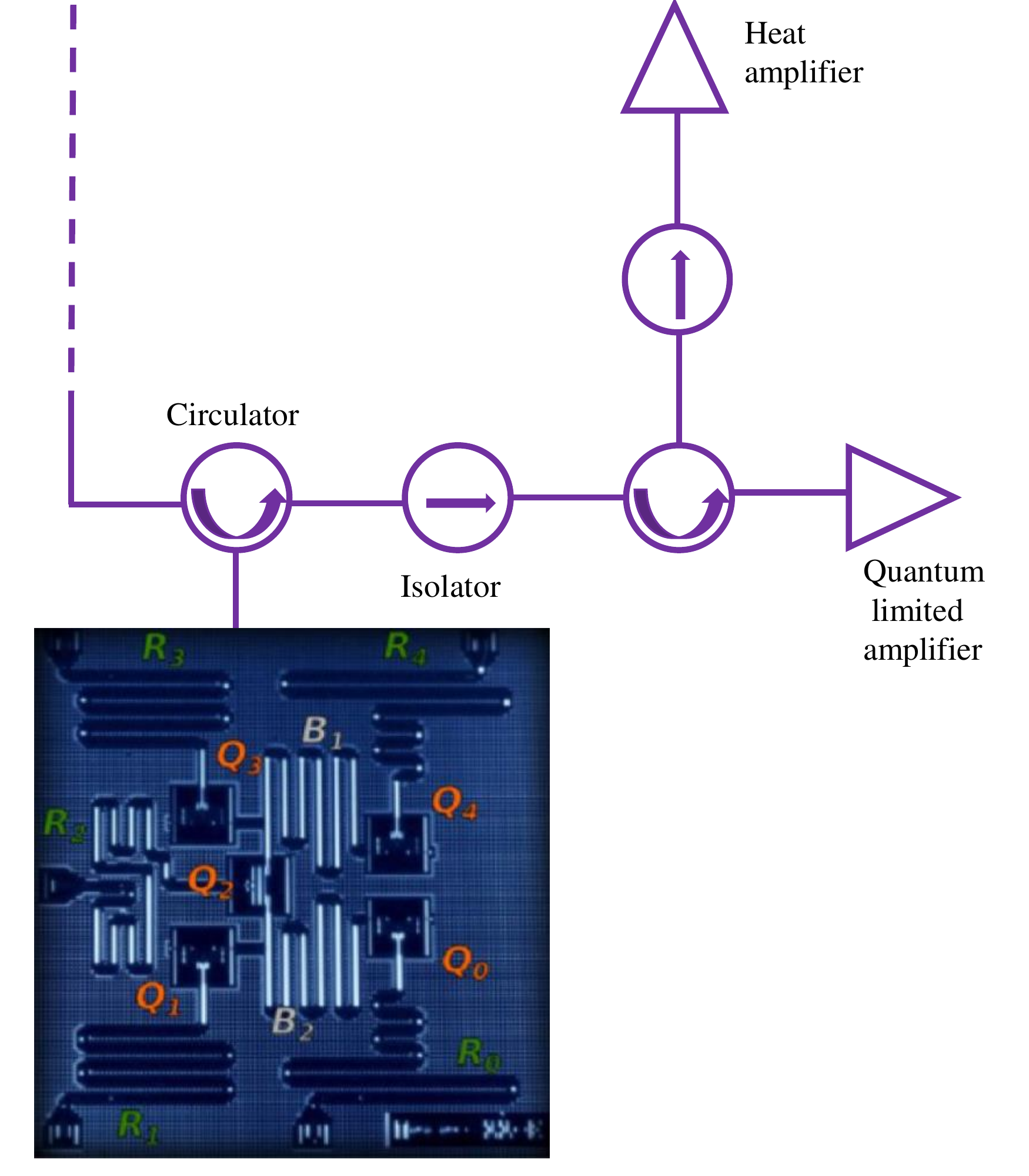}}
\par\end{centering}
\textcolor{black}{\caption{\textcolor{red}{\label{fig:architecture-1}}\textcolor{black}{The
architecture used in IBM five-qubit quantum computer. The upper trace
in the figure helps in control and read out. The left arm is used
for input, and the right arm is used for output. The lower trace is
the chip layout, in which five-qubits are positioned. A dedicated
coplanar waveguide resonator is used to control and read-out the individual
qubits, transmission lines for the purpose are shown in dark color.
The coupling between two qubits can be realized via coupled microwave
resonator shown by white lines.}}
}
\end{figure}

\textcolor{black}{}
\textcolor{black}{In general, there are several types of superconducting
qubits that can be used for realizing quantum information tasks. In
the IBM quantum processors, the qubits used are transmon qubits, which
are charged qubits designed to reduce the charge noise \cite{malkoc2013quantum}.
The arrangement of five superconducting qubits (q{[}0{]}, q{[}1{]},
q{[}2{]}, q{[}3{]}, q{[}4{]}) and their control mechanism as provided
in IBM quantum experience website \cite{IBMQE} including chip layout,
is shown in  Figure \ref{fig:architecture-1}. Each qubit can be controlled
and read out by a dedicated coplanar waveguide resonator, shown in
the figure by black transmission lines. Also, the qubit-qubit coupling
realized by coplanar microwave resonator  and is shown by white transmission
lines. These coplanar wave guides have also been used for read out
purpose. It is evident from the qubit topology, as given in \cite{IBMQE}
that qubits q{[}0{]}, q{[}1{]} and q{[}2{]} are interconnected, but
only qubit q{[}2{]} is connected to qubits q{[}3{]} and q{[}4{]},
in addition qubits q{[}3{]} and q{[}4{]} are also interconnected.
Such a topological set-up limits the applicability of two-qubit gates
between the possible pairs. Details about experimental parameters,
obtained from the IBM quantum experience website \cite{architectureurl}
are given in Table \ref{tab:exp-1}. Two relaxation time scales namely,
longitudinal relaxation time (T1) and transverse relaxation time (T2)
for qubits (q{[}0{]}, q{[}1{]}, q{[}2{]}, q{[}3{]}, q{[}4{]}) are
given in another Table \ref{tab:crosstalk-1} as well as the coupling
strengths between pair of connected qubits are provided through the
crosstalk matrix (see Table \ref{tab:crosstalk-1}).}

\subsection{\textcolor{black}{IBM QX4 \label{subsec:IBM-QX4}}}

\begin{figure}
\begin{centering}
\textcolor{black}{\includegraphics[scale=1.2]{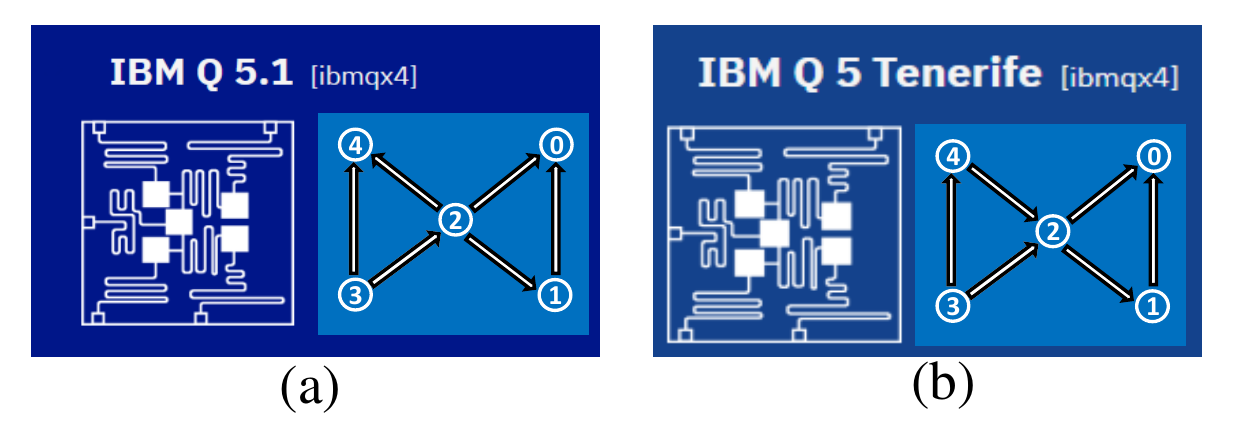}}
\par\end{centering}
\centering{}\textcolor{black}{\caption{\label{fig:IBM QX4}Figures (a) and (b) are depicting chip layout
and the topology of the old and new versions of IBM QX4 processor.}
}
\end{figure}

\textcolor{black}{IBM QX4 is another type of five-qubit IBM quantum
computer. This processor also has two versions. In Figures \ref{fig:IBM QX4}
(a) and (b) the topology of the new and old versions of IBM QX4 processor
are shown respectively. Similar to IBM QX2, in IBM QX4, the directly
implementable CNOT operations among the five-qubits in old and new
version of IBM QX4 are different. Allowed CNOT operations between
in five-qubits for old version are q{[}2{]} $\rightarrow$ {} q{[}0{]},
q{[}2{]} $\rightarrow$ {} q{[}1{]}, q{[}2{]} $\rightarrow$ {} q{[}4{]},
q{[}3{]} $\rightarrow$ {} q{[}2{]}, q{[}3{]} $\rightarrow$ {} q{[}4{]},
q{[}1{]} $\rightarrow$ {} q{[}0{]} and those for new version are
q{[}2{]} $\rightarrow$ {} q{[}0{]}, q{[}2{]} $\rightarrow$ {} q{[}1{]},
q{[}4{]}$\rightarrow$ {} q{[}2{]}, q{[}3{]} $\rightarrow$ {} q{[}2{]},
q{[}3{]} $\rightarrow$  q{[}4{]}, q{[}2{]} $\rightarrow$ {} q{[}1{]}.
}
\textcolor{black}{}
\begin{table}
\textcolor{black}{\caption{\textcolor{red}{\label{tab:exp-1}}\textcolor{black}{Details about
experimental parameters used in IBM quantum computer architecture
which are available on the website. The first column is qubit index
q in IBM quantum computer. The second column shows resonance frequencies
$\omega^{R}$ of corresponding read-out resonators. The qubit frequencies
$\omega$ are given in the third column. Anhormonicity $\delta$,
a measure of information leakage out of the computational space, is
the difference between two subsequent transition frequencies. Also,
$\chi$ and $\kappa$ given in the fifth and sixth columns are qubit-cavity
coupling strengths and coupling of the cavity to the environment for
corresponding qubits.}}
}
\centering{}\textcolor{black}{}%
\begin{tabular}{|>{\centering}p{1.5cm}|>{\centering}p{2.2cm}|>{\centering}p{2cm}|>{\centering}p{2cm}|>{\centering}p{2cm}|>{\centering}p{2cm}|}
\hline 
\textcolor{black}{Qubit number q} & \textcolor{black}{$\omega^{R}/2\pi$(GHz)} & \textcolor{black}{$\omega/2\pi$(GHz)} & \textcolor{black}{$\delta/2\pi$(MHz)} & \textcolor{black}{$\chi/2\pi$(kHz)} & \textcolor{black}{$\kappa/2\pi$(kHz)}\tabularnewline
\hline 
\textcolor{black}{q{[}0{]}} & \textcolor{black}{6.530350} & \textcolor{black}{5.2723} & \textcolor{black}{-330.3} & \textcolor{black}{476} & \textcolor{black}{523}\tabularnewline
\hline 
\textcolor{black}{q{[}1{]}} & \textcolor{black}{6.481848} & \textcolor{black}{5.2145} & \textcolor{black}{-331.9} & \textcolor{black}{395} & \textcolor{black}{489}\tabularnewline
\hline 
\textcolor{black}{q{[}2{]}} & \textcolor{black}{6.436229} & \textcolor{black}{5.0289} & \textcolor{black}{-331.2} & \textcolor{black}{428} & \textcolor{black}{415}\tabularnewline
\hline 
\textcolor{black}{q{[}3{]}} & \textcolor{black}{6.579431} & \textcolor{black}{5.2971} & \textcolor{black}{-329.4} & \textcolor{black}{412} & \textcolor{black}{515}\tabularnewline
\hline 
\textcolor{black}{q{[}4{]}} & \textcolor{black}{6.530225} & \textcolor{black}{5.0561} & \textcolor{black}{-335.5} & \textcolor{black}{339} & \textcolor{black}{515}\tabularnewline
\hline 
\end{tabular}
\end{table}
\textcolor{black}{}
\begin{table}
\textcolor{black}{\caption{\textcolor{red}{\label{tab:crosstalk-1}}\textcolor{black}{Details
about experimental parameters used in IBM quantum computer architecture
which are available on the website. The first row and the first column
shows qubit index q{[}i{]}. Entries of the matrix are the couplings
between corresponding qubits. Last two columns depict values of} longitudinal
relaxation time (T1) and transverse relaxation time (T2). used in
IBM quantum computer. Rest of the entries in this table have the same
meaning as was described in the the previous table.}
}
\centering{}\textcolor{black}{}%
\begin{tabular}{|>{\centering}p{1.5cm}|>{\centering}p{1cm}|>{\centering}p{1cm}|>{\centering}p{1cm}|>{\centering}p{1cm}|>{\centering}p{1cm}|>{\centering}p{2cm}|>{\centering}p{2cm}|}
\hline 
\textcolor{black}{$\zeta_{ij}$ (kHz)} & \textcolor{black}{q{[}0{]}} & \textcolor{black}{q{[}1{]}} & \textcolor{black}{q{[}2{]}} & \textcolor{black}{q{[}3{]}} & \textcolor{black}{q{[}4{]}} & \textcolor{black}{T$_{1}$ ($\mu$s)} & \textcolor{black}{T$_{2}$ ($\mu$s)}\tabularnewline
\hline 
\textcolor{black}{q{[}0{]}} &  & \textcolor{black}{-43} & \textcolor{black}{-83} &  &  & \textcolor{black}{53.04 (0.64)} & \textcolor{black}{48.50 (2.63)}\tabularnewline
\hline 
\textcolor{black}{q{[}1{]}} & \textcolor{black}{-45} &  & \textcolor{black}{-25} &  &  & \textcolor{black}{63.94 (1.06)} & \textcolor{black}{35.07 (0.59)}\tabularnewline
\hline 
\textcolor{black}{q{[}2{]}} & \textcolor{black}{-83} & \textcolor{black}{-27} &  & \textcolor{black}{-127} & \textcolor{black}{-38} & \textcolor{black}{52.08 (0.58)} & \textcolor{black}{89.73 (1.82)}\tabularnewline
\hline 
\textcolor{black}{q{[}3{]}} &  &  & \textcolor{black}{-127} &  & \textcolor{black}{-97} & \textcolor{black}{51.78 (0.55)} & \textcolor{black}{60.93 (1.09)}\tabularnewline
\hline 
\textcolor{black}{q{[}4{]}} &  &  & \textcolor{black}{-34} & \textcolor{black}{-97} &  & \textcolor{black}{55.80 (0.95)} & \textcolor{black}{84.18 (2.41)}\tabularnewline
\hline 
\end{tabular}
\end{table}
\textcolor{black}{}

\subsection{\textcolor{black}{Sixteen-qubit IBM quantum computer \label{subsec:Sixteen-qubit-IBM-quantum}}}

\begin{figure}
\begin{centering}
\textcolor{black}{\includegraphics[scale=1.1]{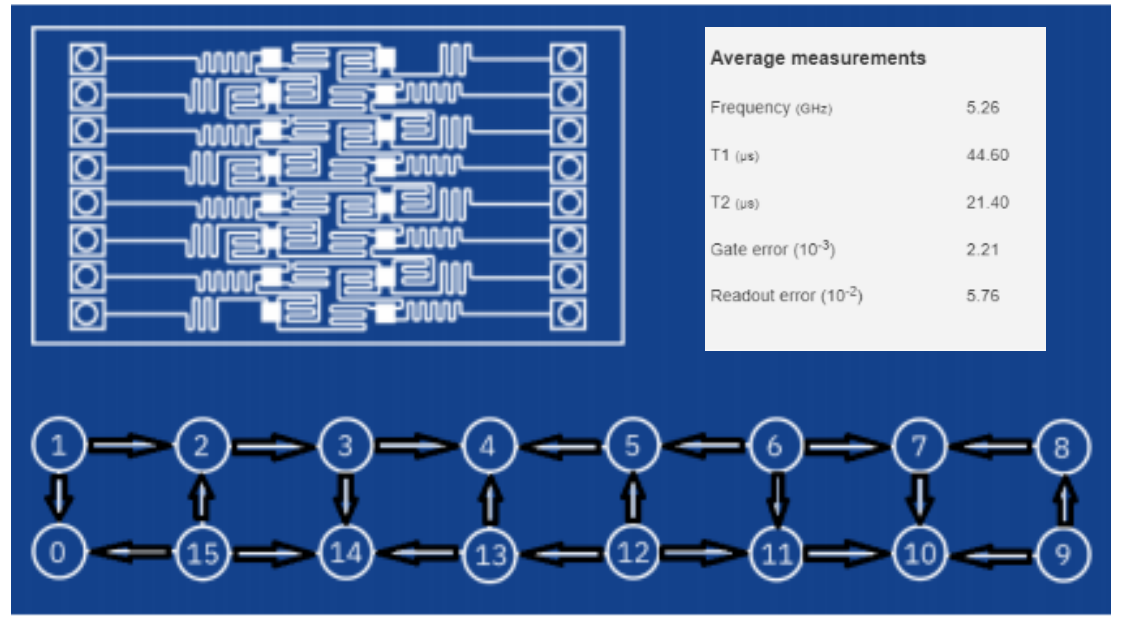}}
\par\end{centering}
\textcolor{black}{\caption{\label{fig:16-qubit}\textcolor{black}{Chip layout and qubit topology
of IBM QX5 quantum processor.}}
}
\end{figure}

\textcolor{black}{Figure \ref{fig:16-qubit} shows the chip layout
and qubit topology of the sixteen-qubit quantum processor IBM QX5.
Allowed CNOT operations among the sixteen qubits are q{[}0{]} $\rightarrow$ {}
q{[}1{]}, q{[}0{]} $\rightarrow$ {} q{[}2{]}, q{[}2{]}$\rightarrow$ {}
q{[}3{]}, q{[}3{]} $\rightarrow$  q{[}14{]}, q{[}3{]} $\rightarrow$ {}
q{[}4{]}, q{[}5{]} $\rightarrow$ {} q{[}4{]}, q{[}6{]} $\rightarrow$ {}
q{[}11{]}, q{[}6{]} $\rightarrow$ {} q{[}7{]}, q{[}6{]} $\rightarrow$ {}
q{[}5{]}, q{[}8{]} $\rightarrow$ {} q{[}7{]}, q{[}7{]} $\rightarrow$ {}
q{[}10{]}, q{[}9{]} $\rightarrow$ {} q{[}8{]}, q{[}9{]} $\rightarrow$ {}
q{[}10{]}, q{[}11{]} $\rightarrow$ {} q{[}10{]}, q{[}12{]} $\rightarrow$ {}
q{[}13{]}, q{[}12{]} $\rightarrow$ {} q{[}11{]}, q{[}12{]} $\rightarrow$ {}
q{[}5{]}, q{[}13{]} $\rightarrow$ {} q{[}14{]}, q{[}13{]} $\rightarrow$ {}
q{[}4{]}, q{[}15{]} $\rightarrow$ {} q{[}14{]}, q{[}15{]} $\rightarrow$ {}
q{[}2{]}, q{[}15{]} $\rightarrow$ {} q{[}0{]}. }
\textcolor{black}{IBM has also introduced a twenty-qubit quantum processor
\cite{IBMQE2} and has proposed to launch a fifty-qubit quantum processor.
However, we restrict ourselves from describing those processors as
those are not used in this thesis.}

\section{\textcolor{black}{Basic idea of quantum state tomography (QST) \label{sec:Basic-idea-of}}}

\textcolor{black}{Quantum state tomography is the process of reconstructing
the density matrix of a state. In the context of this thesis, to obtain
the full picture, we need to reconstruct the density matrix of the
teleported state using QST \cite{chuang1998bulk}. Till date, a large
number of advanced protocols have been proposed for quantum state
tomography \cite{james2001measurement,hebenstreit2017compressed,alsina2016experimental,rundle2016quantum,filipp2009two,schmied2016quantum}.
An advanced protocol suppresses the requirement of repeated preparation
of the state to be tomographed and hence allows state characterization
in dynamical environment using only one experiment \cite{shukla2013ancilla}.
Quantum state tomography requires extraction of information from the
experiments and the subsequent use of that information in the reconstruction
process of the experimental density matrix. An experimental density
matrix in the Pauli basis can be expressed as $\rho^{E}=\frac{1}{2^{N}}\stackrel[i=1]{2^{N}}{\sum}c_{i...N}\sigma_{i}\otimes.....\otimes\sigma_{N}$,
where c$_{i...N}$ can be obtained as $\langle\sigma_{i}\otimes....\otimes\sigma_{N}\rangle$
with $\sigma_{i...N}\in\left\{ I,X,Y,Z\right\} $. For single-qubit
case the density matrix is, $\rho=\frac{1}{2}\left(\begin{array}{cc}
1+\langle\sigma_{Z}\rangle & \langle\sigma_{X}\rangle-i\langle\sigma_{Y}\rangle\\
\langle\sigma_{X}\rangle+i\langle\sigma_{Y}\rangle & 1-\langle\sigma_{Z}\rangle
\end{array}\right)$. Availability of the expectation value $\langle\sigma_{i}\rangle$
in an experiment depends on whether $\sigma_{i}$ is a direct observable
or an indirect observable. In the IBM architecture, $Z$ is a direct
observable hence from the experimental outcomes, i.e., available probabilities
of $|0\rangle$ and $|1\rangle$, the value of $\langle Z\rangle$
can be directly calculated (without the use of any tomographic gate),
while $X$ and $Y$ being indirect observables, calculation of $\langle X\rangle$
and $\langle Y\rangle$ would require the application of $H$ and
$HS^{\dagger}$ gates, respectively \cite{hebenstreit2017compressed}.
In the present thesis, we have discussed QST in relatively more detail in
Chapter \ref{cha:resourceopt} and \ref{cha:BellIBM} for two and
three qubit states.}

\section{\textcolor{black}{A brief summary of this chapter and the structure
of the rest of the thesis \label{sec:A-brief-summary}}}

This thesis is focused on quantum communication protocols and in this
chapter we have provided an introduction to the field of quantum communication,
with an appropriate stress on its applicability in our daily life.
In the beginning of this chapter, we have described the history of
quantum communication and the basic concepts related to this field
with a specific attention on quantum teleportation and quantum cryptography.
Some quantum communication schemes which are rigorously studied in
the subsequent chapters of this thesis are also discussed. A short
introduction to quantum noise, QST, quantum process tomography, quantum
gates and realization of quantum gates with some optical elements
have also been provided in this chapter. Further, the power and limitations
of the SQUID-based IBM quantum computer are discussed in this chapter.

\textcolor{black}{The structure of the rest of the thesis can be summarized
as follows. Chapters \ref{cha:resourceopt}-\ref{cha:nonorthogonal}
are dedicated to the quantum teleportation schemes using different
type of quantum resources (entangled orthogonal state and entangled
nonorthogonal state) and their experimental realization using superconductivity-based
IBM quantum computer. In Chapter \ref{cha:resourceopt}, an explicit
scheme (quantum circuit) is designed for the teleportation of an $n$-qubit
quantum state. It is established that the proposed scheme requires
an optimal amount of quantum resources, whereas larger amount of quantum
resources have been used in a large number of recently reported teleportation
schemes for the quantum states which can be viewed as special cases
of the general n-qubit state considered here. A trade-off between
our knowledge about the quantum state to be teleported and the amount
of quantum resources required for the same is observed. A proof-of-principle
experimental realization of the proposed scheme (for a two-qubit state)
is also performed using five-qubit SQUID-based IBM quantum computer.
The experimental results show that the state has been teleported with
high fidelity. Relevance of the proposed teleportation scheme has
also been discussed in the context of controlled, bidirectional, and
bidirectional controlled state teleportation.}

\textcolor{black}{In Chapter \ref{cha:comment} of this thesis, we
explicitly show that a teleportation protocol reported by Zhao et
al., \cite{zhao2018quantum} for the teleportation of an eight-qubit
state utilizing a six-qubit state can actually be implemented by using
the scheme proposed in the previous chapter. Thus, the use of six-qubit
state as the teleportation channel can be circumvented. Specifically,
in this chapter, we have esta}blished that there is a conceptual mistake
in the work of Zhao et al., \cite{zhao2018quantum} and the teleportation
task that they have performed can be realized using any two Bell states
(i.e., without using the multi-partite entangled state used by them).
Further, it is mentioned that the applicability of the observations
of this Chapter is not limited to the work of Zhao et al., rather
it's applicable to a large set of recent proposals for the teleportation
of multi-qubit states.

\textcolor{black}{Chapter \ref{cha:nonorthogonal} aims to investigate
the effect of nonorthogonality of an entangled nonorthogonal state-based
quantum channel in detail in the context of the teleportation of a
qubit. Specifically, average fidelity, minimum fidelity and minimum
assured fidelity (MASFI) are computed for teleportation of a single-qubit
state using all the Bell-type entangled nonorthogonal states known
as quasi-Bell states. Using Horodecki criterion, it is shown that
the teleportation scheme obtained by replacing the quantum channel
(Bell state) of the usual teleportation scheme by a quasi-Bell state
is optimal. Further, the performance of various quasi-Bell states
as teleportation channel is compared in an ideal situation (i.e.,
in the absence of noise) and under different noise models (e.g., AD
and PD). It is observed that the best choice of the quasi-Bell state
depends on the amount of nonorthogonality, both in noisy and noiseless
cases. A specific quasi-Bell state, which was found to be maximally
entangled in the ideal conditions, is shown to be less efficient as
a teleportation channel compared to other quasi-Bell states in particular
cases when subjected to noisy channels. It has also been observed
that usually the value of average fidelity falls with an increase
in the number of qubits exposed to noisy channels (viz., Alice\textquoteright s,
Bob\textquoteright s and to be teleported qubits), but the converse
may be observed in some particular cases. Chapter \ref{cha:resourceopt}-\ref{cha:nonorthogonal}
are primarily focused on a particular aspect of quantum communication,
teleportation.}

\textcolor{black}{So, in Chapter \ref{cha:BellIBM}, we moved our
attention to a scheme for distributed quantum measurement which plays
an important role in deciding which strategies/steps are to be avoided
in designing schemes for quantum cryptography. The scheme for distributed
quantum measurement studied here allows nondestructive or indirect
Bell measurement which was proposed by Gupta et al., \cite{NDBSD}.
The scheme is experimentally realized here using the five-qubit IBM
SQUID-based quantum computer. The experiment confirmed that the Bell
state can be constructed and measured in a nondestructive manner with
a reasonably high fidelity. A comparison of the outcomes of this study
and the results obtained earlier in an NMR-based experiment (Samal
et al., (2010) \cite{samal2010non}) has also been performed. The
study indicates that to make a scalable SQUID-based quantum computer,
errors introduced by the gates (in the present technology used by
IBM) have to be reduced considerably.}

\textcolor{black}{In Chapter \ref{chap:OPTICAL-DESIGNS-FOR}, we have
presented optical designs for the realization of a set of quantum
cryptography schemes. There are several theoretical schemes for QKD
and other quantum cryptographic tasks (e.g., schemes for secure direct
quantum communication and their controlled version). However, only
a few of them have yet been performed experimentally. Other schemes
which have not yet been performed experimentally include, schemes
for QD \cite{pathak2013elements}, CQD \cite{thapliyal2017quantum},
Kak\textquoteright s three stage scheme \cite{kak2006three,mandal2013multi,thapliyal2018kak}.
This chapter aims to report optical circuits for the realization of
these quantum cryptograhic schemes using available optical elements,
like laser, BS, PBS, HWP, and experimentally realizable quantum states
like single photon states (which represents a single-qubit), two-qubit
and multi qubit entangled states of light (such as GHZ-like state,
W state). Finally, the thesis work is concluded in Chapter \ref{cha:Conclusions-and-Scope}
with a brief discussion on the limitations of the present work and
the scope for future work.}

\chapter[CHAPTER \thechapter \protect\newline DESIGN AND EXPERIMENTAL REALIZATION OF AN OPTIMAL SCHEME FOR TELEPORTATION
OF AN $n$-QUBIT QUANTUM STATE]{DESIGN AND EXPERIMENTAL REALIZATION OF AN OPTIMAL SCHEME FOR TELEPORTATION
OF AN $n$-QUBIT QUANTUM STATE\textsc{\label{cha:resourceopt}}}

{\large{}\lhead{}}{\large\par}

\section{Introduction\label{sec:Introduction-chap2} }

\textcolor{black}{In the previous chapter, we have already introduced
the concept of QT. Here, we may note that perfect QT of an arbitrary
quantum state using a classical channel would require an infinite
amount of classical resources. However, it is possible to teleport
an arbitrary quantum state with unit fidelity using a quantum channel
(a pre-shared entangled state) and a few bits of classical communication.
As perfect teleportation does not have a classical analogue, schemes
for quantum teleportation (QT) and their variants discussed in the
previous chapter drew considerable attention of the quantum communication
community. In addition to the teleportation-based schemes mentioned
in Section \ref{sec:Various-facets-of-insecure}, there exist proposals
to employ teleportation in quantum repeaters to enhance the feasibility
of quantum communication, and in ensuring security against an eavesdropper's
attempt to encroach the private spaces of legitimate users devising
trojan-horse attack \cite{lo1999unconditional}. This wide applicability
of QT and its variants and the fact that quantum resources are costly
(for example, preparation and maintenance of an $n$-qubit entangled
state is extremely difficult for large $n$) have motivated us to
investigate whether the recently proposed schemes \cite{li2016quantum,hassanpour2016bidirectional,da2007teleportation,li2016asymmetric,song2008controlled,cao2005teleportation,muralidharan2008quantum,tsai2010teleportation,nie2011quantum,tan2016deterministic,wei2016comment,li2016quantum1,yu2013teleportation,nandi2014quantum}
are using an optimal amount of quantum resources. If not, how to reduce
the amount of quantum resources to be used? An effort to answer these
questions has led to the present work, where we aim to propose a scheme
for teleportation of an $n$-qubit quantum state using an optimal
amount of quantum resources and to experimentally realize a particular
case of the proposed scheme using five-qubit IBM quantum computer.
Before we proceed to describe the findings of the present work, we
would like to elaborate a bit on what makes it fascinating to work
on teleportation even after almost quarter century of its introduction.}

\textcolor{black}{As a natural generalization of QT schemes, protocols
for teleportation of multi-qubit states have also been proposed. In
2006, Chen et al., proposed a multi-qubit generalization of the standard
QT scheme \cite{chen2006general}, which enabled teleportation of
multi-qubit states using a genuine multi-partite entangled channel
having a general form. Chen et al.'s work also indicated that the
bipartite nature of the channel is sufficient for teleportation of
multi-qubit quantum states. Specifically, it was shown that QT of
an arbitrary $n$-qubit state can be achieved by performing $n$ rounds
of Bennett et al.'s protocol \cite{bennett1993teleporting} for QT
(one for each qubit). Later, this scheme was extended to CT of an
arbitrary $n$-qubit quantum state \cite{man2007genuine}. Along the
same line, a bidirectional state teleportation and a bidirectional
controlled state teleportation schemes may be designed for teleporting
two arbitrary multi-qubit states, one each by Alice and Bob. Specifically,
a quantum state suitable as a quantum channel for a CT (bidirectional
controlled state teleportation) scheme should essentially reduce to
a useful quantum channel for QT (bidirectional state teleportation)
after the controller's measurement (see \cite{thapliyal2015general}
for detail discussion). In the previous Chapter, we have already mentioned
that the original QT protocol of Bennett et al., was experimentally
realized by Bouwmeester et al., \cite{bouwmeester1997experimental}
in 1997 and later on, a number of experimental realizations of single-qubit
QT has been reported \cite{bouwmeester1997experimental,nielsen1998complete,furusawa1998unconditional,zhao2004experimental,riebe2004deterministic,barrett2004deterministic}.
However, hardly any proposal for teleportation of multi-qubit quantum
states have been tested experimentally because the experimental realization
of those schemes would require quantum resources that are costly.
This observation has further motivated us to design a general teleportation
scheme that would require lesser amount of quantum resource and/or
such resources that can be generated and maintained easily using available
technology.}

\textcolor{black}{It may be noted that an optimized scheme for multi-partite
QT must involve optimization of both procedure and resources. Optimization
of the procedure demands use of those states as quantum channel, which
can be prepared easily and are least affected by decoherence; while
the optimization of resources demands that the scheme should exploit/utilize
all the channel qubits that are available for performing QT. The results
of Ref. \cite{chen2006general} can be viewed as an optimization of
the procedure (as Bell states can be easily prepared and are less
prone to decoherence in comparison with the multi-partite entangled
states). The importance of such strategies becomes evident when we
try to realize QT in a quantum network designed for quantum internet
\cite{sun2016quantum}. Naturally, an optimized QT scheme is expected
to improve the performance of such a quantum internet. Another kind
of optimization has been attempted in some recent works. Specifically,
efforts have been made to form teleportation channel (having lesser
number of entangled qubits) suitable for the teleportation of specific
types of unknown quantum states \cite{li2016quantum,hassanpour2016bidirectional,da2007teleportation,li2016asymmetric,song2008controlled,cao2005teleportation,muralidharan2008quantum,tsai2010teleportation,nie2011quantum,tan2016deterministic,wei2016comment,li2016quantum1,yu2013teleportation,nandi2014quantum}.
For example, in \cite{li2016quantum}, the three-qubit and the four-qubit
unknown quantum states have been teleported using four and five-qubit
cluster states, respectively. An extended list of these complex states
and the corresponding channels are given in Table \ref{tab:A-list-of}.
In fact, some of this set of schemes has exploited the fact that some
of the probability amplitudes in the state to be teleported are zero
and a QT scheme essentially transfers the unknown probability amplitudes
to distant qubits. Teleportation of such states has its own significance,
a trivial example that can justify its significance is the teleportation
of entangled quantum states (say, a non-maximally entangled Bell-type
state). We noticed that the quantum resources used in these protocols
are not optimal and most of the cases involve redundant qubits. Keeping
these in mind, here, we set the task for us to minimize the number
of these qubits exploiting the available information regarding the
mathematical structure of the quantum state to be teleported. Specifically,
in what follows, we would propose an efficient and optimal (uses minimum
number of Bell states as quantum channel) scheme to teleport a multi-qubit
state of specific form. In what follows, it will be established that
the bottleneck of our optimal scheme is the application of a unitary
operation which transforms the state to be teleported from the entangled
basis to the computational basis. Such a transformation allows us
to render the information encoded into a smaller number of qubits.
On the other hand, it increases quantum computational resources required
at each node (because of the application of an extra unitary). However,
it is desirable to minimize the number of qubits to be transmitted
at the cost of computational resources as transmitting a rather large
quantum state is much harder than computation. This is so because
the former requires more resources at each step, i.e., in initialization,
transmission, and measurement steps and is also prone to environmental
effects. Further, it would be established that the proposed scheme
is simple in nature and can be extended to corresponding CT and bidirectional
controlled state teleportation schemes.}

\textcolor{black}{Actual relevance of an optimized scheme lies in
the experimental realization of the scheme only. An interesting window
for experimental realization of the schemes of quantum computation
and communication in general and optimized schemes in particular has
been opened up recently, when IBM provided free access to a five-qubit
superconducting quantum computer by placing it in cloud \cite{IBMQE,devitt2016performing}.
This has provided a platform for experimental testing of various proposals
for quantum communication and computation. In the present work, we
have used IBM quantum computer to experimentally realize the optimal
scheme designed here. Specifically, we have successfully implemented
the optimal quantum circuit designed for teleportation of a two-qubit
state. The experiment performed is a proof-of-principle experiment
as the receiver and the sender are not located at two distant places,
but it shows successful teleportation with very high fidelity and
paves the way for future realizations of the proposed scheme using
optical elements.}

\textcolor{black}{The rest of the chapter is organized as follows.
In Section \ref{mun}, we propose a scheme for the teleportation of
a multi-qubit quantum state having $m$ unknown coefficients using
optimal quantum resources. We also discuss a specific case of this
scheme which corresponds to QT of a two-unknown multi-qubit quantum
state using a Bell state as quantum channel. In Section \ref{cbt},
we describe optimal schemes for controlled unidirectional and bidirectional
state teleportation of the quantum states. Further, in Section \ref{sec:Experimental-implementation},
an experimental realization of the proposed scheme is reported using
the five-qubit IBM quantum computer available on cloud. Finally, we
conclude the chapter in Section \ref{sec:Conclusion-2}.}

\section{\textcolor{black}{Teleportation of an $n$-qubit state with $m$
unknown coefficients \label{sec:Teleportation-of-an}}}

\textcolor{black}{\label{mun}}

\textcolor{black}{Consider an $n$-qubit quantum state to be teleported
as}

\textcolor{black}{
\begin{equation}
|\psi\rangle=\sum_{i=1}^{m}\alpha_{i}\ket{x_{i}},\label{eq:psi}
\end{equation}
where $\alpha_{i}$s are the probability amplitudes ensuring normalization
$\sum_{i=1}^{m}\vert\alpha_{i}\vert^{2}=1$. Additionally, $x_{i}$s
are mutually orthogonal to each other, i.e., $\langle x_{i}|x_{j}\rangle=\delta_{ij}\,\forall\,\left(1<i,j<m\right)$.
Therefore, one may note that $x_{i}$s are the elements of an $n$-qubit
orthogonal basis iff $m\leq2^{\left(n\right)}$. For example, we may
consider 3-qubit quantum states $\ket{\xi_{1}}=\alpha_{1}|000\rangle+\alpha_{2}|111\rangle$
and $\ket{\xi_{2}}=\alpha_{1}|000\rangle+\alpha_{2}|011\rangle+\alpha_{3}|100\rangle+\alpha_{4}|111\rangle$,
teleportation schemes for which were proposed in Refs. \cite{yu2013teleportation}
and \cite{wei2016comment}, respectively. For both the states $n=3$,
but we can easily observe that $m=2$ for $\ket{\xi_{1}}$ and $m=4$
for $\ket{\xi_{2}}$. In what follows, we will establish that because
of this difference in the values of $m$, teleportation of $\ket{\xi_{1}}$
would require only one Bell state, whereas that of $\ket{\xi_{2}}$
would require 2 Bell states.}

\textcolor{black}{Here, we set the task as to teleport state $\ket{\psi}:\,m\leq2^{\left(n-1\right)}$
using optimal quantum resources (i.e., using minimum number of qubits
in the multi-qubit entangled state used as quantum channel). To do
so, we will transform the state to be teleported (say, $|\psi\rangle$)
to a quantum state of a preferred form (say, $|\psi^{\prime}\rangle$).
Specifically, the central idea of our scheme lies in finding out a
unitary $U$, which transforms state $\ket{\psi}$ into $\ket{\psi^{\prime}}$,
i.e., $\ket{\psi^{\prime}}=U\ket{\psi}$, such that}

\begin{figure}
\begin{centering}
\textcolor{black}{\includegraphics[scale=0.48]{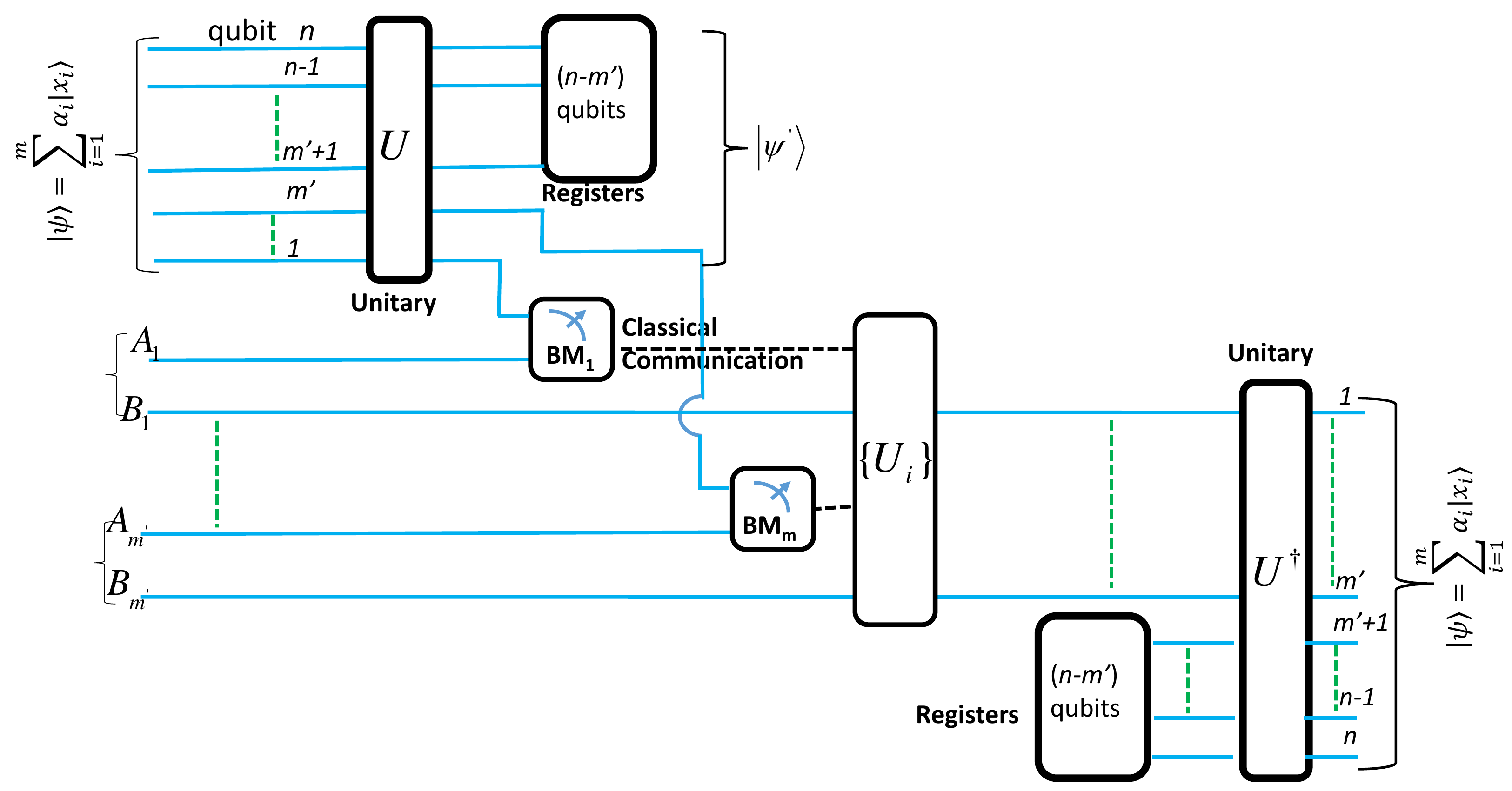} }
\par\end{centering}
\textcolor{black}{\caption{\label{fig:mun}\textcolor{blue}{{} }\textcolor{black}{The circuit designed
for generalized quantum teleportation of an $n$-qubit quantum states
with $m$ unknown coefficients. Here, $m^{\prime}=\left\lceil \log_{2}{m}\right\rceil $
is the number of Bell states. $U_{i}$ is the unitary operation applied
by Bob and BM stands for Bell measurement.}}
}
\end{figure}

\textcolor{black}{
\begin{equation}
|\psi^{\prime}\rangle=\sum_{i=1}^{m}{\alpha_{i}^{\prime}|y_{i}\rangle}.\label{eq:psiD}
\end{equation}
Here, $|\psi^{\prime}\rangle$ is a unique $n$-qubit quantum state
which can be reduced to an $m$-qubit quantum state in the computational
basis $\{y_{i}\}$, after measuring the redundant qubits. Specifically,
the unitary is expected to possess a one-to-one map for each element
of $\{x_{i}\}\rightarrow\{y_{i}\}$. As shown in Figure \ref{fig:mun},
here we prefer $\ket{y_{i}}=\ket{0}^{n-m^{\prime}}\otimes\ket{\widetilde{i}}$,
where $m^{\prime}=\left\lceil \log_{2}{m}\right\rceil $ and $\widetilde{i}$
is the binary equivalent of decimal value $i$ in an $m^{\prime}$-bit
binary string. This step transforms $m$ elements of $\ket{\psi}$
with non-zero projections in Eq. (\ref{eq:psi}) to that of $m$ elements
of $\ket{\psi^{\prime}}$ in Eq. (\ref{eq:psiD}). For example, we
may consider the quantum state $\ket{\xi_{1}}$ or $\ket{\xi_{2}}$
described above as the quantum state to be teleported. As both of
these states are 3-qubit states, $n=3$ for both of them. However,
in the expansion of $\ket{\xi_{1}}$ ($\ket{\xi_{2}}$) there are
2 (4) non-zero coefficients. Consequently, $m=2$ ($m=4$) and $m^{\prime}=\left\lceil \log_{2}{2}\right\rceil =1$
($m^{\prime}=\left\lceil \log_{2}{4}\right\rceil =2$) for the quantum
state $\ket{\xi_{1}}$ ($\ket{\xi_{2}}$). Therefore, the maximum
advantage of quantum resources would be obtained if the quantum state
$\ket{\xi_{1}}$ ($\ket{\xi_{2}}$) is teleported using a scheme that
stores two qubits (one-qubit) in a register.}

\textcolor{black}{However, to design the unitary able to accomplish
such a task the map $\{x_{i}\}\rightarrow\{y_{i}\}$ should exist
between all the elements (both possessing either zero or non-zero
projection in $\ket{\psi}$ and $\ket{\psi^{\prime}})$ in both the
basis. In other words, the unitary $U$ mapping the basis elements
from $\{x_{i}\}$ to $\{y_{i}\}$ required for the desired transformation
would be}

\textcolor{black}{
\begin{equation}
U=\sum_{i=1}^{2^{n}}|y_{i}\rangle\langle x_{i}|.\label{eq:U}
\end{equation}
Being a $2^{n}$ dimensional computational basis $\left\{ y_{i}\right\} $
is already known while state $\ket{\psi}$ reveals only $m$ orthogonal
vectors of basis $\{x_{i}\}$. Therefore, the remaining $\left(2^{n}-m\right)$
elements of basis $\{x_{i}\}$ can be obtained by Gram-Schmidt procedure,
such that the elements of $\{x_{i}\}$ follow the completeness relation,
i.e., $\sum_{i=1}^{2^{n}}{\sum}|x_{i}\rangle\langle x_{i}|=\mathbb{I}$.}

\textcolor{black}{The obtained quantum state $\ket{\psi^{\prime}}$
possesses the first ${n-m^{\prime}}$ qubits in $\ket{0}$, while
the remaining $m^{\prime}$ qubits hold the complete information of
$\ket{\psi}$. Therefore, our task reduces to a teleportation of an
$m^{\prime}$-qubit quantum states using an optimal amount of quantum
resources. An $m^{\prime}$-qubit quantum state can be teleported
either by using at least $2m^{\prime}$-qubit entangled state or $m^{\prime}$
Bell states, one for teleporting each qubit \cite{chen2006general}.}

\textcolor{black}{Preparing a multi-qubit entangled state is relatively
expensive and such a state is more prone to decoherence then a two-qubit
entangled state. For the reason, we prefer the second strategy and
select Bell states as a quantum channel (see Figure \ref{fig:mun}).
Once the quantum state $\ket{\psi^{\prime}}$ is reconstructed at
Bob's port, he would require to perform a unitary operation $U^{\dagger}.$}

\textcolor{black}{For the sake of completeness of the chapter we may
summarize teleportation of an arbitrary $m^{\prime}$-qubit quantum
state in the following steps:}
\begin{enumerate}
\item \textcolor{black}{The $m^{\prime}$-qubit unknown state to be teleported
(whose qubits are indexed by $1,2,\cdots,m^{\prime}$) and the first
qubits of each $m^{\prime}$ Bell states (indexed by $A_{1},A_{2},\cdots,A_{m^{\prime}}$)
are with Alice and all the second qubits (indexed by $B_{1},B_{2},\cdots,B_{m^{\prime}}$)
are with Bob.}
\item \textcolor{black}{Alice performs pairwise Bell measurements on her
qubits ($i,A_{i}$) and finally announces $m^{\prime}$ measurement
outcomes.}
\item \textcolor{black}{Bob applies Pauli operations on each qubit in his
possession depending upon the measurement outcome of Alice (see Table
\ref{tab:tel-U}, which is adapted from Ref. \cite{thapliyal2015applications}).
At the end of this step Bob obtains the $m^{\prime}$-qubit unknown
quantum state that Alice has teleported.}
\end{enumerate}
\textcolor{black}{}
\begin{table}
\textcolor{black}{\caption{\label{tab:tel-U} \textcolor{black}{Unitary applied by Bob to reconstruct
the quantum state for QT. In the table, SMO is the sender's measurement
outcome.}}
}

~
\centering{}\textcolor{black}{}%
\begin{tabular}{|c|c|c|c|c|}
\hline 
 & \multicolumn{4}{c|}{\textcolor{black}{Initial state shared by Alice and Bob}}\tabularnewline
\hline 
 & \textcolor{black}{$|\psi^{+}\rangle$} & \textcolor{black}{$|\psi^{-}\rangle$} & \textcolor{black}{$|\phi^{+}\rangle$} & \textcolor{black}{$|\phi^{-}\rangle$}\tabularnewline
\hline 
\textcolor{black}{SMO} & \textcolor{black}{Receiver} & \textcolor{black}{Receiver} & \textcolor{black}{Receiver} & \textcolor{black}{Receiver}\tabularnewline
\hline 
\textcolor{black}{00} & \textcolor{black}{$I$} & \textcolor{black}{$Z$} & \textcolor{black}{$X$} & \textcolor{black}{$iY$}\tabularnewline
\hline 
\textcolor{black}{01} & \textcolor{black}{$X$} & \textcolor{black}{$iY$} & \textcolor{black}{$I$} & \textcolor{black}{$Z$}\tabularnewline
\hline 
\textcolor{black}{10} & \textcolor{black}{$Z$} & \textcolor{black}{$I$} & \textcolor{black}{$iY$} & \textcolor{black}{$X$}\tabularnewline
\hline 
\textcolor{black}{11} & \textcolor{black}{$iY$} & \textcolor{black}{$X$} & \textcolor{black}{$Z$} & \textcolor{black}{$I$}\tabularnewline
\hline 
\end{tabular}
\end{table}
\textcolor{black}{Once Bob has access to the $m^{\prime}$-qubit unknown
state and knowledge of the unitary $U$ Alice has applied, he prepares
$n-m^{\prime}$ qubits in $\ket{0}$. Finally, he applies $U^{\dagger}$
to reconstruct the unknown quantum state $|\psi\rangle=U^{\dagger}|\psi^{\prime}\rangle.$
Note that the unitary operation defined in Eq. (\ref{eq:U}) is independent
of the unknown parameters of the quantum state to be teleported and
exploits only the available knowledge of the state, i.e., the number
of qubits, bases $\{x_{i}\}$ and $\{y_{i}\}$, and the number of
vanished coefficients.}

\textcolor{black}{}
\textcolor{black}{In Table \ref{tab:A-list-of}, we have given unitary
operations involved in teleportation of various multi-qubit states
with different number of unknowns using our scheme. Teleportation
of these states using relatively fragile and expensive quantum resources
have been reported in the recent past. To be specific, our technique
can be used to teleport any quantum state having two unknown coefficients
\cite{da2007teleportation,cao2005teleportation,tsai2010teleportation,yu2013teleportation,li2016quantum}
using a single Bell state, irrespective of the number of physical
qubits present in the state. In contrast, in the existing literature
(cf. Columns 2 and 3 of Table \ref{tab:A-list-of}) it is observed
that the number of qubits used in the quantum channel increases with
the increase in number of qubits to be teleported. Similarly, a quantum
state having four-unknown coefficients can be teleported only using
two Bell states, unlike the higher dimensional entangled states used
in \cite{wei2016comment,li2016quantum1,tan2016deterministic}. This
clearly establishes that our scheme allows one to circumvent the use
of redundant qubits and complex entangled states that are used until
now, and thus the present proposal increases the possibility of experimental
realization.}

\begin{table}
\caption{\label{tab:A-list-of}\textcolor{black}{A list of quantum states and
the quantum resources used to teleport them. The number of qubits
in the unknown quantum states, the unknown quantum state (i.e., the
state to be teleported) and the resources used to teleport them are
mentioned in the first, second and third column, respectively. Here,
CS stands for cluster state and $a_{i}$ represents the binary value
of decimal number $i$ expanded up to $k$ digits. Corresponding minimum
resources used to teleport that states are given in the forth column.
The unitary required to decrease the number of entangled qubits to
be used as quantum channel is given in the last column.}}

~
\begin{centering}
\textcolor{black}{\footnotesize{}}%
\begin{tabular}{|>{\centering}p{2.6cm}|>{\centering}p{1.3cm}|>{\centering}p{1.8cm}|>{\centering}p{1cm}|>{\centering}p{6.3cm}|}
\hline 
\textbf{\textcolor{black}{\footnotesize{}Quantum state to be teleported}} & \textbf{\textcolor{black}{\footnotesize{}Number of qubits in the state
to be teleporte-d}} & \textbf{\textcolor{black}{\footnotesize{}Quantum state used as quantum
channel and correspondi-ng reference}} & \textbf{\textcolor{black}{\footnotesize{}Num-}}{\footnotesize\par}

\textbf{\textcolor{black}{\footnotesize{}ber of Bell states requir-ed
in our schem-e to telepo-rt the state}} & \textbf{\textcolor{black}{\footnotesize{}Unitary to be applied by
Alice}}\tabularnewline
\hline 
\textcolor{black}{\footnotesize{}$\alpha|a_{0}\rangle+\beta|a_{3}\rangle$} & \textcolor{black}{\footnotesize{}two-qubit} & \textcolor{black}{\footnotesize{}four-qubit CS \cite{da2007teleportation},
3-qubit W class state \cite{cao2005teleportation}} & \textcolor{black}{\footnotesize{}one} & \textcolor{black}{\footnotesize{}$|a_{0}\rangle\langle a_{0}|+|a_{1}\rangle\langle a_{3}|+|a_{2}\rangle\langle a_{1}|+|a_{3}\rangle\langle a_{2}|$}\tabularnewline
\hline 
\textcolor{black}{\footnotesize{}$\alpha|a_{1}\rangle+\beta|a_{2}\rangle$} & \textcolor{black}{\footnotesize{}two-qubit} & \textcolor{black}{\footnotesize{}three-qubit GHZ- like state \cite{tsai2010teleportation}} & \textcolor{black}{\footnotesize{}one}{\footnotesize\par}

~

~ & \textcolor{black}{\footnotesize{}$|a_{0}\rangle\langle a_{1}|+|a_{1}\rangle\langle a_{2}|+|a_{2}\rangle\langle a_{0}|+|a_{3}\rangle\langle a_{3}|$}\tabularnewline
\hline 
\textcolor{black}{\footnotesize{}$\alpha|a_{0}\rangle+\beta|a_{7}\rangle$} & \textcolor{black}{\footnotesize{}three-qubit} & \textcolor{black}{\footnotesize{}four-qubitCS \cite{yu2013teleportation}} & \textcolor{black}{\footnotesize{}one}{\footnotesize\par}

~

~ & \textcolor{black}{\footnotesize{}$\begin{array}{l}
|a_{0}\rangle\langle a_{0}|+|a_{1}\rangle\langle a_{7}|+|a_{2}\rangle\langle a_{1}|+|a_{3}\rangle\langle a_{2}|\\
+|a_{4}\rangle\langle a_{3}|+|a_{5}\rangle\langle a_{4}|+|a_{6}\rangle\langle a_{5}|+|a_{7}\rangle\langle a_{6}|
\end{array}$}\tabularnewline
\hline 
\textcolor{black}{\footnotesize{}$\begin{array}{l}
\alpha(|a_{0}\rangle+|a_{3}\rangle)\\
+\beta(|a_{4}\rangle+|a_{7}\rangle)
\end{array}$} & \textcolor{black}{\footnotesize{}three-qubit} & \textcolor{black}{\footnotesize{}four-qubitCS \cite{li2016quantum}}{\footnotesize\par}

~ & \textcolor{black}{\footnotesize{}one}{\footnotesize\par}

~ & \textcolor{black}{\footnotesize{}$\begin{array}{l}
\frac{1}{\sqrt{2}}(|a_{0}\rangle\langle(|a_{0}+a_{3})|+|a_{1}\rangle\langle a_{4}+a_{7})|\\
+|a_{2}\rangle\langle(a_{1}+a_{2})|+|a_{3}\rangle\langle(a_{5}+a_{6})|\\
+|a_{4}\rangle\langle(a_{0}-a_{3})|+|a_{5}\rangle\langle(a_{4}-a_{7})|\\
+|a_{6}\rangle\langle(a_{1}-a_{2})|+|a_{7}\rangle\langle(a_{5}-a_{6})|)
\end{array}$}\tabularnewline
\hline 
\textcolor{black}{\footnotesize{}$\begin{array}{l}
\alpha(|a_{0}\rangle+|a_{7}\rangle)\\
+\beta(|a_{13}\rangle+|a_{10}\rangle)
\end{array}$} & \textcolor{black}{\footnotesize{}four-qubit} & \textcolor{black}{\footnotesize{}	five-qubit CS \cite{li2016quantum}
	} & \textcolor{black}{\footnotesize{}one} & \textcolor{black}{\footnotesize{}$\begin{array}{l}
\frac{1}{\sqrt{2}}(|a_{0}\rangle\langle|a_{0}+a_{7}|+|a_{1}\rangle\langle(a_{13}+a_{10})|\\
+|a_{2}\rangle\langle(a_{1}+a_{2})|+|a_{3}\rangle\langle(a_{3}+a_{4})|\\
+|a_{4}\rangle\langle(a_{5}+a_{6})|+|a_{5}\rangle\langle(a_{8}+a_{9})|\\
+|a_{6}\rangle\langle(a_{11}+a_{12}|+|a_{7}\rangle\langle(a_{14}+a_{15})|\\
+|a_{8}\rangle\langle(a_{0}-a_{7})|+|a_{9}\rangle\langle(a_{13}-a_{10})|\\
+|a_{10}\rangle\langle(a_{1}-a_{2})|+|a_{11}\rangle\langle(a_{3}-a_{4})|\\
+|a_{12}\rangle\langle(a_{5}-a_{6})|+|a_{13}\rangle\langle(a_{8}-a_{9})|\\
+|a_{14}\rangle\langle(a_{11}-a_{12})|+|a_{15}\rangle\langle(a_{14}-a_{15})|)
\end{array}$}{\footnotesize\par}

~\tabularnewline
\hline 
\end{tabular}{\footnotesize\par}
\par\end{centering}
\end{table}
\begin{center}
\textcolor{black}{}
\begin{table}
\centering{}\textcolor{black}{\footnotesize{}}%
\begin{tabular}{|>{\centering}p{2.6cm}|>{\centering}p{1.3cm}|>{\centering}p{1.8cm}|>{\centering}p{1cm}|>{\centering}p{6.3cm}|}
\hline 
\textbf{\textcolor{black}{\footnotesize{}Quantum state to be teleported}} & \textbf{\textcolor{black}{\footnotesize{}Number of qubits in the state
to be teleporte-d}} & \textbf{\textcolor{black}{\footnotesize{}Quantum state used as quantum
channel and correspondi-ng reference}} & \textbf{\textcolor{black}{\footnotesize{}Num-}}{\footnotesize\par}

\textbf{\textcolor{black}{\footnotesize{}ber of Bell states requir-ed
in our schem-e to telepo-rt the state}} & \textbf{\textcolor{black}{\footnotesize{}Unitary to be applied by
Alice}}\tabularnewline
\hline 
\textcolor{black}{\footnotesize{}$\begin{array}{l}
\alpha|a_{0}\rangle+\beta|a_{3}\rangle\\
+\gamma|a_{4}\rangle+\delta|a_{7}\rangle
\end{array}$} & \textcolor{black}{\footnotesize{}three-qubit} & \textcolor{black}{\footnotesize{}five-qubit state \cite{wei2016comment}} & \textcolor{black}{\footnotesize{}two}{\footnotesize\par}

~

~ & \textcolor{black}{\footnotesize{}$\begin{array}{l}
|a_{0}\rangle\langle a_{0}|+|a_{1}\rangle\langle a_{3}|+|a_{2}\rangle\langle a_{4}|\\
+|a_{3}\rangle\langle a_{7}|+|a_{4}\rangle\langle a_{1}|+|a_{5}\rangle\langle a_{2}|\\
+|a_{6}\rangle\langle a_{5}|+|a_{7}\rangle\langle a_{6}|
\end{array}$}\tabularnewline
\hline 
\textcolor{black}{\footnotesize{}$\begin{array}{l}
\alpha|a_{0}\rangle+\beta|a_{3}\rangle\\
+\gamma|a_{12}\rangle+\delta|a_{15}\rangle
\end{array}$} & \textcolor{black}{\footnotesize{}four-qubit} & \textcolor{black}{\footnotesize{}six-qubit CS \cite{li2016quantum1}} & \textcolor{black}{\footnotesize{}two}{\footnotesize\par}

~

~ & \textcolor{black}{\footnotesize{}$\begin{array}{l}
|a_{0}\rangle\langle a_{0}|+|a_{1}\rangle\langle a_{3}|+|a_{2}\rangle\langle a_{12}|\\
+|a_{3}\rangle\langle a_{15}|+|a_{4}\rangle\langle a_{1}|+|a_{5}\rangle\langle a_{2}|\\
+|a_{6}\rangle\langle a_{4}|+|a_{7}\rangle\langle a_{5}|+|a_{8}\rangle\langle a_{6}|\\
+|a_{9}\rangle\langle a_{7}|+|a_{10}\rangle\langle a_{8}|+|a_{11}\rangle\langle a_{9}|\\
+|a_{12}\rangle\langle a_{10}|+|a_{13}\rangle\langle a_{11}|+|a_{14}\rangle\langle a_{13}|\\
+|a_{15}\rangle\langle a_{14}|
\end{array}$}\tabularnewline
\hline 
\textcolor{black}{\footnotesize{}$\begin{array}{l}
\alpha|a_{0}\rangle+\beta|a_{15}\rangle\\
+\gamma|a_{63}\rangle+\delta|a_{48}\rangle
\end{array}$} & \textcolor{black}{\footnotesize{}six-qubit} & \textcolor{black}{\footnotesize{}six-qubit CS \cite{tan2016deterministic}} & \textcolor{black}{\footnotesize{}two}{\footnotesize\par}

~

~ & \textcolor{black}{\footnotesize{}$\begin{array}{l}
|a_{0}\rangle\langle a_{0}|+|a_{1}\rangle\langle a_{15}|+|a_{2}\rangle\langle a_{63}|\\
+|a_{3}\rangle\langle a_{48}|+|a_{4}\rangle\langle a_{1}|+|a_{5}\rangle\langle a_{2}|\\
+|a_{6}\rangle\langle a_{3}|+|a_{7}\rangle\langle a_{4}|+|a_{8}\rangle\langle a_{5}|\\
+|a_{9}\rangle\langle a_{6}|+|a_{10}\rangle\langle a_{7}|+|a_{11}\rangle\langle a_{8}|\\
+|a_{12}\rangle\langle a_{9}|+|a_{13}\rangle\langle a_{10}|+|a_{14}\rangle\langle a_{11}|\\
+|a_{15}\rangle\langle a_{12}|+|a_{16}\rangle\langle a_{13}|+|a_{17}\rangle\langle a_{14}|\\
+|a_{18}\rangle\langle a_{16}|+|a_{19}\rangle\langle a_{17}|+|a_{20}\rangle\langle a_{18}|\\
+|a_{21}\rangle\langle a_{19}|+|a_{22}\rangle\langle a_{20}|+|a_{23}\rangle\langle a_{21}|\\
+|a_{24}\rangle\langle a_{22}|+|a_{25}\rangle\langle a_{23}|+|a_{26}\rangle\langle a_{24}|\\
+|a_{27}\rangle\langle a_{25}|+|a_{28}\rangle\langle a_{26}|+|a_{29}\rangle\langle a_{27}|\\
+|a_{30}\rangle\langle a_{28}|+|a_{31}\rangle\langle a_{29}|+|a_{32}\rangle\langle a_{30}|\\
+|a_{33}\rangle\langle a_{31}|+|a_{34}\rangle\langle a_{32}|+|a_{35}\rangle\langle a_{33}|\\
+|a_{36}\rangle\langle a_{34}|+|a_{37}\rangle\langle a_{35}|+|a_{38}\rangle\langle a_{36}|\\
+|a_{39}\rangle\langle a_{37}|+|a_{40}\rangle\langle a_{38}|+|a_{41}\rangle\langle a_{39}|\\
+|a_{42}\rangle\langle a_{40}|+|a_{43}\rangle\langle a_{41}|+|a_{44}\rangle\langle a_{42}|\\
+|a_{45}\rangle\langle a_{43}|+|a_{46}\rangle\langle a_{44}|+|a_{47}\rangle\langle a_{45}|\\
+|a_{48}\rangle\langle a_{46}|+|a_{49}\rangle\langle a_{47}|+|a_{50}\rangle\langle a_{49}|\\
+|a_{51}\rangle\langle a_{50}|+|a_{52}\rangle\langle a_{51}|+|a_{53}\rangle\langle a_{52}|\\
+|a_{54}\rangle\langle a_{53}|+|a_{55}\rangle\langle a_{54}|+|a_{56}\rangle\langle a_{55}|\\
+|a_{57}\rangle\langle a_{56}|+|a_{58}\rangle\langle a_{57}|+|a_{59}\rangle\langle a_{58}|\\
+|a_{60}\rangle\langle a_{59}|+|a_{61}\rangle\langle a_{60}|+|a_{62}\rangle\langle a_{61}|\\
+|a_{63}\rangle\langle a_{62}|
\end{array}$}\tabularnewline
\hline 
\end{tabular}
\end{table}
\par\end{center}

\subsection{\textcolor{black}{Teleportation of state of type $\ket{\psi}=\alpha\ket{x_{i}}+\beta\ket{x_{j}}$}}

\textcolor{black}{\label{2un}}

\textcolor{black}{As an explicit example of the proposed scheme, consider
an $n$-qubit state with only two unknowns, i.e., 
\begin{equation}
\ket{\psi}=\alpha\ket{x_{i}}+\beta\ket{x_{j}},\label{eq:}
\end{equation}
such that, $\inpr{x_{i}}{x_{j}}=\delta_{ij}$ and ${\abs{\alpha}}^{2}+{\abs{\beta}}^{2}=1$.
Here, $x_{i}$ and $x_{j}$ are the elements of some $2^{n}$ dimensional
basis set. Our task is to teleport state $\ket{\psi}$ using optimal
quantum resources (i.e., using minimum number of entangled qubits
in quantum channel). As mentioned previously the minimum number of
Bell states required in the quantum channel for this kind of state
would be $\left\lceil \log_{2}{2}\right\rceil =1$.}

\textcolor{black}{Therefore, we will transform state $\ket{\psi}$
into $\ket{\psi^{\prime}}$, such that $U\ket{\psi}=\ket{\psi^{\prime}}$,
such that 
\begin{equation}
\ket{\psi^{\prime}}=\alpha^{\prime}\ket{y_{i}}+\beta^{\prime}\ket{y_{j}},\label{eq:-1}
\end{equation}
with $\inpr{y_{i}}{y_{j}}=\delta_{ij}$ as $y_{i}$ and $y_{j}$ are
the the elements of computational basis. For the simplest choice of
$\ket{\psi^{\prime}}$, we choose $y_{i}={\ket{0}}^{n-1}\otimes\ket{0}$
and $y_{j}={\ket{0}}^{n-1}\otimes\ket{1}$.}

\begin{figure}
\begin{centering}
\textcolor{black}{\includegraphics[scale=0.5]{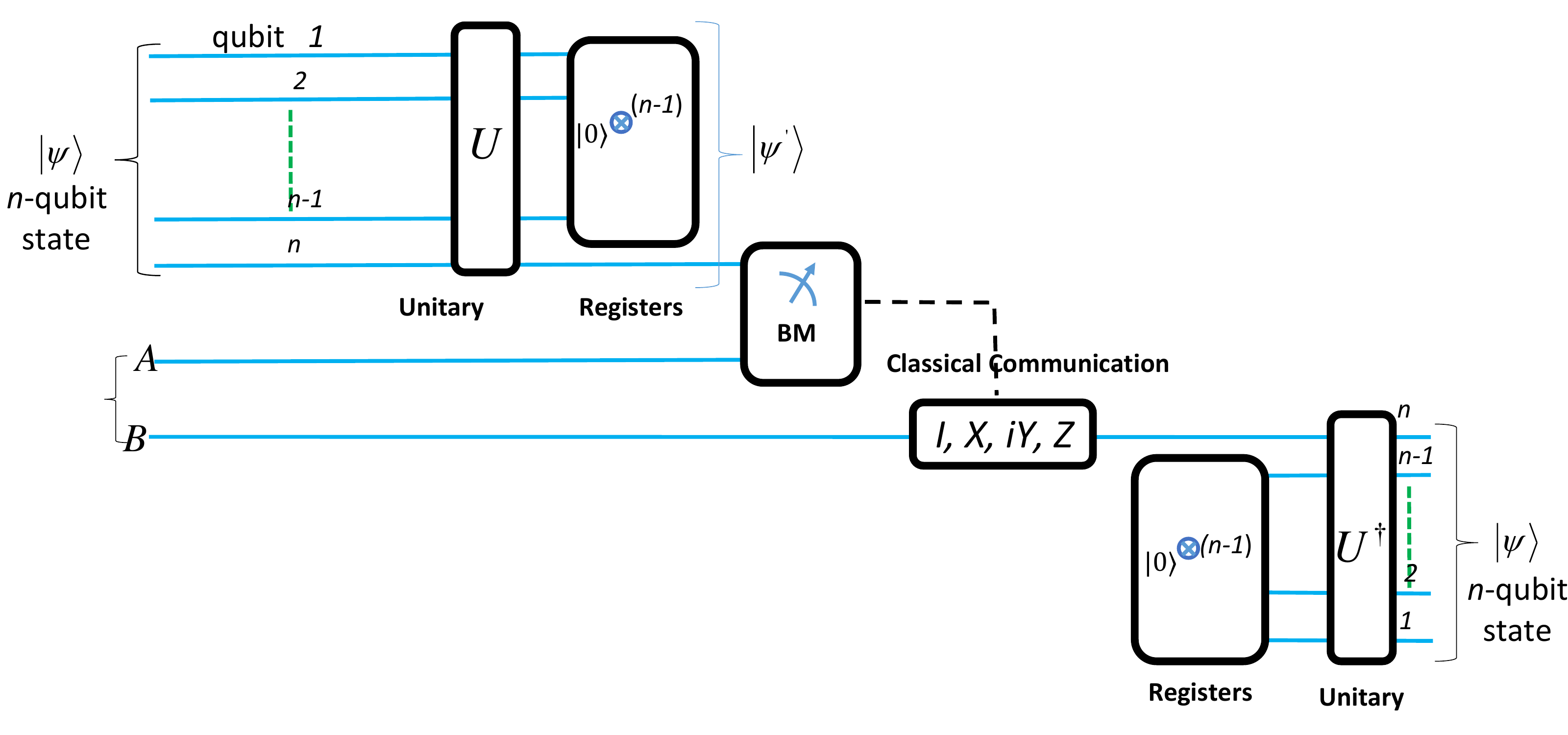}}
\par\end{centering}
\textcolor{black}{\caption{\label{fig:2un} \textcolor{black}{As an explicit example, a quantum
circuit (which uses a single Bell state as quantum channel) for the
the teleportation of an $n$-qubit quantum state having two unknown
coefficients is shown. }}
}
\end{figure}

\textcolor{black}{Now, we will show that it is possible to teleport
state $\ket{\psi^{\prime}}$ using one $e$ bit (Bell state) and classical
resource. The quantum circuit for teleporting state $\ket{\psi}$
is given in Figure \ref{fig:2un}. The first part of the quantum circuit
shows transformation of the state $\ket{\psi}$ to the state $\ket{\psi^{\prime}}$
while the second part of the circuit contains standard scheme for
teleporting an arbitrary single-qubit state. The third part of the
circuit involves application of the unitary $U^{\dagger}$ to transform
the reconstructed state $\ket{\psi^{\prime}}$ into the unknown state
$\ket{\psi}$ to be teleported.}

\textcolor{black}{}

\section{\textcolor{black}{Controlled and bidirectional teleportation with
optimal resource \label{sec:Controlled-and-Bidirectional}}}

\textcolor{black}{\label{cbt} Controlled teleportation of a single-qubit
involves a third party (Charlie) as supervisor and hence instead of
a Bell state we require tripartite entangled state as quantum channel.
As mentioned in Section \ref{sec:Introduction-chap2}, many of the
papers which involve multi-qubit complex states for teleportation
also perform controlled teleportation using those multi-qubit complex
states. Here, we extend our scheme, for optimal QT to optimal CT.
The scheme of CT using optimal resources can be explained along the
same line of the QT scheme as follows.}

\textcolor{black}{To construct an optimal scheme, it is assumed that
Charlie also knows the unitary Alice is using to reduce the size of
the quantum state to be teleported. In other words, he is aware of
the number of entangled qubits Alice and Bob require to perform teleportation.
Suppose the reduced quantum state has $m^{\prime}$ qubits, Charlie
prepares $m^{\prime}$ GHZ states and share the three qubits among
Alice, Bob and himself. Charlie measures his qubit in $\{\ket{+},\ket{-}\}$
basis and withholds the measurement outcome. Independently, Alice
and Bob perform the QT scheme with the only difference that Bob requires
Charlie's measurement disclosure to reconstruct the state. Charlie
announces the required classical information when he wishes Bob to
reconstruct the state.}

\textcolor{black}{Similarly, when both Alice and Bob wish to teleport
a quantum state each to Bob and Alice, respectively, under the supervision
of Charlie, they perform QT schemes independently, while Charlie prepares
the quantum channel in such a way that after his measurement the reduced
state is the product of $2m^{\prime}$ Bell states (half of which
will be used for Alice to Bob, while the remaining half for Bob to
Alice communication). Charlie's disclosure of his measurement outcomes
end both Alice's and Bob's ignorance regarding the quantum channel
they were sharing, and they can subsequently reconstruct the unknown
quantum states teleported to them (see \cite{thapliyal2015general}
for detail). In Ref. \cite{thapliyal2015applications}, it is shown that BCST can also be accomplished solely
using Bell states. Therefore, CT and BCST can also be performed using
only bipartite entanglement. In the absence of the controller, a BCST
scheme can be reduced to a BST scheme. Our results indicate that some
of the recent schemes of CT using four-qubit cluster state \cite{song2008controlled}
and quantum information splitting using four and five-qubit cluster
state \cite{muralidharan2008quantum,nie2011quantum}; BST using three-qubit
GHZ state \cite{hassanpour2016bidirectional}; and six-qubit cluster
state \cite{li2016asymmetric}, can also be performed with reduced
amount of quantum resources (entangled states involving lesser number
of qubits).}

\section{\textcolor{black}{Experimental implementation of the proposed efficient
QT scheme using IBM's real quantum processor \label{sec:Experimental-implementation}}}

\textcolor{black}{In Section \ref{sec:SQUID-based}, we have already
mentioned that recently, IBM corporation has placed a five-qubit superconductivity-based
quantum computer on cloud \cite{IBMQE}, and has provided its access
to everyone. This initiative has enabled the interested researchers
to experimentally realize various proposals for quantum information
processing tasks. Interestingly, a set of superconductivity-based
implementations of quantum computer that are similar to the technology
used in developing IBM's five-qubit quantum computer, have already
been reported \cite{fedortchenko2016quantum,rundle2016quantum,devitt2016performing}.
In Section \ref{sec:SQUID-based}, technical aspects of the IBM quantum
computer has already been reviewed briefly. In addition to what has
already been told in the previous chapter, we may note that currently,
the five-qubit superconductivity-based quantum computers have many
limitations, like available gate library is only approximately universal,
measurement of individual qubits at different time points is not allowed,
limited applicability of CNOT gate, and short decoherence time \cite{IBMDT}.
Further, the real quantum computer (IBM quantum experience) available
at cloud allows a user to perform an experiment using at most five-qubits.
Keeping these limitations in mind, we have chosen a two-qubit two-unknown
quantum state $\ket{\psi}=\alpha(\ket{00}+\ket{11})+\beta(\ket{01}-\ket{10})$
: $2(|\alpha|^{2}+|\beta|^{2})=1$ as the state to be teleported.}

\begin{figure}
\begin{centering}
\textcolor{black}{\includegraphics[scale=0.85]{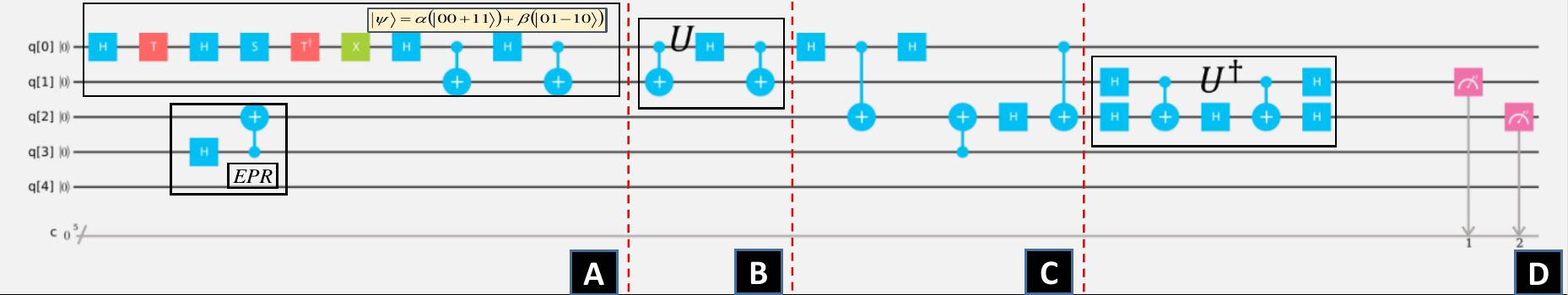}}
\par\end{centering}
\textcolor{black}{\caption{\label{fig:sim} Teleportation circuit used for the teleportation
of two-qubit unknown quantum state $\left(\alpha\left(\ket{00}+\ket{11}\right)+\beta\left(\ket{01}-\ket{10}\right)\right)$
using single Bell state as a quantum channel. This circuit which is
implemented using five-qubit IBM quantum computer, is divided into
four parts. In (A) part state preparation is done, where an EPR channel
and a quantum state $\ket{\psi}$ with $\abs{\alpha}^{2}=0.375$ and
$\abs{\beta}^{2}=0.125$ are prepared. In (B) the decomposition of
the unitary operation $U$ which transforms $\left(\alpha\left(\ket{00}+\ket{11}\right)+\beta\left(\ket{01}-\ket{10}\right)\right)$
to $\left(\alpha\ket{00}+\beta\ket{10}\right)$ in the computational
basis is shown. In (C) teleportation of a single-qubit state is realized,
and In (D) $\ket{\psi}$ is reconstructed from the teleported single-qubit
state by applying $U^{\dagger}$ followed by projective measurement.}
}
\end{figure}

\textcolor{black}{Here, we would like to mention that implementation
of a single-qubit teleportation protocol (which requires only three
qubits) has already been demonstrated using IBM's quantum computer
\cite{fedortchenko2016quantum}. The experimental implementation of
the present QT scheme is relatively complex and can be divided into
four parts as shown in Figure \ref{fig:sim}. Part A involves preparation
of state $\ket{\psi}$ (using qubit q{[}0{]} and q{[}1{]}) and a Bell
state (using qubit q{[}2{]} and q{[}3{]}). The complex circuit comprised
of the quantum gates from Clifford group is shown in Figure \ref{fig:sim}.
Here, it may be noted that the IBM quantum computer accepts quantum
gates from Clifford group only. The state $\ket{\psi}$ in this particular
case is prepared with $|\alpha|^{2}=0.375$ and $|\beta|^{2}=0.125$.
Preparation of the desired two-qubit state by application of specific
quantum gates is detailed in the following.}
\begin{lyxlist}{00.00.0000}
\item [{$\ket{00}\xrightarrow{\hspace{0.2cm}{\mathrm{H}^{1},\,\mathrm{T}^{1}}\hspace{0.2cm}}\left(\frac{\ket{0}+e^{i\frac{\pi}{4}}\ket{1}}{\sqrt{2}}\right)\ket{0}\xrightarrow{\hspace{0.2cm}{\mathrm{H}^{1},\,\mathrm{S}^{1}}\hspace{0.2cm}}e^{i\frac{\pi}{8}}(\cos\left({\frac{\pi}{8}}\right)\ket{0}+\sin\left({\frac{\pi}{8}}\right)\ket{1})\ket{0}\xrightarrow{\hspace{0.2cm}{{\mathrm{T}^{\dagger}}^{1},\,\mathrm{X}^{1},\,\mathrm{H}^{1}}\hspace{0.2cm}}$}]~
\item [{\textcolor{black}{$\sqrt{2}e^{i\frac{\pi}{8}}({\alpha\ket{0}+\beta\ket{1}})\ket{0}$$\xrightarrow{\hspace{0.2cm}{\mathrm{C^{1}-NOT^{2}}}\hspace{0.2cm}}\sqrt{2}e^{i\frac{\pi}{8}}(\alpha\ket{00}+\beta\ket{11})\xrightarrow{\hspace{0.2cm}\mathrm{H}^{1}\hspace{0.2cm}}e^{i\frac{\pi}{8}}\left(\alpha\left(\ket{00}+\ket{10}\right)\right.$}}]~
\item [{\textcolor{black}{$\left.\text{+\ensuremath{\beta\ \left(\ket{01}-\ket{11}\right)}}\right)\xrightarrow{\hspace{0.2cm}\mathrm{C^{1}-NOT^{2}}\hspace{0.2cm}}$$e^{i\frac{\pi}{8}}\left(\alpha\left(\ket{00}+\ket{11}\right)+\beta\left(\ket{01}-\ket{10}\right)\right).$}}]~
\end{lyxlist}
\textcolor{black}{Here, $e^{i\frac{\pi}{8}}$ is the global phase
in the simulated quantum state with $\alpha=\frac{1}{\sqrt{2}}\left(\cos{\frac{\pi}{8}}+e^{-i\frac{\pi}{4}}\sin{\frac{\pi}{8}}\right)$,
and $\beta=\frac{1}{\sqrt{2}}\left(-\cos{\frac{\pi}{8}}+e^{-i\frac{\pi}{4}}\sin{\frac{\pi}{8}}\right)$.
We have explicitly mentioned the qubit-number on which a particular
operation is to be performed by mentioning the qubit-number on the
the superscript of the corresponding unitary operator.}

\textcolor{black}{}
\textcolor{black}{As described in Section \ref{2un}, Part B involves
application of a unitary $U=\left(\mathrm{C^{2}-NOT^{1}}\right)\cdot\left(\mathbb{I}\otimes H\right)\cdot\left(\mathrm{C^{2}-NOT^{1}}\right)$
to transform the state $\ket{\psi}$ from the entangled basis to the
computational basis and is the bottleneck of the protocol. Such a
transformation allows us to render the information encoded into a
smaller number of qubits (in our example it is a single-qubit) and
thus it reduces the amount of resources required. Part C is dedicated
to the teleportation of a single-qubit state. Here, we have used computational
counterpart of teleportation \cite{adami1999quantum}, which can be
performed when both Alice's and Bob's qubits are locally available
for a two-qubit operation. Teleportation of a single-qubit state in
analogy of Ref. \cite{fedortchenko2016quantum} can also be performed.
This part of the circuit can be divided into two sub-parts. The first
one (left aligned), which includes an EPR circuit, entangles qubit
q{[}1{]} to the Bell state while the second part (right aligned) disentangles
Bob's qubit (q{[}3{]}) from Alice's qubits. The need of disentangling
Bob's qubit from Alice's qubit is explained below. In the standard
protocol for QT \cite{bennett1993teleporting}, Alice measures her
qubits and announces measurement outcomes. Depending on the measurement
outcome of Alice, Bob applies a unitary operation and reconstructs
the unknown state. In IBM's quantum computer, simultaneous measurement
of all the qubits is mandatory, which will project Bob's qubit into
a mixed state. Therefore, we preferred to disentangle Bob's qubit
from Alice's qubits before measurement. We would like to mention here
that an optical implementation of the CNOT gate (also Bell measurement)
can only work probabilistically using linear optics, while in contrast
superconducting qubits allow deterministic CNOT operation. The challenges
of experimental implementation of quantum teleportation using optical
qubits are not addressed in the present work. At last, in Part D,
Bob applies the unitary $U^{\dagger}$ followed by the projective
measurement on all qubits, which reveals the state $\ket{\psi}$ teleported
to Bob's qubits. To perform a quantitative analysis of the performance
of the QT scheme under consideration, we would require the density
matrices of the state to be teleported and that of the teleported
state. In a recent implementation of QT on IBM computer only probabilities
of various outcomes were obtained \cite{fedortchenko2016quantum}.
However, to obtain the full picture, we need to reconstruct the density
matrix of the teleported state using QST \cite{chuang1998bulk} which
in turn requires extraction of information from the experiments and
the subsequent use of that information in the reconstruction of the
experimental density matrix. From the method of QST described in Section
\ref{sec:Basic-idea-of}, we can easily recognize that the reconstruction
of two-qubit state requires nine experiments. Using the above method
we have reconstructed the teleported state (using nine rounds of experiments
with 8192 runs of each experiment) as}

\textcolor{black}{
\begin{equation}
\rho^{\prime\prime}=\left[\begin{array}{cccc}
0.41 & 0.013+0.077i & 0.085-0.19i & 0.204-0.054i\\
0.013-0.077i & 0.134 & -0.065-0.016i & -0.021-0.051i\\
0.085+0.19i & -0.065+0.016i & 0.261 & 0.101+0.035i\\
0.204+0.054i & -0.021+0.051i & 0.101-0.035i & 0.195
\end{array}\right],\label{eq:-2}
\end{equation}
whereas theoretically the state prepared for the teleportation is
$\rho=\ket{\Psi}\bra{\Psi}$ with $\ket{\Psi}=$}

\textcolor{black}{$\left\{ \alpha\left(\ket{00}+\ket{11}\right)+\beta\left(\ket{01}-\ket{10}\right)\right\} .$}

\textcolor{black}{During experimental implementation the state prepared
may also have some errors. Keeping this in mind, we have reconstructed
the density matrix of the quantum state (which is to be teleported)
generated in the experiment as}

\textcolor{black}{
\begin{equation}
\rho^{\prime}=\left[\begin{array}{cccc}
0.352 & -0.080+0.104i & 0.18-0.133i & 0.313-0.005i\\
-0.080-0.104i & 0.135 & -0.092-0.017i & -0.099-0.12i\\
0.18+0.133i & -0.092+0.017i & 0.175 & 0.150+0.101i\\
0.313+0.005i & -0.099+0.12i & 0.150-0.101i & 0.338
\end{array}\right].\label{eq:-3}
\end{equation}
}

\begin{figure}
\begin{centering}
\textcolor{black}{\includegraphics[scale=0.65]{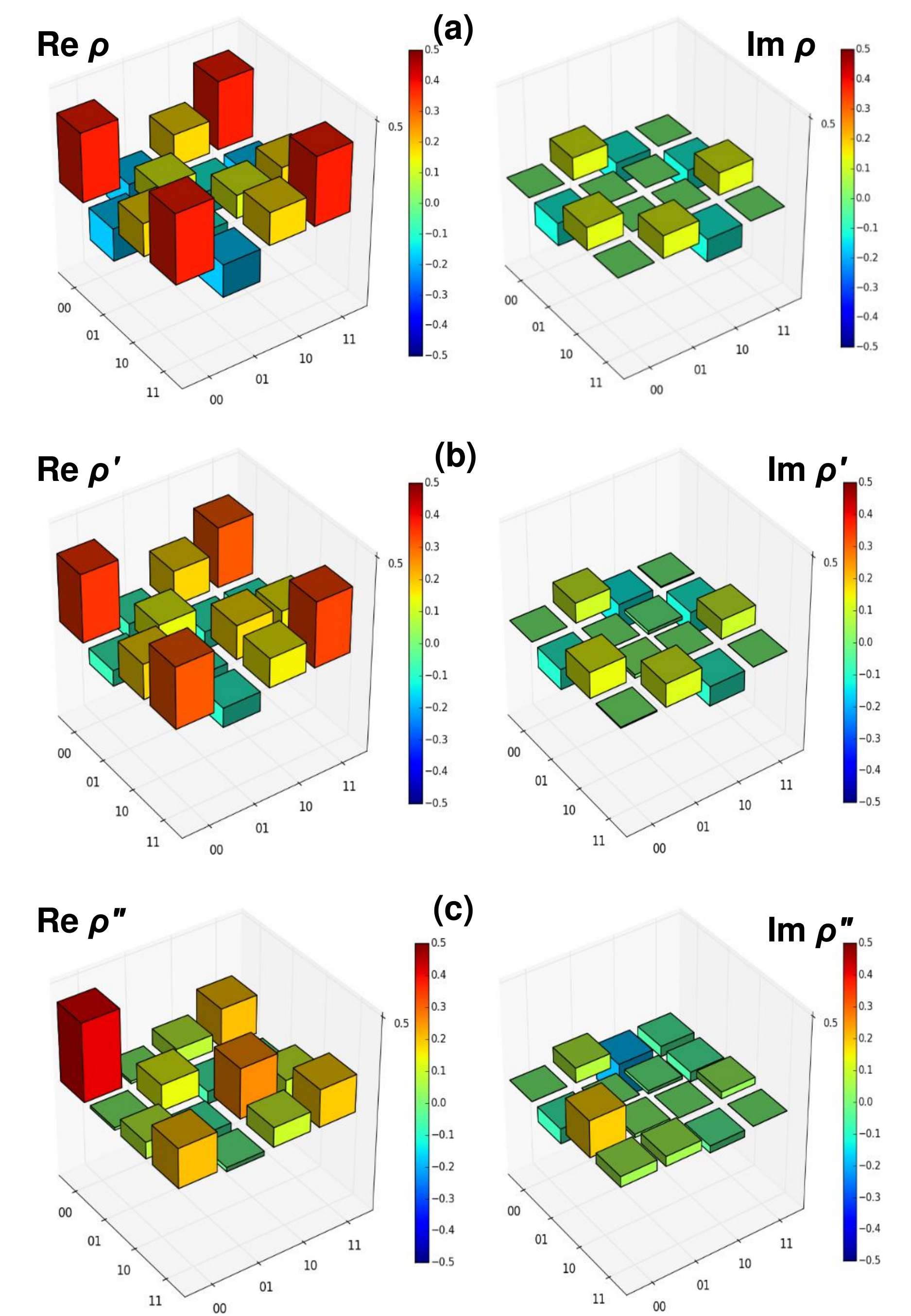}}
\par\end{centering}
\textcolor{black}{\caption{\label{fig:tom} Graphical representation of real (Re) and imaginary
(Im) parts of the density matrices of (a) the theoretical state $\alpha(\ket{00}+\ket{11})+\beta(\ket{01}-\ket{10})$,
(b) the experimentally prepared state, and (c) the reconstructed state
after teleportation.}
}
\end{figure}

\textcolor{black}{Various elements of all the density matrices are
shown pictorially in Figure \ref{fig:tom}. Finally, we would like
to quantize the performance of the QT scheme using a distance-based
measure, i.e. fidelity, which is defined as in Section \ref{subsec:Fidelity-as-a}.
Using this, we calculated the fidelity between the theoretical state
with experimentally generated state (i.e., $\rho^{1}=\rho$ and $\rho^{2}=\rho^{\prime}$)
as 0.9221. The same calculation performed between experimentally generated
and teleported state (i.e., $\rho^{1}=\rho$ and $\rho^{2}=\rho^{\prime\prime}$)
yields a higher value for fidelity (0.9378). Thus, the state preparation
is relatively more erroneous, due to errors in gate implementation
and decoherence. However, the constructed state is found to be teleported
with high fidelity.}

\section{\textcolor{black}{Conclusion\label{sec:Conclusion-2}}}

\textcolor{black}{Teleportation of multi-qubit states with the optimal
amount of quantum resources in terms of the number of entangled qubits
required in the quantum channel has been performed. Specifically,
the amount of quantum resources required to teleport an unknown quantum
state is found to depend (be independent of) on the number of non-zero
probability amplitudes in the quantum state (the number of qubits
in the state to be teleported). Also, the choice of unitary operation
essentially exploits the available information regarding the quantum
state, i.e., only the number of non-zero coefficients and bases $\{x_{i}\}$
and $\{y_{i}\}$, and not on the values of these unknown parameters.
This makes our proposal quite general in nature and manifests its
wide applicability.}

\textcolor{black}{We have explicitly established that the complex
multi-partite entangled states that are used in a large number of
recent works on teleportation (cf. Table \ref{tab:A-list-of}) are
not required for teleportation. Extending this argument one can show
that the complex multi-partite entangled states used for dense-coding
in Refs. \cite{muralidharan2008perfect,tsai2011dense} are not required,
and the task can be performed using an optimal number of Bell states.}

\textcolor{black}{Further, the relevance of the present work is not
restricted to QT. It is also useful in CT, BST and bidirectional controlled
state teleportation schemes. The relevance of the present work also
lies in the fact that the limiting cases of our scheme can perform
the same task with reduced amount of quantum resources in comparison
with the previously achieved counterparts. In fact, for almost all
the existing works reported on teleportation of multi-qubit states
with some non-zero unknowns, we have shown a clear prescription to
optimize the required quantum resources.}

\textcolor{black}{Finally, a proof-of-principle experimental implementation
of the proposed scheme is performed using the IBM quantum computer.
Experimental results are rigorously analyzed. This quantitative analysis
infer that the teleportation circuit implemented here is more efficient
when compared with the state preparation part. This fact establishes
the relevance of the proposed scheme in context of reduction of the
decoherence effects on teleportation, too. As evident from the results
of our four-qubit experiments, the experimental architecture provided
by the IBM quantum experience facility is not sustainable to gate
errors and decoherence. We believe that there exist techniques that
can be used to protect coherence against gate error and decoherence.
For example, IBM may use gates protected by dynamical decoupling to
reduce error; alternatively, in future they may reduce error by using
logic qubits instead of the physical qubits, but that would require
a relatively large physical qubit register. We believe, in order to
provide a reliable quantum computing architecture, incorporation of
these techniques would play an important role.}

\textcolor{black}{We hope our attempt to optimize the resource requirement
for teleportation of multi-qubit quantum states should increase the
feasibility of multi-qubit quantum state teleportation performed in
various quantum systems. This is also expected to impact the teleportation-based
direct secure quantum communication scheme, where resources can be
optimized exploiting the form of the quantum state teleported (e.g.,
\cite{joy2017efficient} and references therein). Along the same line,
optimization of quantum resources in CT, without affecting the controller's
power, will be performed and reported elsewhere.}
\begin{center}
\textbf{\Large{}\newpage}{\Large\par}
\par\end{center}

\chapter[CHAPTER \thechapter \protect\newline QUANTUM TELEPORTATION OF AN EIGHT-QUBIT
STATE USING OPTIMAL QUANTUM RESOURCES]{\textsc{\textcolor{black}{QUANTUM TELEPORTATION OF AN EIGHT-QUBIT
STATE USING OPTIMAL QUANTUM RESOURCES\label{cha:comment}}}}

\textcolor{black}{\large{}\lhead{}}{\large\par}

\section{\textcolor{black}{Introduction\label{sec:Introduction-1}}}

\textcolor{black}{In the previous chapter, we have already mentioned
that the original protocol for quantum teleportation was designed
for the teleportation of a single-qubit state using a Bell state \cite{bennett1993teleporting}.
Subsequently, many schemes have been proposed for the teleportation
of multi-qubit states using various entangled states. Following the
trend, recently, Zhao et al., have proposed a scheme for the teleportation
of the following quantum state}

\textcolor{black}{
\begin{equation}
\begin{array}{lcc}
|\varphi\rangle_{abcdefgh} & = & \left(\alpha|00000000\rangle+\beta|00100000\rangle+\gamma|11011111\rangle+\delta|11111111\rangle\right)_{abcdefgh},\end{array}\label{eq:1}
\end{equation}
where the coefficients $\alpha,\,\beta,\,\gamma,\,\delta$ are unknown
and satisfies $\left|\alpha\right|^{2}+\left|\beta\right|^{2}+\left|\gamma\right|^{2}+\left|\delta\right|^{2}=1$
(cf. Eq. (1) of \cite{zhao2018quantum}). They considered this state
as an eight-qubit quantum state and proposed a scheme for teleportation
of this state using the following six-qubit cluster state described
as}

\textcolor{black}{
\begin{equation}
\begin{array}{lcc}
|\phi\rangle_{123456} & = & \left(\alpha|000000\rangle+\beta|001001\rangle+\gamma|110110\rangle+\delta|111111\rangle\right)_{123456}.\end{array}\label{eq:2-1}
\end{equation}
as the quantum channel (cf. Eq. (2) of \cite{zhao2018quantum}). The
basic conceptual problem with the form of this channel is that this
channel cannot be constituted as the coefficients $\alpha,\,\beta,\,\gamma,\,\delta$
present in the expansion of $|\varphi\rangle_{abcdefgh}$ are unknown.
Further, Eq. (2) of \cite{zhao2018quantum} is not consistent with
Eqs. (5)-(6) of \cite{zhao2018quantum} and thus with the remaining
part of \cite{zhao2018quantum}. To stress on the more important issues
and to continue the discussion, we may consider that Zhao et al.,
actually intended to use 
\begin{equation}
|\phi\rangle_{123456}=\frac{1}{2}\left(|000000\rangle+|001001\rangle+|110110\rangle+|111111\rangle\right),\label{eq:tin}
\end{equation}
which is consistent with Eq. (3) of \cite{zhao2018quantum}. However,
the above mistake does not appear to be a typographical error as the
same error is present in another recent work of the authors (see \cite{zhao2017efficient}).
The more important question is whether we need Eq. (\ref{eq:tin})
for the teleportation of $|\varphi\rangle_{abcdefgh}$, or the task
can be performed using a simpler quantum channel. This is the question,
we wish to address in this chapter. Here, it is important to note
that in a recent work \cite{sisodia2017design}, we have shown that
a quantum state having $n$ unknown coefficients can be teleported
by using $ \left\lceil \log_{2}{n}\right\rceil $ number of 
Bell states. Now, as there are four unknowns in $|\varphi\rangle_{abcdefgh},$
teleportation of this state should require only two Bell states. This
point can be further illustrated by noting that Zhao et al., have
shown that using four CNOT gates (cf. Figure 1 of \cite{zhao2018quantum})
$|\varphi\rangle_{abcdefgh}$ can be transformed to a state $|\varphi^{\prime}\rangle_{abcdefgh}=|\chi\rangle_{abcd}|0000\rangle_{efgh}$,
where}

\textcolor{black}{
\begin{equation}
\begin{array}{lcc}
|\chi\rangle_{abcd} & = & \left(\alpha|0000\rangle+\beta|0010\rangle+\gamma|1101\rangle+\delta|1111\rangle\right)_{abcd},\end{array}\label{eq:3}
\end{equation}
and thus the actual teleportation task reduces to the teleportation
of $|\chi\rangle_{abcd}$. Now, we can introduce a circuit which can
be described as ${\rm CNOT}_{a\rightarrow d}{\rm {\rm CNOT}_{a\rightarrow b}SWAP}_{bc}$,
where ${\rm SWAP}_{ij}$ and ${\rm CNOT}_{i\rightarrow j}$ correspond
to a gate that swaps $i$th and $j$th qubit and a ${\rm CNOT}$ that
uses $i$th qubit as the control qubit and $j$th qubit as the target
qubit, respectively. On application of this circuit, $|\chi\rangle_{abcd}$
would transform to $\left(\alpha|00\rangle+\beta|01\rangle+\gamma|10\rangle+\delta|11\rangle\right)_{ac}|00\rangle_{bd}$
as}

\begin{equation}
\begin{array}{lcl}
{\rm CNOT}_{a\rightarrow d}{\rm {\rm CNOT}_{a\rightarrow b}SWAP}_{bc}|\chi\rangle_{abcd} & = & \left(\alpha|00\rangle+\beta|01\rangle+\gamma|10\rangle+\delta|11\rangle\right)_{ac}|00\rangle_{bd}\\
 & = & |\psi\rangle_{ac}|00\rangle_{bd}
\end{array}.\label{eq:new1}
\end{equation}
\textcolor{black}{Thus, a scheme for teleportation of an arbitrary
two-qubit state $|\psi\rangle_{ac}$ will be sufficient for the teleportation
of $|\varphi\rangle_{abcdefgh}.$ Such a scheme for the QT of an arbitrary
two-qubit state was proposed by Rigolin in 2005 \cite{rigolin2005quantum}.
He proposed sixteen quantum channels that are capable of teleportation
of the arbitrary two-qubit states and referred to those channels as
generalized Bell states. However, soon after the work of Rigolin,
in a comment on it, Deng \cite{deng2005comment} had shown that sixteen
channels introduced by Rigolin were simply the sixteen possible product
states of the 4 Bell states, and thus the work of Deng and Rigolin
established the fact that the product of any two Bell states is sufficient
for the teleportation of an arbitrary two-qubit state. Result of Deng
and Rigolin is consistent with the more general result of ours \cite{sisodia2017design},
which is presented in the previous chapter and that of \cite{thapliyal2015applications,thapliyal2015general,pathak2011efficient}.
This clearly establishes that in addition to the circuit block comprise
of four CNOT gates used in the work of Zhao et al., if one uses the
circuit described above and well known scheme of Rigolin, then one
will be able to teleport so-called an eight-qubit state $|\varphi\rangle_{abcdefgh}$
using only two Bell states. This comment could have been concluded
at this point, but for the convenience of the readers we have added
the following section, where we briefly outline the complete process
of teleportation of $|\varphi\rangle_{abcdefgh}$ using two Bell states.}

\textcolor{black}{}
\begin{figure}
\begin{centering}
\textcolor{black}{\includegraphics[scale=0.6]{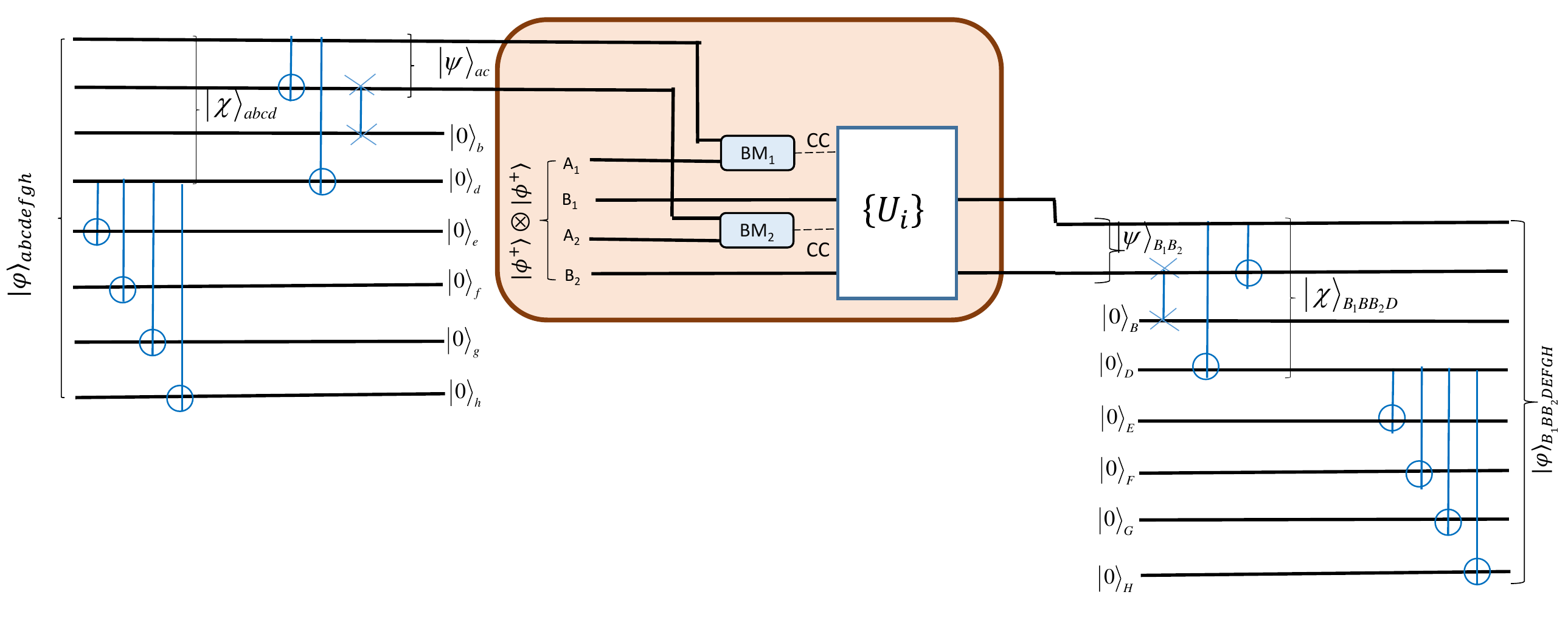} }
\par\end{centering}
\textcolor{black}{\caption{\label{fig:ckt}\textcolor{black}{Teleportation circuit for the teleportation
of an eight-qubit unknown quantum state (i.e., the state described
by Eq. (\ref{eq:1}) using two Bell states (optimal amount of quantum
resource). Here, BM and CC stand for Bell measurement and classical
communication, respectively. $U_{i}$ is the unitary operation required
to reconstruct the quantum state we wish to teleport.}}
}
\end{figure}

\section{\textcolor{black}{Complete teleportation process \label{sec:CTP}}}

\textcolor{black}{Teleportation of $|\varphi\rangle_{abcdefgh}$ is
accomplished in three steps as illustrated in Figure \ref{fig:ckt}.
Firstly, the task of teleportation of the so called eight-qubit quantum
state $|\varphi\rangle_{abcdefgh}$ is reduced to the task of teleporting
a two-qubit quantum state $|\psi\rangle_{ac}$ by transforming $|\varphi\rangle_{abcdefgh}$
into $|\psi\rangle_{ac}|000000\rangle_{bdefgh}$. This is done using
the circuit block (comprised of six CNOT gates and one SWAP gate)
shown in the left side of the complete circuit shown in Figure \ref{fig:ckt}.
The actual teleportation part is the second step which is shown in
the middle block of Figure \ref{fig:ckt} (cf. the rectangular box
in the middle of  Figure \ref{fig:ckt}). In this step, we teleport
$|\psi\rangle_{ac}$ using two Bell states as quantum channel. As
an example, consider. $|\psi^{+}\rangle_{A_{1}B_{1}}\otimes|\psi^{+}\rangle_{A_{2}B_{2}}=$$\frac{|00\rangle+|11\rangle}{\sqrt{2}}\otimes\frac{|00\rangle+|11\rangle}{\sqrt{2}}$
as the quantum channel. Thus, the combined state would be}

\textcolor{black}{
\begin{equation}
\begin{array}{lcc}
|\Phi\rangle_{acA_{1}B_{1}A_{2}B_{2}} & = & |\psi\rangle_{ac}\otimes|\psi^{+}\rangle_{A_{1}B_{1}}\otimes|\psi^{+}\rangle_{A_{2}B_{2}},\end{array}\label{eq:5}
\end{equation}
which can be decomposed as}

\textcolor{black}{
\begin{equation}
{\color{black}{\color{black}\begin{array}{lcl}
|\Phi\rangle_{acA_{1}B_{1}A_{2}B_{2}} & = & {\color{black}{\color{black}{\color{blue}{\color{blue}{\color{black}|\psi^{+}\rangle_{aA_{1}}|\psi^{+}\rangle_{cA_{2}}I\otimes I\,|\psi\rangle_{B_{1}B_{2}}+|\psi^{+}\rangle_{aA_{1}}|\psi^{-}\rangle_{cA_{2}}I\otimes Z\,|\psi\rangle_{B_{1}B_{2}}}}}}}\\
 & + & {\color{black}{\color{black}{\color{black}|{\color{black}\psi^{+}\rangle_{aA_{1}}|\phi^{+}\rangle_{cA_{2}}I\otimes X\,|\psi\rangle_{B_{1}B_{2}}+|\psi^{+}\rangle_{aA_{1}}|\phi^{-}\rangle_{cA_{2}}I\otimes iY\,|\psi\rangle_{B_{1}B_{2}}}}}}\\
 & + & {\color{black}{\color{black}|\psi^{-}\rangle_{aA_{1}}|\psi^{+}\rangle_{cA_{2}}Z\otimes I\,|\psi\rangle_{B_{1}B_{2}}+|\psi^{-}\rangle_{aA_{1}}|\psi^{-}\rangle_{cA_{2}}Z\otimes Z\,|\psi\rangle_{B_{1}B_{2}}}}\\
 & + & {\color{black}{\color{black}|\psi^{-}\rangle_{aA_{1}}|\phi^{+}\rangle_{cA_{2}}Z\otimes X\,|\psi\rangle_{B_{1}B_{2}}+|\psi^{-}\rangle_{aA_{1}}|\phi^{-}\rangle_{cA_{2}}Z\otimes iY\,|\psi\rangle_{B_{1}B_{2}}}}\\
 & + & {\color{black}{\color{black}|\phi^{+}\rangle_{aA_{1}}|\psi^{+}\rangle_{cA_{2}}X\otimes I\,|\psi\rangle_{B_{1}B_{2}}+|\phi^{+}\rangle_{aA_{1}}|\psi^{-}\rangle_{cA_{2}}X\otimes Z\,|\psi\rangle_{B_{1}B_{2}}}}\\
 & + & {\color{black}|\phi^{+}\rangle_{aA_{1}}|\phi^{+}\rangle_{cA_{2}}X\otimes X\,|\psi\rangle_{B_{1}B_{2}}+|\phi^{+}\rangle_{aA_{1}}|\phi^{-}\rangle_{cA_{2}}X\otimes iY\,|\psi\rangle_{B_{1}B_{2}}}\\
 & + & |\phi^{-}\rangle_{aA_{1}}|\psi^{+}\rangle_{cA_{2}}iY\otimes I\,|\psi\rangle_{B_{1}B_{2}}+|\phi^{-}\rangle_{aA_{1}}|\psi^{-}\rangle_{cA_{2}}iY\otimes Z\,|\psi\rangle_{B_{1}B_{2}}\\
 & + & |\phi^{-}\rangle_{aA_{1}}|\phi^{+}\rangle_{cA_{2}}iY\otimes X\,|\psi\rangle_{B_{1}B_{2}}+|\phi^{-}\rangle_{aA_{1}}|\phi^{-}\rangle_{cA_{2}}iY\otimes iY\,|\psi\rangle_{B_{1}B_{2}}.
\end{array}}}\label{eq:6}
\end{equation}
Eq. (\ref{eq:6}) clearly shows that Alice's Bell measurement on qubits
$aA_{1}$ and $cA_{2}$ reduces Bob's qubits $B_{1}B_{2}$ in such
a quantum state that the state $|\psi\rangle_{B_{1}B_{2}}$ can be
obtained by applying local unitary operations. The choice of the unitary
operations depends on the Alice's measurement outcome (as illustrated
in Eq. (\ref{eq:6})). After teleporting $|\psi\rangle$, in the third
step, the initial state to be teleported is reconstructed from $|\psi\rangle_{B_{1}B_{2}}$
by using six ancillary qubits prepared in $\left(|0\rangle^{\otimes6}\right)_{BDEFGH}$
and the circuit block shown in the right side of Figure \ref{fig:ckt}.
At the output of this circuit block we would obtain the so called
eight-qubit state $|\varphi\rangle_{B_{1}BB_{2}DEFGH}$ that we wanted
to teleport.}

\section{\textcolor{black}{Concluding remark \label{sec:Conclusion3}}}

\textcolor{black}{Despite the existence of old results of \cite{rigolin2005quantum,thapliyal2015applications,thapliyal2015general,pathak2011efficient,yang2000multiparticle}
and our recent results \cite{sisodia2017design}, people are frequently
proposing teleportation schemes \cite{zhao2017efficient,li2017quantum,choudhury2017teleportation}
and their variants \cite{choudhury2017asymmetric} using higher amount
of quantum resources, although they know that the preparation and
maintenance of such resources are not easy. Specially, preparation
of multi-partite entanglement is difficult. Keeping that in mind,
it is advisable that while designing new schemes, unnecessary use
of quantum resources should be circumvented. The analysis of Zhao
et al., protocol performed here, and the possible improvement shown
here is only an example. It is not restricted Zhao et al., protocol
only. The main idea of it is applicable to many other schemes (\cite{zhao2017efficient,sisodia2017design,li2017quantum,choudhury2017teleportation,choudhury2017asymmetric}
and references therein) as their amount of quantum resources used
by them are higher than the minimum required amount.}

\chapter[CHAPTER \thechapter \protect\newline TELEPORTATION OF A QUBIT USING ENTANGLED NONORTHOGONAL STATES: A
COMPARATIVE STUDY]{TELEPORTATION OF A QUBIT USING ENTANGLED NONORTHOGONAL STATES: A
COMPARATIVE STUDY\textsc{\label{cha:nonorthogonal}}}

{\large{}\lhead{}}{\large\par}

\section{\textcolor{black}{Introduction \label{sec:Introduction4}}}

\textcolor{black}{In the previous chapters, we have already observed
that one of the most important resources for quantum information processing
is entanglement, which is essential for various tasks of quantum communication
and computation. In Chapter \ref{cha:Introduction1}, we have briefly
mentioned such tasks and have also introduced the notion of entangled
nonorthogonal states. In fact, in Section \ref{subsec:Orthogonal-and-nonorthogonal},
interesting applications of entangled nonorthogonal states have been
reviewed briefly with a focus on the entangled coherent states. In
this chapter, we aim to investigate the effect of non-orthogonality
of an entangled nonorthogonal state-based quantum channel in detail
in the context of the teleportation of a qubit.}

\textcolor{black}{In the context of the studies on the applications
nonorthogonal entangled states in quantum communication, the concept
of minimum assured fidelity (MASFI), which was claimed to corresponds
to the least value of possible fidelity for any given information,
was introduced by Prakash et al. \cite{prakash2007improving}. Subsequently,
in a series of papers (\cite{prakash2007effect,prakash2008effect}
and references therein), they have reported MASFI for various protocols
of quantum communication, and specially for the imperfect teleportation.
Here, it is important to note that Adhikari et al., \cite{adhikari2012quantum}
tried to extend the domain of the standard teleportation protocol
to the case of performing teleportation using entangled nonorthogonal
states. To be precise, they studied teleportation of an unknown quantum
state by using a specific type of entangled nonorthogonal state as
the quantum channel. They also established that the amount of nonorthogonality
present in the quantum channel affects the average fidelity ($F_{ave}$)
of teleportation. However, their work was restricted to a specific
type of entangled nonorthogonal state, and neither the optimality
of the scheme nor the effect of noise on it was investigated by them.
In fact, in their work no effort had been made to perform a comparative
study (in terms of different measures of teleportation quality) among
possible quasi-Bell states that can be used as teleportation channel.
Further, the works of Prakash et al., \cite{prakash2007improving,prakash2007effect,prakash2008effect}
and others (\cite{oh2002fidelity,d2000bell} and references therein)
have established that in addition to $F_{ave},$ minimum assured fidelity
(MASFI) and minimum average fidelity (MAVFI), which we refer here
as minimum fidelity (MFI) can be used as measures of quality of teleportation.}

\textcolor{black}{Keeping these points in mind, in the present chapter,
we have studied the effect of the amount of nonorthogonality on $F_{ave}$,
MFI, and MASFI for teleportation of a qubit using different quasi-Bell
states, which can be used as the quantum channel. We have compared
the performance of these quasi-Bell states as teleportation channel
an ideal situation (i.e., in the absence of noise) and in the presence
of various types of noise (e.g., AD and PD). The relevance of the
choice of these noise models has been well established in the past
(\cite{sharma2016comparative,sharma2016verification,thapliyal2015applications}
and references therein). Further, using Horodecki  et al.'s relation
\cite{horodecki1999general} between optimal fidelity ($F_{opt}$)
and maximal singlet fraction ($f$), it is established that the entangled
nonorthogonal state-based teleportation scheme investigated in the
present work, is optimal for all the cases studied here (i.e., for
all the quasi-Bell states).}

\textcolor{black}{The remaining part of the chapter is organized as
follows. In Section \ref{sec:Entangled-Non-orthngooal-States}, we
briefly describe the mathematical structure of the entangled nonorthogonal
states and how to quantify the amount of entanglement present in such
states using concurrence. In this section, we have restricted ourselves
to very short description as most of the expressions reported here
are well known. However, they are required for the sake of a self-sufficient
description. The main results of the present chapter are reported
in Section \ref{sec:Quantum-Teleportation-using}, where we provide
expressions for ${\rm MASFI}$, $F_{ave}$, and $f$ for all the four
quasi-Bell states and establish $F_{ave}=F_{opt}$ for all the quasi-Bell
states, and deterministic perfect teleportation is possible with the
help of quasi-Bell states. In Section \ref{sec:Effect-of-noise},
effects of AD and PD noise on $F_{ave}$ is discussed for various
alternative situations, and finally the chapter is concluded in Section
\ref{sec:Conclusion4}.}

\section{\textcolor{black}{Entangled nonorthogonal states\label{sec:Entangled-Non-orthngooal-States}}}

\textcolor{black}{Basic mathematical structures of standard entangled
states and entangled nonorthogonal states have been provided in detail
in several papers (\cite{sanders1992entangled,pathak2013elements}
and references therein). Schmidt decomposition of an arbitrary bipartite
state is written as 
\begin{equation}
|\Psi\rangle=\underset{i}{\sum}p_{i}|a_{i}\rangle_{A}\otimes|b_{i}\rangle_{B},
\end{equation}
where $p_{i}$s are the real numbers such that $\underset{i}{\sum}p_{i}^{2}=1$.
Further, $\{|a_{i}\rangle_{A}\}$ $\left(\{|b_{i}\rangle_{B}\}\right)$
is the orthonormal basis of subsystem $A$ $\left(B\right)$ in Hilbert
space $H_{A}$ $\left(H_{B}\right)$. The state $|\Psi\rangle$ is
entangled if at least two of the $p_{i}$s are non-zero. Here, we
may note that a standard bipartite entangled state can be expressed
as 
\begin{equation}
|\psi\rangle=\mu|a\rangle_{A}\otimes|b\rangle_{B}+\nu|c\rangle_{A}\otimes|d\rangle_{B},\label{eq:1-1}
\end{equation}
where $\mu$ and $\nu$ are two complex coefficients that ensure normalization
by satisfying $|\mu|^{2}+|\nu|^{2}=1$ in case of orthogonal states;
$|a\rangle$ and $|c\rangle$ are normalized states of the first system
and $|b\rangle$ and $|d\rangle$ are normalized states of the second
system, respectively. These states of the subsystems satisfy $\langle a|c\rangle=0$
and $\langle b|d\rangle=0$ for the conventional entangled states
of orthogonal states and they satisfy $\langle a|c\rangle\neq0$ and
$\langle b|d\rangle\neq0$ for the entangled nonorthogonal states.
Thus, an entangled state involving nonorthogonal states, which is
expressed in the form of Eq. (\ref{eq:1-1}), has the property that
the overlaps $\langle a|c\rangle$ and $\langle b|d\rangle$ are nonzero,
and the normalization condition would be 
\begin{equation}
|\mu|^{2}+|\nu|^{2}+\mu\nu^{*}\langle c|a\rangle\langle d|b\rangle+\mu^{*}\nu\langle a|c\rangle\langle b|d\rangle=1.\label{eq:2}
\end{equation}
Here and in what follows, for simplicity, we have omitted the subsystems
mentioned in the subscript.}

\textcolor{black}{The two nonorthogonal states of a given}\textcolor{blue}{{}
}system are considered to be linearly independent. They are also assumed
to span a 2D subspace of the Hilbert space. W\textcolor{black}{e may
choose an orthonormal basis $\left\{ |0\rangle,|1\rangle\right\} $
as 
\begin{equation}
|0\rangle=|a\rangle,|1\rangle=\frac{(|c\rangle-p_{1}|a\rangle)}{N_{1}},\label{eq:3a}
\end{equation}
for System $A$, and similarly, $|0\rangle=|d\rangle,{|1\ensuremath{\rangle}=\ensuremath{\frac{(|b\rangle-p_{2}|d\rangle)}{N_{2}}}}$
for System $B$, where $p_{1}=\langle a|c\rangle,$ $p_{2}=\langle d|b\rangle$,
and $N_{i}=\sqrt{1-|p_{i}|^{2}}:\,i\in\{1,2\}.$ Now, we can express
the nonorthogonal entangled state $|\psi\rangle$ described by Eq.
(\ref{eq:1-1}) using the orthogonal basis $\left\{ |0\rangle,|1\rangle\right\} $
as follows}

\textcolor{black}{
\begin{equation}
|\psi\rangle=a'|00\rangle+b'|01\rangle+c'|10\rangle,\label{eq:5-1}
\end{equation}
with $a'=(\mu p_{2}+\nu p_{1})N_{12},\,b'=(\mu N_{2})N_{12},\,c'=(\nu N_{1})N_{12},$
where the normalization constant $N_{12}$ is given by}

\textcolor{black}{
\begin{equation}
N_{12}=[|\mu|^{2}+|\nu|^{2}+\mu\nu^{*}\langle c|a\rangle\langle d|b\rangle+\mu^{*}\nu\langle a|c\rangle\langle b|d\rangle]^{-\frac{1}{2}}.\label{eq:normalization}
\end{equation}
Eq. (\ref{eq:5-1}) shows that an arbitrary entangled nonorthogonal
state can be considered as a state of two logical qubits. Following
standard procedure, the concurrence ($C$) \cite{hill1997entanglement,wootters1998entanglement}
of the entangled state $|\psi\rangle$ can be obtained as \cite{mann1995bell,wang2001bipartite,fu2001maximal}}

\textcolor{black}{
\begin{equation}
C=2|b'c'|=\frac{2|\mu||\nu|\sqrt{(1-|\langle a|c\rangle|^{2})(1-|\langle b|d\rangle|^{2})}}{|\mu|^{2}+|\nu|^{2}+\mu\nu^{*}\langle c|a\rangle\langle d|b\rangle+\mu^{*}\nu\langle a|c\rangle\langle b|d\rangle}.\label{eq:7}
\end{equation}
}

\textcolor{black}{For the entangled state $|\psi\rangle$ to be maximally
entangled, we must have }\textit{\textcolor{black}{$C=1$}}\textcolor{black}{.
Fu et al., \cite{fu2001maximal}, showed that the state $|\psi\rangle$
is maximally entangled state if and only if one of the following conditions
is satisfied: (i) $|\mu|=|\nu|$ for the orthogonal case, and (ii)
$\mu=\nu e^{i\theta}$ and $\langle a|c\rangle=-\langle b|d\rangle^{*}e^{i\theta}$
for the nonorthogonal states, where $\theta$ is a real parameter.}

\textcolor{black}{Before we investigate the teleportation capacity
of the entangled nonorthogonal states, we would like to note that
if we choose $\mu=\nu$ in Eq. (\ref{eq:1-1}), then for the case
of orthogonal basis, normalization condition will ensure that $\mu=\nu=\frac{1}{\sqrt{2}}$,
and the state $|\psi\rangle$ will reduce to a standard Bell state
$|\phi^{+}\rangle=\frac{1}{\sqrt{2}}\left(|01\rangle+|10\rangle\right)$,
and its analogous state under the same condition (i.e., for $\mu=\nu$)
in nonorthogonal basis would be $|\phi_{+}\rangle=N_{+}\left(|a\rangle\otimes|b\rangle+|b\rangle\otimes|a\rangle\right),$
where $N_{+}$ is the normalization constant. In analogy to $|\phi^{+}\rangle$
its analogous entangled nonorthogonal state is denoted as $|\phi_{+}\rangle$
and referred to as quasi-Bell state \cite{hirota2001entangled}. Similarly,
in analogy with the other three Bell states $|\phi^{-}\rangle=\frac{1}{\sqrt{2}}\left(|01\rangle-|10\rangle\right),\,|\psi^{+}\rangle=\frac{1}{\sqrt{2}}\left(|00\rangle+|11\rangle\right),$
and $|\psi^{-}\rangle=\frac{1}{\sqrt{2}}\left(|00\rangle-|11\rangle\right),$
we can obtain entangled nonorthogonal states denoted by $|\phi_{-}\rangle,\,|\psi_{+}\rangle,$
and $|\psi_{-}\rangle$, respectively. In addition to these notations,
in what follows we also use $|\psi^{+}\rangle=|\psi_{1}\rangle,\,|\psi^{-}\rangle=|\psi_{2}\rangle,\,|\phi^{+}\rangle=|\psi_{3}\rangle,\,|\phi^{-}\rangle=|\psi_{4}\rangle.$
Four entangled nonorthogonal states $\left\{ |\psi_{\pm}\rangle,\,|\phi_{\pm}\rangle\right\} $,
which are used in this thesis, are usually referred to as quasi-Bell
states \cite{hirota2001entangled}. They are not essentially maximally
entangled, and they may be expressed in orthogonal basis (see last
column of Table \ref{tab:quasi-bell-states}). Notations used in the
rest of the chapter, expansion of the quasi-Bell states in orthogonal
basis, etc., are summarized in Table \ref{tab:quasi-bell-states},
where we can see that $|\psi_{4}\rangle=|\phi^{-}\rangle$ is equivalent
to $|\phi_{-}\rangle,$ and thus $|\phi_{-}\rangle$ is always maximally
entangled and can lead to perfect deterministic teleportation as can
be done using usual Bell states. So $|\phi_{-}\rangle$ is not a state
of interest in noiseless case. Keeping this in mind, in the next section,
we mainly concentrate on the properties related to the teleportation
capacity of the other three quasi-Bell states. However, in Section
\ref{sec:Effect-of-noise}, we would discuss the effect of noise on
all four quasi-Bell states.}

\textcolor{black}{}
\begin{table}
\textcolor{black}{\caption{\label{tab:quasi-bell-states}Bell states and the corresponding quasi-Bell
states. The table shows that the quasi-Bell states can be expressed
in orthogonal basis and it introduces the notation used in this chapter.
Here, $|\psi^{\pm}\rangle=\frac{1}{\sqrt{2}}\left(|00\rangle\pm|11\rangle\right)\,\,|\phi^{\pm}\rangle=\frac{1}{\sqrt{2}}\left(|01\rangle{\rm {\rm \pm}}|10\rangle\right)$,
$\eta=\frac{2re^{i\theta}}{\sqrt{2(1+r^{2})}},\,\epsilon=\sqrt{\frac{1-r^{2}}{2(1+r^{2})}}$,
$k_{\pm}=\frac{1\pm r^{2}e^{2i\theta}}{\sqrt{2(1\pm r^{2}\cos2\theta)}},\,\,l_{\pm}=\frac{(\sqrt{1-r^{2}})re^{i\theta}}{\sqrt{2(1\pm r^{2}\cos2\theta)}},\,\,m_{\pm}=\frac{1-r^{2}}{\sqrt{2(1\pm r^{2}\cos2\theta)}}$,
$N_{\pm}=\left[2\left(1\pm|\langle a|b\rangle|^{2}\right)\right]^{\frac{-1}{2}},$
and $M_{\pm}=\frac{1}{\sqrt{2(1\pm r^{2}\cos2\theta)}}$ represent
the normalization constant.}
}
\centering{}\textcolor{black}{}%
\begin{tabular}{|>{\centering}p{1cm}|>{\centering}p{2cm}|>{\centering}p{5cm}|>{\centering}p{6.3cm}|}
\hline 
\textcolor{black}{\footnotesize{}S. No.} & \textcolor{black}{\footnotesize{}Bell state} & \textcolor{black}{\footnotesize{}Corresponding quasi-Bell state}{\footnotesize\par}

\textcolor{black}{\footnotesize{}(i.e., Bell-like entangled nonorthogonal
state having a mathematical form analogous to the usual Bell state
given in the 2nd column of the same row)} & \textcolor{black}{\footnotesize{}State in orthogonal basis that is
equivalent to the quasi-Bell state mentioned in the 3rd column of
the same row}\tabularnewline
\hline 
\textcolor{black}{\footnotesize{}1.} & \textcolor{black}{\footnotesize{}$|\psi_{1}\rangle=|\psi^{+}\rangle$} & \textcolor{black}{\footnotesize{}$|\psi_{+}\rangle={\rm M}_{+}(|a\rangle\otimes|a\rangle+|b\rangle\otimes|b\rangle)$} & \textcolor{black}{\footnotesize{}$|\psi_{+}\rangle=k_{+}|00\rangle+l_{+}|01\rangle+l_{+}|10\rangle+m_{+}|11\rangle$}\tabularnewline
\hline 
\textcolor{black}{\footnotesize{}2.} & \textcolor{black}{\footnotesize{}$|\psi_{2}\rangle=|\psi^{-}\rangle$} & \textcolor{black}{\footnotesize{}$|\psi_{-}\rangle={\rm M}_{-}(|a\rangle\otimes|a\rangle-|b\rangle\otimes|b\rangle)$} & \textcolor{black}{\footnotesize{}$|\psi_{-}\rangle=k_{-}|00\rangle-l_{-}|01\rangle-l_{-}|10\rangle-m_{-}|11\rangle$}\tabularnewline
\hline 
\textcolor{black}{\footnotesize{}3.} & \textcolor{black}{\footnotesize{}$|\psi_{3}\rangle=|\phi^{+}\rangle$} & \textcolor{black}{\footnotesize{}$|\phi_{+}\rangle=N_{+}(|a\rangle\otimes|b\rangle+|b\rangle\otimes|a\rangle)$} & \textcolor{black}{\footnotesize{}$|\phi_{+}\rangle=\eta|00\rangle+\epsilon|01\rangle+\epsilon|10\rangle$}\tabularnewline
\hline 
\textcolor{black}{\footnotesize{}4.} & \textcolor{black}{\footnotesize{}$|\psi_{4}\rangle=|\phi^{-}\rangle$} & \textcolor{black}{\footnotesize{}$|\phi_{-}\rangle=N_{-}(|a\rangle\otimes|b\rangle-|b\rangle\otimes|a\rangle)$} & \textcolor{black}{\footnotesize{}$|\phi_{-}\rangle=\frac{1}{\sqrt{2}}\left(|01\rangle-|10\rangle\right)$}\tabularnewline
\hline 
\end{tabular}
\end{table}
\textcolor{black}{In what follows, we aim to perform a rigorous and
comparative investigation of the suitability of using quasi-Bell states
described in Table \ref{tab:quasi-bell-states} as teleportation channel.
Before we do so, it would be apt to note that some studies \cite{adhikari2012quantum,prakash2012minimum,van2001quantum,prakash2007effect,prakash2008effect,prakash2007improving}
have already reported teleportation schemes using quasi-Bell states,
but those studies lack the required rigor as they could not put light
on various important facets of the task. Specifically, we may note
that Adhikari et al., \cite{adhikari2012quantum} showed that it is
possible to perform teleportation using $|\phi_{+}\rangle$. Interestingly,
they showed that the amount of nonorthogonality present in the quantum
channel affects the average fidelity ($F_{ave}$) of teleportation.
However, their work was restricted to the use of $|\phi_{+}\rangle$
as a quantum channel for teleportation. They did not check the suitability
of other quasi-Bell states as quantum channels for teleportation.
Naturally, the study did not lead to a comparison between different
quasi-Bell states. Further, in all realistic situations, it is impossible
to circumvent the noise present in the transmission channel. However,
they did not try to study the effect of noise. In fact, the optimality
of the scheme was not also investigated by them. Such limitations
were also present in some of the earlier works. Specifically, in Ref.
\cite{prakash2012minimum}, MASFI and MFI were calculated for non-maximally
entangled quantum channels that were used to teleport an unknown state
which was reconstructed by the receiver using a set of suitable unitary
operations. The focus of the authors of Ref. \cite{prakash2012minimum}
was to only obtain compact expressions of MASFI and MFI in terms of
concurrence of the non-maximally entangled state in ideal (noiseless)
conditions. They neither tried to compute $F_{ave}$ and use that
to quantify the quality of teleportation, and thus to obtain a comparison
among various possible entangled-nonorthogonal-state-based channels,
nor did they try to establish the optimality of their scheme. We have
used the compact expression of MASFI obtained in Ref. \cite{prakash2012minimum},
but for the computation of MFI we have used Pauli operations in analogy
with the standard single-qubit teleportation scheme that uses Bell
states; this is in contrast to the MFI result reported in Ref. \cite{prakash2012minimum}
using optimized unitary operations. Teleportation of a coherent superposition
state using one of the quasi-Bell states was performed in the past,
and the decay in the amount of entanglement of the quantum channel
was investigated by solving master equation \cite{van2001quantum}.
The same state was teleported using another quasi-Bell state as quantum
channel and considering the channel as a lossy channel in Ref. \cite{prakash2007effect}.
A similar study for two-mode coherent states has been performed using
tripartite entangled coherent state in noiseless \cite{prakash2007improving}
and noisy \cite{prakash2008effect} environments, too. It is relevant
to note here that the focus of all these earlier works was to teleport
a state in ideal or noisy environment using an entangled nonorthogonal
state. However, we wish to perform a comparative study among a set
of entangled nonorthogonal states (quasi-Bell states) in both these
conditions, and also wish to test the optimality of such schemes.}

\section{\textcolor{black}{Teleportation using entangled nonorthogonal state\label{sec:Quantum-Teleportation-using}}}

\textcolor{black}{Let us consider that an arbitrary single-qubit quantum
state}

\textcolor{black}{
\begin{equation}
|I\rangle=\alpha|0\rangle+\beta|1\rangle:\,\,\,\,\,\,\,\,\,\,\,\,\,\,\,|\alpha|^{2}+|\beta|^{2}=1,\label{eq:8}
\end{equation}
is to be teleported using the quasi-Bell state}

\textcolor{black}{
\begin{equation}
|\psi_{\pm}\rangle=N_{\pm}(|a\rangle\otimes|b\rangle\pm|b\rangle\otimes|a\rangle),\label{eq:9}
\end{equation}
where the normalization constant $N_{\pm}=\left[2\left(1\pm|\langle a|b\rangle|^{2}\right)\right]^{\frac{-1}{2}}$.
These quasi-Bell states may be viewed as particular cases of Eq. (\ref{eq:1-1})
with $|d\rangle=|a\rangle,$ $|c\rangle=|b\rangle$, and $\mu=\pm\nu$.
In general, $\langle a|b\rangle$ is a complex number, and consequently,
we can write}

\textcolor{black}{
\begin{equation}
\langle a|b\rangle=re^{i\theta},\label{eq:10}
\end{equation}
where the real parameters }\textit{\textcolor{black}{$r$}}\textcolor{black}{{}
and }\textit{\textcolor{black}{$\theta$}}\textcolor{black}{, respectively,
denote the modulus and argument of the complex number $\langle a|b\rangle$
with $0\leq r\leq1$ and $0\leq\theta\leq2\pi$. As $r=0$ implies,
orthogonal basis, we may consider this parameter as the primary measure
of nonorthogonality. This is so because no value of $\theta$ will
lead to orthogonality condition. Further, for $r\neq0,$ we can consider
$\theta$ as a secondary measure of nonorthogonality. Now, using Eq.
(\ref{eq:10}), and the map between orthogonal and nonorthogonal bases
we may rewrite Eq. (\ref{eq:3a}) as}

\textcolor{black}{
\begin{equation}
|0\rangle=|a\rangle\,{\rm and}\,|1\rangle=\frac{\left[|b\rangle-\langle a|b\rangle a\rangle\right]}{\sqrt{1-r^{2}}}.\label{eq:11.a}
\end{equation}
Thus, we have $|a\rangle=|0\rangle\,{\rm and}\,|b\rangle=\langle a|b\rangle|0\rangle+\sqrt{1-r^{2}}|1\rangle,$
and consequently, $|\phi_{+}\rangle$ can now be expressed as}

\textcolor{black}{
\begin{equation}
|\phi_{+}\rangle=\eta|00\rangle+\epsilon|01\rangle+\epsilon|10\rangle,\label{eq:12}
\end{equation}
where $\eta=\frac{2re^{i\theta}}{\sqrt{2(1+r^{2})}}$ and $\epsilon=\sqrt{\frac{1-r^{2}}{2(1+r^{2})}}$.
This is already noted in Table \ref{tab:quasi-bell-states}, where
we have also noted that if we express $|\phi_{-}\rangle$ in $\{|0\rangle,|1\rangle\}$
basis, we obtain the Bell state $|\phi^{-}\rangle=\frac{1}{\sqrt{2}}\left(|01\rangle-|10\rangle\right)$,
which is maximally entangled and naturally yields unit fidelity for
teleportation. It's not surprising to obtain maximally entangled nonorthogonal
states, as in \cite{wang2001bipartite} it has been already established
that there exists a large class of bipartite entangled nonorthogonal
states that are maximally entangled under certain conditions.}

\textcolor{black}{Using Eq. (\ref{eq:7}), we found the concurrence
of the symmetric state $|\phi_{+}\rangle$ as}

\textcolor{black}{
\begin{equation}
C\left(|\phi_{+}\rangle\right)=\frac{1-|\langle a|b\rangle|^{2}}{1+|\langle a|b\rangle|^{2}}=\frac{1-r^{2}}{1+r^{2}}.\label{eq:13}
\end{equation}
Clearly, $|\phi_{+}\rangle$ is not maximally entangled unless $r=\left|\langle a|b\rangle\right|=0,$
which implies orthogonality. Thus, all quasi-Bell states of the form
$|\phi_{+}\rangle$ are non-maximally entangled. Now, if the state
$|\phi_{+}\rangle$ is used as quantum channel, then following Prakash
et al., \cite{prakash2012minimum} we may express the MASFI for teleportation
of single-qubit state (\ref{eq:8}) as}

\textcolor{black}{
\begin{equation}
\begin{array}{lcl}
\left({\rm MASFI}\right)_{\phi_{+}} & = & \frac{2C\left(|\phi_{+}\rangle\right)}{1+C\left(|\phi_{+}\rangle\right)}=1-r^{2}.\end{array}\label{eq:14}
\end{equation}
}

\textcolor{black}{Since the value of $r$ lies between 0 and 1, the
$\left({\rm MASFI}\right)_{\phi_{+}}$ decreases continuously as $r$
increases. For orthogonal state $r$= $0$, and thus, ${\rm MASFI}=1$.
Thus, we may conclude that the quasi-Bell state $|\phi_{+}\rangle$
will never lead to deterministic perfect teleportation. However, its
Bell state counter part ($r=1$ case) leads to deterministic perfect
teleportation. Here, it would be apt to note that for teleportation
of a single-qubit state using $|\phi_{+}\rangle$ as the quantum channel,
average teleportation fidelity can be obtained as \cite{adhikari2012quantum}}

\textcolor{black}{
\begin{equation}
F_{ave,\phi_{+}}=\frac{3-r^{2}}{3(1+r^{2})}.\label{eq:15}
\end{equation}
This is obtained by computing teleportation fidelity $F^{tel}=\sum_{i=1}^{4}P_{i}\left|\langle I|\zeta_{i}\rangle\right|^{2},$
where $|I\rangle$ is the input state, and $P_{i}=Tr(\langle\Omega|M_{i}|\Omega\rangle)$
with $|\Omega\rangle=|I\rangle\otimes|\psi_{{\rm {channel}}}\rangle$,
and $M_{i}=|\psi_{i}\rangle\langle\psi_{i}|$ is a measurement operator
in Bell basis ($|\psi_{i}\rangle$s are defined in the second column
of Table \ref{tab:quasi-bell-states}), and $|\zeta_{i}\rangle$ is
the teleported state corresponding to $i$th projective measurement
in Bell basis. Interestingly, $F^{tel}$ is found to depend on the
parameters of the state to be teleported (cf. Eq. (11) of Ref. \cite{adhikari2012quantum}).
Thus, if we use Bloch representation and express the state to be teleported
as $|I\rangle=\alpha|0\rangle+\beta|1\rangle=\cos\frac{\theta^{\prime}}{2}|0\rangle+\exp(i\phi^{\prime})\sin\frac{\theta^{\prime}}{2}|1\rangle$,
then the teleportation fidelity $F^{tel}$ will be a function of state
parameters $\theta^{\prime}$ and $\phi^{\prime}$ (here $^{\prime}$
is used to distinguish the state parameter $\theta^{\prime}$ from
the nonorthogonality parameter $\theta$). An average fidelity is
obtained by taking average over all possible states that can be teleported,
i.e., by computing $F_{ave}=\frac{1}{4\pi}\int_{\phi^{\prime}=0}^{2\pi}\int_{\theta^{\prime}=0}^{\pi}F^{tel}\left(\theta^{\prime},\phi^{\prime}\right)\sin(\theta^{\prime})d\theta^{\prime}d\phi^{\prime}$.
This definition of average fidelity is followed in \cite{adhikari2012quantum,henderson2000two}
and in this thesis. However, in the works of Prakash et al., (\cite{prakash2007improving,prakash2007effect,prakash2008effect}
and references therein), $\left|\langle I|\zeta_{i}\rangle\right|^{2}$
was considered as fidelity and $F^{tel}$ as average fidelity. They
minimized $F^{tel}$ over the parameters of the state to be teleported
and referred to the obtained fidelity as the MAVFI. As that notation
is not consistent with the definition of average fidelity used here.
In what follows, we will refer to the minimum value of $F^{tel}$
as MFI, but it would be the same as MAVFI defined by Prakash et al.
Further, we would like to note that in \cite{adhikari2012quantum}
and in the present chapter, it is assumed that a standard teleportation
scheme is implemented by replacing a Bell state by its partner quasi-Bell
state, and as a consequence for a specific outcome of Bell measurement
of Alice, Bob applies the same Pauli operator for teleportation channel
$|\psi_{x}\rangle$ or $|\phi_{x}\rangle$ (which is a quasi-Bell
state) as he used to do for the corresponding Bell state $|\psi^{x}\rangle$
or $|\phi^{x}\rangle,$ where $x\in\{+,-\}.$ However, the expression
of MASFI used here (see Eq. (\ref{eq:14})) and derived in \cite{prakash2012minimum}
are obtained using an optimized set of unitary (cf. discussion after
Eq. (10) in Ref. \cite{prakash2012minimum}) and are subjected to
outcome of Bell measurement of Alice, thus no conclusions should be
made by comparing MASFI with MFI or $F_{ave}$.}

\textcolor{black}{From Eqs. (\ref{eq:14}) and (\ref{eq:15}), we
can see that for a standard Bell state $|\phi^{+}\rangle$ (i.e.,
when $r=0$), ${\rm (MASFI})_{\phi_{+}}=F_{ave}=1$. However, for
$r$ = 1, ${\rm (MASFI})_{\phi_{+}}=0,$ and $F_{ave}=\frac{1}{3}$.
Thus, we conclude that for a standard Bell state both ${\rm MASFI}$
and average teleportation fidelity have the same value. This is not
surprising, as for $r$= 0 the entangled state $|\phi_{+}\rangle$
becomes maximally entangled. However, for $r\neq0,$ this state is
non-maximally entangled, and interestingly, for $r$= 1, we obtain
${\rm MASFI=0}$, whereas $F_{ave}$ is nonzero. We have already noted
that no comparison of ${\rm MASFI}$ and $F_{ave}$ obtained as above
should be made as that may lead to confusing results. Here we give
an example, according to \cite{prakash2007improving,prakash2012minimum},
MASFI is the least possible value of the fidelity, but for certain
values of $r$, we can observe that ${\rm MASFI}>F_{ave}.$ For example,
for $r=0.5$, we obtain ${\rm MASFI}=0.75,$ whereas $F_{ave}=0.733.$
Clearly, minimum found in computation of ${\rm MASFI}$, and the average
found in the computation of $F_{ave}$ is not performed over the same
data set, specifically not using the same teleportation mechanism
(same unitary operations at the receiver's end).}

\textcolor{black}{Now we may check the optimality of the teleportation
scheme by using the criterion introduced by Horodecki et al., in Ref.
\cite{horodecki1999general}. According to this criterion optimal
average fidelity that can be obtained for a teleportation scheme which
uses a bipartite entangled quantum state $\rho$ as the quantum channel
is 
\begin{equation}
F_{opt}=\frac{2f+1}{3},\label{eq:Horodecki}
\end{equation}
where $f$ is the maximal singlet fraction defined as 
\begin{equation}
f=\underset{i}{\max}\langle\psi_{i}|\rho|\psi_{i}\rangle,\label{eq:singlet fraction}
\end{equation}
where $|\psi_{i}\rangle$: $i\in\{1,2,3,4\}$ is Bell state described
above and summarized in Table \ref{tab:quasi-bell-states}. As we
are interested in computing $f$ for quasi-Bell states which are pure
states, we can write $f=\underset{i}{\max}\left|\langle\psi_{i}|\chi\rangle\right|^{2},$
where $|\chi\rangle$ is a quasi-Bell state. A bit of calculation
yields that maximal singlet fraction for the quasi-Bell state $|\phi_{+}\rangle$
is 
\begin{equation}
f_{\phi_{+}}=\frac{1-r^{2}}{1+r^{2}}.\label{eq:singlerfracpsiplus}
\end{equation}
Now using (\ref{eq:15}), (\ref{eq:Horodecki}) and (\ref{eq:singlerfracpsiplus}),
we can easily observe that 
\begin{equation}
F_{opt,\,\phi_{+}}=\frac{2\left(\frac{1-r^{2}}{1+r^{2}}\right)+1}{3}=\frac{3-r^{2}}{3(1+r^{2})}=F_{ave}.\label{eq:optimality}
\end{equation}
Thus, a quasi-Bell state-based teleportation scheme which is analogous
to the usual teleportation scheme, but uses a quasi-Bell state $|\phi_{+}\rangle$
as the quantum channel is optimal. We can also minimize $F_{\phi_{+}}^{tel}(\theta^{\prime},\phi^{\prime})$
with respect to $\theta^{\prime}$ and $\phi^{\prime}$ to obtain
\begin{equation}
{\rm {MFI}}_{\phi_{+}}=\frac{1-r^{2}}{1+r^{2}},\label{eq:mfi-1}
\end{equation}
which is incidentally equivalent to maximal singlet fraction in this
case.}

\textcolor{black}{So far we have reported analytic expressions for
some parameters (e.g., $F_{ave},$ ${\rm MASFI},$ and ${\rm MFI)}$
that can be used as measures of the quality of a teleportation scheme
realized using the teleportation channel $|\phi_{+}\rangle$ and have
shown that the teleportation scheme obtained using $|\phi_{+}\rangle$
is optimal. Among these analytic expressions, $F_{ave,\phi_{+}}$
was already reported in \cite{adhikari2012quantum}. Now, to perform
a comparative study, let us consider that the teleportation is performed
using one of the remaining two quasi-Bell states of our interest (i.e.,
using $|\psi_{+}\rangle$ or $|\psi_{-}\rangle$ described in Table
\ref{tab:quasi-bell-states}) as quantum channel. In that case, we
would obtain the concurrence as}

\textcolor{black}{
\begin{equation}
C\left(|\psi_{\pm}\rangle\right)=2|\pm k_{\pm}m_{\pm}-l_{\pm}^{2}|=\frac{1-r^{2}}{(1\pm r^{2}\cos2\theta)}.\label{eq:19}
\end{equation}
Clearly, in contrast to $C|\phi_{+}\rangle,$ which was only $r$
dependent, the concurrence $C\left(|\psi_{\pm}\rangle\right)$ depends
on both the parameters $r$ and $\theta$. From Eq. (\ref{eq:19})
it is clear that at $\theta=\frac{\pi}{2},\frac{3\pi}{2}$ $\left(\theta=0,\pi\right)$
quasi-Bell state $|\psi_{+}\rangle$ $\left(|\psi_{-}\rangle\right)$
is maximally entangled, even though the states $|a\rangle$ and $|b\rangle$
are nonorthogonal as $r\neq0$. Thus, at these points, states $|\psi_{\pm}\rangle$
are maximally entangled. If quantum state $|\psi_{+}\rangle$ is used
as quantum channel, then ${\rm MASFI}$ for teleportation of an arbitrary
single-qubit information state (\ref{eq:8}) would be}

\textcolor{black}{
\begin{equation}
{\rm (MASFI})_{\psi_{+}}=\frac{2C(|\psi_{+}\rangle)}{1+C(|\psi_{+}\rangle)}=\frac{1-r^{2}}{1-r^{2}\sin^{2}\theta},\label{eq:20}
\end{equation}
and similarly, that for quasi-Bell state $|\psi_{-}\rangle$ would
be}

\textcolor{black}{
\begin{equation}
{\rm (MASFI})_{\psi_{-}}=\frac{1-r^{2}}{1-r^{2}\cos^{2}\theta}.\label{eq:21}
\end{equation}
Thus, the expressions for ${\rm MASFI}$ are also found to depend
on both $r$ and $\theta$. Clearly, at $\theta=\frac{\pi}{2}$ and
$\frac{3\pi}{2},$ ${\rm (MASFI})_{\psi_{+}}=1$, and hence for these
particular choices of $\theta,$ entangled nonorthogonal state $|\psi_{+}\rangle$
leads to the deterministic perfect teleportation of single-qubit information
state. Clearly, for these values of $\theta$, $C(|\psi_{+}\rangle)=1,$
indicating maximal entanglement. However, the entangled state is still
nonorthogonal as $r$ can take any of its allowed values. Similarly,
at $\theta=0$ and $\pi,$ $({\rm MASFI})_{\psi_{-}}=1$, and hence
the entangled state $|\psi_{-}\rangle$ of the nonorthogonal states
$|a\rangle$ and $|b\rangle$ leads to deterministic perfect teleportation
in these conditions. Thus, deterministic perfect teleportation is
possible using quasi-Bell states $|\phi_{-}\rangle$ or $|\psi_{\pm}\rangle$
as quantum channels for teleportation, but it is not possible with
$|\phi_{+}\rangle$ unless it reduces to its orthogonal state counter
part (i.e., $|\phi^{+}\rangle).$ We may now compute the average fidelity
for $|\psi_{\pm}\rangle$, by using the procedure adopted above for
$|\phi_{+}\rangle$ and obtain 
\begin{equation}
F_{ave,\psi_{\pm}}=\frac{3-2r^{2}+r^{4}\mp r^{2}(r^{2}-3)\cos2\theta}{3(1\pm r^{2}\cos2\theta)}.\label{eq:avefidelityphipm}
\end{equation}
}

\textcolor{black}{}
\begin{figure}
\begin{centering}
\textcolor{black}{\includegraphics[scale=0.45]{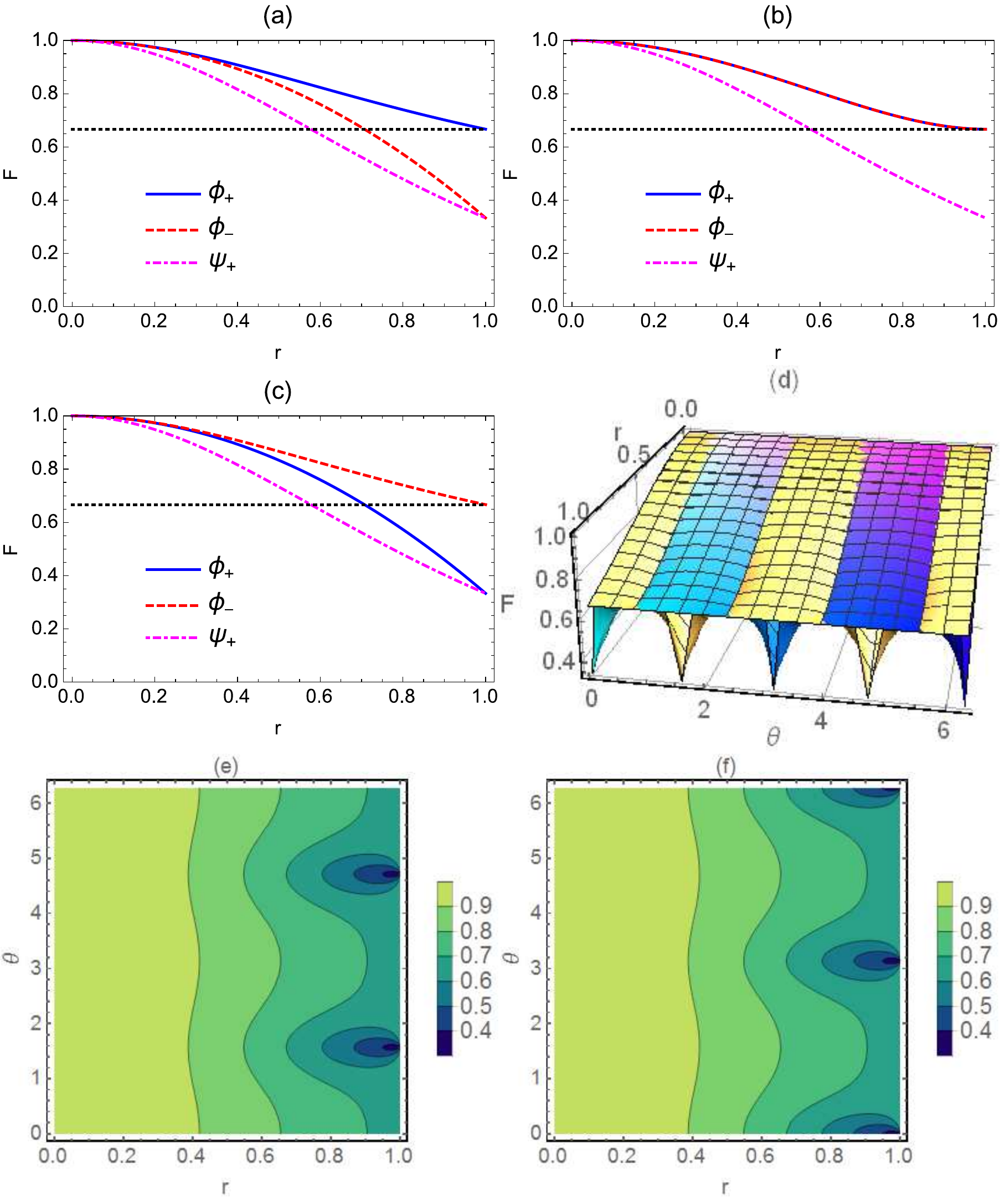}}
\par\end{centering}
\textcolor{black}{\caption{\label{fig:AvFid} \textcolor{black}{The dependence of the average
fidelity on nonorthogonality parameters ($r,$$\theta$) is illustrated
via 2D and 3D plots. Plots (a), (b) and (c) are showing the variation
of $F$ with $r$ for $\theta=0,\frac{\pi}{4},$ and $\frac{\pi}{2}$,
respectively and in these plots the horizontal dotted black line corresponds
to classical fidelity. Similarly, (d) shows the variation of $F$
in 3D for both $\psi_{+}$ and $\psi_{-}$ in light (yellow) and dark
(blue) colored surface plots, respectively. The same fact is illustrated
using contour plots for both of these cases in (e) and (f). Note that
the quantity plotted in this and the following figures is $F_{ave}$,
which is mentioned as $F$ in the $y$-axis. }}
}
\end{figure}
\textcolor{black}{Now, we would like to compare the average fidelity
expressions obtained so far for various values of nonorthogonality
parameters for all the quasi-Bell states. The same is illustrated
in Figure \ref{fig:AvFid}. Specifically, in Figures \ref{fig:AvFid}
(a)-(c) we have shown the dependence of average fidelity on secondary
nonorthogonality parameter ($\theta$) using a set of plots for its
variation with primary nonorthogonality parameter ($r$). This establishes
that the primary nonorthogonality parameter has more control over
the obtained average, fidelity but a secondary parameter is also prominent
enough to change the choice of quasi-Bell state to be preferred for
specific value of primary nonorthogonality parameter. Thus, the amount
of nonorthogonality plays a crucial role in deciding which quasi-Bell
state would provide highest average fidelity for a teleportation scheme
implemented using quasi-Bell state as the teleportation channel. Further,
all these plots also establish that there is always a quasi-Bell state
apart from $|\phi_{-}\rangle$, which has average fidelity more than
classically achievable fidelity $\left(\frac{2}{3}\right)$ for all
values of $r$. We may now further illustrate the dependence of the
average fidelity on both nonorthogonality parameters via 3 D and contour
plots shown in Figures \ref{fig:AvFid} (d), (e) and (f). These plots
establish that the average fidelity of $|\psi_{+}\rangle$ state increases
for the values of $\theta$ for which $|\psi_{-}\rangle$ decreases,
and vice-versa.}

\textcolor{black}{We can now establish the optimality of the teleportation
scheme implemented using $|\psi_{\pm}\rangle$ by computing average
fidelity and maximal singlet fraction for these channels. Specifically,
computing the maximal singlet fraction using the standard procedure
described above, we have obtained}

\textcolor{black}{
\begin{equation}
f_{\psi_{\pm}}=\frac{2-2r^{2}+r^{4}\pm\cos2\theta(2r^{2}-r^{4})}{2(1\pm r^{2}\cos2\theta)}.\label{eq:singlet fractionphipm}
\end{equation}
Using Horodecki et al., criterion (\ref{eq:Horodecki}), and Eq. (\ref{eq:avefidelityphipm})-(\ref{eq:singlet fractionphipm}),
we can easily verify that $F_{ave,\psi_{\pm}}=F_{opt,\psi_{\pm}}$.
Thus, the teleportation scheme realized using any of the quasi-Bell
state are optimal. However, they are not equally efficient for a specific
choice of nonorthogonality parameter as we have already seen in Figure
\ref{fig:AvFid}. This motivates us to further compare the performances
of these quasi-Bell states as a potential quantum channel for teleportation.
For the completeness of the comparative investigation of the teleportation
efficiencies of different quasi-Bell states here we would also like
to report MFI that can be achieved using different quasi-Bell states.
The same can be computed as above, and the computation leads to following
analytic expressions of MFI for $|\psi_{\pm}\rangle$: 
\begin{equation}
\begin{array}{lcl}
{\rm {MFI}}_{\psi_{+}} & = & \frac{1}{2}|k_{+}+m_{+}|^{2}=\frac{1-r^{2}(2-r^{2})\sin^{2}\theta}{1+r^{2}\cos2\theta},\end{array}\label{eq:mavfi}
\end{equation}
and 
\begin{equation}
\begin{array}{lcl}
{\rm {MFI}}_{\psi_{-}} & = & \frac{1}{2}|k_{-}+m_{-}|^{2}=\frac{1-r^{2}(2-r^{2})\cos^{2}\theta}{1-r^{2}\cos2\theta}\end{array}.\label{eq:mavfi}
\end{equation}
}

\textcolor{black}{Interestingly, the comparative analysis performed
with the expressions of MFI using their variation with various parameters
led to quite similar behavior as observed for $F_{ave}$ in Figure
\ref{fig:AvFid}. Therefore, we are not reporting corresponding figures
obtained for MFI.}

\section{\textcolor{black}{Effect of noise on average fidelity \label{sec:Effect-of-noise}}}

\textcolor{black}{In this section, we would like to analyze and compare
the average fidelity obtained for each quasi-Bell state over two well
known Markovian channels, i.e., AD and PD channels. Specifically,
in open quantum system formalism, a quantum state evolving under a
noisy channel can be written in terms of Kraus operators as we have
defined in Section \ref{sec:Effect-of-noise-1} of Chapter \ref{cha:Introduction1}.}

\textcolor{black}{To analyze the feasibility of quantum teleportation
scheme using quasi-Bell states and to compute the average fidelity
we use Eqs. (\ref{eq:Kraus-damping-1}) and (\ref{eq:Kraus-dephasing-1})
in Eq. (\ref{eq:3-2}).} Subsequently, the effect of noise is quantified
by computing \textquotedbl fidelity\textquotedbl{} between the quantum
state which has been actually evolved under the noisy channel under
consideration and the quantum state Alice wished to teleport (say
$\rho=|I\rangle\langle I|$\textcolor{black}{). Mathematically,}

\textcolor{black}{
\begin{equation}
F_{k}=\frac{1}{4\pi}\int_{\phi^{\prime}=0}^{2\pi}\int_{\theta^{\prime}=0}^{\pi}\left(\sum_{i=1}^{4}P_{i}\langle I|\left\{ \rho_{k}\left(\theta^{\prime},\phi^{\prime}\right)\right\} _{i}|I\rangle\right)\sin(\theta^{\prime})d\theta^{\prime}d\phi^{\prime},\label{eq:fidelity}
\end{equation}
which is the square of the conventional fidelity expression, and $\rho_{k}$
is the quantum state recovered at the Bob's port under the noisy channel
$k\in\{AD,PD\}$. Further, details of the mathematical technique adopted
here can be found in some of our group's recent works on secure \cite{sharma2016comparative,sharma2016verification,banerjee2017asymmetric}
and insecure quantum communication \cite{thapliyal2015applications,shukla2017hierarchical}.}

\textcolor{black}{We will start with the simplest case, where we assume
that only Bob's part of the quantum channel is subjected to either
AD or PD noise. The assumption is justified as the quasi-Bell state
used as quantum channel is prepared locally (here assumed to be prepared
by Alice) and shared afterwards. During Alice to Bob transmission
of an entangled qubit, it may undergo decoherence, but the probability
of decoherence is much less for the other qubits that don't travel
through the channel (remain with Alice). Therefore, in comparison
of the Bob's qubits, the Alice's qubits or the quantum state to be
teleported $|I\rangle$, which remain at the sender's end, are hardly
affected due to noise. The effect of AD noise under similar assumptions
has been analyzed for three-qubit GHZ and W states in the recent past
\cite{thapliyal2015quasiprobability}. The average fidelity for all
four quasi-Bell states, when Bob's qubit is subjected to AD channel
while the qubits not traveling through the channel are assumed to
be unaffected due to noise, is obtained as 
\begin{equation}
\begin{array}{lcl}
F_{AD}^{|\psi_{+}\rangle} & = & \frac{-1}{2\left(3+3r^{2}\cos2\theta\right)}\left[-4+r^{2}\left(2+2\sqrt{1-\text{\ensuremath{\eta}}}-3\text{\ensuremath{\eta}}\right)-2\sqrt{1-\text{\ensuremath{\eta}}}\right.\\
 & + & \left.2r^{4}(-1+\eta)+\text{\ensuremath{\eta}}+2r^{2}\left(-2-\sqrt{1-\text{\ensuremath{\eta}}}+r^{2}\sqrt{1-\text{\ensuremath{\eta}}}\right)\cos2\theta\right],\\
F_{AD}^{|\psi_{-}\rangle} & = & \frac{1}{-6+6r^{2}\cos2\theta}\left[-4+r^{2}\left(2+2\sqrt{1-\text{\ensuremath{\eta}}}-3\eta\right)-2\sqrt{1-\text{\ensuremath{\eta}}}\right.\\
 & + & \left.2r^{4}(-1+\text{\ensuremath{\eta}})+\text{\ensuremath{\eta}}-2r^{2}\left(-2-\sqrt{1-\text{\ensuremath{\eta}}}+r^{2}\sqrt{1-\text{\ensuremath{\eta}}}\right)\cos2\theta\right],\\
F_{AD}^{|\phi_{+}\rangle} & = & \frac{4+2\sqrt{1-\text{\ensuremath{\eta}}}-\text{\ensuremath{\eta}}+r^{2}\left(-2\sqrt{1-\text{\ensuremath{\eta}}}+\eta\right)}{6\left(1+r^{2}\right)},\\
F_{AD}^{|\phi_{-}\rangle} & = & \frac{1}{6}\left[\left(4+2\sqrt{1-\eta}-\text{\ensuremath{\eta}}\right)\right].
\end{array}\label{eq:adsq}
\end{equation}
}

\textcolor{black}{Here and in what follows, the subscript of fidelity
$F$ corresponds to noise model and superscript represents the choice
of quasi-Bell state used as teleportation channel. Similarly, all
the average fidelity expressions when Bob's qubit is subjected to
PD noise can be obtained as}

\textcolor{black}{
\begin{equation}
\begin{array}{lcl}
F_{PD}^{|\psi_{+}\rangle} & = & \frac{1}{3}\left[2+\sqrt{1-\eta}+r^{2}\left(-\sqrt{1-\eta}+\frac{-1+r^{2}}{1+r^{2}\cos2\theta}\right)\right],\\
F_{PD}^{|\psi_{-}\rangle} & = & \frac{1}{3}\left[2+\sqrt{1-\text{\ensuremath{\eta}}}-r^{2}\sqrt{1-\text{\ensuremath{\eta}}}+\frac{r^{2}-r^{4}}{-1+r^{2}\cos2\theta}\right],\\
F_{PD}^{|\phi_{+}\rangle} & = & \frac{2+\sqrt{1-\eta}-r^{2}\sqrt{1-\eta}}{3+3r^{2}},\\
F_{PD}^{|\phi_{-}\rangle} & = & \frac{1}{3}\left[2+\sqrt{1-\text{\ensuremath{\eta}}}\right].
\end{array}\label{eq:pdsq}
\end{equation}
}

\textcolor{black}{It is easy to observe that for $\eta=0$ (i.e.,
in the absence of noise) the average fidelity expressions listed in
Eqs. (\ref{eq:adsq}) and (\ref{eq:pdsq}) reduce to the average fidelity
expressions corresponding to each quasi-Bell state reported in Section
\ref{sec:Quantum-Teleportation-using}. This is expected and can also
be used to check the accuracy of our calculation.}

\textcolor{black}{It would be interesting to observe the change in
fidelity when we consider the effect of noise on Alice's qubit as
well. Though, it remains at Alice's port until she performs measurement
on it in suitable basis, but in a realistic situation Alice's qubit
may also interact with its surroundings in the meantime. Further,
it can be assumed that the state intended to be teleported is prepared
and teleported immediately. Therefore, it is hardly affected due to
noisy environment. Here, without loss of generality, we assume that
the decoherence rate for both the qubits is same. Using the same mathematical
formalism adopted beforehand, we have obtained the average fidelity
expressions for all the quasi-Bell states when both the qubits in
the quantum channel are affected by AD noise with the same decoherence
rate. The expressions are}

\textcolor{black}{
\begin{equation}
\begin{array}{lcl}
F_{AD}^{|\psi_{+}\rangle} & = & \frac{1}{3+3r^{2}\cos2\theta}\left[3-2r^{2}(-1+\text{\ensuremath{\eta}})^{2}+r^{4}(-1+\text{\ensuremath{\eta}})^{2}-2\eta\right.\\
 & + & \left.\text{\ensuremath{\eta}}^{2}+r^{2}\left(3+r^{2}(-1+\text{\ensuremath{\eta}1})-\text{\ensuremath{\eta}}\right)\cos2\theta\right],\\
F_{AD}^{|\psi_{-}\rangle} & = & -\frac{1}{-3+3r^{2}\cos2\theta}\left[3-2r^{2}(-1+\text{\ensuremath{\eta}})^{2}+r^{4}(-1+\eta)^{2}+(-2+\text{\ensuremath{\eta}})\text{\ensuremath{\eta}}\right.\\
 & + & \left.r^{2}\left(-3-r^{2}(-1+\text{\ensuremath{\eta}})+\text{\ensuremath{\eta}}\right)\cos2\theta\right],\\
F_{AD}^{|\phi_{+}\rangle} & = & \frac{3-2\eta+r^{2}(-1+2\eta)}{3\left(1+r^{2}\right)},\\
F_{AD}^{|\phi_{-}\rangle} & = & 1-\frac{2\text{\ensuremath{\eta}}}{3}.
\end{array}\label{eq:adbq}
\end{equation}
}

\textcolor{black}{Similarly, the average fidelity expressions when
both the qubits evolve under PD channel instead of AD channel are}

\textcolor{black}{
\begin{equation}
\begin{array}{lcl}
F_{PD}^{|\psi_{+}\rangle} & = & \frac{1}{3}\left[3-\text{\ensuremath{\eta}}+r^{2}\left(-1+\text{\ensuremath{\eta}}+\frac{-1+r^{2}}{1+r^{2}\cos2\theta}\right)\right],\\
F_{PD}^{|\psi_{-}\rangle} & = & \frac{1}{3}\left[3+r^{2}(-1+\eta)-\eta+\frac{r^{2}-r^{4}}{-1+r^{2}\cos2\theta}\right],\\
F_{PD}^{|\phi_{+}\rangle} & = & \frac{3+r^{2}(-1+\text{\ensuremath{\eta}})-\text{\ensuremath{\eta}}}{3\left(1+r^{2}\right)},\\
F_{PD}^{|\phi_{-}\rangle} & = & 1-\frac{\eta}{3}.
\end{array}\label{eq:pdbq}
\end{equation}
}

\textcolor{black}{Finally, it is worth analyzing the effect of noisy
channels on the feasibility of the teleportation scheme, when even
the state to be teleported is also subjected to the same noisy channel.
The requirement of this discussion can be established as it takes
finite time before operations to teleport the quantum state are performed.
Meanwhile, the qubit gets exposed to its vicinity and this interaction
may lead to decoherence. Here, for simplicity, we have considered
the same noise model for the state to be teleported as for the quantum
channel. We have further assumed the same rate of decoherence for
all the three qubits. Under these specific conditions, when all the
qubits evolve under AD channels, the average fidelity for each quasi-Bell
state turns out to be}

~\\
\\
\\

\textcolor{black}{
\begin{equation}
\begin{array}{lcl}
F_{AD}^{|\psi_{+}\rangle} & = & \frac{-1}{2\left(3+3r^{2}\cos2\theta\right)}\left[-4+r^{2}\left(2+2\sqrt{1-\text{\text{\ensuremath{\eta}}}}-3\text{\text{\ensuremath{\eta}}}\right)-2\sqrt{1-\text{\text{\ensuremath{\eta}}}}+2r^{4}(-1+\text{\text{\ensuremath{\eta}}})\right.\\
 & + & \left.\text{\text{\ensuremath{\eta}}}+2r^{2}\left(-2-\sqrt{1-\text{\ensuremath{\eta}}}+r^{2}\sqrt{1-\text{\ensuremath{\eta}}}\right)\cos2\theta\right],\\
F_{AD}^{|\psi_{-}\rangle} & = & \frac{1}{2\left(-3+3r^{2}\cos2\theta\right)}\left[-2\left(2+\sqrt{1-\text{\text{\ensuremath{\eta}}}}\right)+\text{\ensuremath{\eta}}\left(3+2\sqrt{1-\text{\text{\ensuremath{\eta}}}}+2(-2+\text{\text{\ensuremath{\eta}}})\text{\text{\ensuremath{\eta}}}\right)\right.\\
 & - & 2r^{2}(-1+\text{\text{\ensuremath{\eta}}})\left(1+\sqrt{1-\text{\text{\ensuremath{\eta}}}}+\text{\text{\ensuremath{\eta}}}(-3+2\text{\ensuremath{\eta}})\right)+2r^{4}(-1+\text{\text{\ensuremath{\eta}}})^{3}\\
 & + & \left.r^{2}\left(4+2\sqrt{1-\text{\text{\ensuremath{\eta}}}}+2\sqrt{1-\text{\text{\ensuremath{\eta}}}}\left(r^{2}(-1+\text{\ensuremath{\eta}})-\text{\text{\ensuremath{\eta}}}\right)-\text{\text{\ensuremath{\eta}}}\right)\cos2\theta\right],\\
F_{AD}^{|\phi_{+}\rangle} & = & \frac{1}{6\left(1+r^{2}\right)}\left[2\left(2+\sqrt{1-\text{\text{\ensuremath{\eta}}}}\right)+\text{\text{\ensuremath{\eta}}}\left(-3-2\sqrt{1-\text{\ensuremath{\eta}}}+2\text{\ensuremath{\eta}}\right)\right.\\
 & + & \left.r^{2}\left(-2\sqrt{1-\text{\text{\ensuremath{\eta}}}}+\left(5+2\sqrt{1-\text{\ensuremath{\eta}}}-2\text{\text{\ensuremath{\eta}}}\right)\text{\text{\ensuremath{\eta}}}\right)\right],\\
F_{AD}^{|\phi_{-}\rangle} & = & \frac{1}{6}\left[4+2\sqrt{1-\text{\text{\ensuremath{\eta}}}}-3\text{\text{\ensuremath{\eta}}}-2\sqrt{1-\text{\ensuremath{\eta}}}\text{\text{\ensuremath{\eta}}}+2\text{\text{\ensuremath{\eta}}}^{2}\right].
\end{array}\label{eq:adaq}
\end{equation}
}

\textcolor{black}{}
\begin{figure}
\begin{centering}
\textcolor{black}{\includegraphics[scale=0.75]{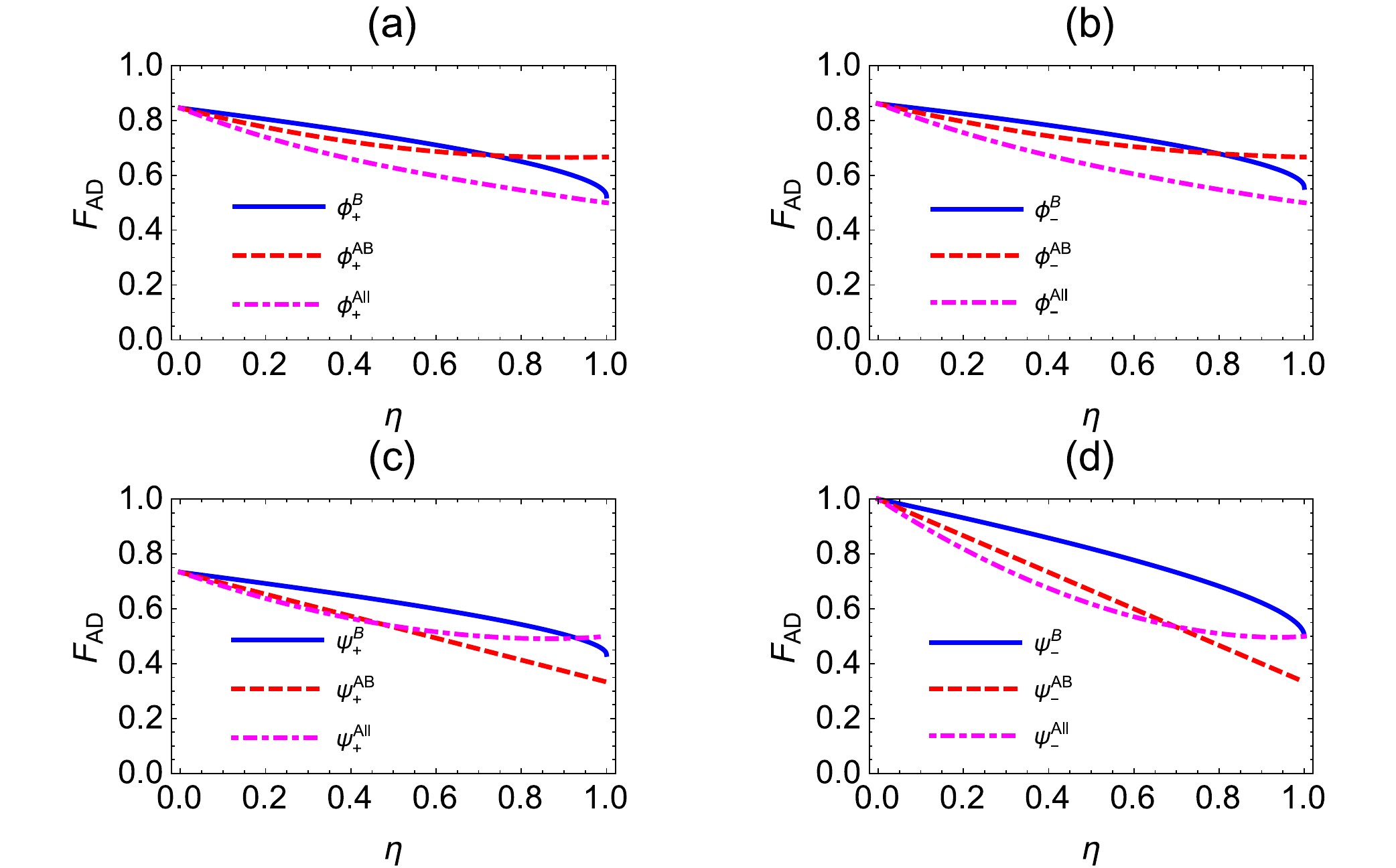}}
\par\end{centering}
\textcolor{black}{\caption{\label{fig:AvFid-AD}\textcolor{black}{The dependence of the average
fidelity on the number of qubits exposed to AD channels is illustrated
for $r=\frac{1}{2}$ and $\theta=\frac{\pi}{3}$. The choice of the
initial Bell states in each case is mentioned in plot legends, where
the superscript B, AB, and All corresponds to the cases when only
Bob's, both Alice's and Bob's, and all three qubits were subjected
to the noisy channel. The same notation is adopted in the following
figures. Amplitude damping noise effect on Bob's, both Alice's and
Bob's, and all three qubits (superscript B, AB, and All is mentioned
in the following figures) is shown with the variation of the average
fidelity ($F_{AD}$) with noise parameter ($\eta)$ for the specific
values of nonorthogonality parameters $r=\frac{1}{2}$ and $\theta=\frac{\pi}{3}$.}}
}
\end{figure}
\textcolor{black}{}
\begin{figure}
\begin{centering}
\textcolor{black}{\includegraphics[scale=0.7]{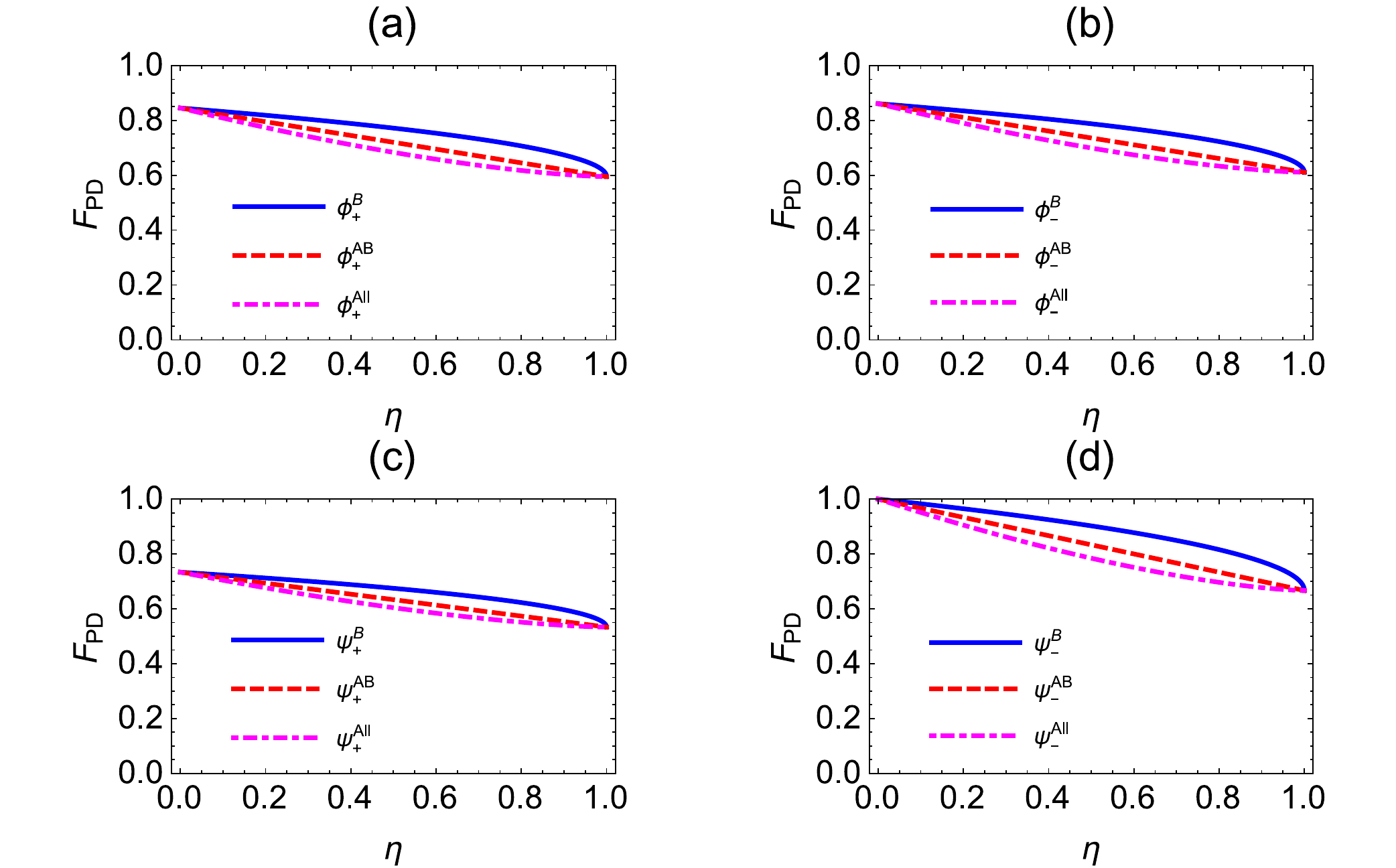}}
\par\end{centering}
\textcolor{black}{\caption{\label{fig:AvFid-PD} \textcolor{black}{The effect of phase damping
noise is shown for all the quasi-Bell states through the variation
of the average fidelity for the specific values of $r=\frac{1}{2}$
and $\theta=\frac{\pi}{3}$.}}
}
\end{figure}
\textcolor{black}{Similarly, when all three qubits are subjected to
PD noise with the same decoherence rate, the analytic expressions
of the average fidelity are obtained as}

\textcolor{black}{
\begin{equation}
\begin{array}{lcl}
F_{PD}^{|\psi_{+}\rangle} & = & \frac{1}{\left(3+3r^{2}\cos2\theta\right)}\left[2+r^{4}+\sqrt{1-\text{\ensuremath{\eta}}}-\sqrt{1-\text{\ensuremath{\eta}}}\text{\ensuremath{\eta}}+r^{2}\left(-1-\sqrt{1-\text{\ensuremath{\eta}}}+\sqrt{1-\text{\ensuremath{\eta}}}\eta\right)\right.\\
 & + & \left.r^{2}\left(2+\sqrt{1-\text{\ensuremath{\eta}}}-r^{2}(1-\text{\ensuremath{\eta}})^{3/2}-\sqrt{1-\text{\ensuremath{\eta}}}\text{\ensuremath{\eta}}\right)\cos2\theta\right],\\
F_{PD}^{|\psi_{-}\rangle} & = & \frac{1}{3}\left[2+\sqrt{1-\text{\ensuremath{\eta}}}-r^{2}\sqrt{1-\text{\ensuremath{\eta}}}-\sqrt{1-\text{\ensuremath{\eta}}}\text{\ensuremath{\eta}}+r^{2}\sqrt{1-\text{\ensuremath{\eta}}}\text{\text{\ensuremath{\eta}}}+\frac{r^{2}-r^{4}}{-1+r^{2}\cos2\theta}\right],\\
F_{PD}^{|\phi_{+}\rangle} & = & \frac{2+\sqrt{1-\text{\text{\ensuremath{\eta}}}}+\sqrt{1-\text{\text{\ensuremath{\eta}}}}\left(r^{2}(-1+\text{\ensuremath{\eta}})-\text{\text{\ensuremath{\eta}}}\right)}{3\left(1+r^{2}\right)},\\
F_{PD}^{|\phi_{-}\rangle} & = & \frac{2+(1-\text{\text{\ensuremath{\eta}}})^{3/2}}{3}.
\end{array}\label{eq:pdaq}
\end{equation}
}

\textcolor{black}{It is interesting to note that in the ideal conditions
$|\phi_{-}\rangle$ is the unanimous choice of quasi-Bell state to
accomplish the teleportation with highest possible fidelity. However,
from the expressions of fidelity obtained in Eqs. (\ref{eq:adsq})-(\ref{eq:pdaq}),
it appears that it may not be the case in the presence of noise. For
further analysis, it would be appropriate to observe the variation
of all the fidelity expressions with various parameters. In what follows,
we perform this analysis.}

\textcolor{black}{Figure \ref{fig:AvFid-AD}, illustrates the dependence
of the average fidelity on the number of qubits exposed to AD channel
for each quasi-Bell state using Eqs. (\ref{eq:adsq}), (\ref{eq:adbq}),
and (\ref{eq:adaq}). Unlike the remaining quasi-Bell states, the
average fidelity for $|\phi_{-}\rangle$ state starts from 1 at $\eta=0$.
Until a moderate value (a particular value that depends on the choice
of quasi-Bell state) of decoherence rate is reached, the decay in
average fidelity completely depends on the number of qubits interacting
with their surroundings. However, at the higher decoherence rate,
this particular nature was absent. Further, Figures \ref{fig:AvFid-AD}
(a) and (b) show that best results compared to remaining two cases
can be obtained for the initial state $|\psi_{\pm}\rangle$, while
both the channel qubits are evolving under AD noise; whereas the same
case turns out to provide the worst results in case of $|\phi_{\pm}\rangle$.
A similar study performed over PD channels instead of AD channels
reveals that the decay in average fidelity solely depends on the number
of qubits evolving over noisy channels (cf. Figure \ref{fig:AvFid-PD}).}

\textcolor{black}{}
\begin{figure}
\centering{}\textcolor{black}{\includegraphics[scale=0.45]{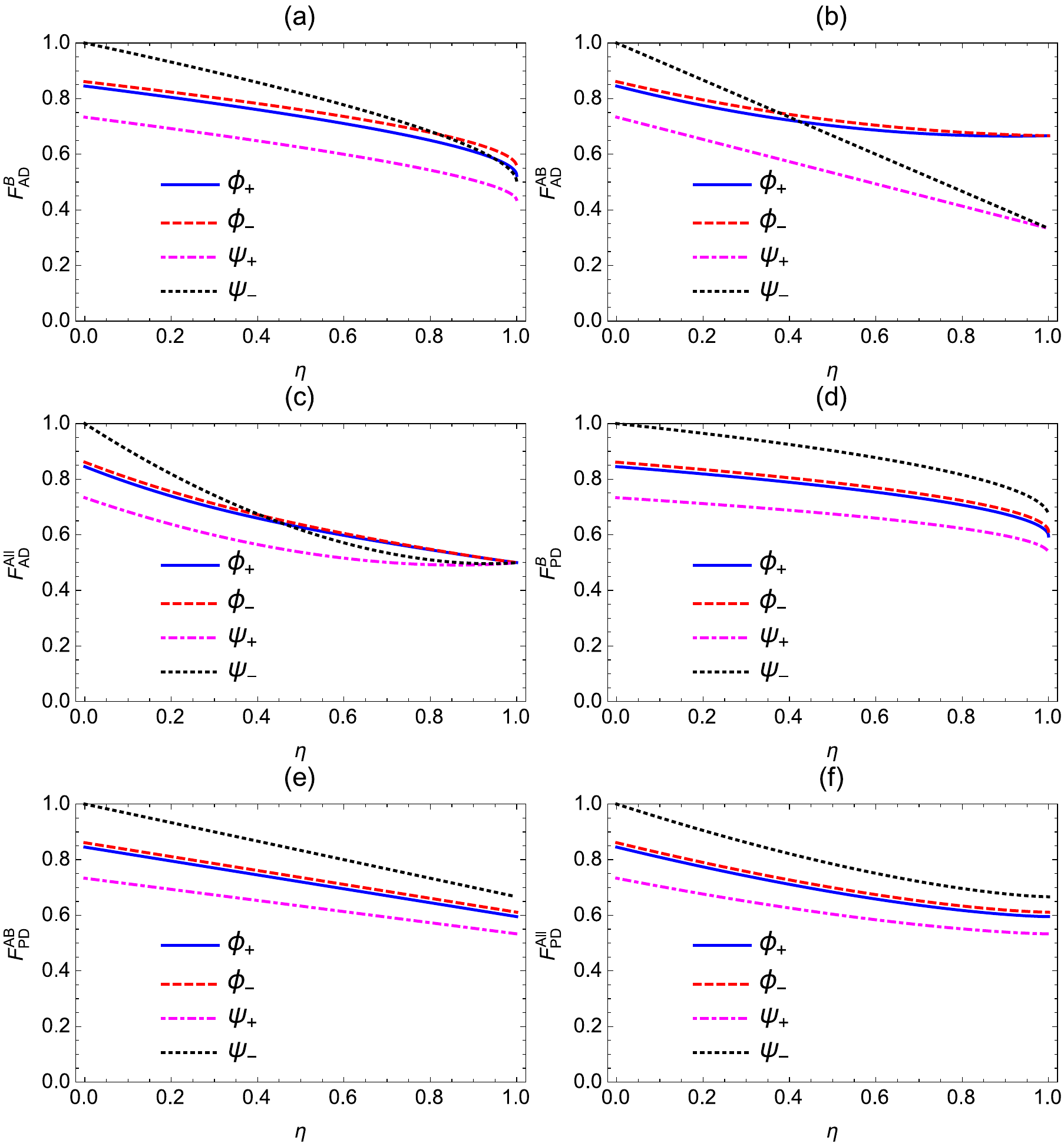}\caption{\label{fig:AvFid-AD-PD}The variation of average fidelity is illustrated
for all the possible cases by considering each quasi-Bell state as
quantum channel.\textcolor{black}{{} Average fidelity is plotted for
AD ((a)-(c)) and PD ((d)-(f)) channels with $r=\frac{1}{2}$ and $\theta=\frac{\pi}{3}$. }}
}
\end{figure}
\textcolor{black}{Finally, it is also worth to compare the average
fidelity obtained for different quasi-Bell states when subjected to
noisy environment under similar condition. This would reveal the suitable
choice of initial state to be used as a quantum channel for performing
teleportation. In Figure \ref{fig:AvFid-AD-PD} (a), the variation
of average fidelity for all the quasi-Bell states is demonstrated,
while only Bob's qubit is exposed to AD noise. It establishes that
although in ideal case and small decoherence rate $|\phi_{-}\rangle$
state is the most suitable choice of quantum channel, which do not
remain true at higher decoherence rate. While all other quasi-Bell
states follow exactly the same nature for decay of average fidelity
and $|\phi_{+}\rangle$ appears to be the worst choice of quantum
channel. A quite similar nature can be observed for the remaining
two cases over AD channels in Figures \ref{fig:AvFid-AD-PD} (b) and
(c). Specifically, $|\phi_{-}\rangle$ remains the most suitable choice
below moderate decoherence rate, while $|\psi_{\pm}\rangle$ may be
preferred for channels with high decoherence, and $|\phi_{+}\rangle$
is inevitably the worst choice.}

\textcolor{black}{A similar study carried out over PD channels and
the obtained results are illustrated in Figures \ref{fig:AvFid-AD-PD}
(d) and (f). From these plots, it may be inferred that $|\phi_{-}\rangle$
undoubtedly remains the most suitable and $|\phi_{+}\rangle$ the
worst choice of quantum channel. The investigation on the variation
of the average fidelity with nonorthogonality parameters over noisy
channels yields a similar nature as was observed in ideal conditions
(cf. Figure \ref{fig:AvFid}). Therefore, we have not discussed it
here, but a similar study can be performed in analogy with the ideal
scenario.}

\textcolor{black}{The results obtained so far are summarized in Tables
\ref{tab:noise-tab1} and \ref{tab:noise-tab2}. In Table \ref{tab:noise-tab1},
we have shown that the present study may be used to select quasi-Bell
state suitable as a quantum channel if the channel is characterized
for noise. The table also summarizes that this choice varies with
the noisy conditions (i.e., number of qubits exposed to noise and
type of noise model) and decoherence rate.}

\textcolor{black}{The study also reveals that for each choice of quasi-Bell
state, the average fidelity obtained over noisy channels falls below
the classical limit for different values of decoherence rate. This
fact may be used to obtain the tolerable decoherence rate for a potential
quantum channel. These limits on the tolerable decoherence rates for
a specific case are explicitly given in Table \ref{tab:noise-tab2}.}

\textcolor{black}{In fact, if one wishes to quantify only the effect
of noise on the performance of the teleportation scheme using a nonorthogonal
state quantum channel, the inner product may be taken with the teleported
state in the ideal condition instead of the state to be teleported.
The mathematical procedure adopted here is quite general in nature
and would be appropriate to study the effect of generalized amplitude
damping \cite{sharma2016comparative,thapliyal2015quasiprobability},
squeezed generalized amplitude damping \cite{sharma2016comparative,thapliyal2015quasiprobability},
bit flip \cite{sharma2016comparative,shukla2017hierarchical}, phase
flip \cite{sharma2016comparative,shukla2017hierarchical}, and depolarizing
channel \cite{shukla2017hierarchical,sharma2016comparative}. This
discussion can further be extended to a set of non-Markovian channels
\cite{thapliyal2017quantum}, which will be carried out in the future
and reported elsewhere.}

\textcolor{black}{}
\begin{table}
\textcolor{black}{\caption{\label{tab:noise-tab1} \textcolor{black}{The selection of the best
and worst quasi-Bell state to be used as a quantum channel for quantum
teleportation depends on the value of decoherence rate. A specific
case ($r=\frac{1}{2}$ and $\theta=\frac{\pi}{3}$) is shown here
for both AD and PD channels. The results observed for PD channel are
mentioned in brackets with corresponding results for AD channel without
bracket. The same notation is used in the next table, too.}}
}
\centering{}\textcolor{black}{\small{}}%
\begin{tabular}{|>{\centering}p{1.8cm}|>{\centering}p{1.9cm}|>{\centering}p{1.8cm}|>{\centering}p{1.9cm}|>{\centering}p{1.8cm}|>{\centering}p{1.9cm}|>{\centering}p{1.8cm}|}
\hline 
\textcolor{black}{\footnotesize{}Qubits} & \multicolumn{1}{c}{} & \multicolumn{1}{c}{} & \multicolumn{1}{c}{\textcolor{black}{\footnotesize{}Decoherence}} & \multicolumn{1}{c}{\textcolor{black}{\footnotesize{}rate}} & \multicolumn{1}{c|}{} & \tabularnewline
\cline{2-7} 
\textcolor{black}{\footnotesize{}exposed} & \multicolumn{1}{c|}{\textcolor{black}{\footnotesize{}Low}} & \textcolor{black}{\footnotesize{}(below $\frac{1}{3}$)} & \multicolumn{1}{c|}{\textcolor{black}{\footnotesize{}Moderate}} & \textcolor{black}{\footnotesize{}(between $\frac{1}{3}$ and $\frac{2}{3}$)} & \multicolumn{1}{c|}{\textcolor{black}{\footnotesize{}High}} & \textcolor{black}{\footnotesize{}(above $\frac{2}{3}$)}\tabularnewline
\cline{2-7} 
\textcolor{black}{\footnotesize{}to noise} & \textcolor{black}{\footnotesize{}Best state} & \textcolor{black}{\footnotesize{}Worst state} & \textcolor{black}{\footnotesize{}Best state} & \textcolor{black}{\footnotesize{}Worst state} & \textcolor{black}{\footnotesize{}Best state} & \textcolor{black}{\footnotesize{}Worst state}\tabularnewline
\hline 
\textcolor{black}{\footnotesize{}Bob's} & \textcolor{black}{\footnotesize{}$|\phi_{-}\rangle$ $\left(|\phi_{-}\rangle\right)$} & \textcolor{black}{\footnotesize{}$|\phi_{+}\rangle$ $\left(|\phi_{+}\rangle\right)$} & \textcolor{black}{\footnotesize{}$|\phi_{-}\rangle$ $\left(|\phi_{-}\rangle\right)$} & \textcolor{black}{\footnotesize{}$|\phi_{+}\rangle$ $\left(|\phi_{+}\rangle\right)$} & \textcolor{black}{\footnotesize{}$|\psi_{-}\rangle$ $\left(|\phi_{-}\rangle\right)$} & \textcolor{black}{\footnotesize{}$|\phi_{+}\rangle$ $\left(|\phi_{+}\rangle\right)$}\tabularnewline
\hline 
\textcolor{black}{\footnotesize{}Alice's and Bob's} & \textcolor{black}{\footnotesize{}$|\phi_{-}\rangle$ $\left(|\phi_{-}\rangle\right)$} & \textcolor{black}{\footnotesize{}$|\phi_{+}\rangle$ $\left(|\phi_{+}\rangle\right)$} & \textcolor{black}{\footnotesize{}$|\psi_{-}\rangle$ $\left(|\phi_{-}\rangle\right)$} & \textcolor{black}{\footnotesize{}$|\phi_{+}\rangle$ $\left(|\phi_{+}\rangle\right)$} & \textcolor{black}{\footnotesize{}$|\psi_{-}\rangle$ $\left(|\phi_{-}\rangle\right)$} & \textcolor{black}{\footnotesize{}$|\phi_{+}\rangle$ $\left(|\phi_{+}\rangle\right)$}\tabularnewline
\hline 
\textcolor{black}{\footnotesize{}All three} & \textcolor{black}{\footnotesize{}$|\phi_{-}\rangle$ $\left(|\phi_{-}\rangle\right)$} & \textcolor{black}{\footnotesize{}$|\phi_{+}\rangle$ $\left(|\phi_{+}\rangle\right)$} & \textcolor{black}{\footnotesize{}$|\psi_{-}\rangle$ $\left(|\phi_{-}\rangle\right)$} & \textcolor{black}{\footnotesize{}$|\phi_{+}\rangle$ $\left(|\phi_{+}\rangle\right)$} & \textcolor{black}{\footnotesize{}$|\psi_{-}\rangle$ $\left(|\phi_{-}\rangle\right)$} & \textcolor{black}{\footnotesize{}$|\phi_{+}\rangle$ $\left(|\phi_{+}\rangle\right)$}\tabularnewline
\hline 
\end{tabular}{\small\par}
\end{table}
\textcolor{black}{}
\begin{table}
\textcolor{black}{\caption{\label{tab:noise-tab2} Even in the presence of noise, different choices
of quasi-Bell states (as quantum teleportation channel) may yield
fidelity higher than the maximum achievable classical fidelity in
the noiseless situation ($\frac{2}{3}$). Here, a specific case for
$r=\frac{1}{2}$ and $\theta=\frac{\pi}{3}$ over both AD and PD channels
is shown. }
}
\centering{}\textcolor{black}{}%
\begin{tabular}{|c|c|c|c|}
\hline 
\textcolor{black}{Quasi-Bell} & \multicolumn{1}{c}{\textcolor{black}{Qubits}} & \multicolumn{1}{c}{\textcolor{black}{exposed}} & \textcolor{black}{to noise}\tabularnewline
\cline{2-4} 
\textcolor{black}{state} & \textcolor{black}{Bob's} & \textcolor{black}{Alice's and Bob's} & \textcolor{black}{All three}\tabularnewline
\hline 
\textcolor{black}{$|\psi_{+}\rangle$} & \textcolor{black}{$0.728$ $\left(0.906\right)$} & \textcolor{black}{$0.728$ $\left(0.692\right)$} & \textcolor{black}{$0.348$ $\left(0.541\right)$}\tabularnewline
\hline 
\textcolor{black}{$|\psi_{-}\rangle$} & \textcolor{black}{$0.803$ $\left(0.933\right)$} & \textcolor{black}{$0.858$ $\left(0.746\right)$} & \textcolor{black}{$0.372$ $\left(0.586\right)$}\tabularnewline
\hline 
\textcolor{black}{$|\phi_{+}\rangle$} & \textcolor{black}{$0.285$ $\left(0.505\right)$} & \textcolor{black}{$0.146$ $\left(0.282\right)$} & \textcolor{black}{$0.110$ $\left(0.206\right)$}\tabularnewline
\hline 
\textcolor{black}{$|\phi_{-}\rangle$} & \textcolor{black}{$0.803$ $\left(1.00\right)$} & \textcolor{black}{$0.490$ $\left(0.981\right)$} & \textcolor{black}{$0.387$ $\left(0.903\right)$}\tabularnewline
\hline 
\end{tabular}
\end{table}

\section{\textcolor{black}{Conclusion\label{sec:Conclusion4}}}

\textcolor{black}{In the present study, it has been established that
all the quasi-Bell states, which are entangled nonorthogonal states
may be used for quantum teleportation of a single-qubit state. However,
their teleportation efficiencies are not the same, and it also depends
on the nature of noise present in the quantum channel. Specifically,
we have considered here four quasi-Bell states as teleportation channel,
and computed average and minimum fidelity that can be obtained by
replacing a Bell state quantum channel in a teleportation scheme by
its nonorthogonal counterpart (i.e., corresponding quasi-Bell state).
The results can be easily reduced to that obtained using usual Bell
state in the limits of vanishing nonorthogonality parameter. Specifically,
there are two real parameters $r$ and $\theta$, considered here
as primary and secondary measures of nonorthogonality.}

\textcolor{black}{Here, $F_{ave}$ and MFI are used as quantitative
measures of the quality of the teleportation scheme which utilizes
a quasi-Bell state instead of usual Bell state as quantum channel.
Therefore, during this discussion, it has been assumed that Bob performs
a Pauli operation corresponding to each Bell state measurement outcome
as in the standard teleportation scheme. However, we have used another
quantitative measure of quality of teleportation performance, MASFI,
which is computed considering an optimal unitary operation to be applied
by Bob. For a few specific cases, the calculated MASFI was found to
be unity. In those cases, concurrence for entangled nonorthogonal
states were found to be unity, which implied maximal entanglement.
However, for these sets of maximally entangled nonorthogonal states,
we did not observe unit average fidelity and minimum fidelity as the
unitary operations performed by Bob were not the same as was in computation
of MASFI.}

\textcolor{black}{The performance of the teleportation scheme using
entangled nonorthogonal states has also been analyzed over noisy channels.
This study yield various interesting results. The quasi-Bell state
$|\phi_{-}\rangle$, which was shown to be maximally entangled in
an ideal situation, remains most preferred choice as quantum channel
while subjected to PD noise as well. However, in AD noise, it is observed
that the preferred choice of the quasi-Bell state depends on the nonorthogonality
parameter and the number of qubits exposed to noisy environment. We
hope the present study will be useful for experimental realization
of teleportation schemes beyond usual entangled orthogonal state regime,
and will also provide a clear prescription for future research on
applications of entangled nonorthogonal states.}
\begin{center}
\textbf{\Large{}\newpage}{\Large\par}
\par\end{center}

\chapter[CHAPTER \thechapter \protect\newline EXPERIMENTAL REALIZATION OF NONDESTRUCTIVE DISCRIMINATION OF BELL
STATES USING A FIVE-QUBIT QUANTUM COMPUTER]{EXPERIMENTAL REALIZATION OF NONDESTRUCTIVE DISCRIMINATION OF BELL
STATES USING A FIVE-QUBIT QUANTUM COMPUTER\textsc{\label{cha:BellIBM}}}

{\large{}\lhead{}}{\large\par}

\section{Introduction \label{sec:Introduction5}}

\textcolor{black}{In the last three chapters, we have mostly discussed
ideas related to QT. However, the domain of quantum communication
is much broader, and in both quantum communication and computation,
discrimination of orthogonal entangled states play a very crucial
role. There exist many proposals for realizing such discrimination
(see \cite{gupta2007general,NDBSD,wang2013nondestructive,zheng2016complete}
and references therein). A particularly important variant of state
discrimination schemes is nondestructive discrimination of entangled
states \cite{gupta2007general,NDBSD,wang2013nondestructive,zheng2016complete},
in which the state is not directly measured. The measurement is performed
over some ancilla qubits/qudits coupled to entangled state and the
original state remains unchanged. Proposals for such nondestructive
measurements in optical quantum information processing using Kerr
type nonlinearity have been discussed in Refs. \cite{li2000non,wang2013nondestructive}
and references therein. Such optical schemes of nondestructive discrimination
are important as they are frequently used in designing entanglement
concentration protocols \cite{banerjee2015maximal}. In the same line,
a scheme for generalized orthonormal qudit Bell state discrimination
and an explicit quantum circuit for the task was provided in \cite{NDBSD}.
In Ref. \cite{NDBSD}, it was also established that the use of distributed
measurement (i.e., nondestructive measurement where the measurement
task is distributed or outsourced to ancilla qubits) have useful applications
in reducing quantum communication complexity under certain conditions.
The work was further generalized in \cite{gupta2007general}, where
the relevance of the quantum circuits for nondestructive discrimination
was established in the context of quantum error correction, }quantum
network and distributed quantum computing. \textcolor{black}{Subsequently,
applications of the nondestructive discrimination of entangled states
have been proposed in various other works, too. Specifically, in \cite{jain2009secure},
a scheme for two-way secure direct quantum communication, which was
referred to as quantum conversation, was developed using the nondestructive
discrimination scheme. In a more general scenario, Luo et al., proposed
a scheme for multi-party quantum private comparison based on $d$-dimensional
entangled states \cite{luo2014multi}, where all the participants
are required to perform nondestructive measurement.}

\textcolor{black}{These applications and the fact that the scheme
proposed in \cite{NDBSD} has been experimentally implemented for
Bell state discrimination using an NMR-based three-qubit quantum computer
\cite{samal2010non}, have motivated us to perform nondestructive
Bell state discrimination using another experimental platform. Specifically,
in this work, we aim to realize nondestructive Bell state discrimination
using a five-qubit superconductivity-(SQUID)-based quantum computer
\cite{IBMQE,devitt2016performing}, which has been recently placed
in cloud by IBM Corporation. This quantum computer was placed in the
cloud in 2016. It immediately drew considerable attention of the quantum
information processing community, and several quantum information
tasks have already been performed using this quantum computer on cloud.
Specifically, in the domain of quantum communication, properties of
different quantum channels that can be used for quantum communication
have been studied experimentally \cite{wei2018efficient} and experimental
realizations of a quantum analog of a bank cheque \cite{behera2017experimental}
that is claimed to work in a banking system having quantum network,
and teleportation of single-qubit \cite{fedortchenko2016quantum}
and two-qubit quantum states using optimal resources \cite{sisodia2017design},
have been reported; in the field of quantum foundation, violation
of multi-party Mermin inequality has been realized for 3, 4, and 5
parties \cite{alsina2016experimental}; an information theoretic version
of uncertainty and measurement reversibility has also been implemented
\cite{berta2016entropic}; in the area of quantum computation, a comparison
of two architectures using demonstration of an algorithm has been
performed \cite{linke2017experimental}, and a quantum permutation
algorithm \cite{yalccinkaya2017optimization}, an algorithm for quantum
summation \cite{majumder2017experimental} and a Deutsch-Jozsa like
algorithm \cite{gangopadhyay2018generalization} have been implemented
recently. Further, a non-abelian braiding of surface code defects
\cite{wootton2017demonstrating} and a compressed simulation of the
transverse field one-dimensional Ising interaction (realized as a
four-qubit Ising chain that utilizes only two qubits) \cite{hebenstreit2017compressed}
have also been demonstrated. Thus, we can see that IBM quantum computer
has already been successfully used to realize various tasks belonging
to different domains of quantum information processing. However, to
the best of our knowledge, IBM quantum computer is not yet used to
perform nondestructive discrimination of the orthogonal entangled
states, and the performance of IBM quantum computer is not yet properly
compared with the performance of the quantum computers implemented
using liquid NMR technology. In the present work, we aim to perform
such a comparison, subject to a specific task. To be precise, we wish
to compare the performance of IBM quantum computer with an NMR-based
quantum computer with respect to the nondestructive discrimination
of Bell states. There is another important reason for testing fundamentally
important quantum circuits (in our case, quantum circuit for nondestructive
discrimination of Bell states) using the IBM quantum computer and/or
a similar platform- it is now understood that liquid NMR-based technology
is not scalable \cite{jones2001quantum}, and it will not lead to
a real scalable quantum computer \cite{jones2001quantum}. However,
it is widely believed that a SQUID-based quantum computer can be made
scalable in future \cite{paauw2009tuning}. In fact, an ideal quantum
information processor should satisfy Di-Vincenzo's criteria \cite{divincenzo1997topics}.
One of these criteria requires the realization of a large quantum
register, which is still a prime technological challenge for experimental
quantum information processing \cite{jones2001quantum}. A ray of
hope is generated in the recent past after the introduction of the
relatively new architectures, like solid-state spin systems (nitrogen
vacancy centers in diamond and phosphorous vacancy centers in silicon)
and superconducting-qubits based systems \cite{cottet2002implementation}
that have the potential to become scalable. Among these technologies,
owing to the scalability and functionality of superconducting-qubit
registers, they have emerged as the best candidate for quantum information
processing. Currently, various types of basic superconducting-qubits,
namely Josephson-junction qubits, Phase qubits, Transmon qubits and
Potential qubits are used \cite{clarke2008superconducting,wendin2007quantum}.
SQUID-based quantum information processing architectures have not
only attained the popularity among the researchers, but have also
led to the path for commercialization of quantum computers. Although
a universal quantum computer with large qubit register is still a
distant hope, large qubit registers to perform specific tasks have
been devised. For example, quantum computers with register size of
512 qubits, 1000 qubits, and 2000 qubits were sold by D-Wave \cite{dwavenews}.
The machine with 51 two qubits has been deployed to tackle classification
problems that are useful in image-recognition \cite{QCvsCC}.}

\textcolor{black}{As mentioned above, the potential scalability of
the SQUID-based systems has also motivated us to perform experimental
realization of the Bell state discrimination circuit using a SQUID-based
five-qubit quantum computer.}

\textcolor{black}{Rest of the chapter is organized as follows. In
Section \ref{sec:Method}, we have described the quantum circuits
(both theoretical and experimental) used here to perform Bell state
discrimination and the method adopted here to perform QST. In Section
\ref{sec:Results}, the results of the experimental realization of
the circuits described in the previous section are reported and analyzed.
Finally, the chapter is concluded in Section \ref{sec:Conclusion5}.}

\section{\textcolor{black}{Quantum circuits and method used for nondestructive
discrimination of Bell states \label{sec:Method}}}

\begin{figure}
\begin{centering}
\textcolor{black}{\includegraphics[scale=0.52]{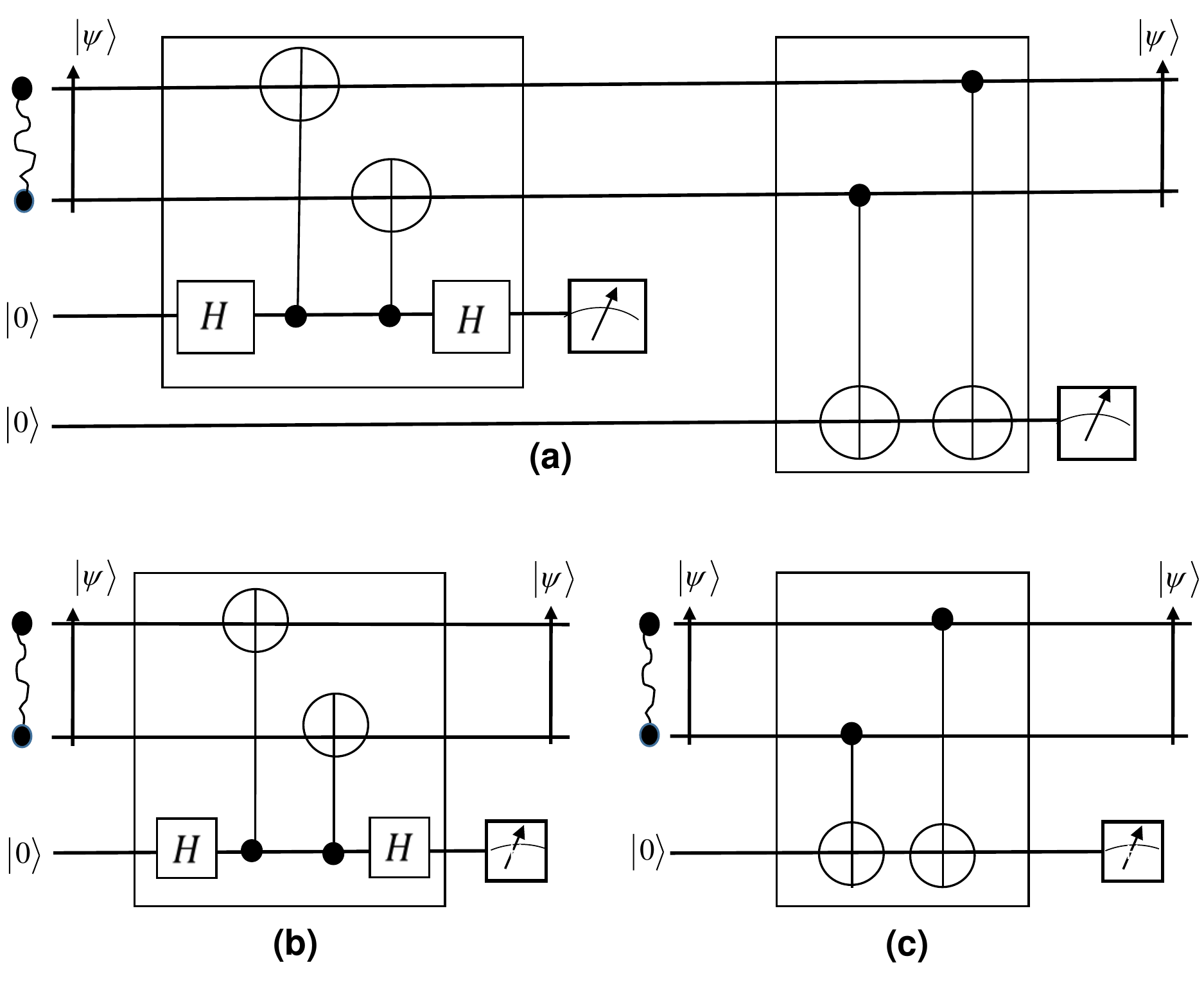}}
\par\end{centering}
\textcolor{black}{\caption{\label{fig:main-circuit} \textcolor{black}{The circuit for nondestructive
discrimination of all the four Bell states. (a) A four-qubit circuit
with two system-qubits (used for Bell-state preparation) and two ancilla-qubits
(subsequently used for phase detection (left block) and parity detection
(right block)). Here, (b) and (c) are the divided part of (a), whereas,
(b) is the phase checking circuit and (c) is the parity checking circuit.}}
}
\end{figure}

\textcolor{black}{We have already mentioned that in Refs. \cite{NDBSD,gupta2007general},
a quantum circuit for the nondestructive discrimination of generalized
orthonormal qudit Bell states was designed by some of the present
authors (cf. Figure 3 of \cite{NDBSD} or Figure 4 of \cite{gupta2007general}).
As a special case of these circuits, one can easily obtain a circuit
for Bell state discrimination, as shown in Figure \ref{fig:main-circuit}
(a) here and in Figure 2 of \cite{NDBSD}. This circuit involves four-qubits,
in which the measurement on the first ancilla qubit would reveal the
phase information, whereas the second measurement would reveal the
parity information. Thus, to discriminate all four Bell states in
a single experiment, we would require a four-qubit system, allowing
non-local operations between system qubits (qubits of the Bell state)
and the ancilla qubits. Apparently, this circuit should have been
implemented as it is in the five-qubit IBM quantum computer, but the
restriction on the application of CNOT gates, restrict us to implement
this circuit as a single circuit (without causing considerable increase
in the gate-count and decrease in the performance). Circumventing,
the increase in circuit complexity (gate-count), initially, we have
implemented the phase checking circuit and the parity checking circuit
separately, as shown in Figures \ref{fig:main-circuit} (b) and (c),
respectively. This is consistent with the earlier NMR-based implementation
of the Bell state discrimination circuit \cite{samal2010non}, where
a three-qubit quantum computer was used and naturally, parity checking
part and the phase checking part was performed via 2 independent experiments.
In fact, in the NMR-based implementation of the nondestructive discrimination
of Bell states, an ensemble of $^{13}CHFBr_{2}$ molecule was used
to perform the quantum computing, as the number of independent Larmor
frequency of that was 3, the quantum computer was a three-qubit one.
Specifically, Samal et al., used three nuclear spins, namely $^{1}H,\,^{13}C$
and $^{19}F$ of $^{13}CHFBr_{2}$ which mimics a three-qubit system
\cite{samal2010non}. In their experiment, they used $^{13}C$ as
ancilla qubit and rest as system qubits. The availability of single
ancilla qubit prohibited nondestructive discrimination of the Bell
states in one shot. Due to the trade-off between the available quantum
resources and number of experiments, they used two experiments to
realize the complete protocol, one for obtaining parity information
and the other one for phase information. The circuit shown in Figure
\ref{fig:main-circuit} (c) is actually used for parity checking and
it is easy to observe that for Bell states having even parity (i.e.,
for $|\psi^{\pm}\rangle=\frac{|00\rangle\pm|11\rangle}{\sqrt{2}}$),
the measurement on the ancilla qubit would yield $|0\rangle$ and
for the other two Bell states (i.e., for odd parity states $|\phi^{\pm}\rangle=\frac{|01\rangle\pm|10\rangle}{\sqrt{2}}$
) it would yield $|1\rangle.$ In a similar fashion, the quantum circuit
shown in Figure \ref{fig:main-circuit} (b) would determine the relative
phase of the Bell state as the measurement on ancilla in Figure \ref{fig:main-circuit}
(b) would yield $|0\rangle$ for $+$ states, i.e., for $|\psi^{+}\rangle$
and $|\phi^{+}\rangle$ and it would yield $|1\rangle$ for $-$ states
$|\psi^{-}\rangle$ and $|\phi^{-}\rangle$ (see third and forth column
of Table \ref{tab:one} for more detail).}

\textcolor{black}{We have already mentioned that the limitations of
the available quantum resources restricted Samal et al., \cite{samal2010non}
to realize the circuits shown in Figures \ref{fig:main-circuit} (b)
and (c) instead of the whole circuit shown in Figure \ref{fig:main-circuit}
(a) in their liquid NMR-based work.}

\textcolor{black}{}
\textcolor{black}{The same situation prevailed in the subsequent work
of the same group \cite{Anil-Kumar-paper2}. In contrast, the IBM's
five-qubit quantum computer IBM QX2}\textbf{\textcolor{black}{{} }}\textcolor{black}{is
SQUID-based, and does not face scalability issues encountered by NMR-based
proposals. Before reporting our experimental observations, it would
be apt to briefly describe the characteristic features of the technology
used in IBM quantum computer, with a focus on its architecture, control
fields for manipulation and readout of qubit register, important experimental
parameters, and the functioning of IBM QX2. The type of superconducting
qubits used in IBM architecture are transmon qubits. Transmon qubits
are charge qubits and can be designed to minimize charge noise which
is a major source of relaxation in charge qubits. The arrangement
of five superconducting qubits (q{[}0{]}, q{[}1{]}, q{[}2{]}, q{[}3{]},
q{[}4{]}) and their control mechanism as given in \cite{architectureurl}
is already described in Section \ref{sec:SQUID-based} (see Figure
\ref{fig:architecture-1} and Table \ref{tab:crosstalk-1} for details).
}
\begin{figure}
\begin{centering}
\textcolor{black}{\includegraphics[scale=0.6]{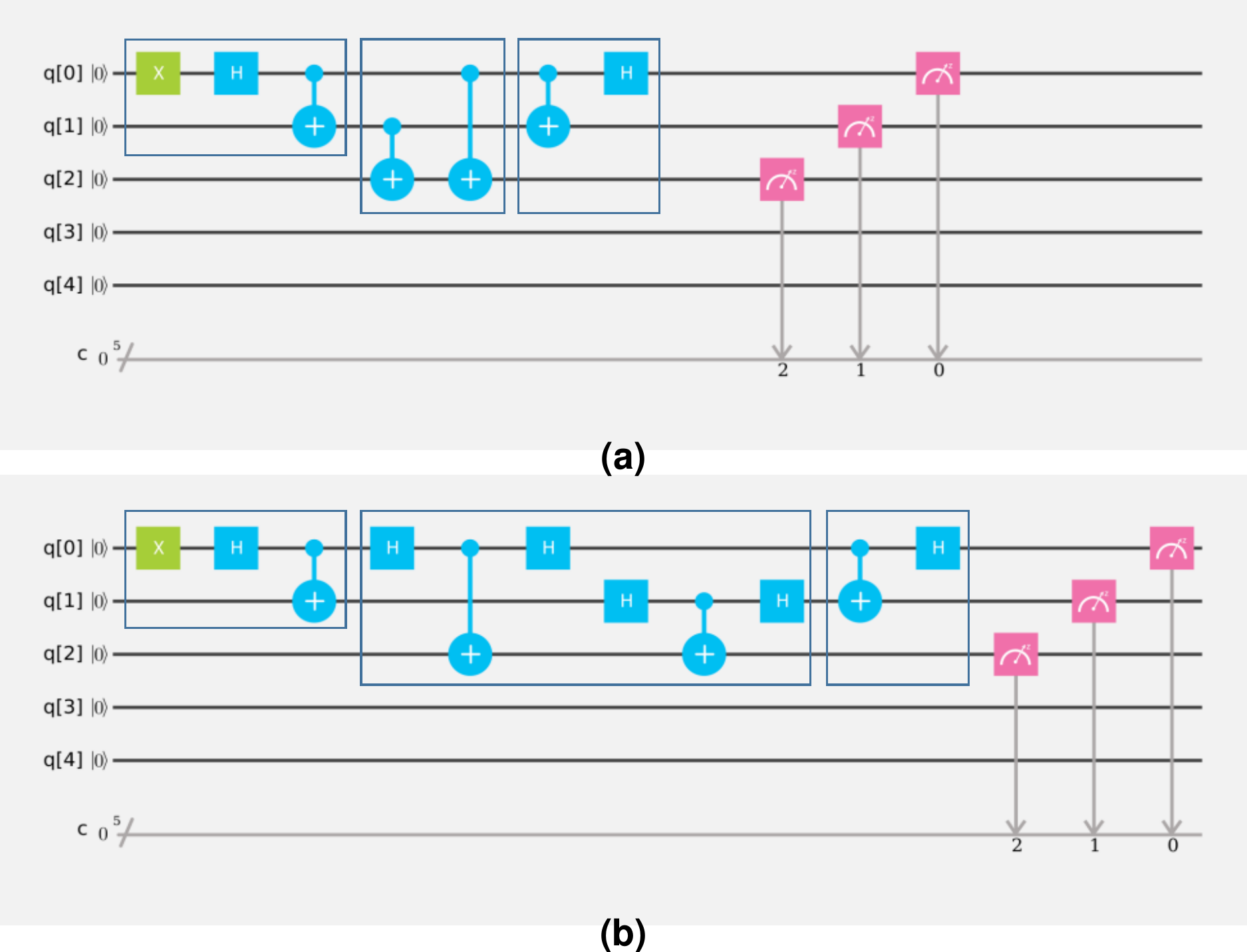}}
\par\end{centering}
\textcolor{black}{\caption{\label{fig:circuit-in-IBM} \textcolor{black}{Experimental implementation
of parity and phase checking circuits using five-qubit IBM quantum
computer for the Bell state $|\psi^{-}\rangle$. Qubits $\text{q[0]}$,
$\text{q[1]}$ are system-qubits, and $\text{q[2]}$ mimics ancilla-qubits.
Here, (a) and (b) correspond to the parity and phase checking circuits,
respectively. In (a) and (b), the left block prepares desired Bell
states, the middle block corresponds to a parity and phase checking
circuit, respectively and the right block is used for reverse EPR
circuit.}}
}
\end{figure}
\textcolor{black}{{} }
\begin{figure}
\begin{centering}
\textcolor{black}{\includegraphics[scale=0.6]{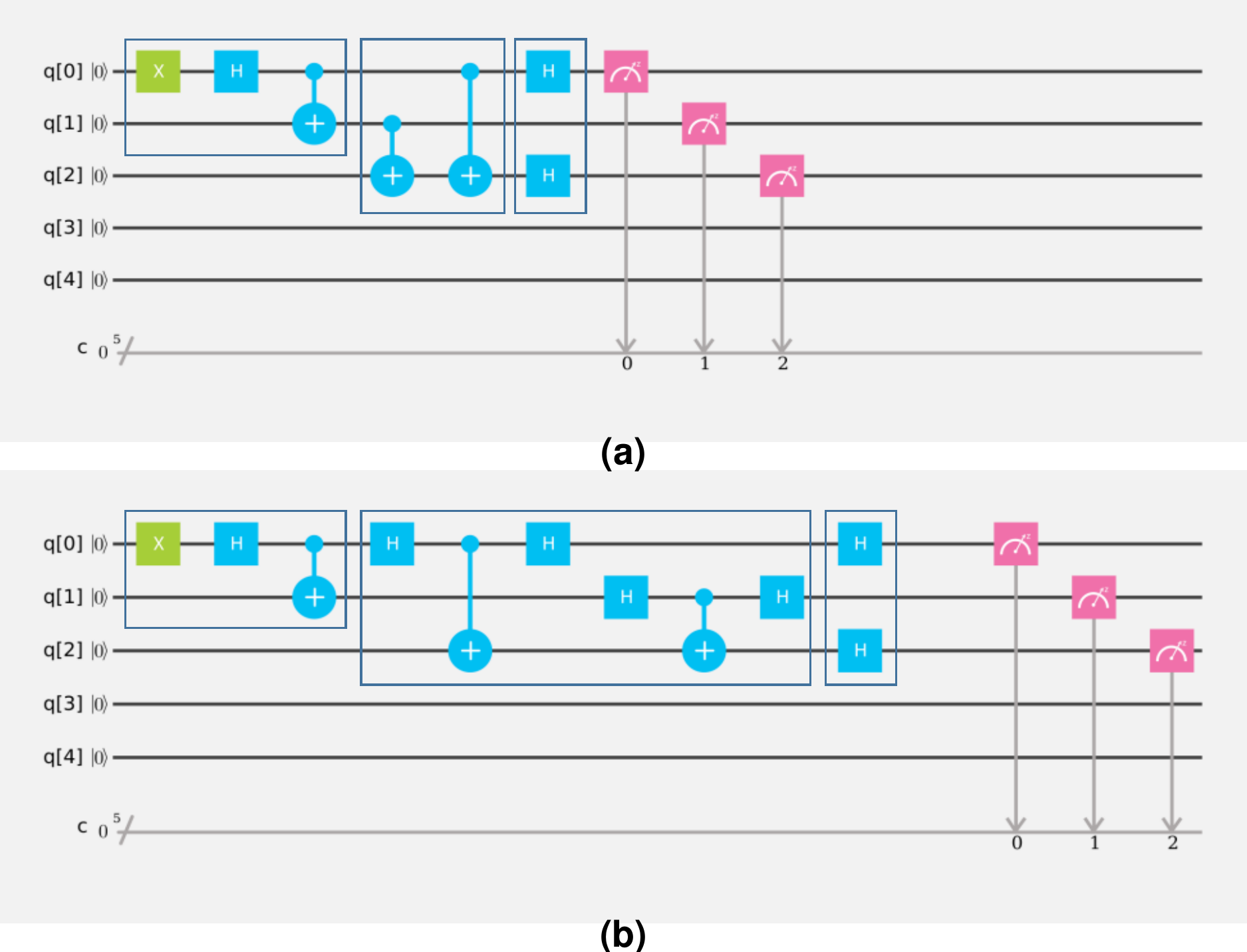}}
\par\end{centering}
\textcolor{black}{\caption{\label{fig:state-tomography}\textcolor{blue}{{} }\textcolor{black}{Experimentally
implementable (a) parity and (b) phase checking circuits on IBM QX2
followed by tomography block (involves measurement in different basis).
Here, configuration implements measurement in X- basis is shown.}}
}
\end{figure}
\textcolor{black}{As evident from the Table \ref{tab:crosstalk-1}
in IBM QX2, couplings between all pairs are not present. Absence of
couplings provides restriction on the applicability of CNOT gates.
In brief, it does not allow us to perform controlled operation between
any two qubits, and consequently restricts us to perform the nondestructive
Bell state discrimination circuit either using two independent experiments
as was done in \cite{samal2010non} (cf. Figure 1 of \cite{samal2010non})
or using a circuit having considerably high gate-count (implementation
of such a circuit will be described in the next section). Unfortunately,
IBM quantum experience does not even allow us to implement the circuits
shown in Figure \ref{fig:main-circuit} (b) in its actual form. We
need to make some modifications to obtain equivalent circuits. In
Figures \ref{fig:circuit-in-IBM} (a) and (b), we have shown the actual
circuits prepared in IBM quantum computer for parity checking and
phase information checking. In both the circuits, left-most box contains
an EPR circuit used for preparation of the Bell state $|\psi^{-}\rangle$(similarly
other states were prepared and measured). The second box from the
left in Figures \ref{fig:circuit-in-IBM} (a) and (b) provide equivalent
circuits for the circuits given in Figures \ref{fig:main-circuit}
(c) and (b), respectively, and one can easily observe that the circuit
for obtaining the phase information required decomposition for successful
implementation in IBM quantum computer. In the third box, a reverse
EPR circuit is inserted to establish that the measurement on ancilla
does not destroy the Bell state. To obtain further information about
the output states of a given circuit, QST is done. In Figure \ref{fig:state-tomography},
we have shown a circuit that can be used to perform QST and thus to
yield each element of the density matrix of the output state before
measurement is performed. Specifically, in Figure \ref{fig:state-tomography},
third block from the left, two hadamard gates are applied. This is
operated to perform state tomography. To be precise, application of
a hadamard gate transforms the measurement basis from computational
basis $\{|0\rangle,|1\rangle\}$ to diagonal basis $\{|+\rangle,|-\rangle\}$
and thus yields $\langle X\rangle X$ element of the single-qubit
density matrix. Here it would be apt to briefly describe the method
adopted here for performing state tomography and measuring fidelity
with an example. Theoretically obtained density matrix of $|\psi^{+}\rangle|0\rangle$
is,}

\textcolor{black}{
\begin{equation}
\begin{array}{lcc}
\rho^{T} & = & |\psi^{+}0\rangle\langle\psi^{+}0|,\end{array}\label{eq:3-1}
\end{equation}
where superscript $T$ indicates a theoretical (ideal) density matrix.
To check how nicely this state is prepared in experiment, we need
to reconstruct the density matrix of the output state using QST by
following the method adopted in Refs. \cite{chuang1998bulk,james2001measurement,hebenstreit2017compressed,alsina2016experimental,rundle2016quantum,filipp2009two,shukla2013ancilla}.
Characterization of a three-qubit experimental density matrix requires
extraction of information from the experiments and then using that
information to reconstruct experimental density matrix. In the Pauli
basis an experimental density matrix can be written as $\rho^{E}=\frac{1}{8}\underset{i,j,k}{\sum}c_{ijk}\sigma_{i}\otimes\sigma_{j}\otimes\sigma_{k},$
where $c_{ijk}=\langle\sigma_{i}\otimes\sigma_{j}\otimes\sigma_{k}\rangle$
and $\sigma_{ijk}=I,\,X,\,Y,\,Z$, and superscript $E$ is used to
indicate experimental density matrix. Reconstruction of this $8\times8$
density matrix of three-qubit state requires knowledge of 63 unknown
real parameters. The evaluation of co-efficient $c_{ijk}$ requires
$\langle\sigma_{i}\rangle$ where $\sigma_{i}=I,\,X,\,Y,\,Z$ for
each qubit. Since $\langle I\rangle$ can be obtain by the experiment
done on $Z$ basis, instead of four we need only three measurement
(measurement in $X$, $Y$, and $Z$ basis) on each qubit, thus requiring
total 27 measurements to tomograph each density matrix. In Chapter
\ref{cha:Introduction1}, we have already mentioned that in IBM, the
only available basis for performing the measurements is $Z$ basis.
Consequently, to realize a measurement in $X$ and $Y$ basis we have
to apply $H$ and $HS^{\dagger}$ gates, respectively prior to the
$Z$ basis measurement. Subsequently, to reconstruct the three-qubit
state $|\psi^{+}\rangle|0\rangle$ and to check how well the state
is prepared, we have to perform 27 experiments, each of which would
run 8192 times. At a later stage of the investigation, following the
same strategy, we would obtain the density matrices of the retained
state after measuring the ancilla qubits to discriminate the Bell
states}

\textcolor{black}{Once $\rho^{E}$ is obtained through QST, the same
may be used to obtain fidelity and thus a quantitative feeling about
the accuracy of the experimental implementation can be obtained. Here
it would be apt to mention that the fidelity is obtained using Eq.
(\ref{eq:fidelity-1}) of Chapter \ref{cha:Introduction1}, which
described in detail in Eq. (\ref{eq:fidelity-1}). As an example,
if we consider that the desired state is $|\psi^{+}0\rangle$ then
$\rho^{1}=|\psi^{+}0\rangle\langle\psi^{+}0|,$ and $\rho^{2}$ would
be the density matrix of the state obtained experimentally by performing
state tomography in a manner described above.}

\textcolor{black}{}
\begin{table*}
\textcolor{black}{\caption{\label{tab:one}\textcolor{blue}{{} }\textcolor{black}{Expected outcomes
after parity and phase checking circuit for ancilla-Bell-state combined
system for all Bell states. Expected obtained results of ancilla qubits
in the same cases are also given. Column 1 illustrates Bell states
to be examined. In Column 2, states of the Bell-state-ancilla composite
system is shown. Outcomes of measurements on ancilla for two cases
is shown in Columns 3 and 4 and outcomes of measuring composite states
in the computational basis is shown in Columns 5 and 6 (from left
to right).}}
}
\centering{}\textcolor{black}{}%
\begin{tabular}{|>{\centering}p{2.7cm}|>{\centering}p{2.3cm}|>{\centering}p{1.5cm}|>{\centering}p{1.5cm}|>{\centering}p{2.5cm}|>{\centering}p{2.5cm}|}
\hline 
\textcolor{black}{\small{}Bell state to be discriminated/identified
by nondestructive measurement} & \textcolor{black}{\small{}Bell-state-ancilla composite state}{\small\par}

\textcolor{black}{\small{}(considered as three-qubit state as implemented
in the experiment performed here)} & \textcolor{black}{\small{}Outcome of measurement on ancilla qubit
used for parity checking} & \textcolor{black}{\small{}Outcome of measurement on ancilla qubit
used for revealing phase information} & \textcolor{black}{\small{}Outcome of three-qubit measurement of parity
checking circuit when the output Bell state is measured after passing
through a reverse EPR circuit (cf. Figures\ref{fig:all qubits measured}
(a)-(d))} & \textcolor{black}{\small{}Outcome of three-qubit measurement of relative
phase checking circuit when the output Bell state is measured after
passing through a reverse EPR circuit (cf. Figures\ref{fig:all qubits measured}
(e)-(h))}\tabularnewline
\hline 
\textcolor{black}{\small{}$|\psi^{+}\rangle=\frac{|00\rangle+|11\rangle}{\sqrt{2}}$} & \textcolor{black}{\small{}$|\psi^{+}\rangle|0\rangle$} & \textcolor{black}{\small{}$|0\rangle$} & \textcolor{black}{\small{}$|0\rangle$} & \textcolor{black}{\small{}$|000\rangle$} & \textcolor{black}{\small{}$|000\rangle$}\tabularnewline
\hline 
\textcolor{black}{\small{}$|\psi^{-}\rangle=\frac{|00\rangle-|11\rangle}{\sqrt{2}}$} & \textcolor{black}{\small{}$|\psi^{-}\rangle|0\rangle$} & \textcolor{black}{\small{}$|0\rangle$} & \textcolor{black}{\small{}$|1\rangle$} & \textcolor{black}{\small{}$|100\rangle$} & \textcolor{black}{\small{}$|101\rangle$}\tabularnewline
\hline 
\textcolor{black}{\small{}$|\phi^{+}\rangle=\frac{|01\rangle+|10\rangle}{\sqrt{2}}$} & \textcolor{black}{\small{}$|\phi^{+}\rangle|0\rangle$} & \textcolor{black}{\small{}$|1\rangle$} & \textcolor{black}{\small{}$|0\rangle$} & \textcolor{black}{\small{}$|011\rangle$} & \textcolor{black}{\small{}$|010\rangle$}\tabularnewline
\hline 
\textcolor{black}{$|\phi^{-}\rangle=\frac{|01\rangle-|10\rangle}{\sqrt{2}}$} & \textcolor{black}{$|\phi^{-}\rangle|0\rangle$} & \textcolor{black}{$|1\rangle$} & \textcolor{black}{$|1\rangle$} & \textcolor{black}{$|111\rangle$} & \textcolor{black}{$|111\rangle$}\tabularnewline
\hline 
\end{tabular}
\end{table*}

\section{\textcolor{black}{Results\label{sec:Results}}}

\textcolor{black}{Initially, we prepare four Bell states (using EPR
circuit, i.e., a hadamard followed by CNOT) and an ancilla in state
$|0\rangle$. It is well known that an EPR circuit transforms input
states $|00\rangle,|01\rangle,|10\rangle,$ $|11\rangle$ into $|\psi^{+}\rangle=\frac{|00\rangle+|11\rangle}{\sqrt{2}}$,$\,|\phi^{+}\rangle=\frac{|01\rangle+|10\rangle}{\sqrt{2}},\,|\psi^{-}\rangle=\frac{|00\rangle-|11\rangle}{\sqrt{2}}$
and $|\phi^{-}\rangle=\frac{|01\rangle-|10\rangle}{\sqrt{2}}$, respectively.
Default initial state in IBM quantum experience is $|0\rangle$ for
each qubit line. However, the input states required by an EPR circuit
to generate different Bell states can be prepared by placing ${\rm NOT}$
gate(s) in appropriate positions before the EPR circuit (see how $|\psi^{-}\rangle$
is prepared in left most block of Figure \ref{fig:circuit-in-IBM}
). This is how Bell states are prepared here. To check the accuracy
of the states prepared in the experiments, QST of the experimentally
obtained density matrices are performed. For this purpose, we have
followed the method described in the previous section. Density matrix
for the experimentally obtained state corresponding to a particular
case (for the expected state $|\psi^{+}0\rangle,$ i.e., for $\text{\ensuremath{\rho}}^{T}=\rho^{1}=|\psi^{+}0\rangle\langle\psi^{+}0|),$
is obtained as 
\begin{equation}
\rho_{|\psi^{+}0\rangle}^{E}={\rm Re}\left[\rho_{|\psi^{+}0\rangle}^{E}\right]+i{\rm \,Im}\left[\rho_{|\psi^{+}0\rangle}^{E}\right],\label{eq:rho1}
\end{equation}
where a subscript is added to uniquely connect the experimental density
matrix with the corresponding ideal state and }\textcolor{black}{\footnotesize{}
\begin{equation}
\begin{array}{lcc}
{\rm Re}\left[\rho_{|\psi^{+}0\rangle}^{E}\right] & = & \mbox{\mbox{\ensuremath{\left(\begin{array}{cccccccc}
0.44 & 0.003 & 0.011 & -0.0102 & 0.006 & -0.005 & 0.365 & 0.007\\
0.003 & 0.002 & 0.014 & -0.005 & 0.005 & 0.005 & 0.011 & 0.001\\
0.011 & 0.014 & 0.074 & 0.003 & 0.005 & -0.004 & 0.006 & 0.002\\
-0.010 & -0.005 & 0.003 & 0.002 & -0.001 & -0.005 & -0.001 & 0\\
0.006 & 0.005 & 0.005 & -0.005 & 0.073 & 0.001 & 0.0035 & -0.004\\
-0.005 & 0.005 & -0.004 & -0.005 & 0.001 & 0.002 & 0.011 & -0.005\\
0.365 & 0.011 & 0.006 & -0.002 & 0.003 & 0.011 & 0.408 & 0.01\\
0.007 & 0.001 & 0.002 & 0 & -0.004 & -0.005 & 0.01 & 0.001
\end{array}\right),}}}\end{array}\label{eq:Repsi0}
\end{equation}
}\textcolor{black}{and }\textcolor{black}{\footnotesize{}
\begin{equation}
\begin{array}{lcc}
{\rm Im}\left[\rho_{|\psi^{+}0\rangle}^{E}\right] & = & \left(\begin{array}{cccccccc}
0 & -0.018 & -0.024 & -0.027 & -0.034 & -0.003 & -0.030 & -0.018\\
0.018 & 0 & -0.003 & 0 & -0.002 & 0 & 0.021 & 0\\
0.024 & 0.003 & 0 & 0.018 & 0.030 & -0.002 & -0.01 & 0.006\\
0.027 & 0 & -0.018 & 0 & 0.004 & -0.005 & 0.0032 & 0\\
0.034 & 0.002 & -0.030 & -0.004 & 0 & -0.005 & -0.003 & -0.023\\
0.003 & 0 & 0.002 & 0.005 & 0.005 & 0 & -0.007 & 0\\
0.030 & -0.021 & 0.01 & -0.003 & 0.003 & 0.007 & 0 & -0.003\\
0.018 & 0 & -0.006 & 0 & 0.023 & 0 & 0.003 & 0
\end{array}\right).\end{array}\label{eq:Impsi0}
\end{equation}
}{\footnotesize\par}

\begin{figure}
\begin{centering}
\textcolor{black}{\includegraphics[scale=0.65]{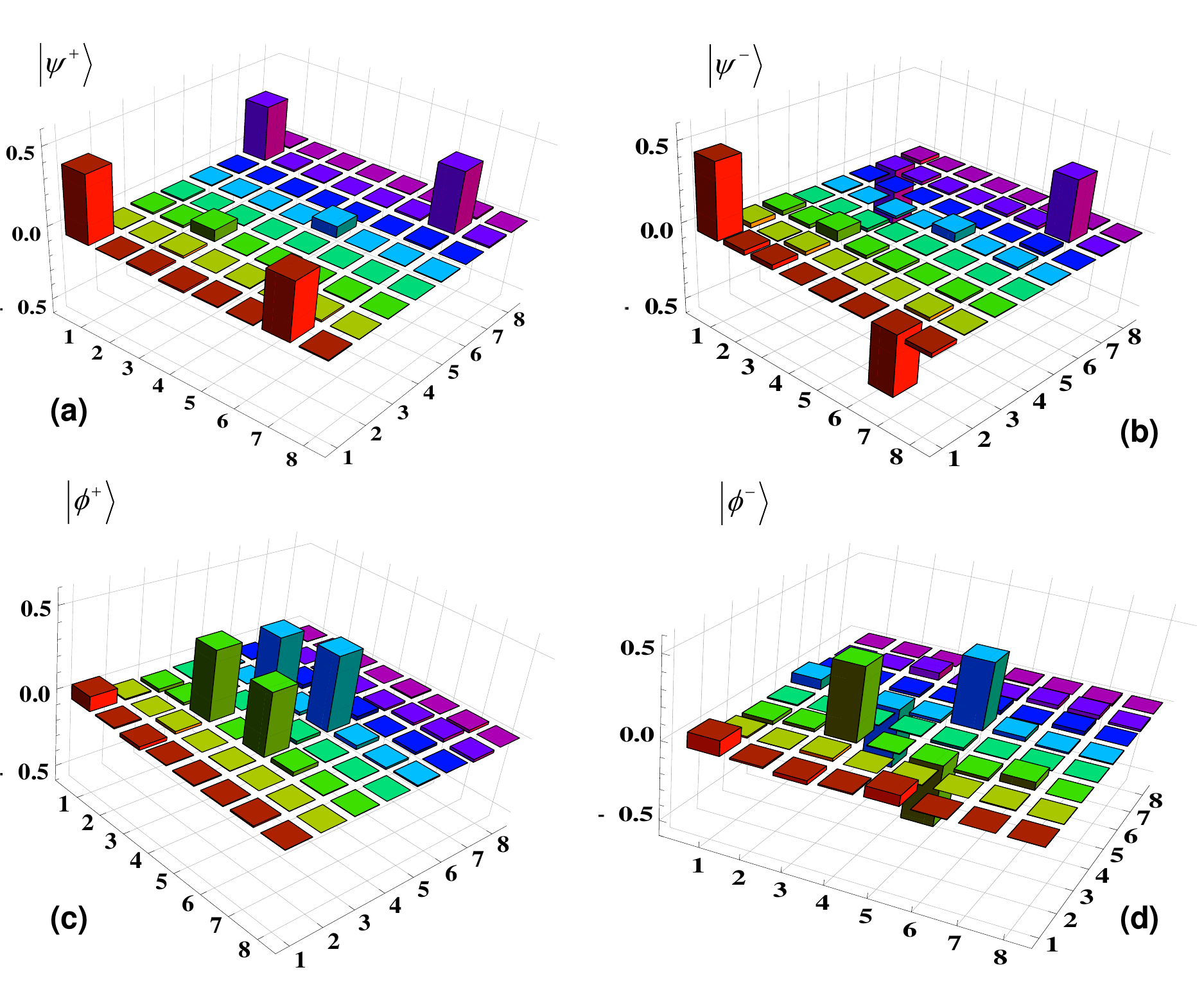}}
\par\end{centering}
\textcolor{black}{\caption{\label{fig:bell-reconstruct}\textcolor{blue}{{} }Reconstructed Bell
states on Bell-state-ancilla composite system corresponding to ideal
states (a) $|\psi^{+}0\rangle$, (b) $|\psi^{-}0\rangle$, (c) $|\phi^{+}0\rangle$,
(d) and $|\phi^{-}0\rangle.$ In each plot, the states $|000\rangle$,\,$|001\rangle$,\,$|010\rangle$,\,$|011\rangle$,\,$|100\rangle,$
$\text{\,}|101\rangle,\text{\,}|110\rangle$ and $|111\rangle$ are
labeled as 1\textendash 8 consecutively in X and Y axis.}
}
\end{figure}

\textcolor{black}{Real part of this density matrix is illustrated
in Figure \ref{fig:bell-reconstruct} (a). Figure \ref{fig:bell-reconstruct}
also illustrates the real part of density matrices of the experimentally
prepared Bell-state-ancilla composite in the other three cases. Corresponding
density matrices are provided below (see Eq. (\ref{eq:repsi-0})-(\ref{eq:Imphi0})).}

\textcolor{black}{\footnotesize{}
\begin{equation}
\begin{array}{lcc}
{\rm Re}\left[\rho_{|\psi^{-}0\rangle}^{E}\right] & = & \left(\begin{array}{cccccccc}
0.476 & 0.029 & 0.029 & -0.001 & -0.007 & -0.005 & -0.375 & -0.024\\
0.029 & 0.003 & 0.022 & 0.001 & 0.002 & 0.0 & -0.021 & -0.003\\
0.029 & 0.022 & 0.066 & -0.02 & -0.015 & -0.001 & -0.012 & 0.002\\
-0.001 & 0.001 & -0.02 & 0.001 & 0.003 & -0.001 & -0.001 & 0.0\\
-0.007 & 0.002 & -0.015 & 0.003 & 0.058 & 0.005 & 0.018 & 0.002\\
-0.005 & 0.0 & -0.0013 & -0.001 & 0.005 & 0.0 & 0.020 & 0.0\\
-0.375 & -0.021 & -0.012 & -0.001 & 0.018 & 0.020 & 0.393 & 0.006\\
-0.024 & -0.003 & 0.002 & 0.0 & 0.002 & 0.0 & 0.006 & 0.003
\end{array}\right),\end{array}\label{eq:repsi-0}
\end{equation}
}{\footnotesize\par}

\textcolor{black}{\footnotesize{}
\begin{equation}
\begin{array}{lcc}
{\rm Im}\left[\rho_{|\psi^{-}0\rangle}^{E}\right] & = & \left(\begin{array}{cccccccc}
0 & -0.022 & -0.01 & -0.019 & -0.009 & 0 & -0.007 & 0.019\\
0.022 & 0 & 0.002 & 0.0 & -0.001 & -0.001 & -0.019 & 0.001\\
0.01 & -0.002 & 0 & 0.023 & -0.013 & -0.003 & -0.012 & -0.001\\
0.019 & 0 & -0.023 & 0 & 0.004 & 0 & -0.001 & 0\\
0.009 & 0.001 & 0.013 & -0.004 & 0 & -0.003 & -0.012 & -0.015\\
0 & 0.001 & 0.003 & 0 & 0.003 & 0.0 & -0.004 & -0.001\\
0.007 & 0.019 & 0.012 & 0.001 & 0.012 & 0.004 & 0 & 0.001\\
-0.019 & -0.001 & 0.001 & 0 & 0.015 & 0.001 & -0.001 & 0
\end{array}\right),\end{array}\label{eq:63}
\end{equation}
}{\footnotesize\par}

\textcolor{black}{\footnotesize{}
\begin{equation}
\begin{array}{lcc}
{\rm Re}\left[\rho_{|\phi^{+}0\rangle}^{E}\right] & = & \left(\begin{array}{cccccccc}
0.089 & 0.007 & -0.021 & 0.004 & -0.006 & -0.004 & 0.008 & -0.002\\
0.007 & 0.0 & 0.016 & 0.0 & -0.002 & 0.0 & -0.001 & 0.0\\
-0.021 & 0.016 & 0.429 & -0.001 & 0.382 & 0.024 & 0.004 & 0.002\\
0.004 & 0.0 & -0.001 & 0.001 & 0.02 & 0.003 & 0.002 & 0.0\\
-0.006 & -0.002 & 0.382 & 0.02 & 0.459 & 0.024 & -0.011 & 0.008\\
-0.004 & 0.0 & 0.024 & 0.003 & 0.024 & 0.002 & 0.015 & -0.001\\
0.008 & -0.001 & 0.004 & 0.002 & -0.011 & 0.015 & 0.02 & -0.017\\
-0.002 & 0.0 & 0.002 & 0.0 & 0.008 & -0.001 & -0.017 & 0.0
\end{array}\right),\end{array}\label{eq:64}
\end{equation}
}{\footnotesize\par}

\textcolor{black}{\footnotesize{}
\begin{equation}
\begin{array}{lcc}
{\rm Im}\left[\rho_{|\phi^{+}0\rangle}^{E}\right] & = & \left(\begin{array}{cccccccc}
0 & -0.007 & -0.016 & -0.016 & -0.03 & 0.001 & 0.001 & 0.003\\
0.007 & 0.0 & 0.001 & 0 & 0.001 & -0.001 & -0.003 & -0.001\\
0.016 & -0.001 & 0 & -0.009 & -0.022 & -0.024 & -0.017 & 0.001\\
0.016 & 0 & 0.009 & 0 & 0.019 & 0.001 & 0.005 & 0.001\\
0.03 & -0.001 & 0.022 & -0.019 & 0 & -0.024 & -0.011 & -0.011\\
-0.001 & 0.001 & 0.025 & -0.001 & 0.024 & 0 & 0.003 & -0.001\\
-0.001 & 0.003 & 0.017 & -0.005 & 0.011 & -0.0025 & 0 & 0.013\\
-0.003 & 0.001 & -0.001 & -0.001 & 0.011 & 0.0005 & -0.013 & 0
\end{array}\right),\end{array}\label{eq:65}
\end{equation}
}{\footnotesize\par}

\textcolor{black}{\footnotesize{}
\begin{equation}
\begin{array}{lcc}
{\rm Re}\left[\rho_{|\phi^{-}0\rangle}^{E}\right] & = & \left(\begin{array}{cccccccc}
0.092 & 0.009 & -0.023 & -0.008 & 0.063 & 0.002 & 0 & -0.001\\
0.009 & 0.001 & 0.018 & 0 & 0.004 & 0 & 0.005 & -0.002\\
-0.023 & 0.017 & 0.454 & 0.005 & -0.38 & -0.017 & 0.047 & -0.001\\
-0.008 & 0 & 0.005 & 0.005 & -0.017 & -0.003 & 0.001 & 0.001\\
0.063 & 0.004 & -0.384 & -0.017 & 0.42 & 0.026 & -0.009 & -0.005\\
0.002 & 0 & -0.017 & -0.002 & 0.025 & 0.003 & 0.025 & 0\\
0 & 0.004 & 0.047 & 0.001 & -0.009 & 0.025 & 0.025 & -0.013\\
-0.001 & -0.002 & -0.001 & 0.001 & -0.005 & 0 & -0.013 & 0
\end{array}\right),\end{array}\label{eq:66}
\end{equation}
}and

\textcolor{black}{\footnotesize{}
\begin{equation}
\begin{array}{lcc}
{\rm Im}\left[\rho_{|\phi^{-}0\rangle}^{E}\right] & = & \left(\begin{array}{cccccccc}
0 & -0.006 & -0.017 & -0.027 & -0.022 & -0.006 & -0.012 & -0.007\\
0.006 & 0 & -0.006 & 0 & 0.008 & -0.001 & -0.007 & -0.001\\
0.017 & 0.006 & 0 & 0.001 & 0.026 & 0.015 & -0.025 & 0\\
0.027 & 0 & -0.001 & 0 & -0.012 & 0.001 & -0.004 & 0\\
0.022 & -0.008 & -0.026 & 0.012 & 0 & -0.018 & -0.009 & -0.029\\
0.006 & 0.001 & -0.015 & -0.001 & 0.018 & 0 & -0.004 & 0.001\\
0.012 & 0.007 & 0.025 & 0.004 & 0.009 & 0.004 & 0 & 0.014\\
0.007 & 0.001 & 0 & 0 & 0.029 & -0.001 & -0.014 & 0
\end{array}\right).\end{array}\label{eq:Imphi0}
\end{equation}
}{\footnotesize\par}

\textcolor{black}{Figure \ref{fig:bell-reconstruct} clearly shows
that the Bell states are prepared with reasonable amount of accuracy,
but does not provide any quantitative measure of accuracy. So we have
calculated the ``average absolute deviation'' $\langle\Delta x\rangle$
and ``maximum absolute deviation'' $\Delta x_{max}$ of the experimental
density matrix from the theoretical one by using this formulae}

\textcolor{black}{
\begin{equation}
\begin{array}{lcc}
\langle\Delta x\rangle & = & \frac{1}{N^{2}}\stackrel[i,j=1]{N}{\sum|x_{i,j}^{T}}-x_{i,j}^{E}|,\\
\Delta x_{max} & = & Max|x_{i,j}^{T}-x_{i,j}^{E}|,\\
\forall i,\,j & \in & \left\{ 1,\,N\right\} .
\end{array}\label{eq:5-2}
\end{equation}
where, $x_{ij}^{T}$ and $x_{ij}^{E}$ are the theoretical and experimental
elements. Putting the values in this equation from Eq. (\ref{eq:Repsi0})
and Eq. (\ref{eq:Impsi0}) we find the ``average absolute deviation''
and the ``maximum absolute deviation'' of the Bell state $|\psi^{+}\rangle|0\rangle$
which is $1.8\%$ and $13.7\%$.}

\textcolor{black}{Similarly, we have also calculated ``average absolute
deviation'' and ``maximum absolute deviation'' for other Bell states
too, these values are $|\psi^{-}\rangle|0\rangle$ is $1.8\%$ and
$12.5\%$, $|\phi^{+}\rangle$ is $1.8\%$ and $11.9\%$ and $|\phi^{-}\rangle$
is $2\%$ and $11.8\%$, respectively. It may be noted that in NMR-based
experiment \cite{samal2010non}, the values for ``average absolute
deviation'' and ``maximum absolute deviation'' is reported as $\sim1\%$
and $\sim4\%$, respectively. This probably indicates, that Bell states
were prepared in NMR-based experiment, with higher accuracy, but the
measure used there was not universal. So we also compute fidelity
of the reconstructed state with respect to the desired state and obtained
following values of the fidelity for various Bell-state-ancilla composite
system: $F_{|\psi^{+}0\rangle}=0.8890,\,F_{|\psi^{-}0\rangle}=0.8994,\,F_{|\phi^{+}0\rangle}=0.9091,$
and $F_{|\phi^{-}0\rangle}=0.9060,$ where $F_{|i\rangle}$ denotes
the fidelity at which the desired quantum state $|i\rangle$ is prepared
in the IBM quantum computer. The computed values of $F_{|i\rangle}$
clearly show that the Bell-state-ancilla composite system has been
prepared with reasonable accuracy in the present experiment. The values
of fidelity obtained here cannot be considered very high in comparison
to the fidelity with which standard quantum states are prepared in
other technologies. For example, in Ref. \cite{benedetti2013experimental}
various states (including a set of phase-dampled Bell states and $X$
states belonging to the families of Werner states) were prepared with
fidelity $\geq0.96$, and in some cases, fidelity was as high as $0.998\pm0.002$
(cf. Table 1 of Ref. \cite{benedetti2013experimental}). Similarly,
in the NMR-based implementation, some of the  authors had earlier
prepared states with an average fidelity of 0.99 \cite{shukla2013ancilla,joshi2014estimating}.
For example, quantum states $\rho^{1}=\frac{1}{2}(\sigma_{z}^{1}+\sigma_{z}^{2})$
and $\rho^{2}=\frac{1}{2}(\sigma_{x}^{1}+\sigma_{x}^{2})-\frac{1}{2}(\sigma_{y}^{1}+\sigma_{y}^{2})+\frac{1}{\sqrt{2}}(\sigma_{z}^{1}+\sigma_{z}^{2})$
were prepared with fidelities $0.997$ and $0.99$, respectively \cite{shukla2013ancilla}.
Before, we proceed further and continue our analysis using fidelity
as a measure of figure of merit that quantify the similarity (closeness)
between two quantum states, we must note that in Ref. \cite{bina2014drawbacks},
it was established that high fidelities may be achieved by pairs of
quantum states with considerably different physical properties. Further,
in \cite{benedetti2013experimental}, it was shown for the families
of depolarized or phase-damped states that two states having high
fidelity may have largely different values of discord. Thus, one needs
to be extra cautious while using fidelity as a quantitative measure
of closeness. However, it is very frequently used in the works related
to quantum information processing in general and in particular, it
has also been used in a set of works \cite{majumder2017experimental,sisodia2017design,gangopadhyay2018generalization}
reporting interesting results using IBM quantum experience. In this
work, we also use fidelity as a quantitative measure of closeness.}

\begin{figure}
\begin{centering}
\textcolor{black}{\includegraphics[scale=0.66]{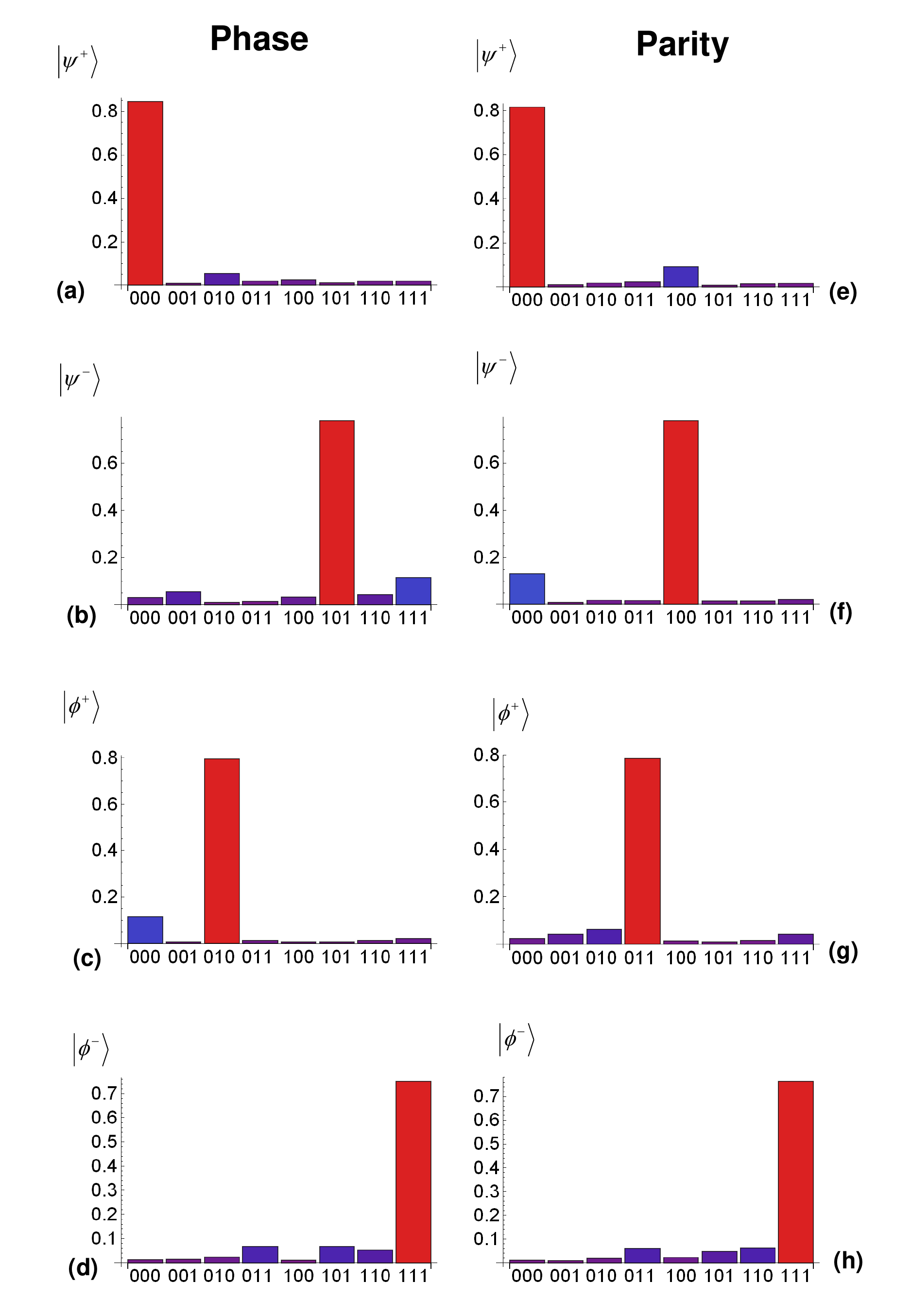}}
\par\end{centering}
\textcolor{black}{\caption{\label{fig:all qubits measured} \textcolor{black}{Experimental results
obtained by implementing the circuits shown in Figure \ref{fig:circuit-in-IBM}.
Measurement results after implementing phase checking and parity checking
circuits are shown in (a)-(d) and (e)-(h), respectively.}}
}
\end{figure}

\textcolor{black}{After verifying that the Bell-state-ancilla composite
system are prepared successfully, we perform measurements on the ancilla
to perform nondestructive discrimination of the Bell state. After
the measurement of ancilla, the Bell state is expected to remain unchanged,
to test that a reverse EPR circuit is applied to the system qubits,
and subsequently the system qubits are measured in computational basis.
The reverse EPR circuit actually transforms a Bell measurement into
a measurement in the computational basis. }
\textcolor{black}{Outcomes of these measurements are shown in Figure
\ref{fig:all qubits measured}, within the experimental error, these
results are consistent with the expected theoretical results shown
in Column 4 and 5 of Table \ref{tab:one}. Thus, nondestructive discrimination
is successfully performed in IBM quantum computer.}

\textcolor{black}{Figure \ref{fig:all qubits measured} definitely
indicates a successful implementation of nondestructive discrimination
of Bell states. However, it does not reveal the whole picture. To
obtain the full picture, we perform QST after implementation of parity
checking circuit and phase information checking circuits. In the following,
we provide experimental density matrices ${\normalcolor {\left[\rho^{E}{}_{|\psi^{+}0\rangle}\right]}}$
for both the cases (phase and parity) corresponding to the ideal state
$|\psi^{+}0\rangle$ as}

\textcolor{black}{}
\textcolor{black}{}
\begin{figure}[h]
\begin{centering}
\textcolor{black}{\includegraphics[scale=0.6]{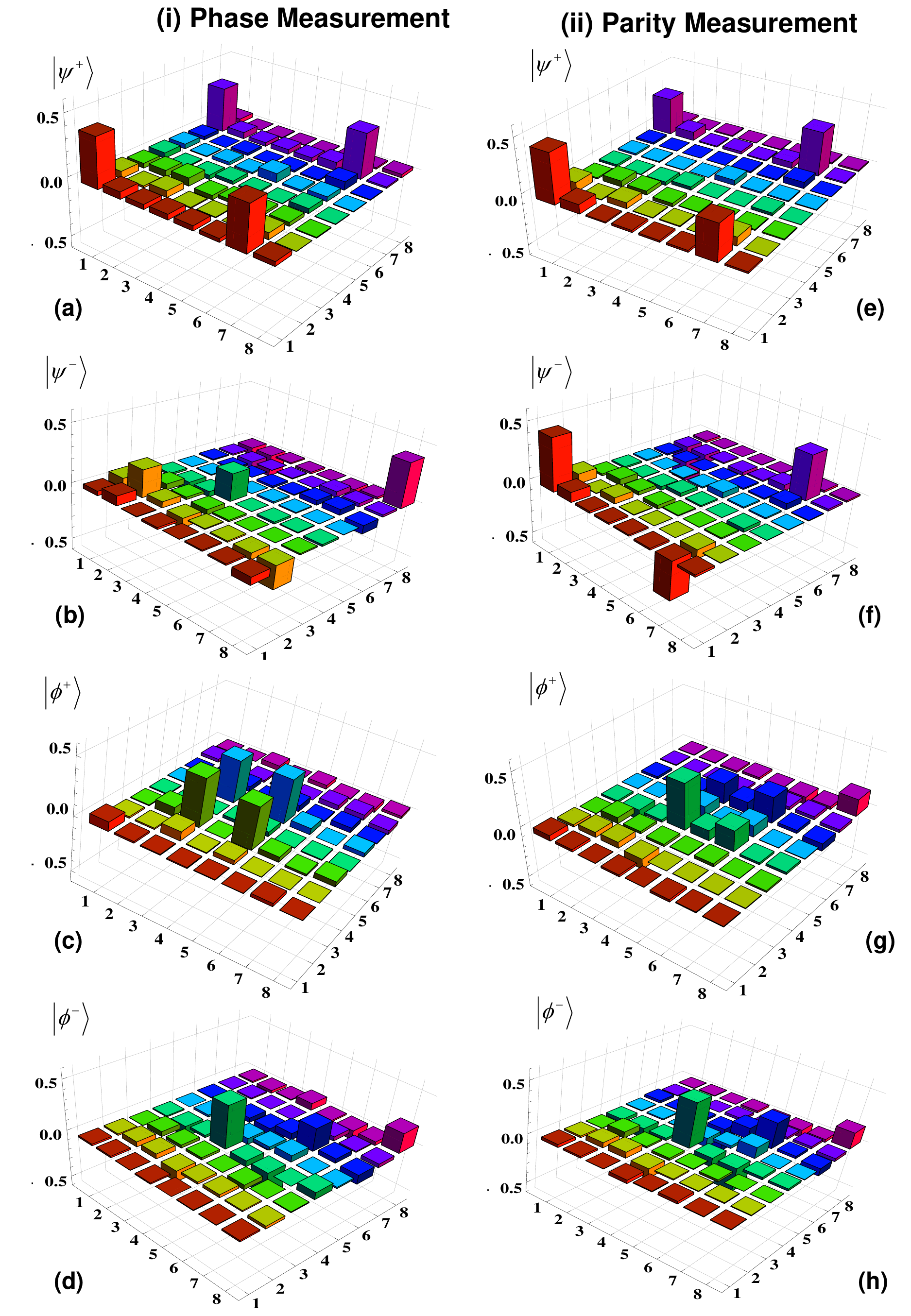}}
\par\end{centering}
\textcolor{black}{\caption{\label{fig:Final-fig} \textcolor{black}{Reconstructed density matrices
of various states of Bell-state-ancilla composite system obtained
after the implementation of phase and parity checking circuits shown
in Figure \ref{fig:state-tomography}. The first column (a)-(d) illustrates
the density matrices corresponding to ideal quantum states $|\psi^{+}0\rangle$,
$|\psi^{-}1\rangle$, $|\phi^{+}0\rangle$, and $|\phi^{-}1\rangle$,
respectively. The density matrices in the second column correspond
to ideal states (e)-(h) $|\psi^{+}0\rangle$, $|\psi^{-}0\rangle$,
$|\phi^{+}1\rangle$, and $|\phi^{-}1\rangle,$ respectively.}}
}
\end{figure}
\textcolor{black}{}
\begin{figure}[h]
\begin{centering}
\textcolor{black}{\includegraphics[scale=0.7]{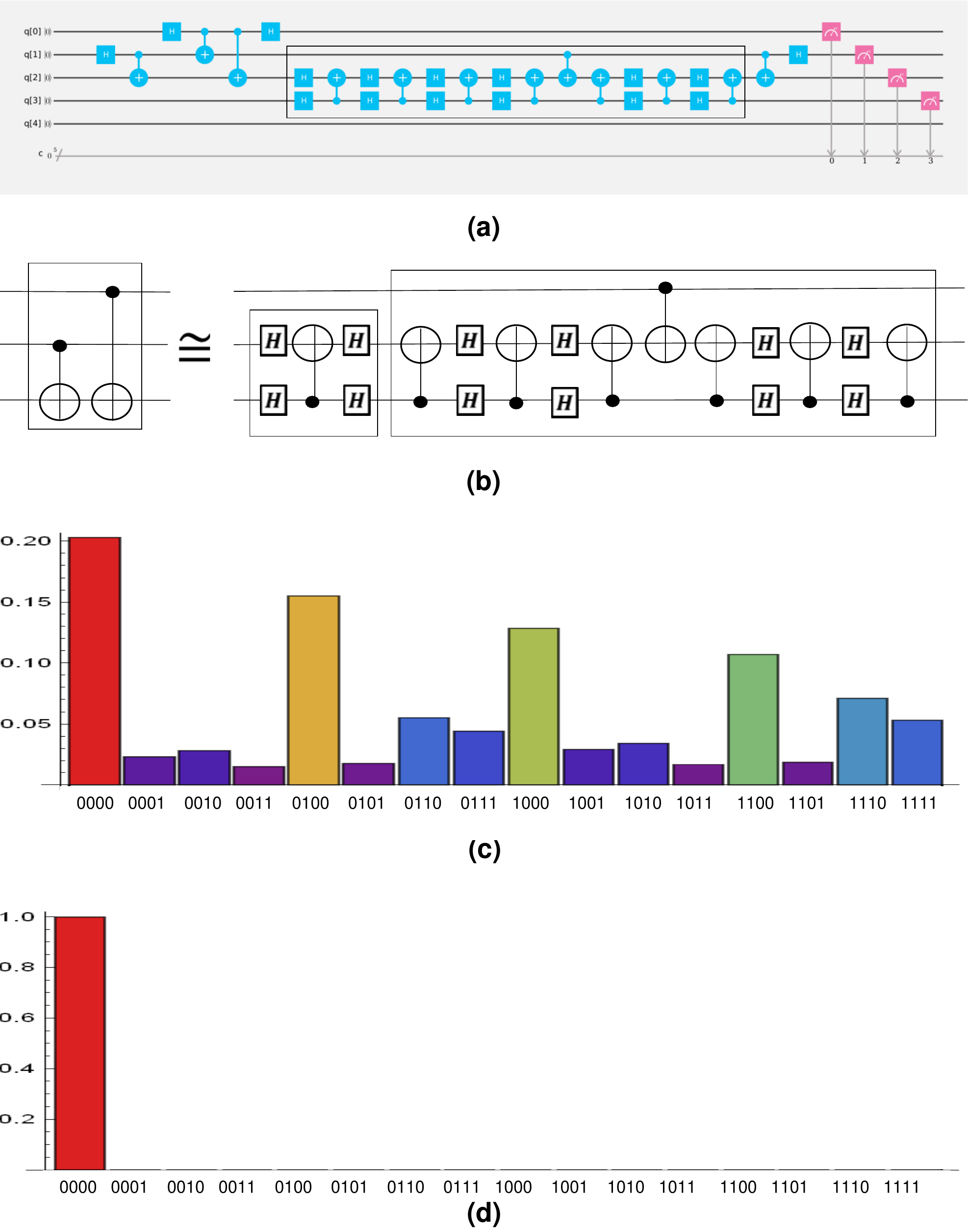}}
\par\end{centering}
\textcolor{black}{\caption{\label{fig:fig4}\textcolor{black}{(a) For the Bell state $|\psi^{+}\rangle,$
actual implementation of the combined (phase and parity checking)
four-qubit quantum circuit shown in Figure \ref{fig:main-circuit}
a using five-qubit IBM quantum computer. (b) To circumvent the constraints
of IBM quantum computer, the implemented circuit utilizes this circuit
identity. (c) Probability of various measurement outcomes obtained
in actual quantum computer after running the circuit 8192 times. (d)
Simulated outcome after running the circuit on IBM simulator. This
outcome coincides exactly with the expected theoretical value, but
the experimental outcome shown in (c) is found to deviate considerably
from it.}}
}
\end{figure}
\textcolor{black}{\footnotesize{}
\begin{equation}
\begin{array}{lcc}
{\rm Re}\left[\rho_{|\psi^{+}0\rangle}^{E}\right]_{parity} & = & \left(\begin{array}{cccccccc}
0.462 & 0.084 & 0.011 & 0.019 & 0.016 & 0.017 & 0.34 & 0.015\\
0.084 & 0.005 & 0.0537 & -0.005 & 0.008 & 0.001 & 0.079 & -0.002\\
0.011 & 0.053 & 0.036 & -0.032 & -0.005 & 0.001 & 0.009 & 0.009\\
0.019 & -0.005 & -0.032 & 0.029 & -0.002 & 0.002 & 0.024 & 0.002\\
0.016 & 0.008 & -0.005 & -0.002 & 0.046 & 0.012 & 0.009 & 0.006\\
0.017 & 0.001 & 0.001 & 0.002 & 0.012 & 0.010 & 0.057 & -0.003\\
0.34 & 0.079 & 0.009 & 0.024 & 0.009 & 0.056 & 0.399 & 0.012\\
0.015 & -0.002 & 0.009 & 0.002 & 0.006 & -0.003 & 0.012 & 0.012
\end{array}\right),\end{array}\label{eq:reparitypsi0}
\end{equation}
}{\footnotesize\par}

\textcolor{black}{\footnotesize{}
\begin{equation}
\begin{array}{lcc}
{\rm Im}\left[\rho_{|\psi^{+}0\rangle}^{E}\right]_{parity} & = & \left(\begin{array}{cccccccc}
0 & -0.054 & -0.007 & -0.057 & -0.011 & -0.011 & -0.148 & -0.083\\
0.054 & 0 & 0.007 & 0.006 & -0.005 & -0.001 & 0.006 & -0.002\\
0.007 & -0.007 & 0 & 0.046 & -0.011 & -0.006 & -0.009 & 0.005\\
0.057 & -0.006 & -0.046 & 0 & 0 & 0.001 & -0.009 & -0.004\\
0.011 & 0.005 & 0.011 & 0 & 0 & -0.011 & -0.012 & -0.039\\
0.011 & 0.001 & 0.006 & -0.002 & 0.011 & 0 & -0.003 & 0.005\\
0.148 & -0.006 & 0.009 & 0.009 & 0.012 & 0.003 & 0 & -0.031\\
0.083 & 0.002 & -0.005 & 0.004 & 0.039 & -0.005 & 0.031 & 0
\end{array}\right),\end{array}\label{eq:Imparitypsi0}
\end{equation}
}{\footnotesize\par}

\textcolor{black}{\footnotesize{}
\begin{equation}
\begin{array}{lcc}
{\rm Re}\left[\rho_{|\psi^{+}0\rangle}^{E}\right]_{phase} & = & \left(\begin{array}{cccccccc}
0.409 & 0.058 & 0.038 & 0.0437 & 0.0385 & 0.028 & 0.366 & 0.035\\
0.058 & 0.014 & 0.052 & -0.007 & 0.0055 & 0 & 0.0523 & 0.001\\
0.038 & 0.052 & 0.064 & -0.034 & 0.024 & 0.019 & 0.03 & -0.006\\
0.043 & -0.007 & -0.034 & 0.026 & 0.011 & -0.004 & 0.038 & -0.004\\
0.038 & 0.005 & 0.024 & 0.011 & 0.073 & 0.015 & 0.037 & -0.004\\
0.028 & 0 & 0.019 & -0.004 & 0.015 & 0.009 & 0.067 & -0.005\\
0.366 & 0.052 & 0.03 & 0.038 & 0.037 & 0.067 & 0.373 & 0.012\\
0.035 & 0.001 & -0.006 & -0.004 & -0.004 & -0.005 & 0.012 & 0.027
\end{array}\right),\end{array}\label{eq:rephasepsi0}
\end{equation}
}and

\textcolor{black}{\footnotesize{}
\begin{equation}
\begin{array}{lcc}
{\rm Im}\left[\rho_{|\psi^{+}0\rangle}^{E}\right]_{phase} & = & \left(\begin{array}{cccccccc}
0 & -0.040 & 0.039 & -0.034 & 0.019 & 0.005 & -0.021 & -0.080\\
0.040 & 0 & 0.015 & 0.005 & 0.009 & 0 & 0.022 & 0.002\\
-0.039 & -0.015 & 0 & 0.043 & -0.002 & -0.014 & -0.054 & 0.005\\
0.034 & -0.005 & -0.043 & 0 & 0.013 & 0.003 & -0.02 & -0.01\\
-0.019 & -0.009 & 0.002 & -0.013 & 0 & -0.009 & -0.059 & -0.054\\
-0.005 & 0 & 0.014 & -0.003 & 0.008 & 0 & -0.016 & 0.007\\
0.021 & -0.022 & 0.054 & 0.02 & 0.059 & 0.016 & 0 & -0.043\\
0.080 & -0.002 & -0.005 & 0.01 & 0.054 & -0.007 & 0.043 & 0
\end{array}\right).\end{array}\label{eq:Imphasepsi0}
\end{equation}
}These density matrices are obtained through QST.

\textcolor{black}{The subscripts ``parity'' and ``phase'' denotes
the experiment for which the experimental density matrix is obtained
via QST.}\textcolor{black}{\footnotesize{} }\textcolor{black}{Corresponding
density matrices for the other Bell-state-ancilla composites are reported
below (Eq. (\ref{eq:rephasepsi-1})-(\ref{eq:Imparity})).}

\textcolor{black}{\footnotesize{}
\begin{equation}
\begin{array}{lcc}
{\rm Re}\left[\rho_{|\psi^{-}1\rangle}^{E}\right]_{phase} & = & \left(\begin{array}{cccccccc}
0.037 & 0.081 & -0.003 & -0.015 & -0.002 & 0.009 & -0.011 & -0.062\\
0.081 & 0.232 & 0.049 & -0.077 & 0.007 & -0.009 & -0.054 & -0.192\\
-0.003 & 0.049 & 0.020 & -0.043 & -0.013 & -0.002 & 0.001 & 0.009\\
-0.015 & -0.077 & -0.043 & 0.233 & 0.001 & -0.016 & 0.007 & -0.015\\
-0.002 & 0.007 & -0.013 & 0.001 & 0.013 & 0.008 & 0.001 & 0.002\\
0.008 & -0.009 & -0.002 & -0.016 & 0.008 & 0.0391 & 0.048 & -0.07\\
-0.011 & -0.054 & 0.001 & 0.007 & 0.001 & 0.048 & 0.030 & 0.018\\
-0.062 & -0.192 & 0.009 & -0.015 & 0.002 & -0.07 & 0.018 & 0.395
\end{array}\right),\end{array}\label{eq:rephasepsi-1}
\end{equation}
}{\footnotesize\par}

\textcolor{black}{\footnotesize{}
\begin{equation}
\begin{array}{lcc}
{\rm Im}\left[\rho_{|\psi^{-}1\rangle}^{E}\right]_{phase} & = & \left(\begin{array}{cccccccc}
0 & -0.085 & 0.003 & -0.062 & -0.001 & 0.009 & 0.007 & 0.075\\
0.085 & 0 & -0.012 & 0.052 & -0.001 & 0.023 & -0.022 & 0.03\\
-0.003 & 0.012 & 0 & 0.040 & -0.002 & -0.006 & -0.003 & 0.004\\
0.062 & -0.052 & -0.040 & 0 & 0.002 & 0.006 & -0.015 & -0.051\\
0.001 & 0.001 & 0.002 & -0.002 & 0 & -0.011 & 0 & -0.048\\
-0.009 & -0.023 & 0.006 & -0.006 & 0.011 & 0 & 0.016 & 0.119\\
-0.007 & 0.021 & 0.003 & 0.015 & 0 & -0.016 & 0 & 0.016\\
-0.075 & -0.03 & -0.004 & 0.051 & 0.048 & -0.119 & -0.016 & 0
\end{array}\right)\end{array},\label{eq:74}
\end{equation}
}{\footnotesize\par}

\textcolor{black}{\footnotesize{}
\begin{equation}
\begin{array}{lcc}
{\rm Re}\left[\rho_{|\phi^{+}0\rangle}^{E}\right]_{phase} & = & \left(\begin{array}{cccccccc}
0.079 & 0.013 & 0.007 & -0.008 & 0.004 & -0.0022 & 0.026 & -0.002\\
0.013 & 0.010 & 0.078 & -0.007 & 0.0362 & 0.002 & -0.001 & 0\\
0.007 & 0.078 & 0.418 & 0.001 & 0.37 & 0.0427 & -0.016 & 0.042\\
-0.008 & -0.007 & 0.001 & 0.024 & 0.039 & -0.002 & -0.007 & 0.001\\
0.004 & 0.036 & 0.37 & 0.039 & 0.387 & 0.053 & 0 & 0.034\\
-0.002 & 0.002 & 0.042 & -0.002 & 0.053 & 0.0098 & 0.041 & -0.005\\
0.026 & -0.001 & -0.016 & -0.007 & 0 & 0.041 & 0.046 & -0.037\\
-0.002 & 0 & 0.042 & 0.001 & 0.034 & -0.006 & -0.037 & 0.020
\end{array}\right)\end{array},\label{eq:75}
\end{equation}
}{\footnotesize\par}

\textcolor{black}{\footnotesize{}
\begin{equation}
\begin{array}{lcc}
{\rm Im}\left[\rho_{|\phi^{+}0\rangle}^{E}\right]_{phase} & = & \left(\begin{array}{cccccccc}
0 & -0.015 & -0.065 & -0.056 & -0.055 & -0.01 & 0.002 & 0.002\\
0.015 & 0 & -0.023 & 0.006 & -0.021 & 0.005 & 0.007 & 0.002\\
0.065 & 0.023 & 0 & -0.024 & 0.002 & -0.045 & 0.045 & 0.022\\
0.056 & -0.005 & 0.024 & 0 & 0.064 & 0.005 & 0.002 & 0.005\\
0.055 & 0.021 & -0.001 & -0.064 & 0 & -0.043 & 0.056 & -0.024\\
0.01 & -0.005 & 0.045 & -0.005 & 0.043 & 0 & 0.006 & 0.005\\
-0.002 & -0.007 & -0.045 & -0.002 & -0.056 & -0.006 & 0 & 0.035\\
-0.002 & -0.005 & -0.022 & -0.004 & 0.024 & -0.005 & -0.035 & 0
\end{array}\right)\end{array},\label{eq:76}
\end{equation}
}{\footnotesize\par}

\textcolor{black}{\footnotesize{}
\begin{equation}
\begin{array}{ccc}
{\rm Re}\left[\rho_{|\phi^{-}1\rangle}^{E}\right]_{phase} & = & \left(\begin{array}{cccccccc}
0.015 & 0.014 & -0.001 & -0.010 & 0 & 0.0055 & 0.0032 & -0.013\\
0.014 & 0.05 & 0.048 & -0.107 & 0.020 & 0.028 & 0.011 & -0.027\\
-0.001 & 0.048 & 0.029 & 0.022 & -0.002 & -0.045 & -0.001 & 0.008\\
-0.010 & -0.107 & 0.022 & 0.431 & -0.058 & -0.231 & 0.006 & 0.069\\
0 & 0.020 & -0.002 & -0.058 & 0.032 & 0.064 & -0.002 & -0.013\\
0.005 & 0.028 & -0.045 & -0.231 & 0.064 & 0.23 & 0.042 & -0.091\\
0.0032 & 0.011 & -0.001 & 0.006 & -0.002 & 0.0422 & 0.01 & -0.037\\
-0.013 & -0.027 & 0.008 & 0.069 & -0.013 & -0.091 & -0.037 & 0.203
\end{array}\right),\end{array}\label{eq:77}
\end{equation}
}{\footnotesize\par}

\textcolor{black}{\footnotesize{}
\begin{equation}
\begin{array}{lcc}
{\rm Im}\left[\rho_{|\phi^{-}1\rangle}^{E}\right]_{phase} & = & \left(\begin{array}{cccccccc}
0 & -0.008 & 0.0015 & -0.036 & -0.052 & -0.016 & 0.006 & -0.011\\
0.008 & 0 & 0.0157 & 0.14 & -0.004 & -0.094 & 0.0026 & 0.033\\
-0.001 & -0.015 & 0 & 0.006 & -0.001 & 0.035 & -0.051 & 0.022\\
0.036 & -0.14 & -0.006 & 0 & -0.056 & -0.003 & -0.003 & -0.031\\
0.052 & 0.004 & 0.001 & 0.056 & 0 & -0.08 & 0.001 & -0.048\\
0.016 & 0.094 & -0.035 & 0.003 & 0.08 & 0 & -0.004 & 0.026\\
-0.006 & -0.002 & 0.051 & 0.002 & -0.001 & 0.004 & 0 & 0.035\\
0.011 & -0.033 & -0.022 & 0.031 & 0.048 & -0.026 & -0.035 & 0
\end{array}\right)\end{array},\label{eq:78}
\end{equation}
}{\footnotesize\par}

\textcolor{black}{\footnotesize{}
\begin{equation}
\begin{array}{lcc}
{\rm Re}\left[\rho_{|\psi^{-}0\rangle}^{E}\right]_{parity} & = & \left(\begin{array}{cccccccc}
0.462 & 0.085 & 0.016 & 0.013 & 0.010 & -0.012 & -0.334 & -0.017\\
0.085 & 0.003 & 0.052 & -0.004 & 0.006 & 0 & -0.0776 & -0.003\\
0.016 & 0.052 & 0.030 & -0.031 & 0.007 & 0.0018 & 0.01 & 0.001\\
0.013 & -0.004 & -0.031 & 0.029 & 0.010 & -0.007 & -0.0295 & 0.003\\
0.010 & 0.006 & 0.007 & 0.011 & 0.048 & 0.011 & 0.0105 & 0.004\\
-0.012 & 0 & 0.001 & -0.007 & 0.011 & 0.009 & 0.0522 & -0.005\\
-0.334 & -0.077 & 0.01 & -0.029 & 0.010 & 0.052 & 0.4008 & 0.016\\
-0.017 & -0.003 & 0.001 & 0.003 & 0.004 & -0.005 & 0.0165 & 0.012
\end{array}\right)\end{array},\label{eq:79}
\end{equation}
}{\footnotesize\par}

\textcolor{black}{\footnotesize{}
\begin{equation}
\begin{array}{lcc}
{\rm Im}\left[\rho_{|\psi^{-}0\rangle}^{E}\right]_{parity} & = & \left(\begin{array}{cccccccc}
0 & -0.051 & -0.020 & -0.066 & -0.011 & 0 & 0.158 & 0.071\\
0.051 & 0 & 0.003 & 0.01 & 0.001 & 0 & -0.010 & 0\\
0.020 & -0.003 & 0 & 0.048 & -0.01 & -0.006 & -0.009 & -0.002\\
0.066 & -0.01 & -0.048 & 0 & 0.006 & -0.001 & 0.006 & -0.005\\
0.011 & -0.001 & 0.01 & -0.006 & 0 & -0.011 & -0.018 & -0.041\\
0 & 0 & 0.006 & 0.001 & 0.011 & 0 & -0.004 & 0.003\\
-0.158 & 0.010 & 0.009 & -0.006 & 0.018 & 0.004 & 0 & -0.027\\
-0.071 & 0 & 0.002 & 0.005 & 0.041 & -0.003 & 0.027 & 0
\end{array}\right)\end{array},\label{eq:60}
\end{equation}
}{\footnotesize\par}

\textcolor{black}{\footnotesize{}
\begin{equation}
\begin{array}{lcc}
{\rm Re}\left[\rho_{|\phi^{+}1\rangle}^{E}\right]_{parity} & = & \left(\begin{array}{cccccccc}
0.05 & 0.020 & -0.002 & -0.017 & -0.002 & -0.020 & 0.007 & -0.005\\
0.020 & 0.046 & 0.054 & -0.093 & 0.002 & 0.003 & 0.005 & -0.005\\
-0.002 & 0.054 & 0.021 & -0.021 & 0.006 & 0.026 & 0.003 & -0.014\\
-0.017 & -0.093 & -0.021 & 0.427 & 0.092 & 0.189 & -0.018 & 0.019\\
-0.002 & 0.002 & 0.006 & 0.092 & 0.03 & 0.102 & -0.003 & -0.004\\
-0.020 & 0.003 & 0.026 & 0.189 & 0.102 & 0.239 & 0.0282 & -0.067\\
0.007 & 0.005 & 0.003 & -0.018 & -0.003 & 0.028 & 0.029 & -0.026\\
-0.005 & -0.005 & -0.014 & 0.019 & -0.004 & -0.067 & -0.026 & 0.158
\end{array}\right)\end{array},\label{eq:81}
\end{equation}
}{\footnotesize\par}

\textcolor{black}{\footnotesize{}
\begin{equation}
\begin{array}{lcc}
{\rm Im}\left[\rho_{|\phi^{+}1\rangle}^{E}\right]_{parity} & = & \left(\begin{array}{cccccccc}
0 & -0.011 & 0.004 & -0.054 & -0.001 & -0.002 & -0.002 & -0.013\\
0.011 & 0 & 0.004 & 0.091 & 0.006 & -0.008 & -0.005 & 0.038\\
-0.004 & -0.004 & 0 & 0 & -0.003 & -0.044 & -0.002 & 0.007\\
0.054 & -0.091 & 0 & 0 & 0.055 & -0.032 & 0.013 & -0.019\\
0.001 & -0.006 & 0.003 & -0.055 & 0 & -0.066 & 0.004 & -0.04\\
0.002 & 0.008 & 0.044 & 0.032 & 0.066 & 0 & 0.007 & 0.076\\
0.002 & 0.005 & 0.002 & -0.013 & -0.004 & -0.007 & 0 & 0.034\\
0.013 & -0.038 & -0.007 & 0.019 & 0.04 & -0.076 & -0.034 & 0
\end{array}\right)\end{array},\label{eq:82}
\end{equation}
}{\footnotesize\par}

\textcolor{black}{\footnotesize{}
\begin{equation}
\begin{array}{lcc}
{\rm Re}\left[\rho_{|\phi^{-}1\rangle}^{E}\right]_{parity} & = & \left(\begin{array}{cccccccc}
0.028 & 0.015 & 0.002 & -0.032 & 0.006 & 0.024 & -0.005 & -0.004\\
0.015 & 0.049 & 0.058 & -0.094 & 0.01 & 0.006 & 0.007 & 0.006\\
0.002 & 0.058 & 0.015 & -0.022 & -0.005 & -0.019 & -0.005 & 0.002\\
-0.032 & -0.094 & -0.022 & 0.439 & -0.083 & -0.226 & 0.022 & 0.019\\
0.006 & 0.01 & -0.005 & -0.083 & 0.026 & 0.108 & -0.002 & -0.003\\
0.024 & 0.006 & -0.019 & -0.226 & 0.108 & 0.256 & 0.022 & -0.072\\
-0.005 & 0.007 & -0.005 & 0.022 & -0.002 & 0.022 & 0.022 & -0.029\\
-0.004 & 0.006 & 0.002 & 0.019 & -0.003 & -0.072 & -0.029 & 0.165
\end{array}\right)\end{array},\label{eq:83}
\end{equation}
}and

\textcolor{black}{\footnotesize{}
\begin{equation}
\begin{array}{lc}
{\rm Im}\left[\rho_{|\phi^{-}1\rangle}^{E}\right]_{parity}= & \left(\begin{array}{cccccccc}
0 & -0.007 & 0.001 & -0.052 & 0.003 & 0.002 & 0.002 & 0\\
0.007 & 0 & 0.006 & 0.097 & 0.002 & -0.003 & 0.007 & 0.016\\
-0.001 & -0.006 & 0 & 0.004 & 0.007 & 0.037 & -0.005 & 0.007\\
0.052 & -0.097 & -0.004 & 0 & -0.044 & -0.007 & -0.003 & -0.01\\
-0.003 & -0.002 & -0.007 & 0.044 & 0 & -0.055 & -0.002 & -0.028\\
-0.002 & 0.003 & -0.037 & 0.007 & 0.055 & 0 & 0.018 & 0.071\\
-0.002 & -0.007 & 0.005 & 0.003 & 0.002 & -0.08 & 0 & 0.032\\
0 & -0.015 & -0.007 & 0.01 & 0.028 & -0.071 & -0.032 & 0
\end{array}\right)\end{array}.\label{eq:Imparity}
\end{equation}
}{\footnotesize\par}

\textcolor{black}{Real part of the density matrices obtained through
the parity checking and phase information checking circuits are shown
in Figure \ref{fig:Final-fig}. The results illustrated through these
plots clearly show that the Bell state discrimination has been realized
appropriately. Further, the obtained density matrices allows us to
quantitatively establish this fact through the computation of fidelity,
and analogy of Figure \ref{fig:Final-fig} with Figure 6 of Ref. \cite{samal2010non}
allows us to compare the NMR-based results with the SQUID-based results.
However, the nonavailability of the exact density matrices for the
NMR-based results, restricts us from a quantitative comparison. The
obtained fidelities for the realization of phase and parity information
checking circuits are given below. The corresponding cases can be
identified by superscript phase and parity. $F_{|\psi^{+}0\rangle}^{phase}$=$0.8707$,
$F_{|\psi^{-}1\rangle}^{phase}$=$0.7114$, $F_{|\phi^{+}0\rangle}^{phase}$=$0.8794$,
$F_{|\phi^{-}1\rangle}^{phase}$=$0.7493$, $F_{|\psi^{+}0\rangle}^{parity}$=$0.8751$,
$F_{|\psi^{-}0\rangle}^{parity}$=$0.8751$, $F_{|\phi^{+}1\rangle}^{parity}$=$0.7224$,
and $F_{|\phi^{-}1\rangle}^{parity}$=$0.7576$, here the ideal state
is given in subscript and superscript \textquotedbl phase\textquotedbl{}
and \textquotedbl parity\textquotedbl{} corresponds to phase discrimination
and parity discrimination realized by phase checking circuit in Figure
\ref{fig:state-tomography} (a) and parity checking circuit in Figure
\ref{fig:state-tomography} (b). Obtained fidelities are reasonably
good, but to make a SQUID-based scalable quantum computer, it is necessary
to considerably improve the quality of quantum gates. Specifically,
we can see that the fidelities of the constructed Bell states were
much higher than the fidelities obtained after phase information or
parity information of the given Bell state is obtained through the
distributed measurement. Clearly, increase in the circuit complexity
has resulted in the reduction of fidelity. To illustrate this point,
we would now report implementation of the circuit shown in Figure
\ref{fig:main-circuit} (a), i.e., implementation of a four-qubit
circuit for nondestructive discrimination of Bell state, where phase
information and parity information will be revealed in a single experiment.
A four-qubit quantum circuit corresponding to the circuit shown in
Figure \ref{fig:main-circuit} (a) with initial Bell state $|\psi^{+}\rangle$
is implemented using IBM quantum computer and the same is shown in
Figure \ref{fig:fig4} (a), where we have used a circuit theorem shown
in Figure \ref{fig:fig4} (b). Although, the circuit and the corresponding
results shown here are for $|\psi^{+}\rangle$, but we have performed
experiments for all possible Bell states and have obtained similar
results (which are not illustrated here). Due to the restrictions
provided by the IBM computer, the left (right) CNOT gate shown in
LHS of Figure \ref{fig:fig4} (b) is implemented using the gates shown
in the left (right) rectangular box shown in RHS of Figure \ref{fig:fig4}
(b). In fact, the right most rectangular box actually swaps qubits
2 and 3, apply a CNOT with control at first qubit and target at the
second qubit and again swaps qubit 2 and 3. The use of the circuit
identity Figure \ref{fig:fig4} (b), allows us to implement the circuit
shown in Figure \ref{fig:main-circuit} (a), but it causes 10 fold
increase in gate count (from 2 CNOT gates to a total of 20 gates)
for the parity checking circuit. As a consequence of the increase
in gate count, the success probability of the experiment reduces considerably,
and that can be seen easily by comparing the outcome of the real experiment
illustrated in Figure \ref{fig:fig4} (c) with the outcome of the
simulation (expected state in the ideal noise-less situation) shown
in Figure \ref{fig:fig4} (d). This comparison in general and the
outcome observed in Figure \ref{fig:fig4} (c), clearly illustrate
that until now the technology used in IBM quantum computer is not
good enough for the realization of complex quantum circuits. This
fact is also reflected in the low fidelity (as low as $47.64$) reported
in Ref. \cite{majumder2017experimental} in the context of a quantum
circuit (having gate count of 11) that implements quantum summation
(cf. Figure 6 (d) of \cite{majumder2017experimental}). A relatively
low value of fidelity (57.03) for a circuit implementing Deutsch-Jozsa
algorithm (having a gate count of 18) has also been reported in \cite{gangopadhyay2018generalization}
(cf. Figure 7 of Ref. \cite{gangopadhyay2018generalization})}\footnote{\textcolor{black}{Gate count (number of elementary quantum gates)
is not an excellent measure for circuit cost here as the error introduced
by different gates are different and even the error introduced by
the same gate placed in different qubit lines are different. However,
reduction of fidelity with this idealized circuit cost (gate count)
provides us a qualitative feeling about the problems that may restrict
the scalability of the technology used in IBM quantum experience.}}\textcolor{black}{. However, we cannot be conclusive about the fidelity
reported in \cite{gangopadhyay2018generalization} as the definition
of fidelity used there (cf. the definition of fidelity given above
Eq. (23) of \cite{gangopadhyay2018generalization} is $F(\rho^{E},\rho^{T})=Tr\sqrt{\rho^{E}\rho^{T}\rho^{E}}$
which is not consistent with the standard definition of fidelity ($F(\rho^{E},\rho^{T})=Tr\sqrt{\sqrt{\rho^{E}}\rho^{T}\sqrt{\rho^{E}}}$)
which is used here. In a similar manner, nothing conclusive can be
obtained from the extremely high fidelity values reported in \cite{wei2018efficient}.
This is so because, the procedure followed to obtain the fidelity
was not described in \cite{wei2018efficient} and extremely high fidelity
were even reported when the experimental results were found to considerably
mismatch with the theoretically expected results (cf. Figure 3 of
\cite{wei2018efficient}). In contrast, in this work a clear prescription
for computation of fidelity has been provided and fidelity is computed
rigorously. We hope this would help others to compute fidelity for
various circuits implemented using IBM quantum experience. The fact
that most of the gates implemented in IBM quantum experience introduce
more error in comparison to the error introduced by the same gates
implemented using some of the other technologies used in quantum computing
leads to relatively low fidelity. To illustrate this, one would require
to perform quantum process tomography and obtain gate fidelity, average
purity, and entangling capability (where it is applicable) and these
quantities reflecting gate performance are to be computed for arbitrary
input states \cite{obrien2004quantum}. In \cite{hebenstreit2017compressed},
an effort has been made to obtain gate fidelities for the quantum
gates used in IBM quantum experience. However, the obtained results
don't provide us the desired gate fidelity as they are obtained only
for input state $|0\rangle$ or $|00\rangle$, and as QST was performed
instead of quantum process tomography. We will further elaborate this
point in a future work.}

\section{\textcolor{black}{Conclusion\label{sec:Conclusion5}}}

\textcolor{black}{We have already noted that nondestructive discrimination
of Bell states have wide applicability. Ranging from quantum error
correction to measurement-based quantum computation, and }quantum
communication in a network to distributed quantum computing. \textcolor{black}{Keeping
that in mind, here, we report an experimental realization of a scheme
for nondestructive Bell state discrimination. Due to the limitations
of the available quantum resources (IBM quantum computer being a five-qubit
quantum computer with few restrictions) this study is restricted to
the discrimination of Bell states only, but our earlier theoretical
proposal is valid in general for discrimination of generalized orthonormal
qudit Bell states. }We hope that the work reported in this chapter
will be generalized in the near future and will be used for the experimental
discrimination \textcolor{black}{of more complex entangled states.
Further, as the work provides a clean prescription for using IBM quantum
experience to experimentally realize quantum circuits that may form
building blocks of a real quantum computer, a similar approach may
be used to realize a set of other important circuits. Finally, the
comparison with the NMR-based technology, reveals that this SQUID-based
quantum computer's performance is comparable to that of the NMR-based
quantum computer as far as the discrimination of Bell states is concerned.
As the detail of the density matrix obtained (through QST) in earlier
works \cite{samal2010non,Anil-Kumar-paper2} was not available, fidelity
of NMR-based realization earlier and the SQUID-based realization reported
here could not be compared. However, the fidelity computed for the
states prepared and retained after the nondestructive discrimination
operation is reasonably high and that indicate the accuracy of the
IBM quantum computer. Further, it is observed that all the density
matrices produced here through the state tomography are mixed state
(i.e., for all of them $Tr(\rho^{2})$<1). This puts little light
on the nature of noise present in the channel and/or the errors introduced
by the gates used. However, it can exclude certain possibilities.
For example, it excludes the possibility that the combined effect
of noise/error present in the circuit can be viewed as bit flip and/or
phase flip error as such errors would have kept the state as pure.
More on characterization of noise present in IBM quantum experience
will be discussed elsewhere.}

\chapter[CHAPTER \thechapter \protect\newline OPTICAL DESIGNS FOR THE REALIZATION OF A SET OF SCHEMES FOR QUANTUM
CRYPTOGRAPHY]{OPTICAL DESIGNS FOR THE REALIZATION OF A SET OF SCHEMES FOR QUANTUM
CRYPTOGRAPHY\textsc{\label{chap:OPTICAL-DESIGNS-FOR}}}

{\large{}\lhead{}}{\large\par}

\section{Introduction \label{sec:Introduction6}}

\textcolor{black}{In the previous chapters, we have proposed and analyzed
a set of protocols for insecure quantum communication with entangled
orthogonal and entangled nonorthogonal state based quantum channels.
Specifically, schemes for QT have been studied and their performance
has been analyzed over Markovian channels. Above studies have motivated
us to look at the most fascinating and useful aspect of the quantum
communication, i.e., quantum cryptography, where various secure communication
tasks can be performed with unconditional security. In Chapter \ref{cha:Introduction1},
we have already mentioned that the first unconditionally secure protocol
for QKD (BB84 protocol) was proposed by Bennett and Brassard in 1984
\cite{bennett1984quantum}. Since then it has been strongly established
that the quantum cryptography can provide unconditional security,
which is a clear advantage over its classical counterparts \cite{shor2000simple,renner2008security}.
Because of this advantage, various schemes of QKD and other cryptographic
tasks have been proposed (we have already mentioned about many of
them in Chapter \ref{cha:Introduction1}). Some of them have also
been realized experimentally \cite{duplinskiy2017low,mavromatis2018experimental,hu2016experimental,zhang2017quantum,zhu2017experimental}.
Interestingly, the quantum resources and the experimental techniques
used in these successful experiments are not the same. To stress on
this particular point, we may note that the QKD entered into the experimental
era with the pioneering experimental work of Bennett et al., in 1992
\cite{bennett1992experimental}. In this work, randomly prepared polarization
states of single photons were used, but as there does not exist any
on demand single photon source, they used faint laser beams as approximate
single photon source. If such a source is used, to circumvent photon
number splitting (PNS) attack, it expected that decoy qubits are to
be used. In 1992 work of Bennett et al., no decoy qubit was used.
However, in later experiments, decoy states are frequently used. For
example, in 2006, Zhao et al., had realized a decoy state based QKD
protocol \cite{zhao2006experimental} using the acousto-optic modulators
(AOMs) to achieve polarization insensitive modulation. In the absence
of on demand single photon sources, various strategies have been used
to realize single photon based QKD schemes, like BB88 and B92. Some
of the experimentalists used weak coherent pulse (WCP) as an approximate
single photon source \cite{diamanti2016practical,duplinskiy2017low,kiktenko2017demonstration,koashi2006efficient,korzh2015provably,lo2014secure,wang2016experimental,xu2015experimental,gleim2016secure}.
Others used heralded single photon source \cite{soujaeff2007quantum,wang2008experimental}.
In the 27 years, a continuous progress has been observed in the experimental
QKD. It started from the experimental realization of a single photon
based QKD scheme using WCP, but as time passes many other facets of
QKD have been experimentally realized. For example, in one hand MDI-QKD
has been realized using untrusted source \cite{liu2016polarization,lo2012measurement},
and heralded single photon source \cite{zhang2018biased,zhou2016measurement}.
On the other hand, soon it was realized that continuous variable QKD
can be used to circumvent the need of single photon sources and thus
to avoid several attacks. Naturally, some of the continuous variable
QKD schemes have been realized in the recent past \cite{grosshans2002continuous,gottesman2001secure,ralph1999continuous}.
Beyond this, to address the concerns of the end users, over the time
the devices used have become portable (say, a silicon photonic transmitter
is designed for polarization-encoded QKD \cite{ma2016silicon,ding2017high},
and chip-based QKD systems have been realized \cite{sibson2017chip});
QKD has been realized using erroneous source \cite{xu2015experimental};
key generation rate over noisy channel has been increased (e.g., in
\cite{zhang2018experimental}, a key generation rate of 1.3 Gbit/s
was achieved over a 10-dB-loss channel); distance over which a key
can be securely distributed has been increased, for example, in \cite{boaron2018secure}
QKD is performed over 421 km in optical fiber and in the last 2-3
years couple of QKD experiments have been performed using satellites
\cite{khan2018satellite,liao2018satellite} - the one which needs
special mention is the quantum communication between the ground stations
located at China and Austria at distance of 7600 kilometers \cite{liao2018satellite}.
Furthermore, various commercial products like Clavis 2 and Clavis
3 of ID Quantique \cite{IDQ} and MagiQ QPN of MagiQ \cite{mgQ}  have
also been marketed.}

\textcolor{black}{From the discussion above it seems that the experimental
QKD is now a matured area. However, the same is not true for other
aspects of quantum cryptography, i.e., for the schemes beyond QKD
(e.g., schemes for  QD, QSDC, DSQC, CQD). Only a handful of experiments
have yet been performed. Specifically, QSDC has been realized with
entangled photons \cite{zhu2017experimental,zhang2017quantum}, and
single photons \cite{hu2016experimental}. On top of that, quantum
secret sharing has also been demonstrated \cite{hai2013experimental,schmid2005experimental}
and extended to multiparty scenario as well \cite{smania2016experimental}.
However, our discussion is focused on direct communication schemes
and quantum secret sharing is beyond the scope of the present chapter. }

\textcolor{black}{The above status of the experimental works have
motivated us to investigate possibilities of experimental realization
of quantum cryptographic schemes, such as QD \cite{pathak2013elements,nguyen2004quantum},
CQD \cite{thapliyal2015applications}, Kak\textquoteright s three
stage scheme inspired direct communication scheme \cite{kak2006three},
controlled DSQC with entanglement swapping \cite{pathak2015efficient}
which have not been experimentally realized so far. In the process,
to design optical schemes for the realization of these schemes, we
have realized that the implementation requires some modifications
of the original schemes. Keeping this point in mind, in the following
sections of this chapter, we have modified the original schemes which
remains operationally equivalent to the original scheme/without compromising
with the security and have designed optical circuits for the above
mentioned quantum cryptograhic schemes, which are based on single
photon, two-qubit and multi-qubit entangled states (such as GHZ-like
state, W state) using available optical elements, like laser, BS,
PBS, HWP.}

\textcolor{black}{The rest of the chapter is organized as follows.
In Section \ref{sec:Quantum-cryptographic-protocols}, we have presented
the designs of optical circuits for various quantum cryptographic
tasks. Each circuit and the protocol it implements are also described
in the section. Finally, the Chapter is concluded in Section \ref{sec:Conclusions6}.}

\section{\textcolor{black}{Quantum cryptographic protocols \label{sec:Quantum-cryptographic-protocols}}}

\textcolor{black}{In the previous section and in Chapter \ref{cha:Introduction1},
we have already mentioned that there exist unconditionally secure
protocols for various quantum communication tasks and a good number
of experiments have been done. However, until the recent past, experimental
works on secure quantum communication were restricted to the experimental
realizations of different protocols of QKD. Only recently (in 2016),
a protocol of QSDC was realized experimentally by Hu et al. \cite{hu2016experimental}.
Specifically, Hu et al., realized DL04 protocol \cite{deng2004secure}
using single-photon frequency coding. This pioneering work was a kind
of proof-of-principle table-top experiment. In this work, the requirement
of quantum memory was circumvented by delaying the photonic qubits
in the fiber coils. However, soon after Hu et al.'s pioneering work,
Zhang et al., \cite{zhang2017quantum} reported another experimental
realization of QSDC protocol through a table top experiment. Zhang
et al.'s experiment was fundamentally different from Hu et al.'s experiment
in two aspects- firstly Zhang et al., used entangled states and secondly
they used quantum memories based on atomic ensembles. Almost immediately
after the Zhang et al.'s experiment, Zhu et al., reported experimental
realization of a QSDC scheme over a relatively longer distance in
2017 \cite{zhu2017experimental}. With these three experiments, experimental
quantum cryptography arrived at a stage beyond QKD, where a set of
schemes of two-party one-way secure direct communication can be experimentally
realized using the available technologies. However, there exist many
multi-party schemes of secure direct quantum communication, some of
which are also two-way schemes. For example, any scheme for QD would
require two-way communication, whereas any scheme of controlled quantum
communication involves at least three parties (say, a scheme for CQD).
No such protocol has yet been realized experimentally. In what follows,
we will see that many of these protocols can be realized experimentally
using the existing technology. However, to do so, some of the protocols
would require some modifications, which are needed for experimental
realizations. In this chapter, we would concentrate on such suitably
modified protocols, and the optical circuits which can be used to
experimentally realize those schemes. The optical designs proposed
here can also be used for experimental implementation of some other
DSQC, for instance, MDI-QSDC scheme using two-qubit entanglement and
single photon source as well as Bell measurement to accomplish required
teleportation \cite{niu2018measurement}.}

\textcolor{black}{In the following section, we will briefly describe
a protocol of CQD and how to implement that using the existing optical
technology. To do so, we will be very precise and restrict ourselves
from the detailed discussion of the protocols or their security proof
as those are available elsewhere and those are not of the interest
of the present thesis. Specifically, we will briefly describe a protocol
in a few steps which are essential. We will also provide a clear schematic
diagram of the optical setup that can be used to realize the protocol,
and will provide a step-wise description of the working of the setup.
The same strategy will be followed in describing the other protocols,
too.}

\subsection{\textcolor{black}{Controlled quantum dialogue}}

\textcolor{black}{To begin with, we may note that CQD is a three party
scheme. In this scheme, Alice and Bob want to exchange their secret
messages simultaneously to each other with the help of a third party
Charlie (controller). In what follows, we will first summarize a set
of CQD schemes of our interest \cite{thapliyal2017quantum,thapliyal2015applications}
and the bottleneck present in the implementation of the theoretical
schemes. After that we will explicitly show that it's possible to
design optical circuits to experimentally realize CQD with entangled
photons and single photon (in more than one way).}

\subsubsection{\textcolor{black}{CQD with single photons}}

\textcolor{black}{CQD scheme based on single-qubit is summarized in
the following steps:}
\begin{description}
\item [{\textcolor{black}{CQDS\_1:}}] \textcolor{black}{Charlie prepares
a random string of single qubits prepared in one of four states $|0\rangle,$
$|1\rangle,$ $|+\rangle$ and $|-\rangle$.}
\item [{\textcolor{black}{CQDS\_2:}}] \textcolor{black}{Charlie sends the
string to Alice.}
\item [{\textcolor{black}{CQDS\_3:}}] \textcolor{black}{After Alice confirms
that she has received the string she randomly measures half of the
qubits either in $\left\{ |0\rangle,|1\rangle\right\} $ or $\left\{ |+\rangle,|-\rangle\right\} $
basis and announces her measurement basis and results with corresponding
position for checking the eavesdropping. Then Charlie compares the
Alice's measurement result with that of state preparation by using
classical communication (CC) in all the cases where they have chosen
the same basis}\footnote{\textcolor{black}{Throughout this chapter, transmission of qubits
is performed along the same line after concatenation of a randomly
prepared string of decoy qubits followed by permutation of qubits
in the enlarged string. Subsequently, the error estimation on the
transmitted decoy qubits provides an upper bound of the errors introduced
during transmission on the remaining message qubits, which can be
solely attributed to the disturbance caused due to an eavesdropping
attempt for the sake of simplicity and attaining utmost security.
The choice of decoy qubits could be the single-qubit states used in
BB84 protocol or entangled states (see \cite{sharma2016verification}
for a detailed discussion).}}\textcolor{black}{.}
\item [{\textcolor{black}{CQDS\_4:}}] \textcolor{black}{Alice encodes her
message on the one-half of the remaining qubits by using Pauli operations
$I$ or $iY$ to encode 0 or 1, respectively.}
\item [{\textcolor{black}{CQDS\_5:}}] \textcolor{black}{Alice sends the
message encoded and decoy qubits to Bob.}
\item [{\textcolor{black}{CQDS\_6:}}] \textcolor{black}{Bob receives the
encoded photons (along with the remaining half to be used as decoy
qubits).}
\item [{\textcolor{black}{CQDS\_7:}}] \textcolor{black}{To ensure the absence
of Eve, Alice discloses the positions of the decoy qubits and Charlie
announces corresponding choice of basis.}
\item [{\textcolor{black}{CQDS\_8:}}] \textcolor{black}{Now, Bob encodes
his message on the same qubits as used by Alice to encode his message.
Charlie announces the basis information for the message encoded qubits
when he wishes the task to be accomplished. After knowing the basis
of the initial state from Charlie, Bob measures the corresponding
qubits in that basis and announces the measurement outcome. Using
the measurement outcome both Alice and Bob can decode each other's
messages.}
\end{description}
\textcolor{black}{}
\begin{figure}
\begin{centering}
\textcolor{black}{\includegraphics[scale=0.43]{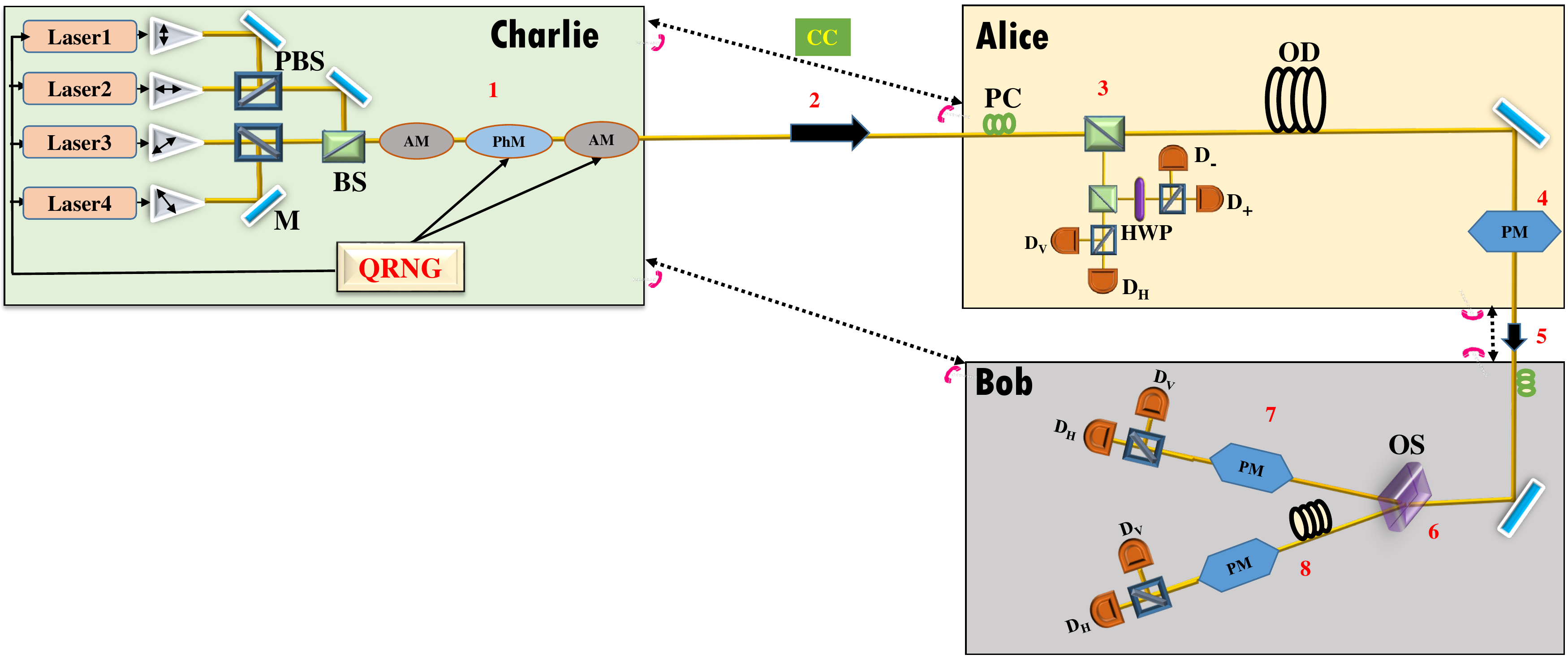}}
\par\end{centering}
\textcolor{black}{\caption{\label{fig:CQDS}Schematic diagram for CQD scheme based on polarization
qubit. Four lasers are used to prepare the polarization states of
photons. In the proposed optical design, BS stands for symmetric (50:50)
beam splitter , PBS is polarizing BS, M is mirror, AM is amplitude
modulator, PhM is phase modulator, QRNG is quantum random number generator,
PC is polarization controller, HWP is half wave plate, OD is optical
delay, PM is polarization modulator, OS is optical switch, $D_{i}$
represents the detector; whereas $D_{H}$ and $D_{V}$ correspond
to detectors used for measurements in horizontal and vertical basis
and similarly $D_{+}$ and $D_{-}$ correspond to measurements in
the diagonal basis; and CC is classical communication.}
}
\end{figure}
\textcolor{black}{The aforementioned CQD scheme has not yet been realized
experimentally due to unavailability of quantum memory, difficulty
in building on demand single photon sources, and limitation in performing
the scheme over scalable distance due to the complexity of the task.
However, there is a silver lining that the CQD protocol using single
photons can be realized using the existing technology, such as using
frequency encoding. To stress on this point a schematic diagram of
the optical setup that can be used to realize the above protocol using
polarization qubits is illustrated in Fig. \ref{fig:CQDS} and in
what follows, the same is elaborated in a few steps.}

\subsubsection*{\textcolor{black}{Optical design for CQD protocol using single photons
(polarization qubits)}}
\begin{description}
\item [{\textcolor{black}{CQDS-Op\_1:}}] \textcolor{black}{Charlie uses
four lasers to generate the polarization state of the photons and
two PBS, three mirrors, one symmetric (50:50) BS, two AM, one PhM,
to generate the random string of single photons in one of the four
polarization states $|H\rangle,$ $|V\rangle,$ $\frac{|H\rangle+|V\rangle}{\sqrt{2}}$
and $\frac{|H\rangle-|V\rangle}{\sqrt{2}}$. Whereas, the first AM
is used to generate the decoy photons, the second AM is used to control
the intensity of light and the global phase of each photon is modulated
by PhM \cite{lo2014secure}. This is not the unique method for the
preparation of polarization qubit, which can also be generated by
heralding one outputs of the SPDC outputs.}
\item [{\textcolor{black}{CQDS-Op\_2:}}] \textcolor{black}{Charlie sends
the string of single photons to Alice through optical fiber, open-air,
through satellite, or under-water in case of maritime cryptography.}
\item [{\textcolor{black}{CQDS-Op\_3:}}] \textcolor{black}{Alice receives
the string of photons and randomly selects one-half of the incoming
photons using a BS to check any eavesdropping attempt. She randomly
measures all the reflected photons either in $\left\{ |H\rangle,\,|V\rangle\right\} $
or $\left\{ \frac{|H\rangle+|V\rangle}{\sqrt{2}},\frac{|H\rangle-|V\rangle}{\sqrt{2}}\right\} $
basis (again using a BS) and announces her measurement basis and results
with corresponding position for checking the eavesdropping. Then Charlie
compares Alice's measurement result with that of state preparation.
While the eavesdropping checking between Charlie and Alice, she uses
an optical delay (serving as a quantum memory) for the rest of the
photons.}\\
\textcolor{black}{Suppose Charlie prepares the photon from laser1,
i.e., in state $|V\rangle$ then detector $D_{V}$ is expected click
in the ideal case, but if the detector $D_{H}$ clicks, then it will
be registered as bit error. However, if the detectors $D_{+}$ or
$D_{-}$ clicks then these cases will be discarded.}
\item [{\textcolor{black}{CQDS-Op\_4:}}] \textcolor{black}{Alice encodes
her message on one-half of the transmitted photons by using a PM or
a set of a half-wave plate sandwiched between two quarter-wave plates.}
\item [{\textcolor{black}{CQDS-Op\_5:}}] \textcolor{black}{Alice sends
encoded and decoy photons to Bob.}
\item [{\textcolor{black}{CQDS-Op\_6:}}] \textcolor{black}{Bob receives
the encoded photons along with the decoy photons and keeps the received
photons in an optical delay. Subsequently, Alice discloses the positions
of the decoy qubits and Bob passes the string of photons through an
optical switch which sends the encoded photons and decoy photons on
different paths.}
\item [{\textcolor{black}{CQDS-Op\_7:}}] \textcolor{black}{To ensure the
absence of Eve, Bob chooses the basis of the decoy qubits by using
a PM and measures them to compute the error rate. They proceed if
the errors are below threshold.}
\item [{\textcolor{black}{CQDS-Op\_8:}}] \textcolor{black}{Bob encodes
his message using a PM on the same photons used by Alice to encode
his message. Subsequently, Charlie reveals the basis information of
state preparation. After knowing the basis choice of the initial state
from Charlie, he measures the message encoded photons using two single
photon detectors and a PM to choose the basis of the states to be
measured and announces his result. In fact, Bob can perform the same
task using only one PM if he delays his encoding till Charlie reveals
the basis information. From the measurement outcomes both Alice and
Bob will be able to decode each other's messages.}
\end{description}

\subsubsection{\textcolor{black}{Kak\textquoteright s three-stage scheme inspired
five-stage scheme of CQD with single photons}}

\begin{figure}
\begin{centering}
\textcolor{black}{\includegraphics[scale=0.47]{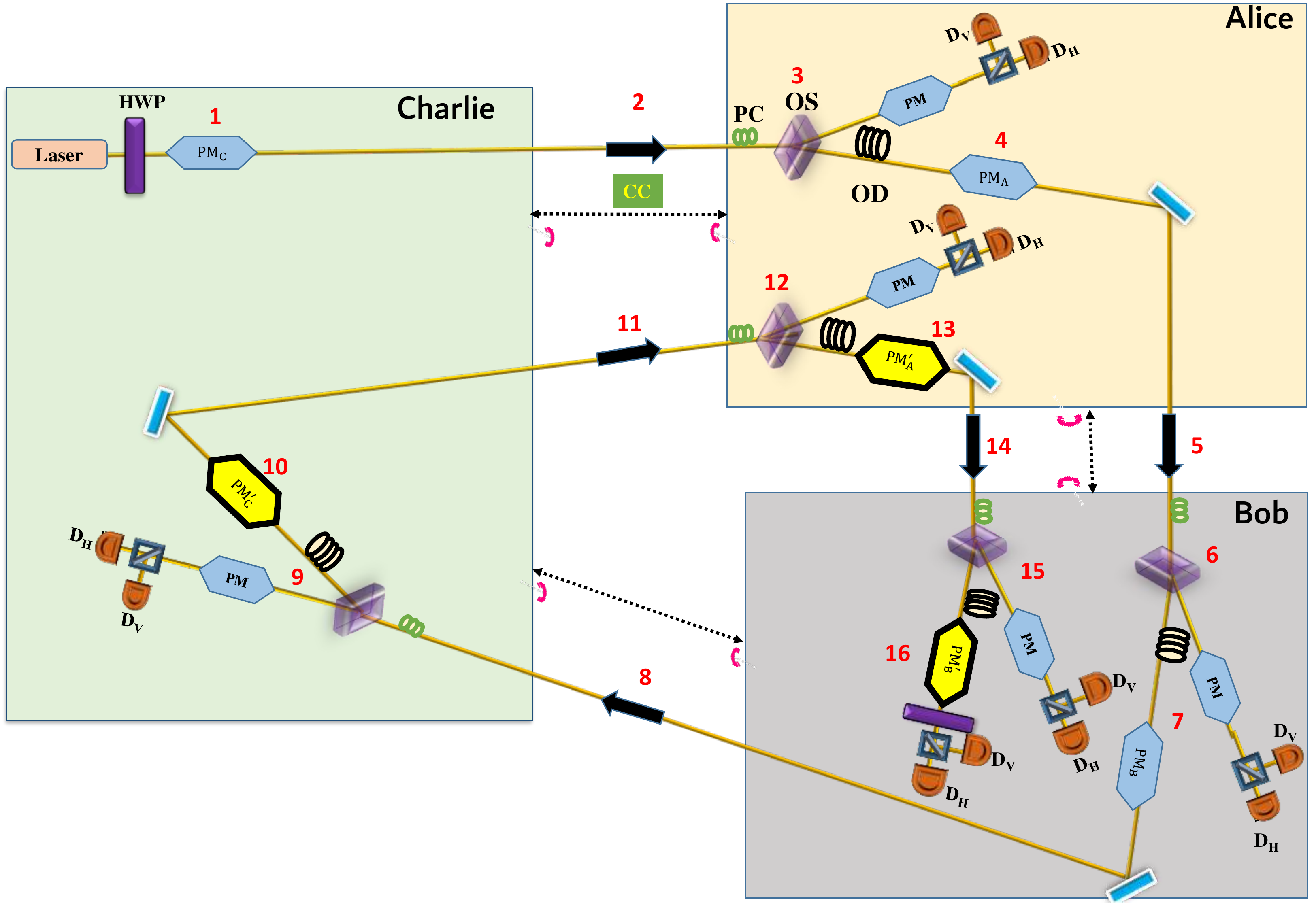}}
\par\end{centering}
\textcolor{black}{\caption{\label{fig:KAK_MODIFIED}A proposed optical implementation using polarization
qubit of the CQD scheme inspired from the three stage scheme. Laser
is used to prepare the photons. In the all lab's, ${\rm PM_{A,B,C}}$
is polarization modulator used to implement a rotation operator, and
${\rm PM_{A,B,C}^{\prime}}$ performs corresponding inverse rotation
operator.}
}
\end{figure}

\textcolor{black}{A three stage QKD scheme proposed in the past \cite{kak2006three}
was shown recently able to perform direct communication. Here, we
propose a three-stage scheme inspired CQD protocol, which can be viewed
as a five-stage protocol of CQD. The protocol is summarized in the
following steps:}
\begin{description}
\item [{\textcolor{black}{CQD-K\_1:}}] \textcolor{black}{Charlie prepares
a string of single qubits in the computational basis. Subsequently,
he applies random unitary operators on each qubit and keeps the corresponding
information with himself.}
\item [{\textcolor{black}{CQD-K\_2:}}] \textcolor{black}{Same as}\textbf{\textcolor{black}{{}
CQDS\_2.}}
\item [{\textcolor{black}{CQD-K\_3:}}] \textcolor{black}{After Alice confirms
that she has received the qubits she measures one-half of the received
qubits to check eavesdropping chosen by Charlie, who also disclose
corresponding rotation operator and the initial state.}
\item [{\textcolor{black}{CQD-K\_4:}}] \textcolor{black}{Alice applies
a random rotation operator on all the qubits.}
\item [{\textcolor{black}{CQD-K\_5:}}] \textcolor{black}{Same as}\textbf{\textcolor{black}{{}
CQDS\_5}}\textcolor{black}{.}
\item [{\textcolor{black}{CQD-K\_6:}}] \textcolor{black}{Same as}\textbf{\textcolor{black}{{}
CQD-K\_3}}\textcolor{black}{, here Bob measures one-half of the received
qubits with the help of information of rotation operators by Charlie
and Alice as well as the initial state revealed by Charlie.}
\item [{\textcolor{black}{CQD-K\_7:}}] \textcolor{black}{Same as}\textbf{\textcolor{black}{{}
CQD-K\_4}}\textcolor{black}{, but Bob applies his rotation operator.}
\item [{\textcolor{black}{CQD-K\_8:}}] \textcolor{black}{Bob sends all
the qubits to Charlie.}
\item [{\textcolor{black}{CQD-K\_9:}}] \textcolor{black}{After Charlie
confirms that he has received the qubits he measures one half of the
received qubits to check eavesdropping with the help of Alice's, Bob's,
and his own rotation operators.}
\item [{\textcolor{black}{CQD-K\_10:}}] \textcolor{black}{Charlie applies
an inverse of his rotation operator applied in }\textbf{\textcolor{black}{CQD-K\_1}}\textcolor{black}{.}
\item [{\textcolor{black}{CQD-K\_11:}}] \textcolor{black}{Same as}\textbf{\textcolor{black}{{}
CQDS\_2}}\textcolor{black}{.}
\item [{\textcolor{black}{CQD-K\_12:}}] \textcolor{black}{Same as}\textbf{\textcolor{black}{{}
CQD-KS\_3,}}\textcolor{black}{{} but here Alice requires information
of the rotation operator from Bob.}
\item [{\textcolor{black}{CQD-K\_13:}}] \textcolor{black}{Alice applies
inverse of the rotation operator applied in }\textbf{\textcolor{black}{CQD-K\_4}}\textcolor{black}{.
Subsequently, she also encodes her message on one-half of the remaining
qubits using Pauli operations $I$ or $X$ to send 0 or 1, respectively.}
\item [{\textcolor{black}{CQD-K\_14:}}] \textcolor{black}{Same as}\textbf{\textcolor{black}{{}
CQDS\_5}}\textcolor{black}{.}
\item [{\textcolor{black}{CQD-K\_15:}}] \textcolor{black}{Same as}\textbf{\textcolor{black}{{}
CQD-KS\_6,}}\textcolor{black}{{} but Bob needs only his rotation operator.}
\item [{\textcolor{black}{CQD-K\_16:}}] \textcolor{black}{Bob applies inverse
of his rotation operator and encodes his message on all the qubits.
He subsequently measures the transformed qubits in the computational
basis and announces the measurement outcome. Finally, Charlie reveals
the initial state when he wishes them to accomplish the task. With
the help of the initial and final states both Alice and Bob can decode
each other's messages.}
\end{description}
\textcolor{black}{To complete two rounds, first for locking and second
for unlocking, between three parties the qubits should travel five
times through the lossy transmission channel which sets limitations
on the experimental implementation of the present scheme. However,
to remain consistent with the theme of the present work, where reduction
of complex quantum cryptographic tasks to obtain the solutions of
simpler secure communication tasks, we now discuss the optically implementable
scheme. In principle, the protocol described here can also be realized
using available optical elements and a schematic diagram for that
is shown in Figure \ref{fig:KAK_MODIFIED}, and the same is described
below in a few steps.}

\textcolor{black}{}

\subsubsection*{\textcolor{black}{Optical design for five-stage scheme using single
photons}}
\begin{description}
\item [{\textcolor{black}{CQD-Op-K\_1:}}] \textcolor{black}{Charlie prepares
a random string of horizontal and vertical polarized single photons
and uses ${\rm PM_{C}}$ to rotate the polarization of light randomly.}
\item [{\textcolor{black}{CQD-Op-K\_2:}}] \textcolor{black}{Same as}\textbf{\textcolor{black}{{}
CQDS-Op\_2.}}
\item [{\textcolor{black}{CQD-Op-K\_3:}}] \textcolor{black}{Same as}\textbf{\textcolor{black}{{}
CQDS-Op\_3, }}\textcolor{black}{but here Charlie informs the decoy
qubits by revealing the positions and corresponding random unitary
operation using which Alice measures the photons in the computational
basis with the help of optical switch, polarization modulator and
single photon detectors. They proceed only if fewer than the threshold
bit-flip errors are observed. }
\item [{\textcolor{black}{CQD-Op-K\_4:}}] \textcolor{black}{Alice applies
a random rotation on the rest of the qubits using ${\rm PM_{A}}$.}
\item [{\textcolor{black}{CQD-Op-K\_5:}}] \textcolor{black}{Same as}\textbf{\textcolor{black}{{}
CQDS-Op\_5.}}
\item [{\textcolor{black}{CQD-Op-K\_6:}}] \textcolor{black}{Same as}\textbf{\textcolor{black}{{}
CQD-Op-K\_3}}\textcolor{black}{, here Alice and Bob perform eavesdropping
checking with the help of Charlie.}
\item [{\textcolor{black}{CQD-Op-K\_7:}}] \textcolor{black}{Same as}\textbf{\textcolor{black}{{}
CQD-Op-K\_4}}\textcolor{black}{, but Bob applies the random operation.}
\item [{\textcolor{black}{CQD-Op-K\_8:}}] \textcolor{black}{Bob sends all
the qubits to Charlie.}
\item [{\textcolor{black}{CQD-Op-K\_9:}}] \textcolor{black}{Same as}\textbf{\textcolor{black}{{}
CQD-Op-K\_3}}\textcolor{black}{, here Charlie needs assistance of
both Alice and Bob to perform eavesdropping checking.}
\item [{\textcolor{black}{CQD-Op-K\_10:}}] \textcolor{black}{Charlie applies
the inverse of his rotation operator}\textbf{\textcolor{black}{{} }}\textcolor{black}{${\rm PM_{C}^{\prime}}$.}
\item [{\textcolor{black}{CQD-Op-K\_11-12:}}] \textcolor{black}{Same as}\textbf{\textcolor{black}{{}
CQDS-Op\_2-3.}}
\item [{\textcolor{black}{CQD-Op-K\_13:}}] \textcolor{black}{Alice applies
an inverse operation of her rotation operator}\textbf{\textcolor{black}{{}
}}\textcolor{black}{using ${\rm PM_{A}^{\prime}}$. She also encodes
her message in this step.}
\item [{\textcolor{black}{CQD-Op-K\_14-15:}}] \textcolor{black}{Same as}\textbf{\textcolor{black}{{}
CQDS-Op\_5-6.}}
\item [{\textcolor{black}{CQD-Op-K\_16:}}] \textcolor{black}{Bob applies
his inverse rotation operator ${\rm PM_{B}^{\prime}}$ and encodes
his message. Then he measures the polarization of the transformed
qubits and announces the result. When Charlie wishes them to complete
the task, he reveals the initial choice of polarization of his qubit.}
\end{description}
\textcolor{black}{So far, we have discussed schemes of CQD using single
photons. Extending the idea, in what follows, we will discuss the
CQD scheme using entangled states and subsequently discuss the optical
implementation of such scheme.}

\subsubsection{\textcolor{black}{CQD protocol with entangled photons}}

\textcolor{black}{CQD scheme based on entangled qubits is summarized
in the following steps:}
\begin{description}
\item [{\textcolor{black}{CQDE\_1:}}] \textcolor{black}{Charlie prepares
a string of three qubit GHZ-like state $|\psi\rangle=|\psi^{+}0\rangle+|\psi^{-}1\rangle$.
He uses first and second qubit as a travel qubit and third qubit as
a home qubit.}
\item [{\textcolor{black}{CQDE\_2:}}] \textcolor{black}{Charlie keeps all
the third qubits and sends the strings of first and second qubits
to Alice and Bob, respectively after inserting some decoy qubits in
each string.}
\item [{\textcolor{black}{CQDE\_3:}}] \textcolor{black}{Same as }\textbf{\textcolor{black}{CQDS\_3,}}\textcolor{black}{{}
but here both Alice and Bob perform eavesdropping checking on the
strings received from Charlie.}
\item [{\textcolor{black}{CQDE\_4:}}] \textcolor{black}{Same as }\textbf{\textcolor{black}{CQDS\_4,}}\textcolor{black}{{}
but Alice can use dense coding and encode 2 bits of message using
all four Pauli operations.}
\item [{\textcolor{black}{CQDE\_5:}}] \textcolor{black}{Same as }\textbf{\textcolor{black}{CQDS\_5}}\textcolor{black}{,
but here Alice randomly inserts freshly prepared decoy qubits before
sending message encoded qubits to Bob.}
\item [{\textcolor{black}{CQDE\_6:}}] \textcolor{black}{Same as }\textbf{\textcolor{black}{CQDS\_6}}\textcolor{black}{.}
\item [{\textcolor{black}{CQDE\_7:}}] \textcolor{black}{Same as }\textbf{\textcolor{black}{CQDS\_7}}\textcolor{black}{,
but Charlie need not disclose anything. }
\begin{figure}
\begin{centering}
\textcolor{black}{\includegraphics[scale=0.43]{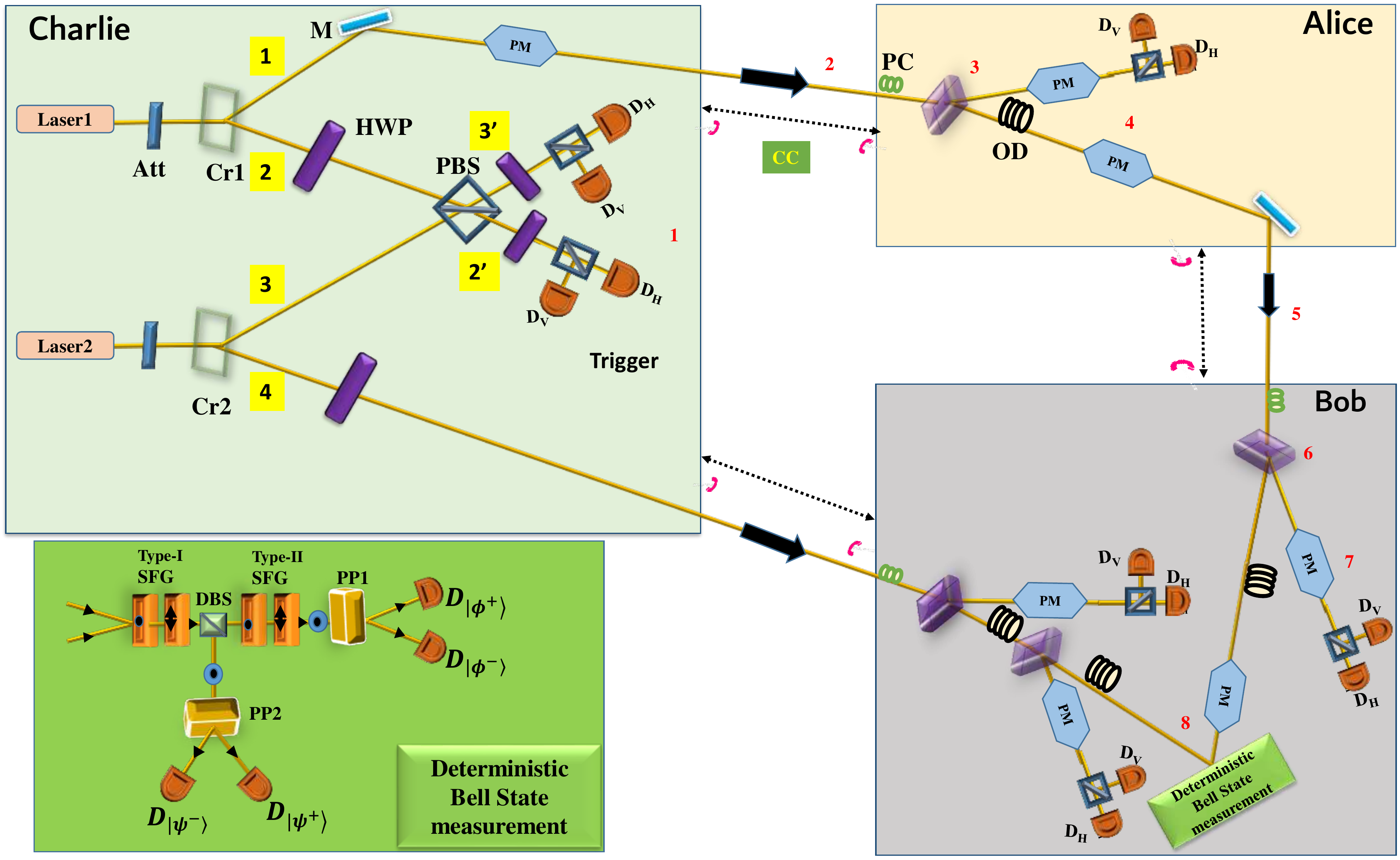}}
\par\end{centering}
\textcolor{black}{\caption{\label{fig:CQD_entangled}Optical design of the CQD scheme with a
complete Bell state measurement (BSM). In Charlie's lab Cr is a nonlinear
crystal which is used to generate the entangled photon. Sum frequency
generation (SFG) type-I and type-II are nonlinear interactions, which
are used to perform the BSM. DBS is dichroic BS and PP1 and PP2 are
$45^{0}$ projector. Attenuator (Att) is used to control the intensity
of light so that a single Bell pair can be generated. This can also
be controlled by changing the pump power. BSM is shown in the box.}
}
\end{figure}
\item [{\textcolor{black}{CQDE\_8:}}] \textcolor{black}{Now, Bob encodes
his message on the same qubits which qubits were used by Alice to
encode his message. Subsequently, Bob performs the Bell measurement
on the messages encoded string and the string received from Charlie.
Then he announces the measurement results. Now, Alice and Bob decode
the message with the help of Charlie's measurement results of the
third qubits.}
\end{description}
\textcolor{black}{The CQD protocol using entangled photons described
above can be realized using optical elements. A schematic diagram
for that is shown in Figure \ref{fig:CQD_entangled}, which is described
briefly below in a few steps.}

\subsubsection*{\textcolor{black}{Optical design for CQD protocol using entangled
photons}}
\begin{description}
\item [{\textcolor{black}{CQDE-Op\_1:}}] \textcolor{black}{Charlie uses
two lasers and two non-linear crystals (Cr1 and Cr2) to generate the
two pair of Bell states with the help of SPDC process.
\[
\begin{array}{ccc}
|\psi\rangle_{1} & = & \left(\frac{|HH\rangle+|VV\rangle}{\sqrt{2}}\right)_{12}\otimes\left(\frac{|HH\rangle+|VV\rangle}{\sqrt{2}}\right)_{34},\end{array}
\]
}
\end{description}
\textcolor{black}{where $H$ and $V$ represent the horizontal and
vertical polarizations, respectively.}

\textcolor{black}{Now, $2^{nd}$ and $4^{th}$ photon passes through
the HWP $\left(2\theta=45^{0}\right)$ and the state becomes}

\textcolor{black}{
\begin{equation}
\begin{array}{ccc}
|\psi\rangle_{2} & = & \left(\frac{|H+\rangle+|V-\rangle}{\sqrt{2}}\right)_{12}\otimes\left(\frac{|H+\rangle+|V-\rangle}{\sqrt{2}}\right)_{34}\\
 & = & \frac{1}{2}\left(|H+H+\rangle+|H+V-\rangle+|V-H+\rangle+|V-V-\rangle\right)_{1234}
\end{array},\label{eq:-4}
\end{equation}
where diagonal polarization states are represented by $|\pm\rangle=\frac{\left(|H\rangle\pm|V\rangle\right)}{\sqrt{2}}$.
Subsequently, the $2^{nd}$ and $4^{th}$ photon pass through the
PBS (which transmits the horizontal photon and reflect the vertical
photon). The postselected state after passing through PBS such that
only one photon will be in each output path can be written as after
renormalization}

\textcolor{black}{
\[
\begin{array}{ccc}
|\psi\rangle_{3} & = & \frac{1}{2}\left(|HHH+\rangle+|HVV-\rangle+|VHH+\rangle-|VVV-\rangle\right)_{12^{'}3^{'}4}\end{array}.
\]
Then photon $2^{'}$ passes through a HWP $\left(2\theta=45^{0}\right)$
and state transforms as}

\textcolor{black}{
\begin{equation}
\begin{array}{lcl}
|\psi\rangle_{4} & = & \frac{1}{2}\left(|+\rangle_{2^{'}}\left(|HH+\rangle+|HV-\rangle+|VH+\rangle-|VV-\rangle\right)_{13^{'}4}\right.\\
 & + & \left.|-\rangle_{2^{'}}\left(|HH+\rangle-|HV-\rangle+|VH+\rangle+|VV-\rangle\right)_{13^{'}4}\right)\\
|\psi\rangle_{4} & = & \frac{1}{2}\left(|+\rangle_{2^{'}}\left(|+H+\rangle+|-V-\rangle\right)_{13^{'}4}+|-\rangle_{2^{'}}\left(|+H+\rangle-|-V-\rangle\right)_{13^{'}4}\right)\\
 & = & \frac{1}{2}\left(|+\rangle_{2^{'}}\left(|+\rangle_{3^{'}}|\psi^{+}\rangle_{14}+|-\rangle|\phi^{+}\rangle_{14}\right)+|-\rangle_{2^{'}}\left(|+\rangle_{3^{'}}|\phi^{+}\rangle_{14}+|-\rangle|\psi^{+}\rangle_{14}\right)\right)\\
 & = & \frac{1}{2}\left(|+\rangle_{2^{'}}\left(|\Phi_{1}\rangle\right)+|-\rangle_{2^{'}}\left(|\Phi_{2}\rangle\right)\right)
\end{array}.\label{eq:-2-1}
\end{equation}
}

\textcolor{black}{From the obtained state one can clearly see that
if Charlie measures qubit $2^{'}$ and announces the measurement outcome,
depending upon which all the parties can decide which channel they
are sharing. Otherwise, if Charlie measures $|+\rangle$, then he
will get $|+\rangle_{3^{'}}|\psi^{+}\rangle_{14}+|-\rangle|\phi^{+}\rangle_{14}$,
no need to apply any gate, i.e., identity, if he measures $|-\rangle$,
then he will get $|+\rangle_{3^{'}}|\phi^{+}\rangle_{14}+|-\rangle|\psi^{+}\rangle_{14}$
, he need to apply NOT gate by using on $1$ or $4$, then the state
will be $|+\rangle_{3^{'}}|\psi^{+}\rangle_{14}+|-\rangle|\phi^{+}\rangle_{14}.$
Also note that $|\Phi_{i}\rangle$ are the unitary equivalent of the
state prepared in }\textbf{\textcolor{black}{CQDE\_1}}\textcolor{black}{.
The experimental preparation of three-qubit states using this approach
can be found in Ref. \cite{zhang2006experimental}.}
\begin{description}
\item [{\textcolor{black}{CQDE-Op\_2:}}] \textcolor{black}{Charlie sends
corresponding photons 1 and 4 to Alice and Bob, respectively.}
\item [{\textcolor{black}{CQDE-Op\_3:}}] \textcolor{black}{Both Alice and
Bob receive the photons and choose the same set of photons using an
optical switch to check their correlations with Charlie to check the
eavesdropping. Bob keeps her photons in an optical delay.  Here, it
is worth mentioning that Alice and Bob can also use BS for this task,
but that will reduce qubit efficiency as the cases where one of them
has measured entangled state, but not other will be discarded.}
\item [{\textcolor{black}{CQDE-Op\_4:}}] \textcolor{black}{Same as }\textbf{\textcolor{black}{CQDS-Op\_4,}}\textcolor{black}{{}
but Alice uses dense coding as well.}
\item [{\textcolor{black}{CQDE-Op\_5-6:}}] \textcolor{black}{Same as }\textbf{\textcolor{black}{CQDS-Op\_5}}\textcolor{black}{-}\textbf{\textcolor{black}{6}}\textcolor{black}{.}
\item [{\textcolor{black}{CQDE-Op\_7:}}] \textcolor{black}{Same as }\textbf{\textcolor{black}{CQDS-Op\_7,}}\textcolor{black}{{}
but Charlie has to reveal the measurement outcome for the corresponding
decoy qubits.}
\item [{\textcolor{black}{CQDE-Op\_8:}}] \textcolor{black}{Same as }\textbf{\textcolor{black}{CQDS-Op\_8,
}}\textcolor{black}{but Bob encode his 2 bits of message on each photon
by using PM on the same photons used by Alice. After that, Bob performs
Bell measurement \cite{kim2001quantum} and announces the measurement
outcome. Similarly, Charlie measures his qubit in the diagonal basis
and announces the measurement result when he wishes Alice and Bob
to decode the messages. There are schemes for probabilistic Bell measurement,
but is not desirable in the implementation of direct communication
schemes as it is prone to loss in encoded message. The drawback of
deterministic BSM is small efficiency due to involvement of nonlinear
optics in its implementation.}
\end{description}

\subsection{\textcolor{black}{Controlled direct secure quantum communication}}

\textcolor{black}{Another controlled communication scheme where DSQC
from Alice to Bob is controlled by a controller. Specifically, Alice
can directly transmits the message in a secure manner to Bob with
the help of controller. Controller controls the channel between Alice
and Bob.}

\subsubsection{\textcolor{black}{CDSQC with single photons}}

\textcolor{black}{The CQD schemes discussed in the previous section
can be reduced to CDSQC schemes if Bob does not encode his message
in the last step. Additionally, he need not to announce the measurement
outcome as he is not sending message to Alice in this task.}

\subsubsection{\textcolor{black}{CDSQC with entangled photons}}

\begin{figure}
\begin{centering}
\includegraphics[scale=0.4]{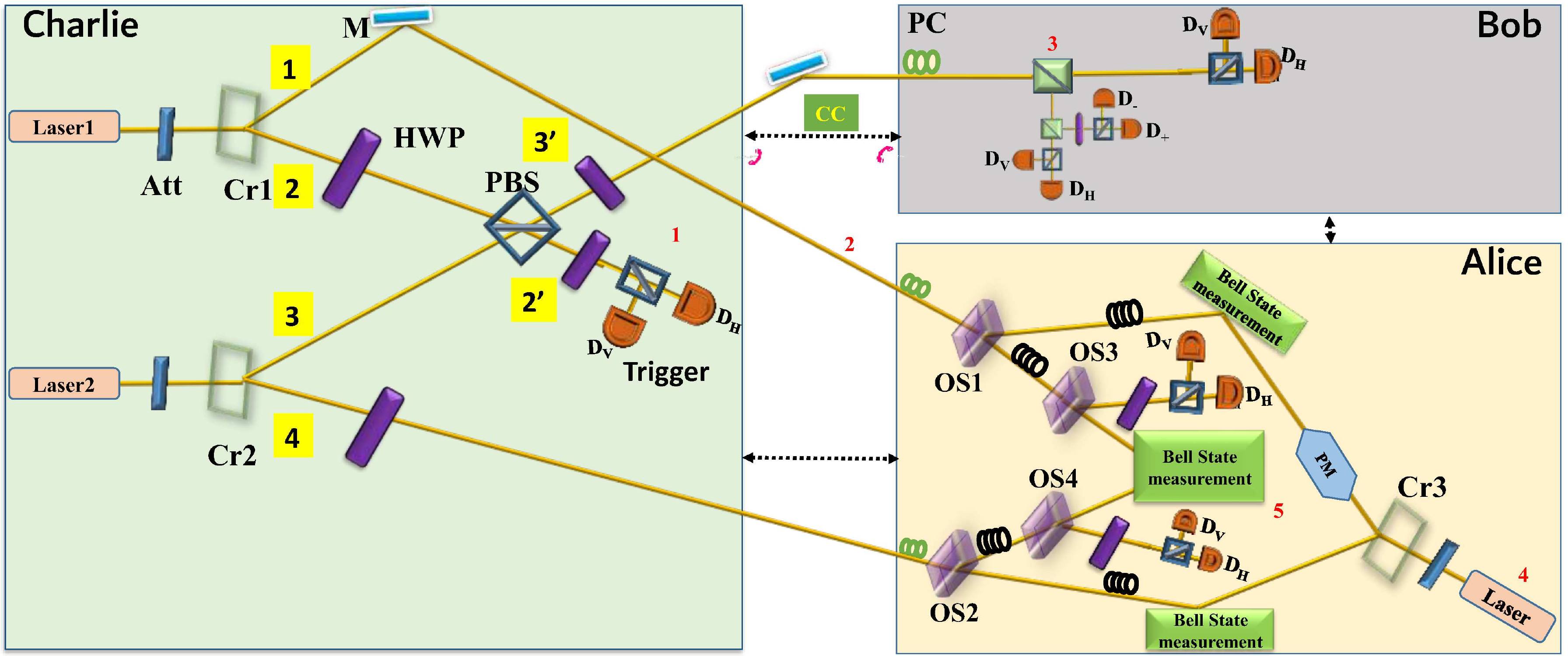}
\par\end{centering}
\caption{\label{fig:CDSQC}Schematic optical design of controlled direct secure
quantum communication with entanglement swapping.}
\end{figure}

\textcolor{black}{The entangled states based CQD scheme can also be
reduced analogous to single photon based scheme to obtain a CDSQC
scheme. Here, we have presented another entangled state based CDSQC
\ref{fig:CDSQC} with entanglem}ent swapping, where message encoded
qubits are not accessible to Eve as those qubits don't travel through
the channel.

\subsubsection{\textcolor{black}{CDSQC with entanglement swapping}}

\textcolor{black}{CDSQC with entanglement swapping is summarized in
the following steps:}
\begin{description}
\item [{\textcolor{black}{CDSQC\_1:}}] \textcolor{black}{Charlie prepares
a four qubit entangled state 
\begin{equation}
|\psi\rangle=\frac{1}{2}\left(|+\rangle_{2^{'}}\left(|0\rangle_{3^{'}}|\psi^{+}\rangle_{14}+|1\rangle_{3^{'}}|\phi^{+}\rangle_{14}\right)+|-\rangle_{2^{'}}\left(|0\rangle_{3^{'}}|\phi^{+}\rangle_{14}+|1\rangle_{3^{'}}|\psi^{+}\rangle_{14}\right)\right),\label{eq:4-qubit state}
\end{equation}
}
\end{description}
\textcolor{black}{where qubit $2^{'}$ corresponds to Charlie, qubits
$1$ and $4$ for Alice and $3^{'}$ for Bob.}
\begin{description}
\item [{\textcolor{black}{CDSQC\_2:}}] \textcolor{black}{Same as }\textbf{\textcolor{black}{CQDE\_2}}\textcolor{black}{.}
\item [{\textcolor{black}{CDSQC\_3:}}] \textcolor{black}{Same as }\textbf{\textcolor{black}{CQDE\_3}}\textcolor{black}{.}
\item [{\textcolor{black}{CDSQC\_4:}}] \textcolor{black}{Alice prepares
$|\psi^{+}\rangle_{A_{1}A_{2}}$ to encode her secret information.
Specifically, she encodes 1 (0) by a}pplying a $X$ ($I)$ gate on
one of the qubits of the Bell state. Thus, the combined state becomes\textcolor{black}{
\begin{equation}
\begin{array}{lcl}
|\psi^{\prime}\rangle & = & \frac{1}{2\sqrt{2}}\left(|\psi^{+}\rangle_{A_{1}A_{2}}|+\rangle_{2^{'}}\left(|0\rangle_{3^{'}}|\psi^{+}\rangle_{14}+|1\rangle_{3^{'}}|\phi^{+}\rangle_{14}\right)\right.\\
 & + & \left.|\psi^{+}\rangle_{A_{1}A_{2}}|-\rangle_{2^{'}}\left(|0\rangle_{3^{'}}|\phi^{+}\rangle_{14}+|1\rangle_{3^{'}}|\psi^{+}\rangle_{14}\right)\right)
\end{array}.\label{eq:-5}
\end{equation}
}
\end{description}
\textbf{\textcolor{black}{CDSQC\_5:}}\textcolor{black}{{} Alice measures
qubits $A_{1}$ and 1 as well as $A_{2}$ and 4 in the Bell basis,
while Bob and Charlie can measure his qubits in the computational
basis and diagonal basis, respectively. Subsequently, Alice and Charlie
announce their measurement outcomes, which should reveal Alice's message
to Bob.}\\
\textcolor{black}{To illustrate this point we can write the state
before Alice's and Bob's measurements while Charlie's measurement
result is $|+\rangle$ when Alice encodes $0$. 
\begin{equation}
\begin{array}{lcl}
|\psi^{\prime}\rangle & = & \frac{1}{2\sqrt{2}}\left(\left\{ |\psi^{+}\rangle_{A_{1}1}|\psi^{+}\rangle_{A_{2}4}+|\phi^{+}\rangle_{A_{1}1}|\phi^{+}\rangle_{A_{2}4}+|\phi^{-}\rangle_{A_{1}1}|\phi^{-}\rangle_{A_{2}4}+|\psi^{-}\rangle_{A_{1}1}|\psi^{-}\rangle_{A_{2}4}\right\} |0\rangle_{3^{'}}\right.\\
 & + & \left.\left\{ |\psi^{+}\rangle_{A_{1}1}|\phi^{+}\rangle_{A_{2}4}+|\psi^{-}\rangle_{A_{1}1}|\phi^{-}\rangle_{A_{2}4}+|\phi^{+}\rangle_{A_{1}1}|\psi^{+}\rangle_{A_{2}4}+|\phi^{-}\rangle_{A_{1}1}|\psi^{-}\rangle_{A_{2}4}\right\} |1\rangle_{3^{'}}\right).
\end{array}\label{eq:msg-1}
\end{equation}
}

\textcolor{black}{Similarly, if Charlie's measurement result is $|-\rangle$
and Alice encodes $1$}

\textcolor{black}{
\begin{equation}
\begin{array}{lcl}
|\psi^{\prime}\rangle & = & \frac{1}{2\sqrt{2}}\left(\left\{ |\psi^{+}\rangle_{A_{1}1}|\phi^{+}\rangle_{A_{2}4}+|\phi^{+}\rangle_{A_{1}1}|\psi^{+}\rangle_{A_{2}4}-|\phi^{-}\rangle_{A_{1}1}|\psi^{-}\rangle_{A_{2}4}-|\psi^{-}\rangle_{A_{1}1}|\phi^{-}\rangle_{A_{2}4}\right\} |1\rangle_{3^{'}}\right.\\
 & + & \left.\left\{ |\psi^{+}\rangle_{A_{1}1}|\psi^{+}\rangle_{A_{2}4}-|\psi^{-}\rangle_{A_{1}1}|\psi^{-}\rangle_{A_{2}4}+|\phi^{+}\rangle_{A_{1}1}|\phi^{+}\rangle_{A_{2}4}-|\phi^{-}\rangle_{A_{1}1}|\phi^{-}\rangle_{A_{2}4}\right\} |0\rangle_{3^{'}}\right).
\end{array}\label{eq:19-1}
\end{equation}
Similarly, if Charlie's measurement result is $|+\rangle$ and Alice
encodes $1$}

\textcolor{black}{
\begin{equation}
\begin{array}{lcl}
|\psi^{\prime}\rangle & = & \frac{1}{2\sqrt{2}}\left(\left\{ |\psi^{+}\rangle_{A_{1}1}|\psi^{+}\rangle_{A_{2}4}-|\psi^{-}\rangle_{A_{1}1}|\psi^{-}\rangle_{A_{2}4}+|\phi^{+}\rangle_{A_{1}1}|\phi^{+}\rangle_{A_{2}4}-|\phi^{-}\rangle_{A_{1}1}|\phi^{-}\rangle_{A_{2}4}\right\} |1\rangle_{3^{'}}\right.\\
 & + & \left.\left\{ |\psi^{+}\rangle_{A_{1}1}|\phi^{+}\rangle_{A_{2}4}-|\psi^{-}\rangle_{A_{1}1}|\phi^{-}\rangle_{A_{2}4}+|\phi^{+}\rangle_{A_{1}1}|\psi^{+}\rangle_{A_{2}4}-|\phi^{-}\rangle_{A_{1}1}|\psi^{-}\rangle_{A_{2}4}\right\} |0\rangle_{3^{'}}\right).
\end{array}\label{eq:-1-1}
\end{equation}
and if Charlie's measurement result is $|-\rangle$ and Alice encodes
$0$}

\textcolor{black}{
\begin{equation}
\begin{array}{lcl}
|\psi^{\prime}\rangle & = & \frac{1}{2\sqrt{2}}\left(\left\{ |\psi^{+}\rangle_{A_{1}1}|\phi^{+}\rangle_{A_{2}4}+|\psi^{-}\rangle_{A_{1}1}|\phi^{-}\rangle_{A_{2}4}+|\phi^{+}\rangle_{A_{1}1}|\psi^{+}\rangle_{A_{2}4}+|\phi^{-}\rangle_{A_{1}1}|\psi^{-}\rangle_{A_{2}4}\right\} |0\rangle_{3^{'}}\right.\\
 & + & \left.\left\{ |\psi^{+}\rangle_{A_{1}1}|\psi^{+}\rangle_{A_{2}4}+|\psi^{-}\rangle_{A_{1}1}|\psi^{-}\rangle_{A_{2}4}+|\phi^{+}\rangle_{A_{1}1}|\phi^{+}\rangle_{A_{2}4}+|\phi^{-}\rangle_{A_{1}1}|\phi^{-}\rangle_{A_{2}4}\right\} |1\rangle_{3^{'}}\right).
\end{array}\label{eq:-3-1}
\end{equation}
}

\subsubsection*{Optical design for CDSQC protocol with entanglement swapping using
single photons:}
\begin{description}
\item [{CDSQC-Op\_1\textcolor{black}{:}}] \textcolor{black}{Same as }\textbf{\textcolor{black}{CQDE-Op\_1.
}}\textcolor{black}{The four qubit state is
\[
|\psi\rangle=\frac{1}{2}\left(|+\rangle_{2^{'}}\left(|+\rangle_{3^{'}}|\psi^{+}\rangle_{14}+|-\rangle|\phi^{+}\rangle_{14}\right)+|-\rangle_{2^{'}}\left(|+\rangle_{3^{'}}|\phi^{+}\rangle_{14}+|-\rangle|\psi^{+}\rangle_{14}\right)\right).
\]
}
\item [{\textcolor{black}{CDSQC-Op\_2:}}] \textcolor{black}{Charlie sends
corresponding photons $3^{'}$ after passing through HWP to Bob and
photons $1$ and $4$ Alice without any operation.}
\item [{\textcolor{black}{CDSQC-Op\_3:}}] \textcolor{black}{Bob randomly
selects the photons by using BS from the received photons to check
the eavesdropping and measures the photons by using single photon
detectors. Same will be happen from Alice's side, but, Alice's photons
will pass through two optical switches OS 1 and OS 2 to choose the
decoy qubits. After that, she chooses a set of qubits (corresponding
to computational basis measurement by Bob) using optical switches
OS 3 and OS 4 to measure the received photons in Bell basis, while
she performs single-qubit measurements on the rest of the qubits (corresponding
to diagonal basis measurement by Bob) to check eavesdropping.}
\item [{\textcolor{black}{CDSQC-Op\_4:}}] \textcolor{black}{Alice prepares
entangled state $|\psi^{+}\rangle_{A_{1}A_{2}}$ to encode her secret
information. Specifically, she applies a PM on one of the qubits of
the Bell state to encode ``1'' and does nothing to send ``0''.
Therefore, the combined state of Alice and Bob before her encoding
is
\[
\begin{array}{lcl}
|\psi^{\prime}\rangle & = & \frac{1}{2\sqrt{2}}\left(|\psi^{+}\rangle_{A_{1}A_{2}}|+\rangle_{2^{'}}\left(|0\rangle_{3^{'}}|\psi^{+}\rangle_{14}+|1\rangle|\phi^{+}\rangle_{14}\right)\right.\\
 & + & \left.|\psi^{+}\rangle_{A_{1}A_{2}}|-\rangle_{2^{'}}\left(|0\rangle_{3^{'}}|\phi^{+}\rangle_{14}+|1\rangle|\psi^{+}\rangle_{14}\right)\right)
\end{array}.
\]
}
\item [{\textcolor{black}{CDSQC-Op\_5:}}] \textcolor{black}{Alice measures
qubits $A_{1}$ and $1$ as well as $A_{2}$ and $2$ in Bell basis,
while Bob can measure his qubits in the computational basis by using
SPD. Subsequently, she announces her measurement outcomes, which should
reveal her message to Bob and Charlie measure his qubit in diagonal
basis.}
\end{description}

\subsection{\textcolor{black}{Quantum Dialogue}}

\textcolor{black}{Quantum dialogue (QD) is a two party scheme, whereas
Alice and Bob as two parties wish to communicate their secret messages
simultaneously to each other. Quantum dialogue can be reduced from
the CQD as shown in \cite{thapliyal2017quantum}. Therefore, we have
presented the feasibility of QD with single photons and entangled
photons. Here, we briefly discuss the changes to be made in the CQD
scheme to obtain the corresponding QD scheme.}

\subsubsection{\textcolor{black}{QD with single photons}}

\begin{figure}
\begin{centering}
\includegraphics[scale=0.43]{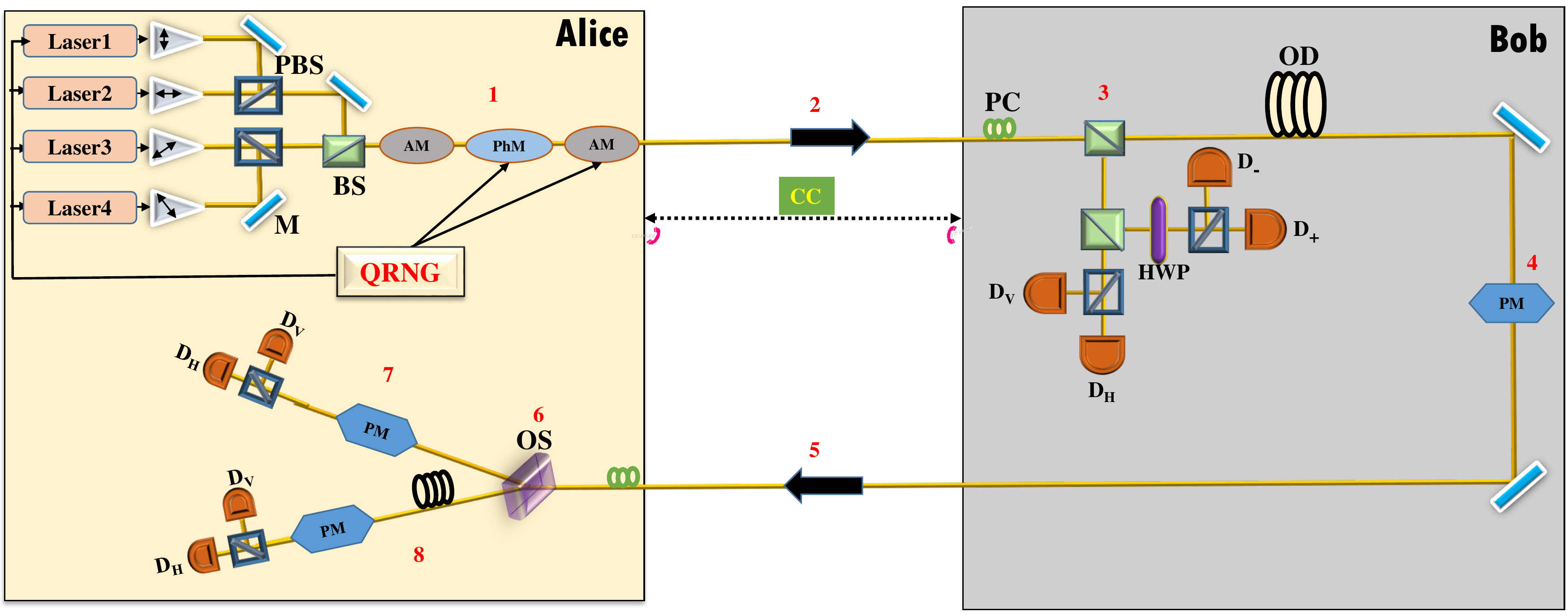}
\par\end{centering}
\centering{}\caption{\label{fig:QD}The proposed optical diagram using linear optics of
the QD scheme which is based on single photons.}
\end{figure}

\textcolor{black}{QD protocol is summarized in the following steps:}
\begin{description}
\item [{\textcolor{black}{QD\_1:}}] \textcolor{black}{Same as }\textbf{\textcolor{black}{CQDS\_1,
}}\textcolor{black}{but here Alice prepares the string.}
\item [{\textcolor{black}{QD\_2:}}] \textcolor{black}{Alice sends the string
to Bob as in }\textbf{\textcolor{black}{CQDS\_2}}\textcolor{black}{.
In the following steps up to }\textbf{\textcolor{black}{CQD\_8}}\textcolor{black}{{}
is same as}\textbf{\textcolor{black}{{} CQD\_8 }}\textcolor{black}{,
but the difference is only here, Alice plays a role of Bob and Bob
plays a role of Alice. }
\item [{\textcolor{black}{QD\_3:}}] \textcolor{black}{Same as }\textbf{\textcolor{black}{CQDS\_3}}\textcolor{black}{,
but Alice and Bob perform eavesdropping checking.}
\item [{\textcolor{black}{QD\_4-7:}}] \textcolor{black}{Same as }\textbf{\textcolor{black}{CQDS\_4}}\textcolor{black}{-}\textbf{\textcolor{black}{7}}\textcolor{black}{.}
\item [{\textcolor{black}{QD\_8:}}] \textcolor{black}{Same as}\textbf{\textcolor{black}{{}
CQDS\_8}}\textcolor{black}{, while Alice prepares the initial string,
so she knows the basis used for its preparation. Therefore, in the
end, Alice announces both initially prepared state and final states
on measurement. Without loss of generality the initial state can be
assumed public knowledge.}
\end{description}
\textcolor{black}{The above summarized QD protocol using single photons
can be realized by optical elements and the optical circuit for that
is illustrated in Figure \ref{fig:QD}, and the same is explained
below in steps.}

\subsubsection*{Optical design for QD protocol using single photons:}
\begin{description}
\item [{QD-Op\_1:}] \textcolor{black}{Same as }\textbf{\textcolor{black}{CQDS-Op\_1}}\textcolor{black}{,
but here Alice prepares the string.}
\item [{\textcolor{black}{QD-Op\_2:}}] \textcolor{black}{Alice sends it
to Bob. In the following steps up to }\textbf{\textcolor{black}{QD-Op\_8}}\textcolor{black}{{}
is same as}\textbf{\textcolor{black}{{} CQDS-Op\_8, }}\textcolor{black}{but
the difference is only here, Alice plays a role of Bob and Bob plays
a role of Alice. }
\item [{\textcolor{black}{QD-Op\_3-7:}}] \textcolor{black}{Same as }\textbf{\textcolor{black}{CQDS-Op\_3}}\textcolor{black}{-}\textbf{\textcolor{black}{7}}\textcolor{black}{.}
\item [{\textcolor{black}{QD-Op\_8:}}] \textcolor{black}{Same as}\textbf{\textcolor{black}{{}
CQDS-Op\_8}}\textcolor{black}{, while Alice prepares the initial string,
so she knows the basis of string.}
\end{description}
\textcolor{black}{Similarly, the rest of the QD schemes using single
photon and entangled states can be reduced from corresponding CQD
schemes. Therefore, here we avoid such repetition and mention only
briefly for the same of completeness.}

\subsection{\textcolor{black}{Quantum secure direct communication/Direct secure
quantum communication}}

\textcolor{black}{In quantum secure direct communication scheme, messages
are transmitted directly in a deterministic and secure manner from
Alice to Bob. A QSDC scheme can be viewed as a quantum dialogue, but
the difference is only here that one party is restricted to encode
the identity only. Similarly, a DSQC scheme can be deduced from a
QSDC or CDSQC scheme where the receiver does not encode his/her message
and requires an additional 1 bit of classical communication from the
sender to decode the message.}

\subsection{\textcolor{black}{Quantum key agreement}}

\textcolor{black}{The proposed optical designs can also be used to
reduce QKA schemes. Specifically, in quantum key agreement, all parties
take part in the key generation process and none can control the key
solely. A QKA scheme has been proposed in the recent past using a
modified version of QSDC/DSQC scheme \cite{shukla2014protocols} which
can be realized experimentally using present optical designs. Precisely,
using QSDC scheme, one party sends his/her raw to another party in
a secure manner, while the other party publicly announces his/her
key and the final key is combined from both raw keys. Therefore, the
optical designs can be used for secure transmission of the first parties
raw key.}

\subsection{\textcolor{black}{Quantum key distribution}}

\textcolor{black}{The optical designs can be used to describe prepare-and-measure
QKD schemes, too. Specifically, decoy qubit based QKD \cite{lo2005decoy,rosenberg2007long}
can be visualized from }\textbf{\textcolor{black}{CQDS-Op\_1-3 }}\textcolor{black}{if
Charlie (Bob) is the sender (receiver) and they share a quantum key
by the end of this scheme. Similarly, the entangled state based CQD
can be used to describe BBM scheme \cite{bennett1992quantumcryptography}.}

\section{\textcolor{black}{Conclusions \label{sec:Conclusions6}}}

\textcolor{black}{In this chapter, we have provided optical circuits
for various quantum cryptographic schemes (CQD, CDSQC with entanglement
swapping, QD, Kak's three stage scheme, etc.) with single photons,
entangled photon (GHZ-like) and also modified version of Kak's protocol.
Interestingly, most of the designed optical circuits can be realized
using both optical fiber based and open air based architectures. Being
a theoretical physics group, we could not realize these optical circuits
in the laboratory, but others having laboratory facility are expected
to find the work reported here as interesting enough to perform experiments
to realize the optical circuits designed in this chapter.}

\newpage{}

\chapter[CHAPTER \thechapter \protect\newline CONCLUSIONS AND SCOPE FOR FUTURE WORK]{\textsc{CONCLUSIONS AND SCOPE FOR FUTURE WORK\label{cha:Conclusions-and-Scope}}}

{\large{}\lhead{}}{\large\par}

\textcolor{black}{This concluding chapter aims to briefly summarize
the results obtained in this thesis work and it also aims to provide
some insights in to the scope of future works. Before we start summarizing
the results, we may note that the domain of quantum communication
is rapidly growing. Several schemes for quantum communication to perform
various cryptographic and non-cryptographic tasks (e.g., QT and its
variants, QKD, QSDC, dense coding) have been introduced and critically
analyzed in the recent past. Many of them have also been experimentally
analyzed. Realizing the importance of this rapidly growing field,
we have tried to contribute something to this field through this thesis
work.}

\textcolor{black}{Among various facets of quantum communication, quantum
teleportation deserves specific attention as it leads to many other
schemes for quantum communication (see Section \ref{sec:Various-facets-of-insecure}).
Motivated by this fact, a major part of the present thesis work is
dedicated to the study of quantum teleportation schemes. In addition,
in this thesis we have also worked on another extremely important
facet of quantum communication- quantum cryptography. Here, we may
note that in this thesis, we have designed two schemes for quantum
communication, but the focus of the thesis is not to design new protocols,
rather it aims to address the following question: Which quantum resources
are essential/sufficient for implementing a particular quantum communication
task? In what follows, we briefly summarize the results obtained in
this work in an effort to answer the above question.}

\section{\textcolor{black}{Conclusions and a brief summary of the work \label{sec:Conclusion-of-the}}}

\textcolor{black}{This thesis is entitled, ``Design and analysis
of communication protocols using quantum resources''. As the title
suggests, this thesis work focused on quantum communication has two
aspects-(a) designing and (b) analysis. In what follows, we will first
summarize the outcome of the works done in the designing part and
subsequently we will summarize the outcome of the analysis done in
this thesis work.}

\textbf{\textcolor{black}{Designing: }}\textcolor{black}{In the designing
part, this work introduces two schemes for teleportation- (i) an entangled
orthogonal state-based scheme that allows us to teleport multi-qubit
quantum states with the optimal quantum resources and (ii) an entangled
nonorthogonal states-based scheme. Further, relevant circuits are
designed for the experimental realization of the first scheme by using
IBM's five-qubit quantum processor (IBMQX2). Quantum circuits for
the implementation of nondestructive discrimination of Bell states
using IBM quantum processor are also designed. In Chapter \ref{chap:OPTICAL-DESIGNS-FOR},
optical designs for the realization of a set of schemes for quantum
cryptography have also been provided. We may briefly summarize designing
part of this thesis work under the following points:}
\begin{enumerate}
\item \textbf{\textcolor{black}{Quantum teleportation scheme: }}\textcolor{black}{The
major part of this thesis is dedicated to the quantum teleportation
schemes (entangled orthogonal-based and entangled nonorthogonal state-based).
As we mentioned in Chapter \ref{cha:resourceopt}, a large number
of quantum teleportation protocols of multi-qubit quantum states have
been proposed in the recent past. Most of them are found to use multi-qubit
entangled orthogonal states which are very difficult to produce and
maintain (cf. Table \ref{tab:A-list-of}). Motivated by this fact,
in the second chapter of this thesis, we have tried to design an entangled
orthogonal state-based quantum teleportation scheme that can teleport
a multi-qubit quantum state using minimum amount of quantum resource,
i.e., the minimum number of Bell states. Specifically, it has been
shown that the amount of quantum resources required to teleport an
unknown quantum state depends only on the number of non-zero probability
amplitudes present in the quantum state to be teleported and is independent
of the number of qubits in the state to be teleported. We have explicitly
established that the complex multi-partite entangled states that are
used in a large number of recent works on teleportation (cf. Table
\ref{tab:A-list-of}) are not essential for teleportation as there
exists a minimum number of Bell states that can be used as a quantum
channel to teleport a multi-qubit state of specific form in an optimal
and efficient manner. This scheme has also been discussed in the context
of dense coding and variants of QT, such as CT and BST. In  Chapter
\ref{cha:comment}, we have elaborated on a specific quantum teleportation
scheme, which can be viewed as a particular case of the scheme proposed
in  Chapter \ref{cha:resourceopt}. Usually, standard entangled states,
which are inseparable states of orthogonal states, are used to realize
schemes for teleportation and its variants (as is done in Chapters
\ref{cha:resourceopt}-\ref{cha:comment}). However, entangled nonorthogonal
states do exist, and they may be used to implement some of the teleportation-based
protocols \cite{adhikari2012quantum}. Specifically, entangled coherent
states \cite{van2001entangled} and Schr{\"o}dinger cat states prepared
using SU(2) coherent states \cite{wang2000entangled} are the typical
examples of entangled nonorthogonal states. In the forth chapter of
this thesis, we have designed a teleportation scheme which can be
realized using entangled nonorthogonal states. Specifically, we have
considered here four quasi-Bell states (Bell-type entangled nonorthogonal
states) as teleportation channel for the teleportation of a single-qubit
state, and computed average and minimum fidelity that can be obtained
by replacing a Bell state quantum channel in a teleportation scheme
by its nonorthogonal counterpart (i.e., corresponding quasi-Bell state). }
\item \textbf{\textcolor{black}{Experimental realization: }}\textcolor{black}{In
this thesis work, to experimentally realize the quantum communication
schemes, we have used a recently introduced experimental platform
(superconducitivity IBM quantum computer) that can be accessed through the
cloud based services. Specifically, in Chapter \ref{cha:resourceopt},
a proof-of-principle experimental realization of our proposed optimal
quantum teleportation scheme has been reported. For the same, we have
designed a quantum circuit on the IBM quantum computer for teleportation
of two-qubit quantum state using maximally entangled orthogonal state,
i.e., Bell state. Further, in Chapter \ref{cha:BellIBM}, we have
discussed discrimination of orthogonal entangled states, which plays
a very crucial role in quantum information processing. There exist
many proposals for realizing such discrimination (\cite{gupta2007general,NDBSD}
and references therein). A particularly important variant of state
discrimination schemes is nondestructive discrimination of entangled
states \cite{gupta2007general,li2000non,wang2013nondestructive},
in which the state is not directly measured. Specifically, we have
realized nondestructive Bell state discrimination using a five-qubit
superconductivity-(SQUID)-based quantum computer \cite{IBMQE,devitt2016performing},
which has been recently placed in cloud by IBM Corporation. We have
designed Clifford+T circuits for nondestructive discrimination of
Bell states in accordance with the restrictions imposed by the particular
architecture of the IBM quantum processor.}
\item \textbf{\textcolor{black}{Optical Implementation: }}\textcolor{black}{We
have also looked at some aspects of quantum cryptography. Quantum
cryptography in general and QKD in particular have drawn considerable
attention of the scientific community, because of its relevance in
our day-to-day life to defense, banking to inter-Government communication.
Naturally, several protocols of quantum communication have been proposed,
but only a few of those quantum cryptographic schemes have yet been
realized experimentally. This fact motivated us to investigate whether
it's possible to realize hitherto unrealized schemes of quantum cryptography
using the available technology (i.e., using the devices available
in a modern optical laboratory and quantum states that can be prepared
in a lab). A designing effort has been made in this direction, and
optical circuits for the realization of various quantum cryptographic
tasks (including QD, CQD, Kak's three stage scheme modified QD, controlled
DSQC with entanglement swapping) have been designed. Optical designs
are provided for fiber-based implementation as well as open-air implementation
of the schemes of quantum cryptography. Being a theoretical physics
group, we could not experimentally implement these optical circuits,
but we hope other groups will realize these circuits soon.}
\end{enumerate}
\textbf{\textcolor{black}{Analysis: }}\textcolor{black}{The analysis
part of the thesis is also performed in Chapters \ref{cha:resourceopt}-\ref{chap:OPTICAL-DESIGNS-FOR}.
Specifically, in the designed second scheme of QT, i.e., Chapter \ref{cha:nonorthogonal},
the effect of noise is studied to reveal that in a noisy environment
the performance of all the quasi-Bell states are not equivalent. Further,
using the IBM quantum computer, experimentally nondestructive discrimination
of Bell states is performed and the results are analyzed to compare
them with those of an earlier experiment done using NMR. The relevance
of this experiment is discussed in the context of quantum cryptography.
Benefits of the schemes and circuits designed here are also quantitatively
analyzed by theoretically computing average fidelity, minimum fidelity
and minimum assured fidelity in the presence of noise and also by
obtaining the fidelity of the experimentally generated quantum states
with the help of QST. We have briefly summarized analysis part of
this thesis work in the following points:}
\begin{enumerate}
\item \textbf{\textcolor{black}{Quantum teleportation scheme:}}\textcolor{black}{{}
The performance of the teleportation scheme using entangled nonorthogonal
states has been analyzed over noisy channels in Chapter \ref{cha:nonorthogonal}
of the present thesis. Specifically, the effect of PD and AD noise
were studied to reveal that the quasi-Bell state $|\phi_{-}\rangle$,
which was shown to be maximally entangled in an ideal situation, remains
the most preferred choice as quantum channel while subjected to PD
noise. However, in the presence of damping effects due to interaction
with an ambient environment (i.e., in AD noise), the choice of the
quasi-Bell state is found to depend on the nonorthogonality parameter
and the number of qubits exposed to noisy environment. We have investigated
the performance of the standard teleportation scheme using $F_{ave}$
and MFI as quantitative measures of the quality of the teleportation
scheme by considering a quasi-Bell state instead of usual Bell state
as quantum channel.}
\item \textbf{\textcolor{black}{Experimental realization: }}\textcolor{black}{In
2010, nondestructive Bell state discrimination scheme has been experimentally
implemented using an NMR-based three-qubit quantum computer \cite{samal2010non}
which have motivated us to perform nondestructive Bell state discrimination
using another experimental platform. So in Chapter \ref{cha:BellIBM},
we have realized this scheme experimentally using IBM quantum computer
and also analyzed the results of experimentally nondestructive discrimination
of Bell states to compare with NMR results.}\\
\textcolor{black}{The analysis of the results obtained using IBM quantum
processors have also revealed that the present technology needs much
improvement to achieve the desired scalability. This is so because
the gate fidelity of the individual gates realized here is relatively
low compared to the same obtained in NMR and a few other more matured
technologies. Consequently, state fidelity of the output of a more
complex circuit would be low.}
\end{enumerate}
\textcolor{black}{We may now conclude this section as well as this
thesis by restressing on the main findings of the thesis by listing
them here below}
\begin{enumerate}
\item \textcolor{black}{The quantum teleportation scheme is generalized
to design a new scheme capable of teleportation of multi-qubit quantum
states with optimal amount of quantum resources. Specifically, our
scheme allows teleportation of a multi-qubit quantum state by using
the minimum number of Bell states. Thus, it circumvents the need for
complex multi-partite entangled quantum states as a teleportation
channel. }
\item \textcolor{black}{It's established that the amount of quantum resources
required to teleport an unknown quantum state depends only on the
number of nonzero probability amplitudes present in the quantum state
and is independent of the number of qubits in the state to be teleported.}
\item \textcolor{black}{The idea behind the designing of the above mentioned
generalized scheme for QT is found to be relevant in the context of
designing corresponding schemes for CT, BST and BCST, too.}
\item \textcolor{black}{In Chapter \ref{cha:nonorthogonal}, it's observed
that if we convert $|\phi_{-}\rangle$ quasi-Bell state in orthogonal
basis $\left\{ |0\rangle,|1\rangle\right\} ,$we obtain the maximally
entangled Bell state $|\phi^{-}\rangle=\frac{1}{\sqrt{2}}\left(|01\rangle+|10\rangle\right)$.
Naturally, this state is found to be the best choice (among quasi-Bell
states) as a teleportation channel.}
\item \textcolor{black}{It's observed that the amount of nonorthogonality
plays a crucial role in deciding which quasi-Bell state would provide
highest average fidelity for a teleportation scheme implemented using
a quasi-Bell state as the teleportation channel.}
\item \textcolor{black}{The quasi-Bell state $|\phi_{-}\rangle$ which was
shown to be maximally entangled in an ideal situation, is found to
remain the most preferred choice as quantum channel while subjected
to PD noise as well.}
\item \textcolor{black}{With appropriate analysis, it's established that
the existing IBM quantum computers are scalable. To construct a larger
and useful quantum computer IBM has to considerably reduce the gate-error
rate. The finding is supported by both quantum state tomography and
quantum process tomography, Further, a comparison of results obtained
using IBM quantum computers with the corresponding results obtained
using NMR-based quantum computers is found to reveal that NMR results
are better as far as the state fidelity and gate fidelity are concerned.}
\item \textcolor{black}{It's shown that the optical implementation of the
other quantum cryptographic schemes which have not yet been performed
experimentally is possible with available technology. Relevant optical
circuits have been designed so that the interested experimental groups
can implement them.}
\end{enumerate}

\section{\textcolor{black}{Limitations of the present work and scope for future
work \label{sec:Limitations-scope}}}

\textcolor{black}{In this thesis, we have studied the effect of Markovian
noise only. In future, the effect of non-Markovian noise can also
be studied. It may be further hoped that the optimal scheme for teleportation
reported in Chapter \ref{cha:resourceopt} will find its application
in designing various other optimal schemes of quantum communication
as many schemes of quantum communications can be viewed as the variant
of teleportation.}

\textcolor{black}{This thesis work was initially, planned to be theoretical.
However, as IBM provided free access to its five-qubit quantum computer,
we have also performed some experiments using IBM quantum processors.
These experiments were only proof-of principle experiments as the
architecture of IBM restricted us to keep Alice and Bob in the same
place. This is a limitation of the present work. However, optical
experiments will not have such a restriction. Keeping this fact in
mind, in Chapter \ref{chap:OPTICAL-DESIGNS-FOR} we have provided
a set of optical circuits for various quantum cryptographic tasks.
Interestingly, most of the designed optical circuits can be realized
using both optical fiber based and open air based architectures. Being
a theoretical physics group, we could not realize these optical circuits
in the laboratory. However, we hope that the teleportation schemes
presented in Chapters \ref{cha:resourceopt} and \ref{cha:comment}
as well as the optical designs presented in Chapter \ref{chap:OPTICAL-DESIGNS-FOR}
will be experimentally realized optically in the near future.}

{\large{}\lhead{}\addcontentsline{toc}{chapter}{REFERENCES}}\renewcommand{\bibname}{References}\renewcommand{\headrulewidth}{0pt}
\begin{center}

\newpage{}
\par\end{center}

{\large{}\addcontentsline{toc}{chapter}{LIST OF PUBLICATIONS DURING
Ph.D. THESIS WORK}}{\large\par}

\chapter*{{\Large{}LIST OF PUBLICATIONS DURING Ph.D. THESIS WORK}}
\begin{flushleft}
\emph{Publications in International Journals}
\par\end{flushleft}
\begin{enumerate}
\item \textbf{\textcolor{black}{Sisodia M.}}\textcolor{black}{, Verma V.,
Thapliyal K.,  Pathak A., }\textcolor{black}{\emph{\textquotedblleft Teleportation
of a qubit using entangled nonorthogonal states: a comparative study\textquotedblright }}\textcolor{black}{,
Quantum Information Processing, vol. 16, p. 76, 2017. (}\textbf{\textcolor{black}{Thomson
Reuters I.F.}}\textcolor{black}{{} = 2.283, }\textbf{\textcolor{black}{h
index}}\textcolor{black}{{} = 38, }\textbf{\textcolor{black}{h5-index}}\textcolor{black}{{}
= 38, }\textbf{\textcolor{black}{Published by}}\textcolor{black}{{}
Springer New York,}\textbf{\textcolor{black}{{} Indexed in }}\textcolor{black}{SCI
and SCOPUS).}
\item \textbf{\textcolor{black}{Sisodia M.}}\textcolor{black}{, Shukla A.,
Thapliyal K., Pathak A., }\textcolor{black}{\emph{\textquotedblleft Design
and experimental realization of an optimal scheme for teleportation
of an n-qubit quantum state\textquotedblright }}\textcolor{black}{,
Quantum Information Processing, vol. 16, p. 292, 2017. (}\textbf{\textcolor{black}{Thomson
Reuters I.F. }}\textcolor{black}{= 2.283, }\textbf{\textcolor{black}{h
index}}\textcolor{black}{{} =38, }\textbf{\textcolor{black}{h5-index}}\textcolor{black}{{}
= 38, }\textbf{\textcolor{black}{Published by}}\textcolor{black}{{}
Springer New York, }\textbf{\textcolor{black}{Indexed in}}\textcolor{black}{{}
SCI and SCOPUS).}
\item \textbf{\textcolor{black}{Sisodia M.}}\textcolor{black}{, Shukla A.,
Pathak A.,}\textcolor{black}{\emph{ \textquotedblleft Experimental
realization of nondestructive discrimination of Bell states using
a five-qubit quantum computer\textquotedblright }}\textcolor{black}{,
Physics Letters A, vol. 381, pp. 3860-3874, 2017. (}\textbf{\textcolor{black}{Thomson}}\textcolor{black}{{}
}\textbf{\textcolor{black}{Reuters I.F. }}\textcolor{black}{= 1.863,
}\textbf{\textcolor{black}{h index }}\textcolor{black}{= 153,}\textbf{\textcolor{black}{{}
h5-index}}\textcolor{black}{{} =41, }\textbf{\textcolor{black}{Published
by }}\textcolor{black}{Elsevier Netherlands, }\textbf{\textcolor{black}{Indexed
in}}\textcolor{black}{{} SCI and SCOPUS).}
\item \textbf{\textcolor{black}{Sisodia M.}}\textcolor{black}{, Pathak A.,
}\textcolor{black}{\emph{\textquotedblleft Comment on \textquotedblleft Quantum
Teleportation of Eight-Qubit State via Six-Qubit Cluster State\textquotedblright }}\textcolor{black}{, International
Journal of Theoretical Physics, vol. 57, pp. 2213-2217, 2018. (}\textbf{\textcolor{black}{Thomson
Reuters I.F.}}\textcolor{black}{{} = 1.121, }\textbf{\textcolor{black}{h
index}}\textcolor{black}{{} =51, }\textbf{\textcolor{black}{h5-index}}\textcolor{black}{{}
=30, }\textbf{\textcolor{black}{Published by }}\textcolor{black}{Springer
New York, }\textbf{\textcolor{black}{Indexed in}}\textcolor{black}{{}
SCI and SCOPUS).}
\item \textbf{Sisodia M.,} Thapliyal K., Pathak A., \emph{\textquotedblleft Optical
designs for a set of quantum cryptographic protocols\textquotedblright },
Communicated, 2019.
\end{enumerate}
\emph{Communicated to International Journals and not included in the
thesis}
\begin{enumerate}
\item Shukla A.,\textbf{ Sisodia M., } Pathak A., \emph{``Complete characterization
of the single-qubit quantum gates used in the IBM quantum processors''}, arxiv:1805.07185,
2018.
\item \textbf{Sisodia M., }Shukla A., de Almeida A.A., Dueck G.W .,
Pathak A., \emph{``Circuit optimization for IBM processors: a way
to get higher fidelity and higher values of nonclassicality witnesses''},
arxiv:1812.11602, 2018.
\item \textbf{Sisodia M.,} Shukla C., Long G-L, \emph{\textquotedblleft Linear
optics based entanglement concentration protocols for Cluster-type
entangled coherent state\textquotedblright{}}, Quantum Information Processing, vol. 18, p. 253,  2019.
\end{enumerate}
\emph{Communicated to International Conference and not included in
the thesis}
\begin{enumerate}
\item \textbf{Sisodia M. }\emph{\textquotedblleft An improved scheme of
quantum teleportation for four-qubit state\textquotedblright{} }, Communicated,
2019.
\end{enumerate}
{\large{}$\thispagestyle{plain}$
}{\large\par}

\textbf{Extended abstracts and short papers in International/National
conferences}
\begin{enumerate}
\item Sisodia, M., Shukla, A., Thapliyal K., Pathak, A., \emph{\textquotedblleft Scheme
for teleporting a multi-qubit state using optimal resource\textquotedblright{}}
Book of Abstract, Young Quantum-2017, Harish-Chandra Research Institute,
Allahabad, February 27- March 1, (2017).
\item Sisodia, M., Shukla, A., Pathak, A., \emph{\textquotedblleft Our experience
with the IBM quantum experience: The story of successful achievements
and failures due to the limitations of the IBM quantum computers \textquotedblright{}}
Book of Abstract, Quantum Frontiers and Fundamentals-2018, Raman Research
Institute, Bengaluru, April 30- May 4, (2018) pp. 99-102.
\item Sisodia, M., Shukla, A., Thapliyal K., Pathak, A., \emph{\textquotedblleft Optical
designs for a set of quantum cryptographic protocols\textquotedblright{}}
Book of Abstract, Student Conference On Optics and Photonics-2018,
Physical Research Laboratory, Ahmedabad, October 4-6, (2018) pp. 47-48.
\item Sisodia, M., Shukla, A., Pathak, A., \emph{``What reduces the accuracy
of the IBM quantum computers: An answer from the perspective of quantum
process tomography''} Book of Abstract, Quantum Information Processing
and Applications (QIPA-2018), Harish-Chandra Research Institute, Allahabad,
December 2-8, (2018).
\item Sisodia, M.\emph{ \textquotedblleft Comment on Improving the Teleportation
Scheme of Five-Qubit State with a Seven-Qubit Quantum Channel\textquotedblright{}}
Book of Abstract, International Conference on Photonics, Metamaterials
and Plasmonics-2019, Jaypee Institute of Information Technology, Noida,
February 14-16, (2019) p. 74.
\end{enumerate}

\end{document}